\pdfoutput=1
\documentclass[PAPER, atlasdraft=false, cernpreprint=true, texlive=2023, UKenglish, texmf, orcidlogo]{atlasdoc}
\usepackage{atlaspackage}[subcaption]
\usepackage{amsmath}
\usepackage{enumitem}
\usepackage{makecell}
\usepackage{mathtools}
\usepackage{multirow}
\usepackage{ragged2e}
\usepackage{siunitx}
\usepackage{tablefootnote}
\usepackage{tabularx}
\usepackage{pdflscape}
\newcolumntype{Y}{>{\centering\arraybackslash}X}

\usepackage{atlasbiblatex}

\usepackage{atlasphysics}
\graphicspath{{logos/}{figures/}}

\usepackage{ANA-HIGG-2023-09-PAPER-defs}

\AtlasTitle{Measurements of the production cross-sections of a Higgs boson in association with a vector boson and decaying into $WW^\ast$ with the ATLAS detector at $\sqrt{s} = 13$ TeV}

\AtlasAbstract{%
Measurements of the total and differential Higgs boson production cross-sections, via $WH$ and $ZH$ associated production using $H\rightarrow WW^\ast\rightarrow\ell\nu\ell\nu$ and $H\rightarrow WW^\ast\rightarrow\ell\nu jj$ decays, are presented. The analysis uses proton--proton events delivered by the Large Hadron Collider at a centre-of-mass energy of 13 TeV and recorded by the ATLAS detector between 2015 and 2018. The data correspond to an integrated luminosity of 140 fb$^{-1}$. The sum of the $WH$ and $ZH$ cross-sections times the $H\rightarrow WW^\ast$ branching fraction is measured to be \makebox{$0.44^{+0.10}_{-0.09}$ (stat.) $^{+0.06}_{-0.05}$ (syst.) pb}, in agreement with the Standard Model prediction. Higgs boson production is further characterised through measurements of the differential cross-section as a function of the transverse momentum of the vector boson and in the framework of Simplified Template Cross-Sections.
}

\author{The ATLAS Collaboration}

\AtlasRefCode{HIGG-2023-09}

\PreprintIdNumber{CERN-EP-2025-037}

\HepDataRecord{https://www.hepdata.net/record/157861}

\AtlasJournal{JHEP}
\AtlasJournalRef{JHEP 08 (2025) 034}
\AtlasDOI{10.1007/JHEP08(2025)034}

\hypersetup{pdftitle={ATLAS document},pdfauthor={The ATLAS Collaboration}}

\begin{document}

\maketitle

\tableofcontents

%
%

% The next lines are included from the .//sections/introduction.tex input file
%
\section{Introduction}

Higgs boson production with a \W or \Z boson, respectively denoted by \WH and \ZH and collectively referred to as \VH associated production, provides direct access to the Higgs boson couplings to weak bosons. In the case of \VH associated production with a subsequent \HWW decay, the Higgs boson couples only to vector bosons at both the production and decay vertices, allowing these couplings to be precisely measured and providing a stringent test of the Standard Model (SM) of particle physics. Statistically significant deviations from their expected values could indicate beyond the SM effects; for example, the presence of additional Higgs bosons~\cite{Lee:1973iz, Branco:2011iw} or new heavy vector bosons~\cite{deBlas:2012qp, Pappadopulo:2014qza}.

Measurements are presented of the total and differential \WH and \ZH production cross-sections in the \HWW decay channel, using proton--proton (\pp) collisions at a centre-of-mass energy of $\sqrt{s} = \qty{13}{\TeV}$. The data correspond to an integrated luminosity of \qty{140}{\ifb} and were recorded by the ATLAS detector~\cite{PERF-2007-01} at the Large Hadron Collider (LHC)~\cite{Evans:2008zzb} between 2015 and 2018. Previous measurements were performed by the ATLAS~\cite{HIGG-2013-15} and CMS~\cite{CMS-HIG-13-023} Collaborations at $\sqrt{s} = 7$ and \qty{8}{\TeV} and more recently by the CMS Collaboration~\cite{CMS-HIG-20-013} at $\sqrt{s} = \qty{13}{\TeV}$ using \qty{137}{\ifb} of data. The latest result from the ATLAS Collaboration used data from 2015 and 2016 with an integrated luminosity of \qty{36.1}{\ifb}~\cite{HIGG-2017-14}. In the \HWW decay channel, measurements of Higgs boson production via gluon--gluon fusion (ggF) and vector-boson fusion (VBF) at $\sqrt{s} = \qty{13}{\TeV}$ have also been performed by the ATLAS~\cite{HIGG-2021-20} and CMS~\cite{CMS-HIG-20-013} Collaborations, and \VH production has been studied in other decay modes including $H \to ZZ\virt \to 4\ell$~\cite{HIGG-2018-28, CMS-HIG-19-001}, $H \to \gamma\gamma$~\cite{HIGG-2020-16, CMS-HIG-19-015}, $H \to b\bar{b}$~\cite{HIGG-2018-51, CMS-HIG-18-016}, and \Htautau~\cite{HIGG-2019-09, CMS-HIG-19-010}.

The analysis is performed using events with two ($2\ell$), three ($3\ell$), or four ($4\ell$) charged leptons (electrons or muons) in the final state, targeting the \WH and \ZH channels. In the $2\ell$ channel, both opposite-sign (OS) and same-sign (SS) lepton pair configurations are considered. Leptonic decays of $\tau$-leptons resulting from the decays of the associated \W/\Z bosons or the \W bosons from the \HWW decay are considered as signal, while no specific selection is performed for events with hadronically decaying $\tau$-leptons in the final state. Events from \VH production with \Htautau are considered as background. Additionally, events from ggF and VBF production are also considered as background.

In each channel, multivariate discriminants are used to maximise the sensitivity to the Higgs boson signal. The distributions of these discriminants are combined in a binned maximum-likelihood fit to extract the signal yield and the background normalisations. The maximum-likelihood fit provides results for the \WH and the \ZH channels separately and for their combination \VH. The \VH fit is performed assuming the SM prediction for the relative cross-sections of the \WH and \ZH production processes. Cross-section measurements are also performed in bins of the transverse momentum, \pt, of the associated vector boson, \ptV, (defined as the ``\ptV scheme'' in the following) and in kinematic fiducial regions defined according to the Simplified Template Cross-Section (STXS) framework~\cite{Andersen:2016qtm, deFlorian:2016spz, Berger:2019wnu} (defined as the ``STXS scheme'' in the following).

The outline of the paper is as follows: Section~\ref{sec:overview} provides an overview of the signal characteristics and the analysis strategy and a description of the STXS framework, Section~\ref{sec:detector} describes the ATLAS detector, Section~\ref{sec:samples} describes the data and the simulated event samples, Section~\ref{sec:reconstruction} describes the event reconstruction, Section~\ref{sec:analyses} details the various selections used to define the signal regions in the analysis, Section~\ref{sec:background} discusses how the backgrounds are estimated and includes the definitions of the control regions, and Section~\ref{sec:systematics} provides commentary on the systematic uncertainties. Finally, Section~\ref{sec:fit_and_results} defines the likelihood fit procedure and presents the results of the analysis, which are summarised in Section~\ref{sec:conclusions}.

% End of text imported from the .//sections/introduction.tex input file

% The next lines are included from the .//sections/overview.tex input file
%
\section{Analysis overview}
\label{sec:overview}

\begin{figure}[htbp]
\centering
\subfloat[]{
\includegraphics[width=0.35\textwidth]{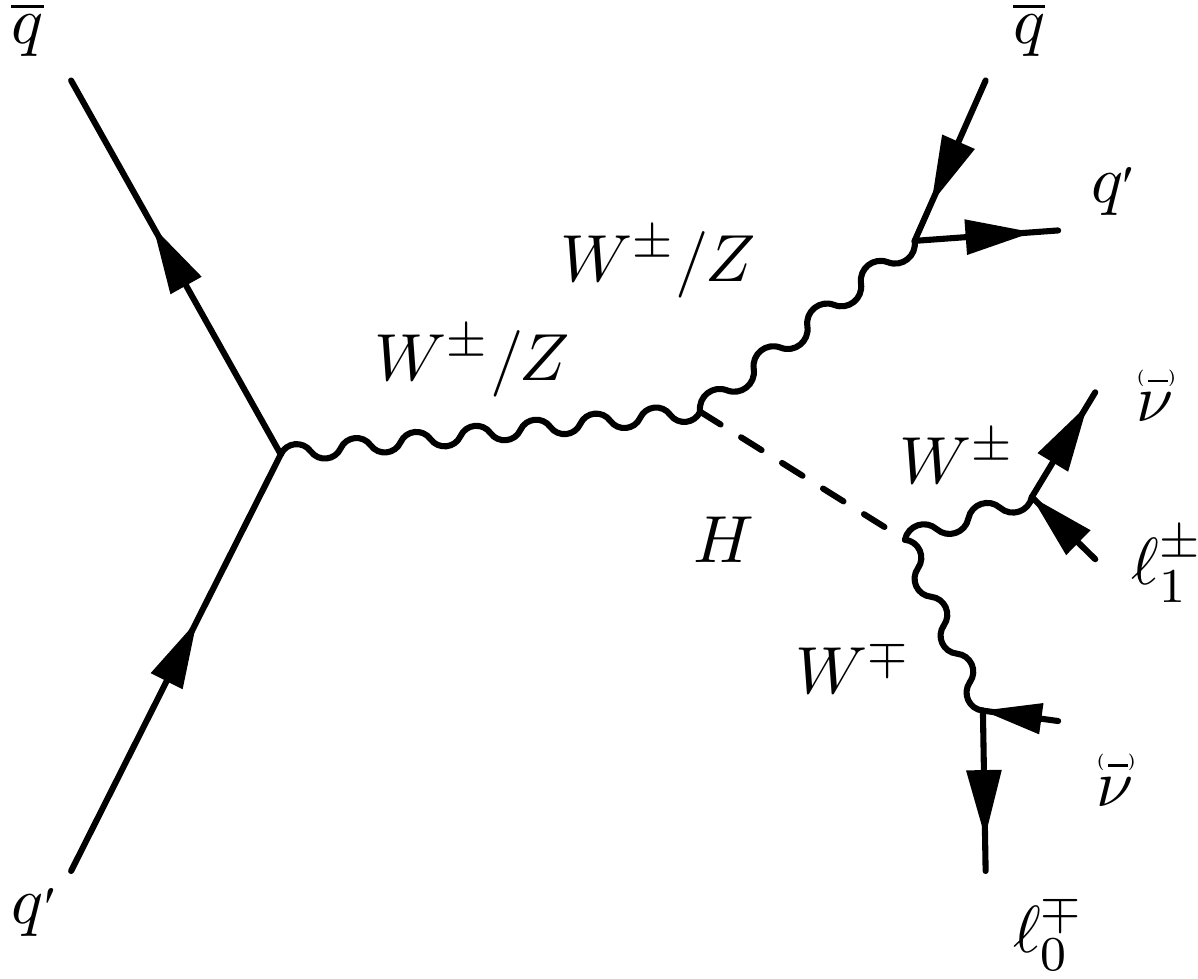}
\label{fig:channels_os}
}
\hspace{2em}
\subfloat[]{
\includegraphics[width=0.35\textwidth]{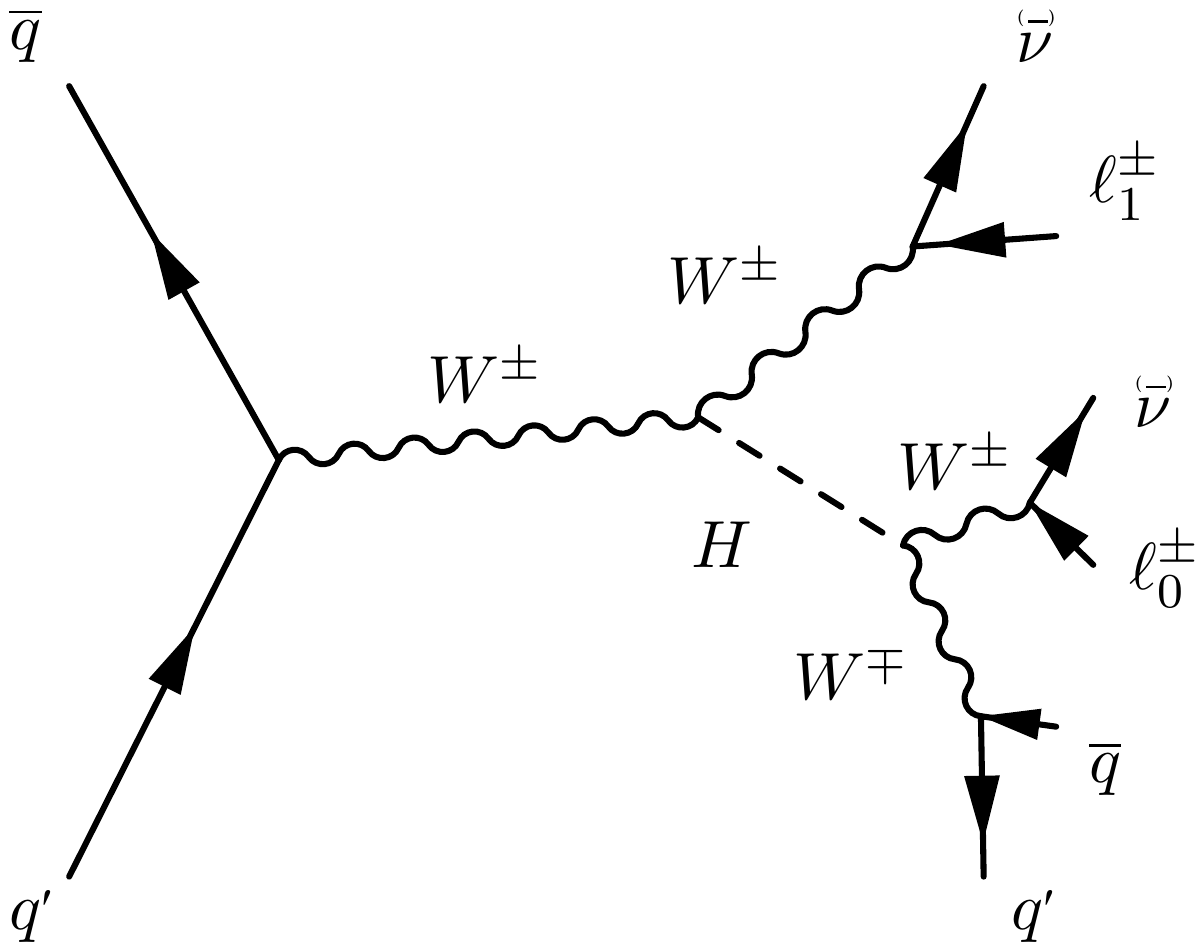}
\label{fig:channels_ss}
} \\
\subfloat[]{
\includegraphics[width=0.35\textwidth]{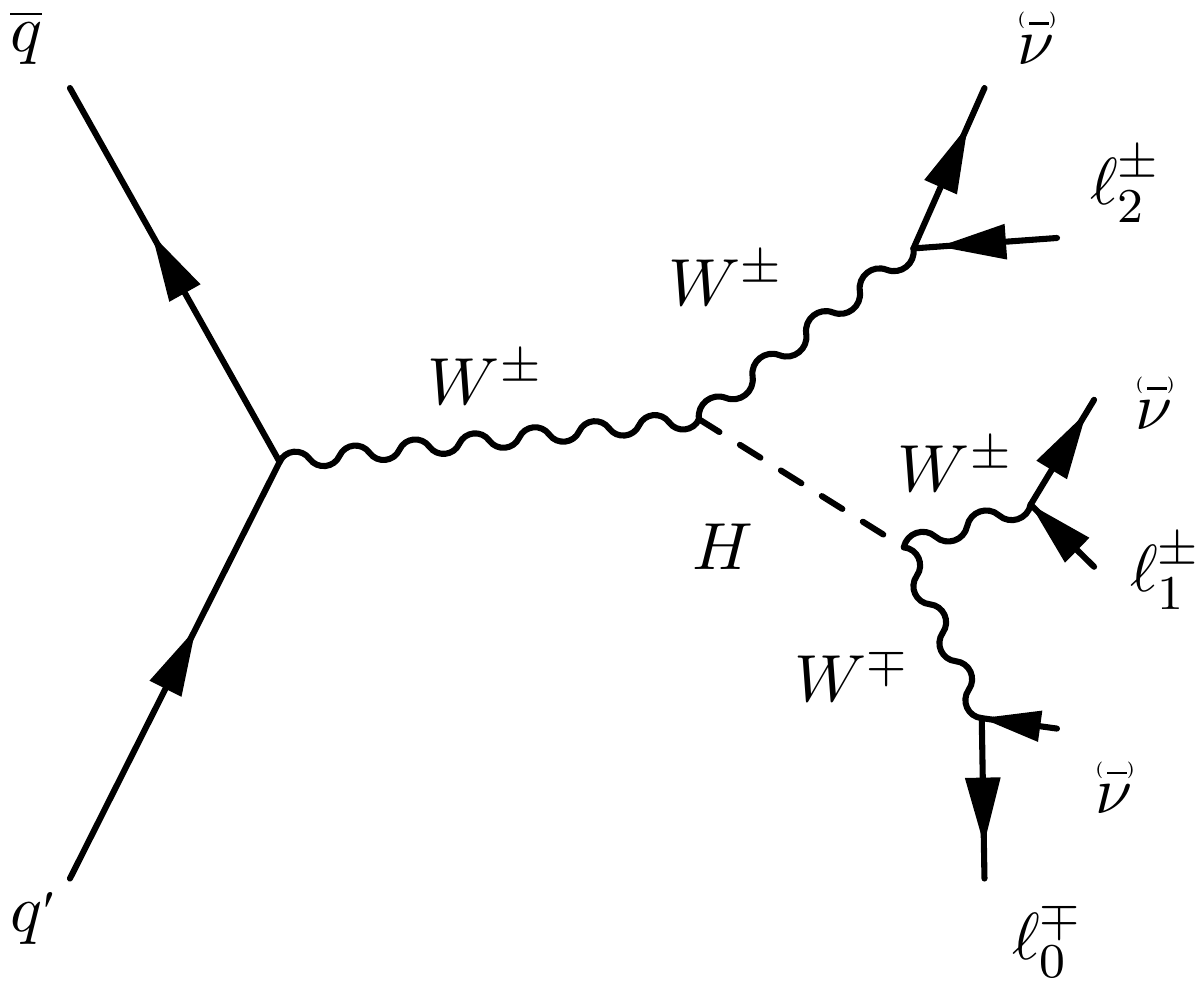}
\label{fig:channels_3l}
}
\hspace{2em}
\subfloat[]{
\includegraphics[width=0.35\textwidth]{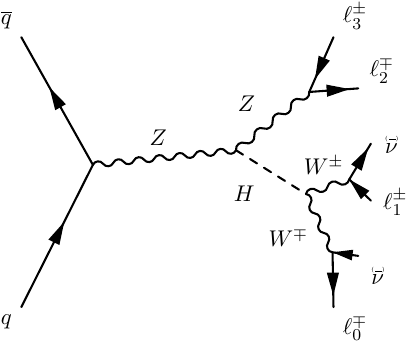}
\label{fig:channels_4l}
} \\
\caption{Leading-order Feynman diagrams for the signal topologies considered by the \VH, \HWW measurement. The subscripts on the leptons are defined for the individual channels in Section~\ref{sec:analyses}: (a) the opposite-sign $2\ell$ channel, (b) the same-sign $2\ell$ channel, (c) the $3\ell$ channel, and (d) the $4\ell$ channel.}
\end{figure}

Higgs boson production with a \W or \Z boson, followed by a \HWW decay, is sought using events with two, three, or four charged leptons in the final state. The analysis is designed to select events which are kinematically consistent with the \WH and \ZH, \HWW processes in order to enhance the signal-to-background ratio. The channels are defined as follows:

\begin{enumerate}[label=(\arabic*)]

\item \textbf{Opposite-sign $2\ell$ channel}: The targeted signal contribution consists of a \VH process in which the associated weak boson \V decays hadronically and produces two energetic jets, while \W bosons from the \HWW decay produce two oppositely charged leptons and two neutrinos. The leading-order Feynman diagram for this process is shown in Figure~\ref{fig:channels_os}. After requiring two leptons of different flavour, the leading backgrounds for this channel are \ttbar and \Wt processes. Other major components are $Z\to\tau\tau$ and \WW production with two associated jets. Final states including \Wjets and multijets may produce mis-identified leptons, contaminating the signal region. Other background sources include \WZg production and other processes involving the Higgs boson, in particular its production through ggF;

\item \textbf{Same-sign $2\ell$ channel}: The targeted signal contribution consists of a \WH process in which the associated \W boson decays leptonically, while for the \HWW decay, the \W boson with the same sign as the associated \W boson decays leptonically and the other \W boson decays hadronically. The final state therefore contains two leptons with the same charge, two neutrinos, and two energetic jets. The leading-order Feynman diagram for this process is shown in Figure~\ref{fig:channels_ss}. Significant backgrounds in this channel include \WZg, \Wy, and \Wjets production; \WW, \Zy, \Zjets, and top-quark processes also contribute to this final state;

\item \textbf{$3\ell$ channel}: The targeted signal contribution consists of a \WH process where all weak bosons decay leptonically producing three charged leptons and three neutrinos in the final state. The leading-order Feynman diagram for this process is shown in Figure~\ref{fig:channels_3l}. The most prominent background to this channel is \WZg production. Non-resonant \WWW production is another significant background with the same final state as the signal. Other important backgrounds are \ZZg, \Zy, \Zjets, \ttbar, and \Wt production;

\item \textbf{$4\ell$ channel}: The targeted signal contribution consists of a \ZH process where all weak bosons decay leptonically producing four charged leptons and two neutrinos in the final state. The leading-order Feynman diagram for this process is shown in Figure~\ref{fig:channels_4l}. The main backgrounds to this channel are non-resonant \ZZg and \WWZ production.

\end{enumerate}

The selections defining the channels described above are mutually exclusive due to the requirements on the respective multiplicity and total charge of the leptons. To maximise the analysis sensitivity to the \VH, \HWW process in each of these event topologies, the data samples for each channel --- except for the opposite-sign $2\ell$ channel --- are further subdivided into several signal regions (SRs). Additional kinematic regions, with orthogonal selection criteria and designated as control regions (CRs), are used to determine normalisation factors that are applied to the major backgrounds in each SR.

A multivariate analysis is performed in all analysis channels to increase the sensitivity. The most suitable type of multivariate analysis, its configuration, and the set of input variables are chosen for each channel. Specifically, the setup providing the best separation of signal from background is selected and less complex setups are preferred if they provide the same performance. The $2\ell$ and $3\ell$ channels use neural networks based on Keras~\cite{chollet2015keras} with the TensorFlow~\cite{Abadi:2016kxi} backend, and the $4\ell$ channel uses boosted decision trees (BDTs) in TMVA~\cite{TMVA}.

The final results are extracted from a fit that simultaneously considers all SRs and CRs. The final results consist of a set of parameters of interest (POIs) which scale the SM expectations for the signal yields or cross-sections to match the observed data. The $3\ell$ and $4\ell$ are the most sensitive channels for the measurements of \WH and \ZH production, respectively, and for the overall measurement of \VH production.

Cross-section measurements of \VH are also conducted using the Stage~1.2 scheme of the STXS framework~\cite{Andersen:2016qtm, deFlorian:2016spz, Berger:2019wnu}, which splits the production modes of the Higgs boson into kinematic fiducial regions or ``categories'' which are theoretically relevant, experimentally accessible, and straightforward to statistically combine. In this framework, the \VH STXS category assumes a leptonic decay of the \V and is further categorised according to its transverse momentum, \ptV; in contrast, the $\qq\to\V(\to\qq)\H$ (along with VBF) and $gg\to\Z(\to\qq)\H$ (along with ggF) topologies are included in the electroweak (EW) \qqH and \ggH STXS categories, respectively. In the STXS measurement scheme, \WH and \ZH are measured independently in fiducial bins of \ptV and $\qq\to\V(\to\qq)H$ is measured as part of EW \qqH in a single fiducial bin of dijet invariant mass, \mjj. In addition, the limited interference between $\qq\to\V(\to\qq)H$ and VBF in the phase space relevant for this analysis motivates the use of the \ptV measurement scheme, which is inclusive in the decay of the \V. In the \ptV scheme, the sensitivity to low values of \ptV is primarily enabled by the $\V(\to\textrm{leptons})\H$ measurements, while the sensitivity to high values of \ptV is primarily enabled by the $\V(\to\qq)\H$ measurements. In both schemes, fiducial bins are merged to provide sensitivity for all measured parameters. This leads to measuring four bins of \VH in the \ptV scheme and three bins for each of \WH and \ZH as well as one bin for EW \qqH in the STXS scheme.

% End of text imported from the .//sections/overview.tex input file

% The next lines are included from the .//sections/detector.tex input file
%
\section{ATLAS detector}
\label{sec:detector}

\newcommand{\AtlasCoordFootnote}{%
ATLAS uses a right-handed coordinate system with its origin at the nominal interaction point (IP) in the centre of the detector and the $z$-axis along the beam pipe. The $x$-axis points from the IP to the centre of the LHC ring, and the $y$-axis points upwards. Cylindrical coordinates $(r,\phi)$ are used in the transverse plane, $\phi$ being the azimuthal angle around the $z$-axis. The pseudorapidity is defined in terms of the polar angle $\theta$ as $\eta = -\ln \tan(\theta/2)$. Angular distance is measured in units of $\Delta R \equiv \sqrt{(\Delta\eta)^{2} + (\Delta\phi)^{2}}$.
}

The ATLAS detector~\cite{PERF-2007-01} at the LHC covers nearly the entire solid angle around the collision point.\footnote{\AtlasCoordFootnote} It consists of an inner tracking detector surrounded by a thin superconducting solenoid, electromagnetic and hadron calorimeters, and a muon spectrometer incorporating three large superconducting air-core toroidal magnets.

The inner-detector system (ID) is immersed in a \qty{2}{\tesla} axial magnetic field and provides charged-particle tracking in the range of $|\eta| < 2.5$. The high-granularity silicon pixel detector covers the vertex region and typically provides four measurements per track, the first hit normally being in the insertable B-layer (IBL)~\cite{ATLAS-TDR-19, PIX-2018-001}. It is followed by the silicon microstrip tracker (SCT), which usually provides eight measurements per track. These silicon detectors are complemented by the transition radiation tracker (TRT), which enables radially extended track reconstruction up to $|\eta| = 2.0$. The TRT also provides electron identification information based on the fraction of hits (typically 30 in total) above a higher energy-deposit threshold corresponding to transition radiation.

The calorimeter system covers the pseudorapidity range $|\eta| < 4.9$. In the region $|\eta|< 3.2$, electromagnetic calorimetry (EM) is provided by barrel and endcap high-granularity lead/liquid-argon (LAr) calorimeters, with an additional thin LAr presampler covering $|\eta| < 1.8$ to correct for energy loss in material upstream of the calorimeters. Hadron calorimetry is provided by the steel/scintillator-tile calorimeter, segmented into three barrel structures within $|\eta| = 1.7$, and two copper/LAr hadron endcap calorimeters. The solid angle coverage is completed with forward copper/LAr and tungsten/LAr calorimeter modules optimised for electromagnetic and hadronic energy measurements respectively.

The muon spectrometer (MS) comprises separate trigger and high-precision tracking chambers measuring the deflection of muons in a magnetic field generated by the superconducting air-core toroidal magnets. The field integral of the toroids ranges between \num{2.0} and \qty{6.0}{\tesla\metre} across most of the detector. Three layers of precision chambers, each consisting of layers of monitored drift tubes, cover the region $|\eta| < 2.7$, complemented by cathode-strip chambers in the forward region, where the background is highest. The muon trigger system covers the range $|\eta| < 2.4$ with resistive-plate chambers in the barrel, and thin-gap chambers in the endcap regions.

Events of interest are selected by the first-level trigger system implemented in custom hardware, followed by selections made by algorithms implemented in software in the high-level trigger~\cite{TRIG-2016-01}. The first-level trigger accepts events from the \qty{40}{\MHz} bunch crossings at a rate below \qty{100}{\kHz}, which the high-level trigger reduces further to record events to disk at about \qty{1}{\kHz}. The full ATLAS Run 2 data sample is used for the analysis, consisting of \pp collision data produced at $\sqrt{s} = \qty{13}{\TeV}$ and recorded between 2015 and 2018. The data are subjected to quality requirements~\cite{DAPR-2018-01}, including the removal of events recorded when relevant detector components were not operating correctly. The total integrated luminosity after this cleaning of the data corresponds to \qty{140}{\ifb}~\cite{DAPR-2021-01}.

An extensive software suite~\cite{SOFT-2022-02_updated} is used in the reconstruction and analysis of real and simulated data, in detector operations, and in the trigger and data acquisition systems of the experiment.

% End of text imported from the .//sections/detector.tex input file

% The next lines are included from the .//sections/samples.tex input file
%
\section{Monte Carlo samples}
\label{sec:samples}

Monte Carlo (MC) generators are used to model the hard \pp scattering matrix element (ME), parton shower (PS) and hadronisation, and underlying event (UE). The generators that are used for modelling signal and background processes are listed in Table~\ref{tab:mcsamples}.

The Higgs boson samples are simulated with the \HWW decay in the four main production channels: ggF, VBF, \VH, and associated production with a top quark pair (\ttH). These samples are simulated using a Higgs boson mass of \qty{125}{\GeV} and then normalised to the cross-sections~\cite{deFlorian:2016spz} computed for a mass of \qty{125.09}{\GeV}~\cite{HIGG-2014-14}. The normalisation of all Higgs boson samples accounts for the decay branching ratio calculated with \textsc{HDECAY}~\cite{Djouadi:1997yw, Spira:1997dg, Djouadi:2006bz} and \PROPHECY~\cite{Bredenstein:2006ha, Bredenstein:2006rh, Bredenstein:2006nk_fixed}. All samples use the PDF4LHC15~\cite{Butterworth:2015oua} parton distribution function (PDF) set and are interfaced to \PYTHIA[8.2]~\cite{Sjostrand:2014zea} for parton showering and hadronisation, with parameters set according to the AZNLO~\cite{STDM-2012-23} (for \VH, ggF, and VBF) or A14~\cite{ATL-PHYS-PUB-2014-021} (for \ttH) set of tuned parameters (tunes). The decays of bottom and charm hadrons are performed using \EVTGEN~\cite{Lange:2001uf}.

The \WH and \ZH events are simulated at next-to-leading order (NLO) in perturbative quantum chromodynamics (QCD) for up to one additional jet using \PowhegvTwo~\MiNLO~\cite{Nason:2004rx, Frixione:2007vw, Nason:2009ai, Alioli:2010xd, Luisoni:2013cuh}. Each sample is normalised to a cross-section calculated at next-to-next-to-leading-order (NNLO) in QCD and at NLO in EW\cite{Ciccolini:2003jy, Brein:2003wg, Brein:2011vx, Brein:2012ne, Denner:2014cla}. The subdominant \ggZH process is simulated at leading-order (LO) in QCD with \PowhegvTwo and normalised to a cross-section calculated at NLO in QCD with next-to-leading logarithmic (NLL) corrections~\cite{Altenkamp:2012sx, Harlander:2014wda}.

The ggF events are simulated using \PowhegvTwo~\NNLOPS~\cite{Hamilton:2013fea}. The simulation achieves NNLO accuracy for arbitrary inclusive $gg\to\H$ observables by reweighting the Higgs boson rapidity spectrum in \textsc{Hj}-\MINLO~\cite{Hamilton:2012np, Campbell:2012am, Hamilton:2012rf} to that of \textsc{HNNLO}~\cite{Catani:2007vq}. The sample is normalised to a cross-section computed at next-to-next-to-next-to-leading-order (N$^3$LO) in QCD with NLO EW corrections~\cite{Actis:2008ug, Actis:2008ts, Harlander:2009bw, Harlander:2009mq, Pak:2009dg, Harlander:2009my, Anastasiou:2015ema, deFlorian:2016spz, Anastasiou:2016cez, Bonetti:2018ukf, Dulat:2018rbf}. The VBF events are simulated using \PowhegvTwo and normalised to a cross-section calculated at NLO in QCD and EW~\cite{Ciccolini:2007jr, Ciccolini:2007ec}, with approximate NNLO in QCD corrections applied~\cite{Bolzoni:2010xr}. The \ttH events are simulated using \PowhegvTwo~\cite{Frixione:2007nw, Hartanto:2015uka} at NLO accuracy in QCD. The sample is normalised to a cross-section computed at NLO accuracy in QCD with NLO EW corrections~\cite{deFlorian:2016spz}.

The main background processes include single boson, diboson, triboson, single top quark, and top quark pair production.

The $Z/\gamma\virt$ (``\Zjets'') processes are simulated using \SHERPA[2.2.1]~\cite{Bothmann:2019yzt} at NLO in QCD for up to two jets and at LO for up to four jets, calculated using the \textsc{Comix}~\cite{Gleisberg:2008fv} and \OPENLOOPS~\cite{Cascioli:2011va, Denner:2016kdg, Buccioni:2019sur} libraries. The single boson samples are normalised to cross-sections computed at NNLO in QCD~\cite{Anastasiou:2003ds}. The \VV and \Vy processes are simulated using \SHERPA[2.2.2] and \SHERPA[2.2.8], respectively, at NLO in QCD for up to one jet and at LO for up to three jets. Throughout the paper, \WZ and \ZZ are assumed to correspond to $\WZ/\W\gamma\virt$ and $\ZZ/\Z\gamma\virt$, respectively (i.e., including the virtual photon contribution). The on-shell \VVV processes are modelled by \SHERPA[2.2.2] at NLO in QCD inclusively and at LO for up two jets. Virtual QCD corrections to the \VV, \Vy, and \VVV processes were provided by the \OPENLOOPS library~\cite{Cascioli:2011va, Denner:2016kdg, Buccioni:2017yxi, Buccioni:2019sur}. The $gg$-initiated diboson processes are modelled by \SHERPA[2.2.2] at LO in QCD for up to one jet and normalised to cross-sections computed at NLO in QCD~\cite{Caola:2015rqy, Grazzini:2018owa}. For all \SHERPA samples, the events are simulated using the \NNPDF[3.0nnlo]~\cite{Ball:2014uwa} PDF set, and the matrix elements are matched with the \SHERPA parton shower~\cite{Schumann:2007mg, Gleisberg:2008fv} using the \MEPSatNLO prescription~\cite{Catani:2001cc, Hoeche:2009rj, Hoeche:2011fd, Hoeche:2012yf} and the set of tuned parameters developed by the \SHERPA authors. EW \WW production in association with two jets ($WWqq$) is simulated using \MGNLO~\cite{Alwall:2014hca} with LO-accurate matrix elements~\cite{Caola:2015rqy} and using the \NNPDF[3.0nlo]~\cite{Ball:2014uwa} PDF set. The events are interfaced to \PYTHIA[8] for parton showering and hadronisation, with parameters set according to the A14 tune.

The top quark pair production is modelled by \PowhegvTwo at NLO in QCD with the \hdamp parameter\footnote{The \hdamp parameter is a resummation damping factor and one of the parameters that controls the matching of \POWHEG matrix elements to the parton shower and thus effectively regulates the high-\pt radiation against which the \ttbar system recoils.} set to $1.5\times{m_\textrm{top}}$~\cite{ATL-PHYS-PUB-2016-020}. The events are normalised to a cross-section computed at NNLO in QCD with next-to-next-to-leading logarithmic (NNLL) corrections~\cite{Beneke:2011mq, Cacciari:2011hy, Baernreuther:2012ws, Czakon:2012zr, Czakon:2012pz, Czakon:2013goa, Czakon:2011xx}. The \Wt process is simulated using \PowhegvTwo~\cite{Re:2010bp} at NLO in QCD using the five-flavour scheme. The diagram removal scheme~\cite{Frixione:2008yi} is used to remove interference and overlap with \ttbar production. The \ttV, \tZ, and \tWZ processes are simulated using \MGNLO at NLO in QCD. For all top processes, events are simulated using the \NNPDF[3.0nlo] PDF set. \EVTGEN is used to model decays of hadrons containing $b$- or $c$-quarks, and the events are interfaced to \PYTHIA[8] for parton showering and hadronisation, with parameters set according to the A14 tune.

All samples are processed through the \GeantFour-based~\cite{Agostinelli:2002hh} ATLAS detector simulation~\cite{SOFT-2010-01} and the standard ATLAS reconstruction software~\cite{SOFT-2022-02}. The effect of pile-up is modelled by overlaying the hard-scattering event with simulated inelastic \pp events simulated with \PYTHIA[8.1]~\cite{Sjostrand:2007gs} using the \NNPDF[2.3lo]~\cite{Ball:2012cx} PDF set and the A3~\cite{ATL-PHYS-PUB-2016-017} tune.

\begin{table}[htbp]
\centering
\caption{MC generators used to model the signal and background processes. Alternative generators or parton shower models --- used to estimate systematic uncertainties --- are shown in parenthesis. In the last column, the prediction order (in QCD, unless specified otherwise) for the total cross-section is shown.}
\label{tab:mcsamples}
\resizebox{\textwidth}{!}{
\begin{tabular}{llll}
\hline\hline
Process                              & ME                  & PS/UE          & Prediction order \\
& (alternative)       & (alternative)  & for total cross-section \\
\hline
\qqWH                                & \PowhegvTwo~\MiNLO  & \PYTHIA[8]     & NNLO QCD + NLO EW~\cite{Ciccolini:2003jy, Brein:2003wg, Brein:2011vx, Brein:2012ne, Denner:2014cla} \\
&                     & (\HERWIG[7]~\cite{Bahr:2008pv, Bellm:2015jjp}) & \\
\qqZH                                & \PowhegvTwo~\MiNLO  & \PYTHIA[8]     & NNLO QCD + NLO EW~\cite{Ciccolini:2003jy, Brein:2003wg, Brein:2011vx, Brein:2012ne, Denner:2014cla} \\
&                     & (\HERWIG[7])   & \\
\ggZH                                & \PowhegvTwo         & \PYTHIA[8]     & NLO + NLL~\cite{Altenkamp:2012sx, Harlander:2014wda} \\
&                     & (\HERWIG[7])   & \\
ggF $H$                              & \PowhegvTwo~\NNLOPS & \PYTHIA[8]     & N$^3$LO QCD + NLO EW~\cite{Actis:2008ug, Actis:2008ts, Harlander:2009bw, Harlander:2009mq, Pak:2009dg, Harlander:2009my, Anastasiou:2015ema, deFlorian:2016spz, Anastasiou:2016cez, Bonetti:2018ukf, Dulat:2018rbf} \\
VBF $H$                              & \PowhegvTwo         & \PYTHIA[8]     & NNLO QCD + NLO EW~\cite{Ciccolini:2007jr, Ciccolini:2007ec, Bolzoni:2010xr} \\
\ttH                                 & \PowhegvTwo         & \PYTHIA[8]     & NLO QCD + NLO EW~\cite{deFlorian:2016spz} \\
\hline\hline
$Z/\gamma\virt$                      & \SHERPA[2.2.1]      & \SHERPA[2.2.1] & NNLO~\cite{Anastasiou:2003ds} \\
& (\MGNLO)            & (\PYTHIA[8])   & \\
$q\bar{q}/g \to WW$                  & \SHERPA[2.2.2]      & \SHERPA[2.2.2] & NLO~\cite{Cascioli:2011va, Denner:2016kdg, Buccioni:2019sur} \\
$WZ/ZZ/V\gamma\virt$                 & \SHERPA[2.2.2]      & \SHERPA[2.2.2] & NLO~\cite{Cascioli:2013gfa} \\
\Vy                                  & \SHERPA[2.2.8]      & \SHERPA[2.2.8] & NLO~\cite{Cascioli:2013gfa} \\
$gg \to WW/ZZ$                       & \SHERPA[2.2.2]      & \SHERPA[2.2.2] & NLO~\cite{Caola:2015rqy, Grazzini:2018owa} \\
\VVV                                 & \SHERPA[2.2.2]      & \SHERPA[2.2.2] & NLO~\cite{Binoth:2008kt} \\
& (\MGNLO)            & (\PYTHIA[8])   & \\
$qq \to WWqq$                        & \MGNLO              & \PYTHIA[8]     & LO \\
\ttbar                               & \PowhegvTwo         & \PYTHIA[8]     & NNLO + NNLL~\cite{Beneke:2011mq, Cacciari:2011hy, Baernreuther:2012ws, Czakon:2012zr, Czakon:2012pz, Czakon:2013goa, Czakon:2011xx} \\
& (\MGNLO)            & (\HERWIG[7])   & \\
\Wt                                  & \PowhegvTwo         & \PYTHIA[8]     & NLO~\cite{Kidonakis:2010ux, Kidonakis:2013zqa} \\
\ttV                                 & \MGNLO              & \PYTHIA[8]     & NLO~\cite{Broggio:2019ewu} \\
\tZ                                  & \MGNLO              & \PYTHIA[8]     & NLO~\cite{Campbell:2013yla} \\
\tWZ                                 & \MGNLO              & \PYTHIA[8]     & NLO~\cite{Maltoni:2019aot} \\
\hline\hline
\end{tabular}
}
\end{table}

% End of text imported from the .//sections/samples.tex input file

% The next lines are included from the .//sections/reconstruction.tex input file
%
\section{Event reconstruction}
\label{sec:reconstruction}

Candidate signal events are selected using triggers that require a single isolated lepton with minimum \pt thresholds ranging from 24 to \qty{26}{\GeV} for electrons and from 20 to \qty{26}{\GeV} for muons~\cite{TRIG-2018-05, TRIG-2018-01}, depending on the data-taking period. At least one of the leptons reconstructed offline is required to have triggered the event. Additionally, the lepton is required to have a reconstructed \pt higher than the nominal trigger threshold by at least \qty{1}{\GeV}, which ensures that the \pt is squarely on the plateau of the trigger's efficiency.

Selected events are required to have at least one primary vertex reconstructed from at least two matched tracks, each with transverse momentum $\pt > \qty{500}{\MeV}$, as described in Ref.~\cite{ATL-PHYS-PUB-2015-026}. If an event has more than one reconstructed primary vertex, the vertex with the largest track $\sum\pt^2$ is selected for the analysis.

Electrons are reconstructed by matching energy clusters in the electromagnetic calorimeter to well-reconstructed tracks that are extrapolated to the calorimeter~\cite{EGAM-2021-01}. Electron candidates are required to satisfy $|\eta| < 2.47$, excluding the transition region $1.37 < |\eta| < 1.52$ between the barrel and end caps of the LAr calorimeter.

Muons are reconstructed from a global fit to matching tracks from the ID and the MS~\cite{MUON-2018-03}. They are required to satisfy $|\eta| < 2.5$.

For the purpose of lepton counting, all leptons with $\pt > \qty{10}{\GeV}$ and passing a set of identification requirements common to all channels are included. This ensures orthogonality between the channels when divided according to an exact lepton multiplicity and total charge.

To suppress particles mis-identified as leptons, several identification, impact parameter, and isolation criteria are applied to electron and muon candidates, shown in Table~\ref{tab:lepton_selections}. For electrons, a likelihood-based identification method~\cite{EGAM-2018-01} is employed. Electron candidates in the $2\ell$ and $3\ell$ channels must satisfy the ``Tight'' likelihood working point, while electron candidates in the $4\ell$ channel must satisfy the ``Medium'' likelihood working point. The ``Tight'' and ``Medium'' working points have identification efficiencies of $\sim$78\% and $\sim$88\%, respectively, for electrons with $\pt \sim \qty{40}{\GeV}$. For muons, a cut-based identification method~\cite{MUON-2018-03} is employed. Muon candidates must satisfy the ``Medium'' working point, which has an identification efficiency of $\sim$96\% for muons with $\pt \sim \qty{40}{\GeV}$. The impact parameter requirements are $|z_0\sin\theta| < \qty{0.5}{mm}$ and $|d_0|/\sigma_{d_0} < 5\ (3)$ for electrons (muons).\footnote{The transverse impact parameter, $d_0$, is defined by the point of closest approach of the track to the beamline in the $r$--$\phi$ plane, its uncertainty being $\sigma_{d_0}$, while the longitudinal impact parameter, $z_0$, is given by the longitudinal distance to the hard-scatter vertex from this same point.} Electron candidates in the $2\ell$ and $3\ell$ channels must also be unambiguously reconstructed as electrons; they cannot be simultaneously reconstructed as photons.

For $2\ell$ and $3\ell$ channels, a newly developed multivariate method based on a BDT, the prompt lepton improved veto (PLIV), is employed. PLIV exploits isolation and lifetime information --- including the presence of secondary vertices --- associated with a track jet matched to the selected lepton candidate. It leads to a substantial improvement in the rejection of mis-identified leptons compared to previously available methods within ATLAS while maintaining high efficiency for selecting prompt leptons. Lepton candidates in the $2\ell$ and $3\ell$ channels must satisfy the ``PLImprovedTight'' working point. For electrons, this working point has prompt and non-prompt electron efficiencies of $\sim$95\% and $\sim$4\%, respectively, for electrons with $\pt \sim \qty{40}{\GeV}$ and $|\eta| < 1.37$; for muons, this working point has prompt and non-prompt muon efficiencies of $\sim$95\% and $\sim$3\%, respectively, for muons with $\pt \sim \qty{40}{\GeV}$. Electron and muon candidates in the $4\ell$ channel must satisfy the cut-based ``FCLoose'' and ``Loose\_VarRad'' working points, respectively, which rely on tracking and calorimeter isolation variables~\cite{EGAM-2018-01, MUON-2018-03}. The ``FCLoose'' working point has an isolation efficiency of $\sim$99\% for electrons with $\pt \sim \qty{40}{\GeV}$ and passing the ``Medium'' identification working point, and the ``Loose\_VarRad'' working point has a prompt muon efficiency of 97\% and an efficiency for muons from heavy-flavour decays of 6.1\% for muons with $20 < \pt < \qty{100}{\GeV}$ and passing the ``Medium'' identification working point. To suppress electron charge mis-identification in the same-sign $2\ell$ channel and the $3\ell$ channel with zero same-flavour, opposite-sign lepton pairs in the final state, another BDT discriminant, the electron charge identification selector (ECIDS)~\cite{EGAM-2018-01}, is used.

\begin{table}[htbp]
\centering
\caption{The electron and muon definitions used by the analysis. Any selection in brackets, $(\ldots)$, is specific to the $4\ell$ channel and is used instead of the preceding selection on that same line of the table; otherwise, the selections are applicable to all channels. $^\ast$ECIDS is only applied to the same-sign $2\ell$ channel and the $3\ell$ channel with zero same-flavour, opposite-sign lepton pairs in the final state.}
\label{tab:lepton_selections}
\begin{tabularx}{0.75\textwidth}{r | Y | Y}
\hline\hline
& \multicolumn{2}{c}{Flavour} \\
Criteria & Electrons & Muons \\
\hline
\pt[$\GeV\xspace$] & \multicolumn{2}{c}{$>10$} \\
$|\eta|$ & $<1.37$ or $\in[1.52,2.47]$ & $<2.5$ \\
Identification & Tight (Medium) & Medium \\
$|z_0\sin{\theta}|$ [mm] & \multicolumn{2}{c}{$<0.5$} \\
$|d_0/\sigma_{d_0}|$ & $<5$ & $<3$ \\
Unambiguous $e/\gamma$ & Yes (No) & -- \\
\multirow{2}{*}{Isolation} & PLImprovedTight & PLImprovedTight \\
& (FCLoose) & (Loose\_VarRad) \\
ECIDS & Yes$^\ast$ (No) & -- \\
\hline\hline
\end{tabularx}
\end{table}

Jets are reconstructed using the anti-$k_t$ algorithm with a radius parameter of $R = 0.4$ and particle-flow objects as input~\cite{Cacciari:2008gp, Fastjet, PERF-2015-09}. The four-momentum of the jets is corrected for the response of the non-compensating calorimeter, signal losses due to noise threshold effects, energy loss in inactive material, and contamination from pile-up (defined as additional \pp interactions in the same and neighbouring bunch crossings)~\cite{JETM-2018-05}. For jets entering the analysis, a kinematic selection of $\pt > \qty{20}{\GeV}$ and $|\eta| < 4.5$ is applied. For the purpose of jet counting, only jets with $\pt > \qty{30}{\GeV}$ are considered. A jet-vertex-tagger multivariate discriminant selection that reduces contamination from pile-up~\cite{PERF-2014-03} is applied to jets with $20 < \pt < \qty{60}{\GeV}$ and $|\eta| < 2.4$, utilising calorimeter and tracking information to separate hard-scatter jets from pile-up jets. Furthermore, to suppress pile-up jets in the forward region, $|\eta| > 2.5$, jet shapes and topological correlations in pile-up interactions are exploited~\cite{PERF-2016-06, ATL-PHYS-PUB-2019-026}.

Jets with $\pt > \qty{20}{\GeV}$ and $|\eta| < 2.5$ containing $b$-hadrons (\bjets) are identified using a neural-network discriminant, DL1r, based on a number of lower-level taggers which utilise relevant quantities such as the associated track impact parameters and information from secondary vertices. The working point that is adopted has an average 85\% \bjet tagging efficiency, as estimated from simulated \ttbar events~\cite{FTAG-2019-07, FTAG-2018-01}.

The missing transverse momentum \pmet, with magnitude \met, is calculated as the negative vector sum of the \pt of all the selected leptons and jets, together with reconstructed tracks that are not associated with these objects but are consistent with originating from the primary vertex~\cite{JETM-2020-03}. It serves as experimental proxy for the momentum carried by undetected particles. An object-based \met significance~\cite{JETM-2020-03}, \metsig, is used to reject events where the \met arises due to the mis-reconstruction of the physics objects entering the calculation.

% End of text imported from the .//sections/reconstruction.tex input file

% The next lines are included from the .//sections/analyses.tex input file
%
\section{Event selection}
\label{sec:analyses}

In this section, the criteria used to define the SR in each channel are described. Table~\ref{tab:SummaryAllSRs} summarises the selections for each channel before applying multivariate techniques. In Table~\ref{tab:SummaryAllSRs} and throughout this section, the minimum lepton \pt selections are smaller than the \pt thresholds of the single lepton triggers. While at least one lepton must have triggered the event and therefore satisfy one of these higher thresholds, all leptons must satisfy minimum \pt selections.

The multivariate techniques applied in each channel are also described in greater detail in this section. The variables used as input by each technique have been confirmed to exhibit good consistency with data in the phase space regions where the technique is trained and deployed.

\begin{table}[htbp]
\centering
\caption{Definition of the event selection for each SR in the analysis before applying multivariate techniques. The selections are defined and motivated for all channels in their respective subsections of Section~\ref{sec:analyses}. ``SFOS'' and ``DFOS'' refer to same- and different-flavour, opposite-sign lepton pairs, respectively. In the context of the $4\ell$ channel, leptons~0 and 1, $\ell_0$ and $\ell_1$, comprise the Higgs boson candidate --- this is described in greater detail in Section~\ref{sec:4l}.}
\label{tab:SummaryAllSRs}
\resizebox{\textwidth}{!}{
\begin{tabular}{ r || r r | r r | r r }
\hline\hline
\multirow{2}{*}{Channel} & \multicolumn{2}{c|}{$2\ell$} & \multicolumn{2}{c|}{$3\ell$} & \multicolumn{2}{c}{$4\ell$} \\
& OS & SS & \Z-dominated & \Z-depleted & 1-SFOS & 2-SFOS \\
\hline
Minimum lepton \pt[$\GeV\xspace$] & 15 & 15 & 15 & 15 & 10 & 10 \\
Number of leptons & 2 & 2 & 3 & 3 & 4 & 4 \\
Total lepton charge & 0 & $\pm 2$ & $\pm 1$ & $\pm 1$ & 0 & 0 \\
Number of SFOS pairs & 0 & -- & 1 or 2 & 0 & 1 & 2 \\
Number of DFOS pairs & 1 & -- & -- & -- & -- & -- \\
Minimum $\Delta{R}_{\ell\ell}$ & 0.1 & 0.4 & 0.1 & 0.1 & \multicolumn{2}{c}{0.2 ($\ell_0\ell_1$)} \\
Minimum \mll[$\GeV\xspace$] & 10 & -- & 12 (all SFOS) & -- & \multicolumn{2}{c}{12 (all SFOS) or 10 (all DFOS)} \\
\hline
Number of jets & $\geq2$ & $\geq1$ & -- & -- & -- & -- \\
Number of $b$-tagged jets & 0 & 0 & 0 & 0 & 0 & 0 \\
\hline
$|\mll - \mZ|$ [$\GeV\xspace$] & -- & $>20$ ($e^{\pm}e^{\pm}$) & $>25$ (all SFOS) & -- & -- & -- \\
\mll[$\GeV\xspace$] & -- & -- & -- & -- & -- & $<50$ ($\ell_0\ell_1$) \\
\met[$\GeV\xspace$] & -- & -- & $>30$ & -- & -- & -- \\
$|\mjj - \qty{85}{\GeV}|$ [$\GeV$] & $<15$ & -- & -- & -- & -- & -- \\
\mjj[$\GeV$] & -- & $<500$ & -- & -- & -- & -- \\
$|\dyjj|$ & $<1.2$ & -- & -- & -- & -- & -- \\
$|\mfl - \qty{122.5}{\GeV}|$ [$\GeV$] & -- & -- & -- & -- & -- & $>7.5$ \\
$\ANNDFOS^{VH}$ & $>0.2$ & -- & -- & -- & -- & -- \\
$\ANNZdep^{\ttbar}$ & -- & -- & -- & $<0.25$ & -- & -- \\
\hline\hline
\end{tabular}
}
\end{table}

% The next lines are included from the .//sections/chan2os.tex input file
%
\subsection{Opposite-sign \texorpdfstring{$2\ell$}{2l} channel}
\label{sec:os2l}

The opposite-sign $2\ell$ channel requires two different-flavour, opposite-sign (DFOS) leptons from the decay of the Higgs boson, labelled $\ell_0$ and $\ell_1$ in decreasing order of \pt, and two or more jets from the decay of the associated vector boson, which can be either a \W or a \Z. The leading and subleading leptons must exceed \pt thresholds of \qty{22}{\GeV} and \qty{15}{\GeV}, respectively, and are required to satisfy the common lepton selections. The invariant mass of the leptons must satisfy $\mll > \qty{10}{\GeV}$ to remove low-mass resonances, and the angular separation between the leptons must satisfy $\Delta{R}_{\ell\ell} > 0.1$ to remove overlapping leptons. To reject top quark production, events with at least one $b$-jet are vetoed. To ensure orthogonality with the ATLAS ggF and VBF production measurements~\cite{HIGG-2021-20}, the leading dijet system is required to have an invariant mass of $|\mjj - \qty{85}{\GeV}| < \qty{15}{\GeV}$ and a rapidity separation of $|\dyjj| < 1.2$.

An artificial neural network (ANN) is trained to simultaneously classify \VH, ggF, VBF, top, \Zjets, and \WW processes; ggF and VBF correspond to the dominant Higgs boson backgrounds, and top, \Zjets, and \WW correspond to the dominant non-Higgs boson backgrounds. The output consists of six distinct output nodes, $\ANNDFOS^i$, each describing the likeliness of the input event to stem from process $i$ ($i=$ \VH, ggF, VBF, top, \Zjets, and \WW) with the condition that $\sum_i\ANNDFOS^i=1$. Different nodes are needed to build dedicated control regions to normalise the relevant background processes, as explained in Section~\ref{sec:background}. In total, 19 variables are used as input --- these are summarised in Table~\ref{tab:inputsTwoLeptonOS} of Appendix~\ref{app:inputs}. Variables including the dilepton invariant mass, \mll, and azimuthal separation, \dphill, are useful for separating Higgs boson production from background processes, and variables including the dijet invariant mass, \mjj, and rapidity separation, $|\dyjj|$, are useful for independently separating \VH and VBF production from all other processes. The \Zjets background is separated from other processes by means of the invariant mass of the $\tau$-lepton pair using the collinear approximation~\cite{Plehn:1999xi}, \mtt, and the top and \WW backgrounds are separated from other processes by means of the transverse mass of the \hww system, \mt:

\begin{equation}
\mt = \sqrt{2\pt^{\ell\ell}\met\left(1 - \cos{\Delta\phi_{\ell\ell,\textrm{miss}}}\right)} \,,
\label{eqn:mT}
\end{equation}

\noindent where $\pt^{\ell\ell}$ is the transverse momentum of the dilepton system and $\Delta\phi_{\ell\ell,\textrm{miss}}$ is the azimuthal separation between the dilepton system and the \met. To construct the SR, the \VH output node of the ANN is required to be greater than 0.2, corresponding to a signal efficiency of 80\% and a background rejection of 77\%.

\begin{figure}[htbp]
\centering
\subfloat[]{
\includegraphics[width=0.46\textwidth]{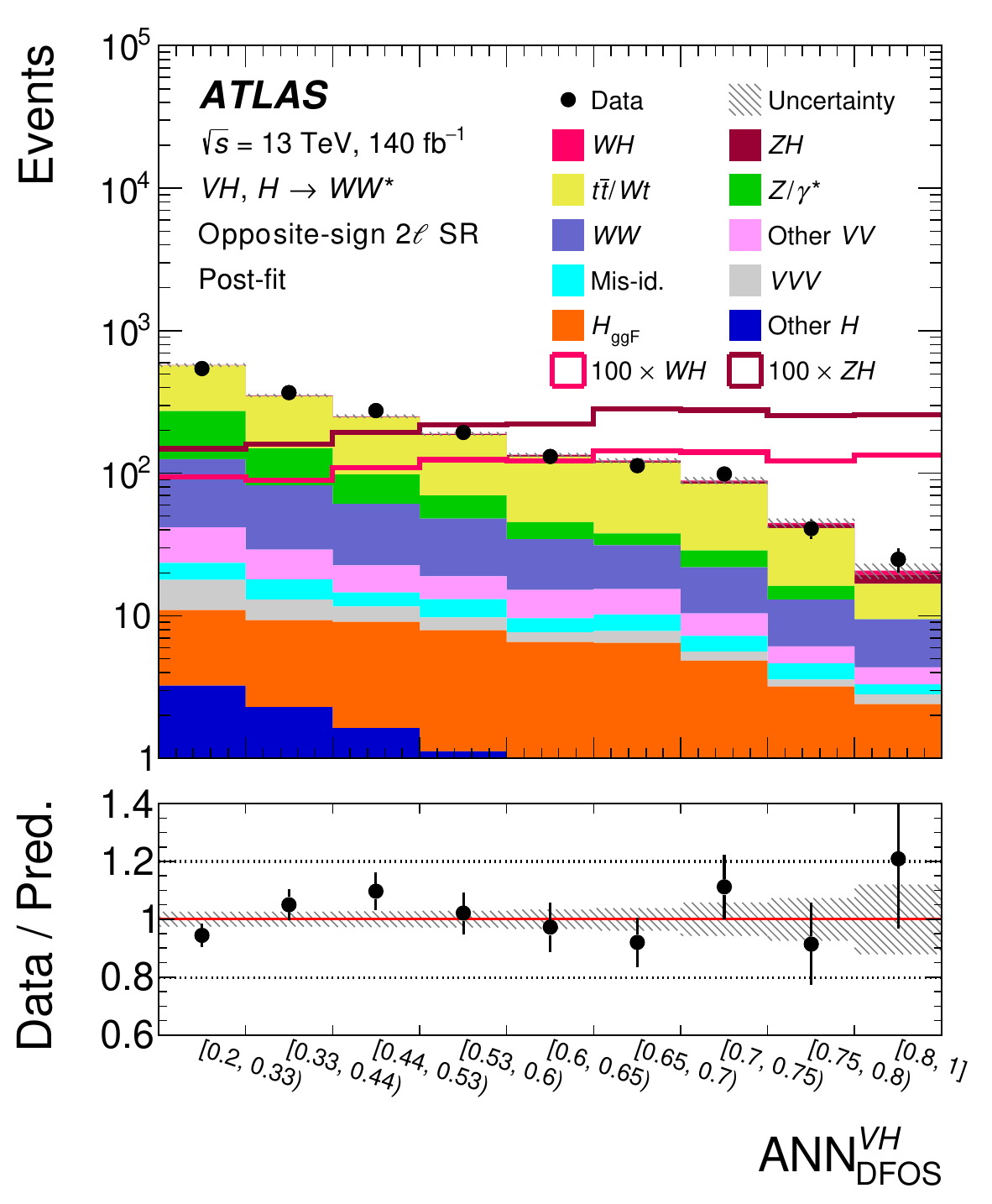}
\label{fig:postfit_MVA_2los_fit}
}
\subfloat[]{
\includegraphics[width=0.51\textwidth]{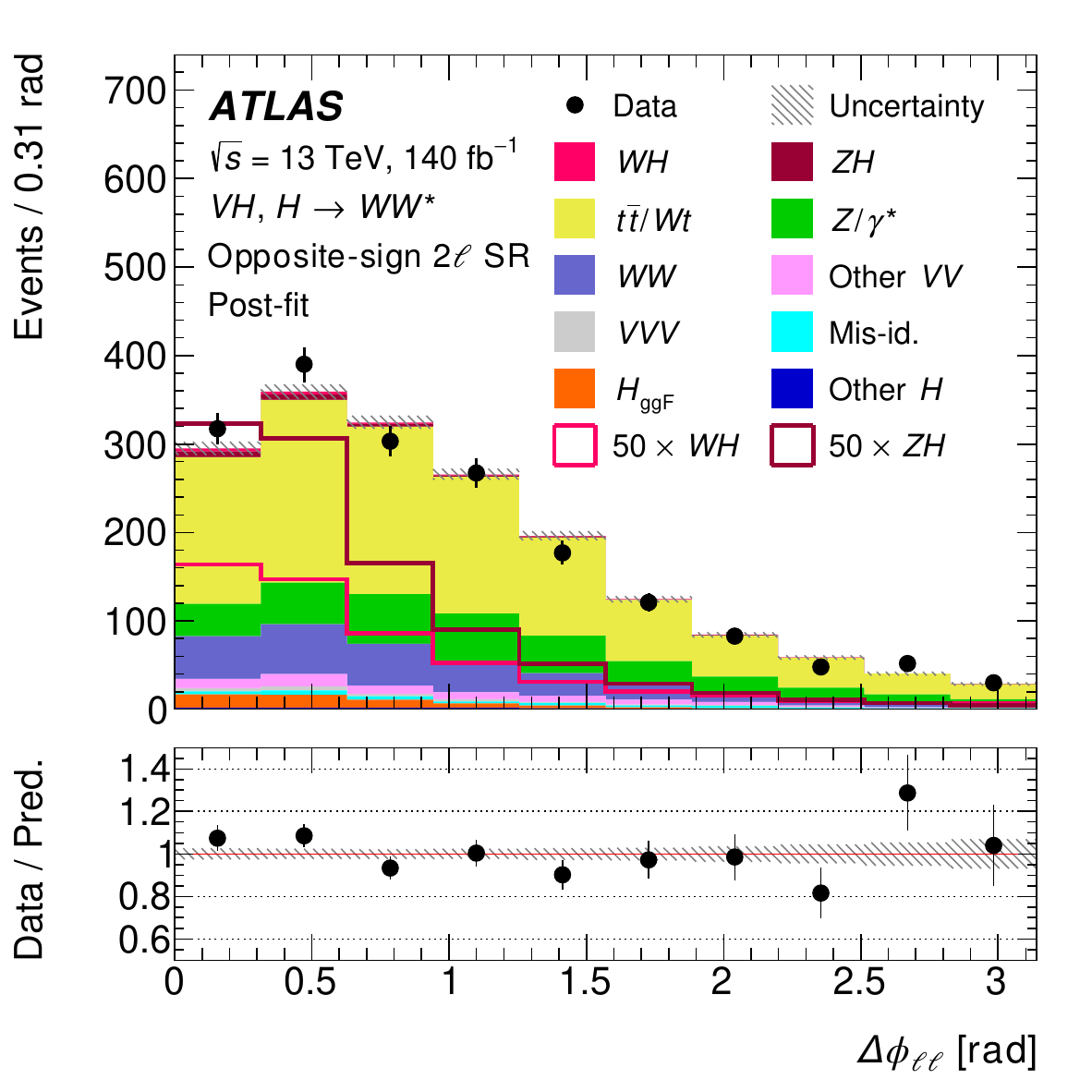}
}
\caption{Post-fit distribution of (a) $\ANNDFOS^{VH}$ and (b) the dilepton azimuthal separation, \dphill, in the opposite-sign $2\ell$ SR. The post-fit \WH and \ZH signal yields are additionally overlaid and upscaled in both distributions. The lower panel shows the ratio of the data to the sum of the fitted signal and background. The hatched band shows the total uncertainty. The post-fit results are obtained from the combined 2-POI fit described in Section~\ref{sec:fit}. In all subsequent post-fit distributions, ``Mis-id.'' is the mis-identified lepton background, as described in Section~\ref{sec:background_fake}.}
\label{fig:postfit_MVA_2los}
\end{figure}

Figure~\ref{fig:postfit_MVA_2los} shows distributions of $\ANNDFOS^{VH}$ and a signal-sensitive kinematic variable in the opposite-sign $2\ell$ SR, as obtained from the fit procedure described in Section~\ref{sec:fit} (hereafter called a ``post-fit'' distribution).

% End of text imported from the .//sections/chan2os.tex input file

% The next lines are included from the .//sections/chan2ss.tex input file
%
\subsection{Same-sign \texorpdfstring{$2\ell$}{2l} channel}

The same-sign $2\ell$ channel selects exactly two same-sign isolated leptons with the leading lepton \pt above \qty{22}{\GeV} and subleading lepton \pt above \qty{15}{\GeV}. The leading and subleading leptons are labelled as $\ell_0$ and $\ell_1$, respectively. Events with at least one reconstructed jet are selected to maintain a high signal efficiency; the choice to require at least one jet instead of at least two jets is made because a significant fraction of simulated signal events, 57\%, have only one reconstructed jet. The invariant mass of the leading dijet system, \mjj, is required to be below \qty{500}{\GeV} for events with two or more jets to suppress contributions from the same-sign \WW process~\cite{STDM-2018-32}. Events containing any $b$-tagged jets are vetoed to reject the top backgrounds. The angular separation between the leptons is required to be $\Delta{R}_{\ell\ell} > 0.4$ to suppress the \Wy events. The events are further split into \SSMM, \SSEE, and \SSDF channels according to the lepton flavour. The \SSEE channel requires two electrons and vetoes events where the invariant mass of the electron pair is within $\pm$\qty{20}{\GeV} of the \Z pole --- this suppresses \Zjets events, where the charge of an electron is reconstructed incorrectly. The \SSMM channel requires two muons, and the \SSDF channel requires one electron and one muon. The selections described above constitute the \SSMM, \SSEE, and \SSDF SRs.

The \WZg process is the dominant background in the same-sign $2\ell$ channel. A recurrent neural network (RNN)~\cite{Hochreiter:1997yld} is trained to distinguish \WH from \WZg, and the same RNN is used in each of the \SSMM, \SSEE, and \SSDF SRs. The RNN has the advantage of being able to learn sequential dependencies for arbitrary-length sequences of objects as input. This feature is used in the analysis where the number of jet variables depend on the jet multiplicity of the events.

The sequences provided to the RNN consist of reconstructed objects placed in order of leptons, \met, and jets (up to a maximum of five jets) with leptons and jets further ordered by \pt. The RNN uses the \pt, $\eta$, and $\phi$ of leptons and jets and the magnitude, \met, and the azimuthal angle, $\phi_\textrm{miss}$, of the missing transverse momentum. A variable containing the information to distinguish between different particle types (leptons, missing transverse momentum, and jets) is also given as input to the RNN. The geometry of the ATLAS detector --- and the physics --- has a rotational symmetry in the $\phi$ direction and a mirror symmetry in the $\eta$ direction. These symmetries are used to pre-process the input variables: the coordinate system of the sequence is rotated to keep the leading lepton with a $\phi$ coordinate of 0 and reflected to keep the leading lepton with a positive $\eta$ coordinate. By doing so, the training data is used in a more efficient way as the RNN does not need to learn these symmetries in the training process.

\begin{figure}[htbp]
\centering
\subfloat[]{
\includegraphics[width=0.31\textwidth]{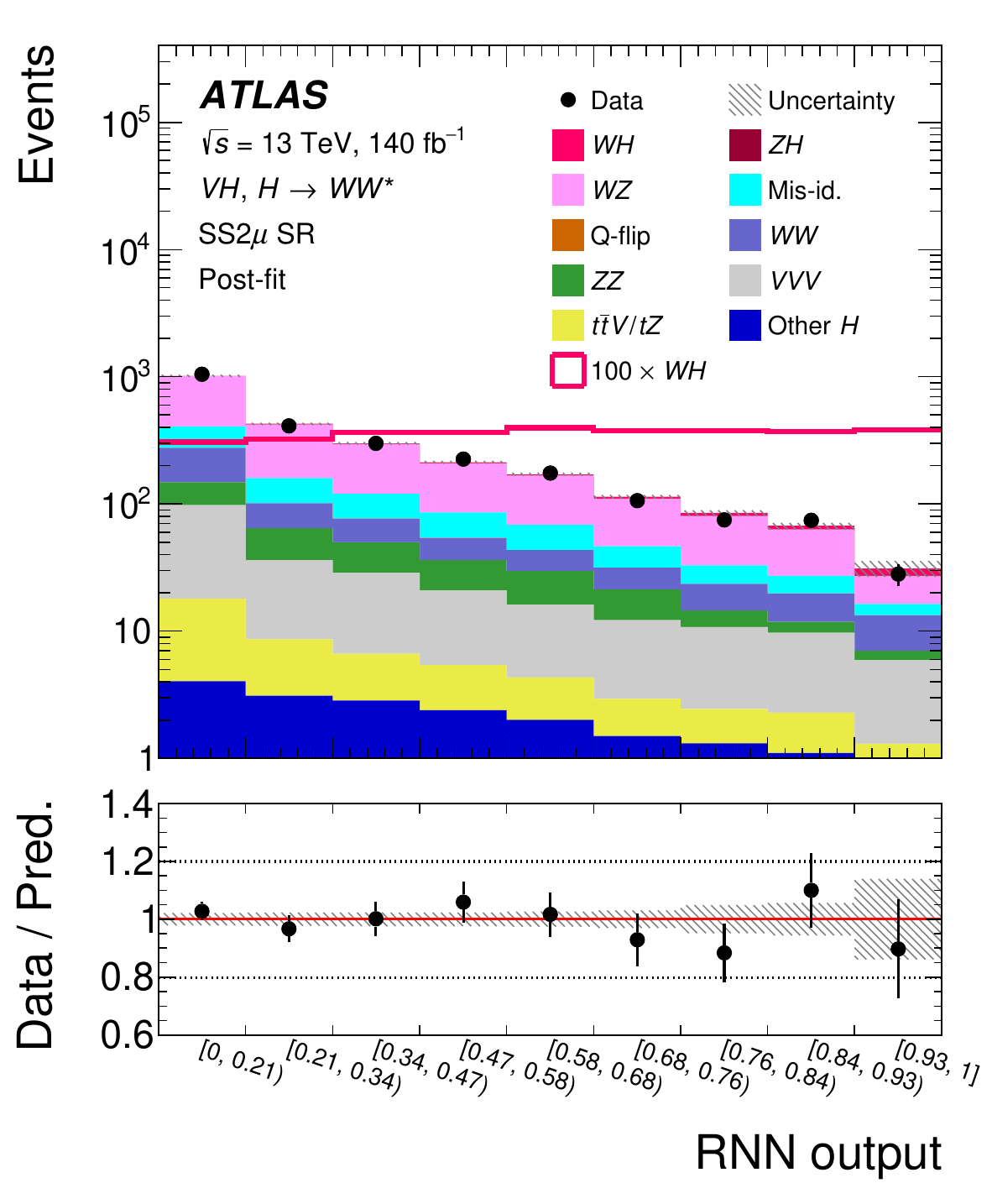}
\label{fig:postfit_MVA_2lss_MM}
}
\subfloat[]{
\includegraphics[width=0.31\textwidth]{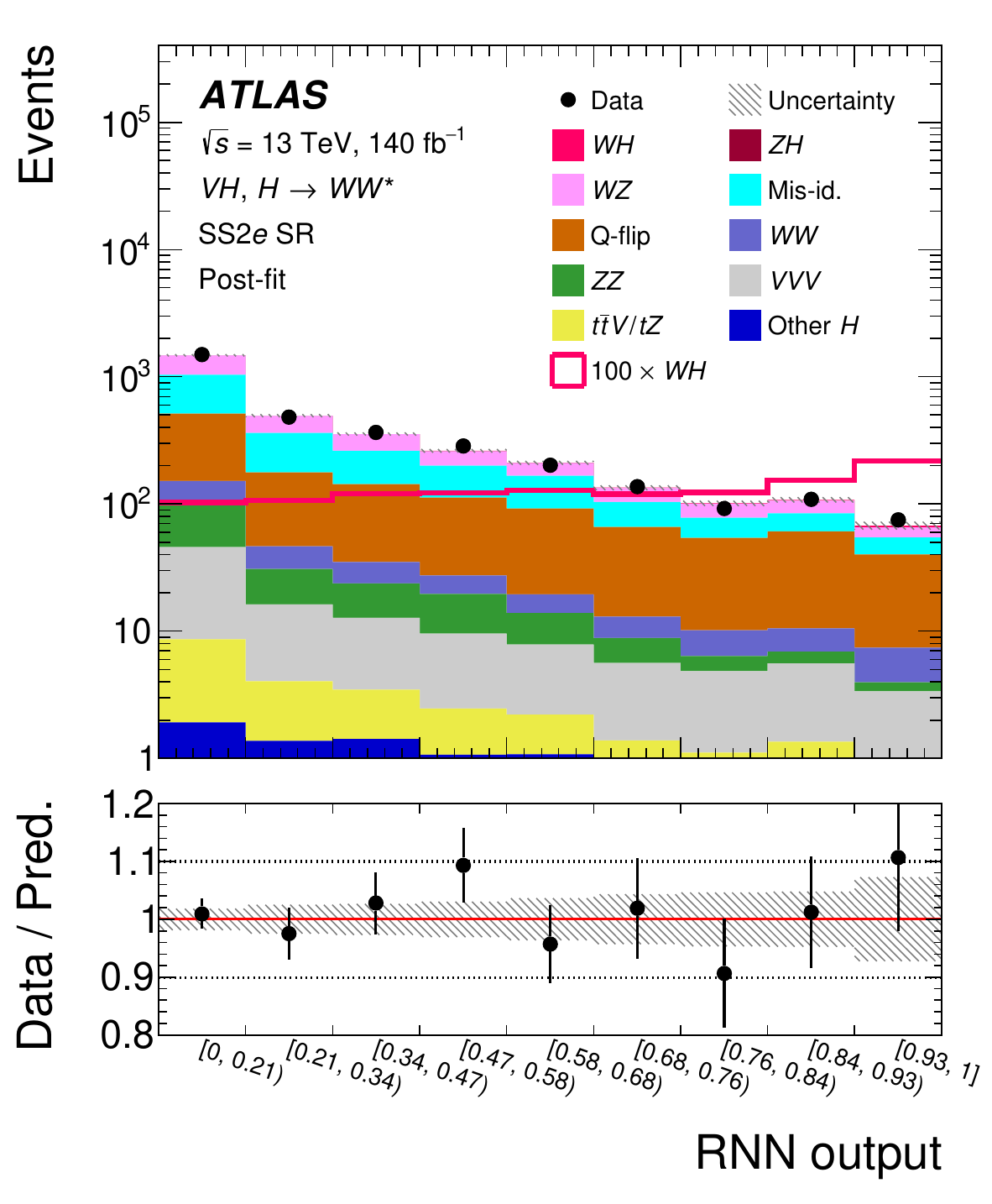}
\label{fig:postfit_MVA_2lss_EE}
}
\subfloat[]{
\includegraphics[width=0.31\textwidth]{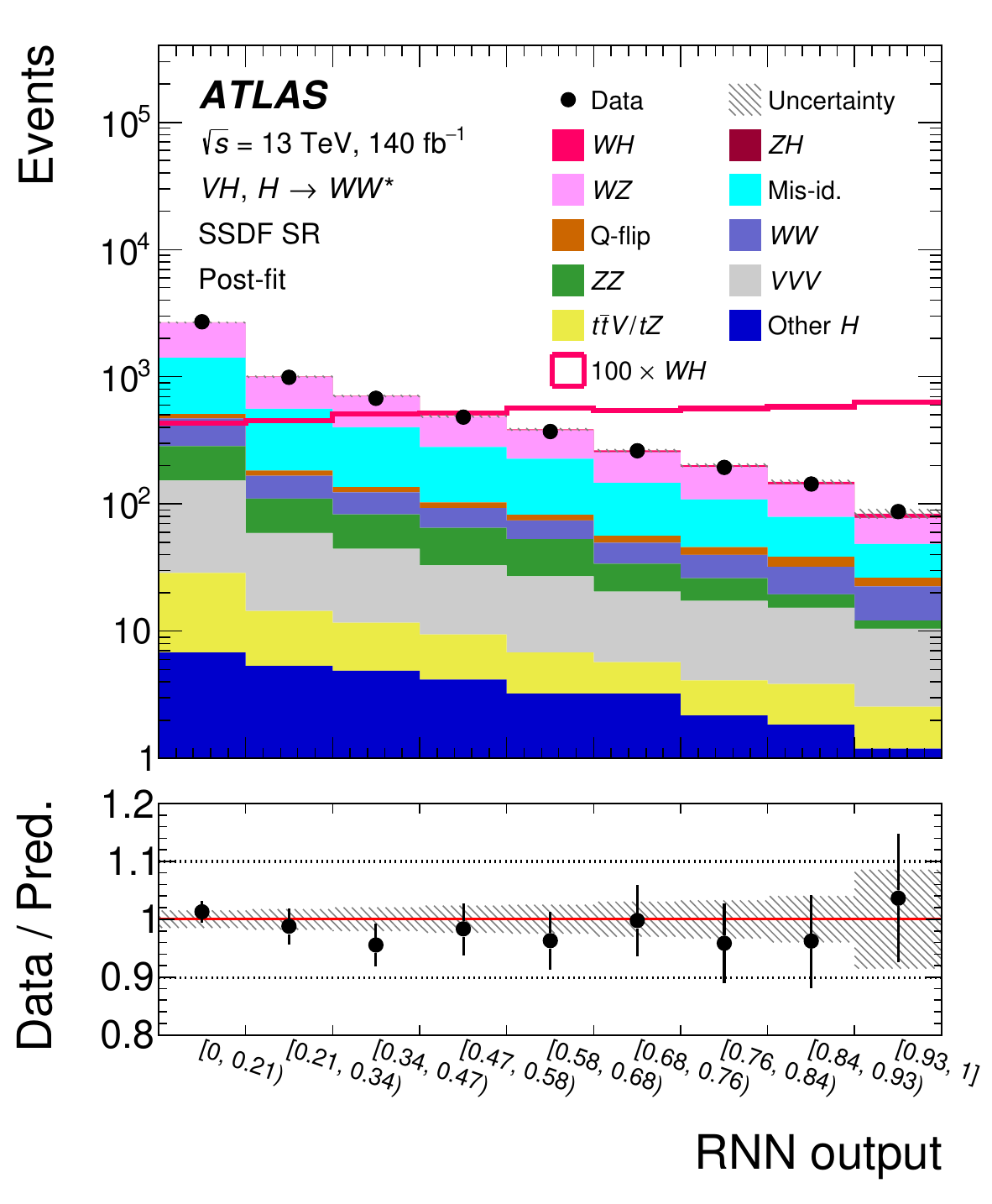}
\label{fig:postfit_MVA_2lss_DF}
} \\
\subfloat[]{
\includegraphics[width=0.32\textwidth]{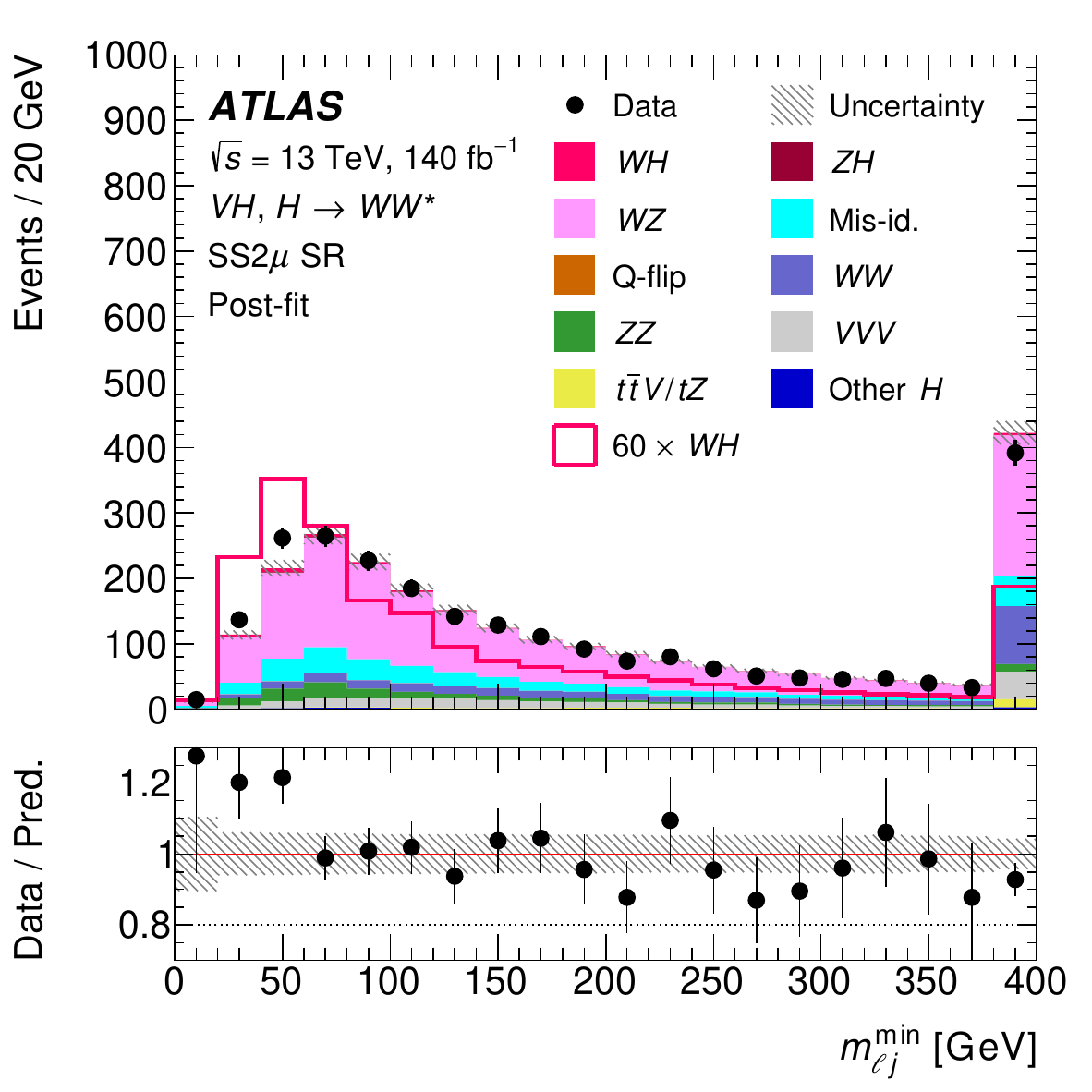}
\label{fig:postfit_mlj_2lss_MM}
}
\subfloat[]{
\includegraphics[width=0.32\textwidth]{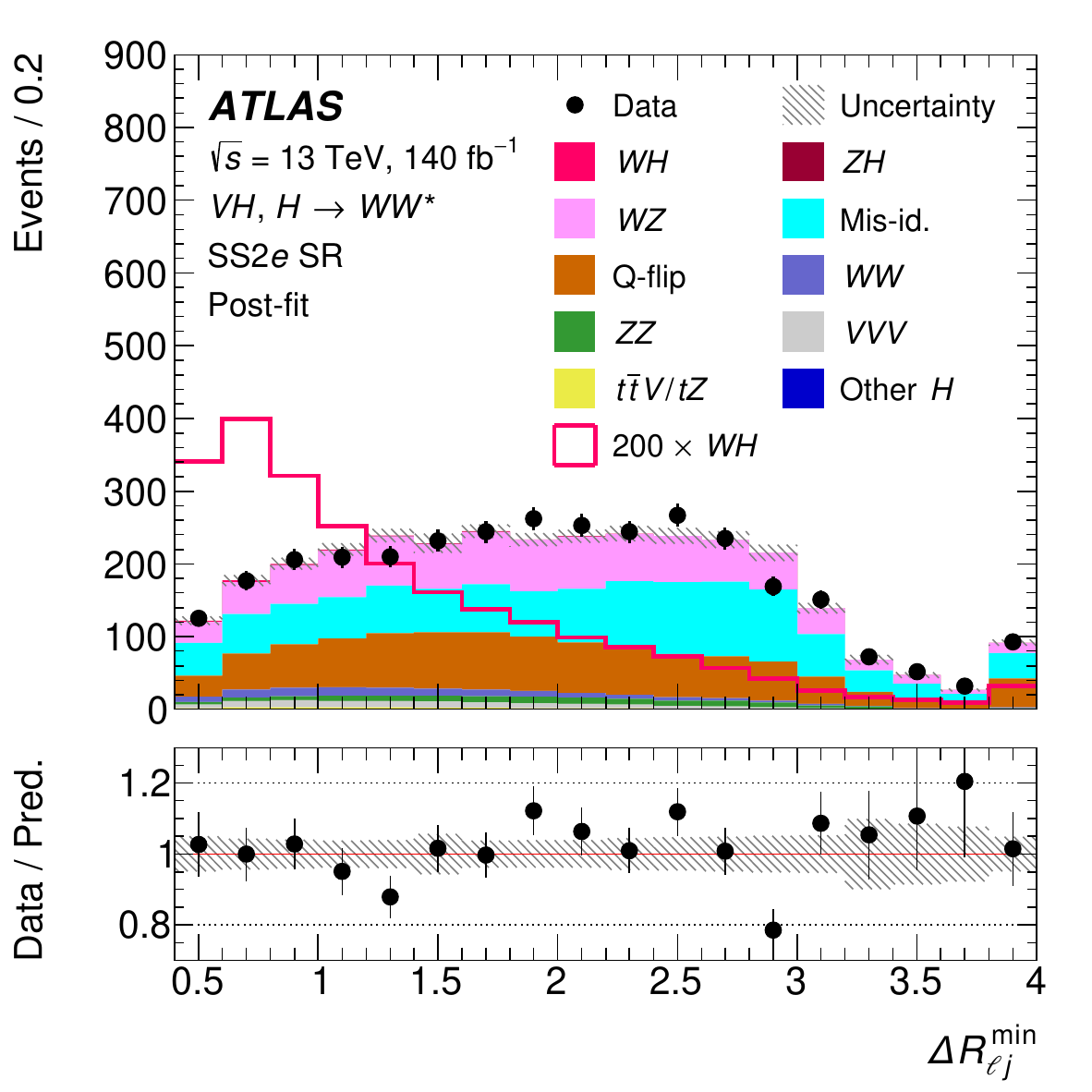}
}
\subfloat[]{
\includegraphics[width=0.32\textwidth]{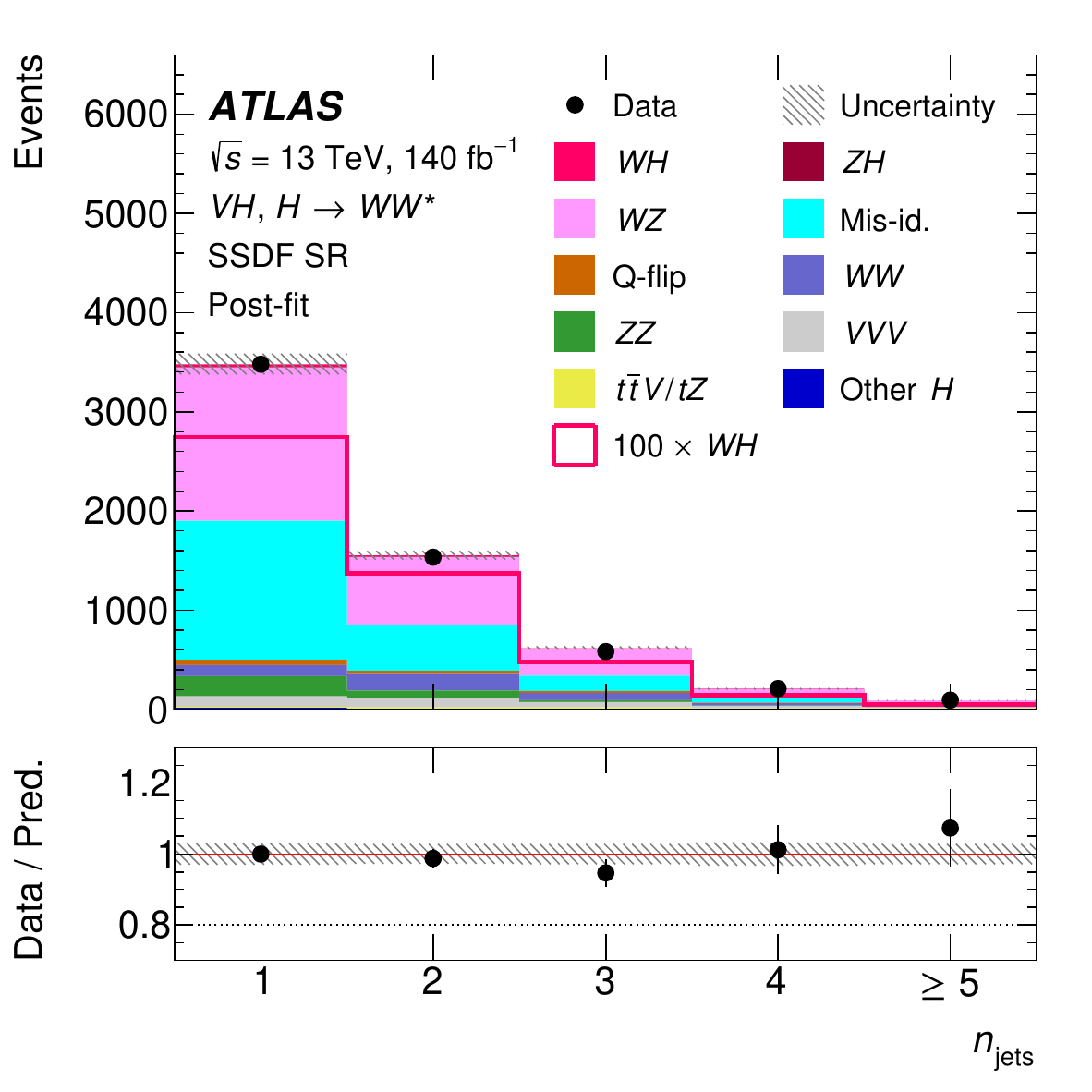}
} \\
\caption{Post-fit distributions of the RNN discriminant in the (a) \SSMM, (b) \SSEE, and (c) \SSDF SRs and (d) the minimum value of the invariant mass of a lepton and jet in the \SSMM SR, (e) the minimum value of the $\Delta{R}$ between a lepton and a jet in the \SSEE SR, and (f) the number of jets in the \SSDF SR. The post-fit \WH signal yields are additionally overlaid and upscaled in all distributions. The lower panel shows the ratio of the data to the sum of the fitted signal and background. The hatched band shows the total uncertainty. The last bin includes overflow, where applicable. The post-fit results are obtained from the combined 2-POI fit described in Section~\ref{sec:fit}. ``Q-flip'' is the electron background with mis-identified charge, as described in Section~\ref{sec:background}.}
\label{fig:postfit_MVA_2lss}
\end{figure}

Figure~\ref{fig:postfit_MVA_2lss} shows the post-fit distributions of the RNN discriminants and signal-sensitive kinematic variables in the three same-sign $2\ell$ SRs. In the \SSMM signal region, an RNN shape mis-modelling in \WZg events, the largest background in that signal region, was observed. A correction to the RNN shape was evaluated using a validation region enriched in the \WZg background and applied to each of the \SSMM, \SSEE, and \SSDF signal regions. A detailed description of this correction is given in Section~\ref{sec:background}. A residual discrepancy between data and simulation is observed in the minimum value of the invariant mass of a lepton and jet, shown in Figure~\ref{fig:postfit_mlj_2lss_MM}. As described in Section~\ref{sec:background}, this variable is sensitive to the mis-modelling in \WZg events; however, unlike the RNN shape, this variable is not explicitly corrected as it does not directly enter the fit model.

% End of text imported from the .//sections/chan2ss.tex input file

% The next lines are included from the .//sections/chan3l.tex input file
%
\subsection{\texorpdfstring{$3\ell$}{3l} channel}
\label{sec:3l}

In the $3\ell$ channel, exactly three isolated leptons with $\pt > \qty{15}{\GeV}$ are required with a total charge of $\pm1$. The lepton with unique charge is labelled $\ell_0$, the lepton closest to $\ell_0$ in angular distance $\Delta{R}$ is labelled $\ell_1$, and the remaining lepton is labelled $\ell_2$. This labelling scheme is demonstrated pictorially in the leading-order Feynman diagram, Figure~\ref{fig:channels_3l}. In signal events, leptons $\ell_0$ and $\ell_1$ are most likely to originate from the \HWW decay with probabilities of 99\% and 85\%, respectively.

The most prominent background processes in the $3\ell$ channel are \WZg production and top quark processes with either three prompt leptons (e.g., \ttV) or two prompt leptons and one non-prompt lepton from a $b$-hadron decay (e.g., \ttbar).

The analysis of the $3\ell$ channel separates events with at least one same-flavour, opposite-sign charge (SFOS) lepton pair from events with zero SFOS lepton pairs, which have different signal-to-background ratios. Due to the presence of background processes with $Z\to\ell\ell$ decays as a dominant background, the former set of events is hereafter referred to as the \Z-dominated channel, while the latter is referred to as the \Z-depleted channel.

In the \Z-dominated channel, the major background processes are those involving \Z bosons. Therefore, the invariant mass, \mll, for each SFOS pair is required to satisfy a \Z-veto selection: $|\mll - \mZ| > \qty{25}{\GeV}$. To suppress background events from heavy-flavour quarkonia, the minimum invariant mass of all SFOS pairs is required to be greater than \qty{12}{\GeV}. Furthermore, the \met is required to be larger than \qty{30}{\GeV} to select final states with neutrinos. To suppress processes containing top quarks, events with at least one $b$-tagged jets are vetoed. To remove overlapping leptons, the angular separation between all lepton pairs must satisfy $\Delta{R}_{\ell\ell} > 0.1$. These selections constitute the \Z-dominated SR.

For both the channels, dedicated ANNs are used to separate signal from background. In the \Z-dominated channel, fifteen variables are used as input --- these are summarised in Table~\ref{tab:inputsZdom} of Appendix~\ref{app:inputs}. The purpose of the single output classifier, \ANNZdom, is to distinguish between the signal and the dominant \WZg background. Accordingly, it is trained against this background process. The output is used as final discriminant in this channel. Particularly useful variables for separating \WH and \WZg production include the invariant mass of leptons~0 and 1, $m_{\ell_0\ell_1}$, as well as the angular distance between leptons~0 and 1, $\Delta{R}_{\ell_0\ell_1}$; also useful are the \met and \met-related variables such as the azimuthal separation between lepton~0 and the \met, $\Delta\phi_{\ell_0,\textrm{miss}}$.

In the \Z-depleted channel, the three dominant background processes are \ttbar, \WZg, and \WWW. To simultaneously separate the \WH signal from these backgrounds, a multi-classifier is used providing four distinct output nodes, $\ANNZdep^i$, each describing the likeliness of the input event to stem from process $i$ ($i=$ \WH, \ttbar, \WZg, \WWW) with the condition that $\sum_i\ANNZdep^i=1$. Table~\ref{tab:inputsZdep} of Appendix~\ref{app:inputs} summarises the input variables used in the \Z-depleted channel. The angular distance between leptons~0 and 1, $\Delta{R}_{\ell_0\ell_1}$, is particularly useful for separating \WH production from background processes, and the \met and \met-related variables are useful for separating \WH, \ttbar, and \WZg production, which all have varying numbers of neutrinos in the final state. The transverse momentum of the leading jet, $\pt^{j_0}$, is useful for separating \ttbar production from other processes, and the compatibility of the event with the $\WZg(\to\tau\tau)$ hypothesis, $F_\alpha$, is useful for separating \WZg production from other processes. The latter variable is described in greater detail in the appendix.

To suppress processes containing top quarks, events with a high $\ANNZdep^{\ttbar}$ score --- greater than 0.25 --- or at least one $b$-tagged jets are vetoed. Additionally, all lepton pairs must satisfy $\Delta{R}_{\ell\ell} > 0.1$ to remove overlapping leptons. These selections constitute the \Z-depleted SR, whose final discriminant, $\ANNZdep^\Delta$, is defined as:

\begin{equation}
\mathrm{ANN}_\textrm{Zdep}^{\Delta} = \mathrm{ANN}_\textrm{Zdep}^{WH} - \mathrm{ANN}_\textrm{Zdep}^{WZ} - \mathrm{ANN}_\textrm{Zdep}^{WWW} \,.
\end{equation}

\begin{figure}[htbp]
\centering
\subfloat[]{
\includegraphics[width=0.48\textwidth]{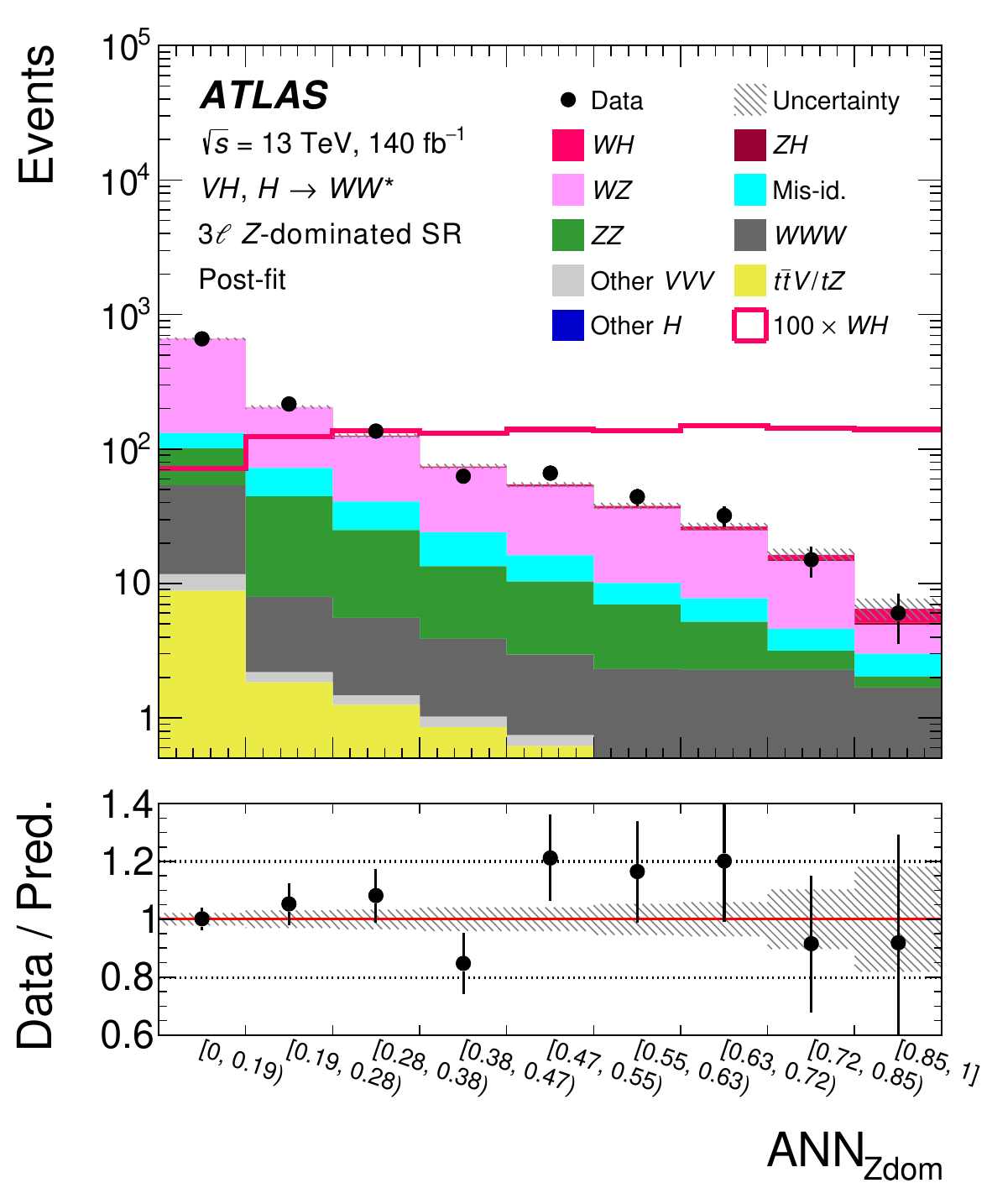}
\label{fig:postfit_MVA_3l_Zdom}
}
\subfloat[]{
\includegraphics[width=0.48\textwidth]{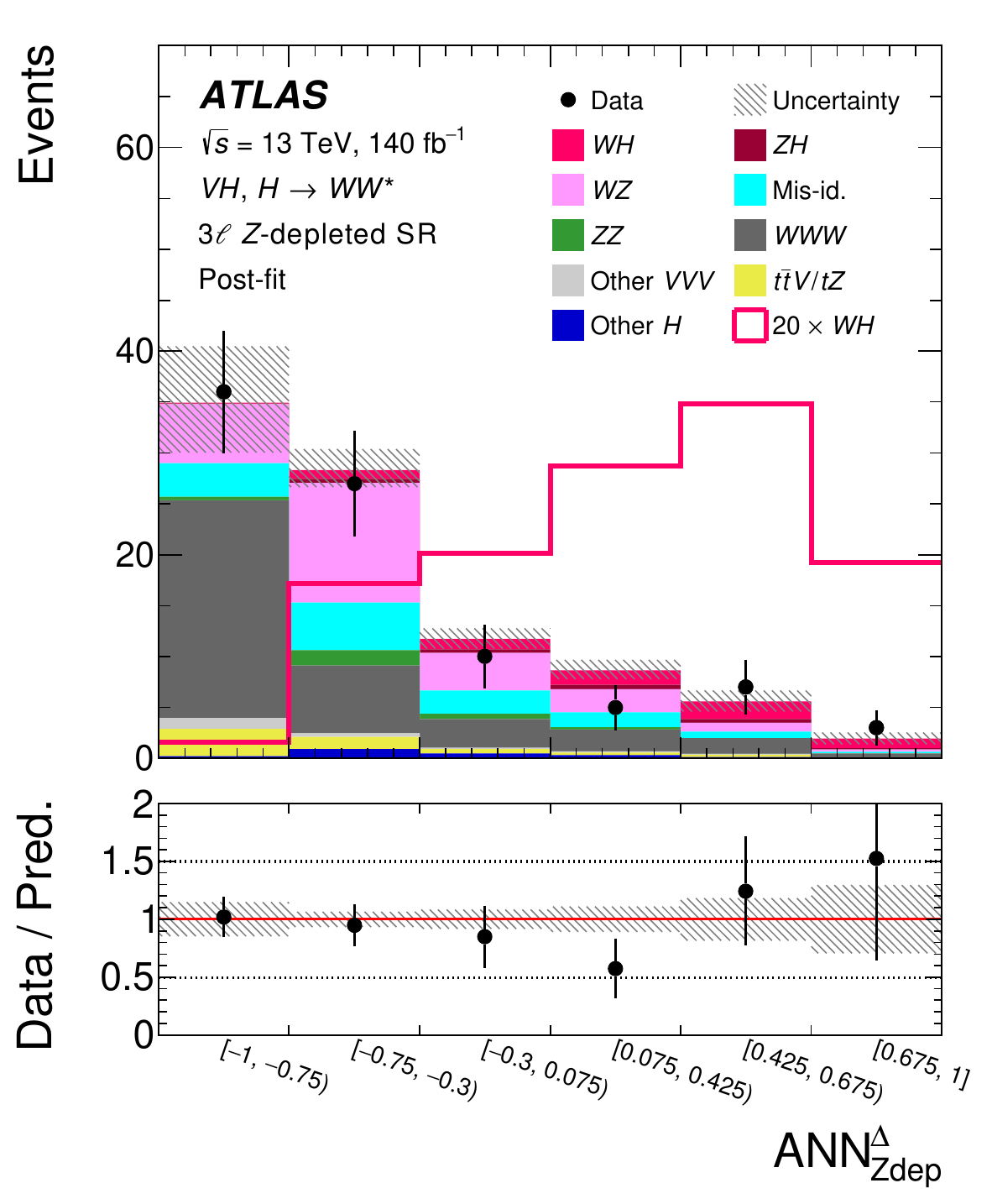}
\label{fig:postfit_MVA_3l_Zdep}
} \\
\subfloat[]{
\includegraphics[width=0.48\textwidth]{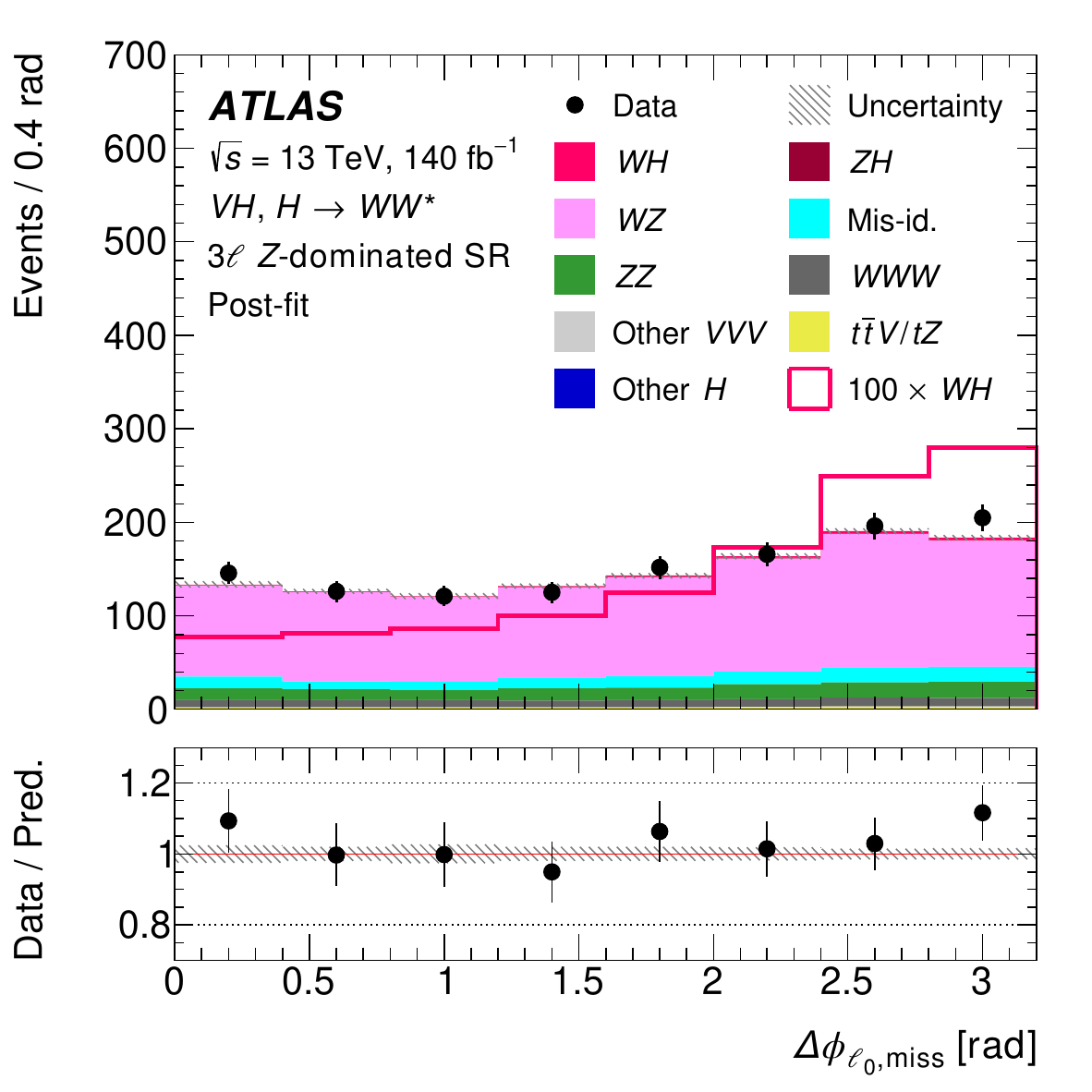}
}
\subfloat[]{
\includegraphics[width=0.48\textwidth]{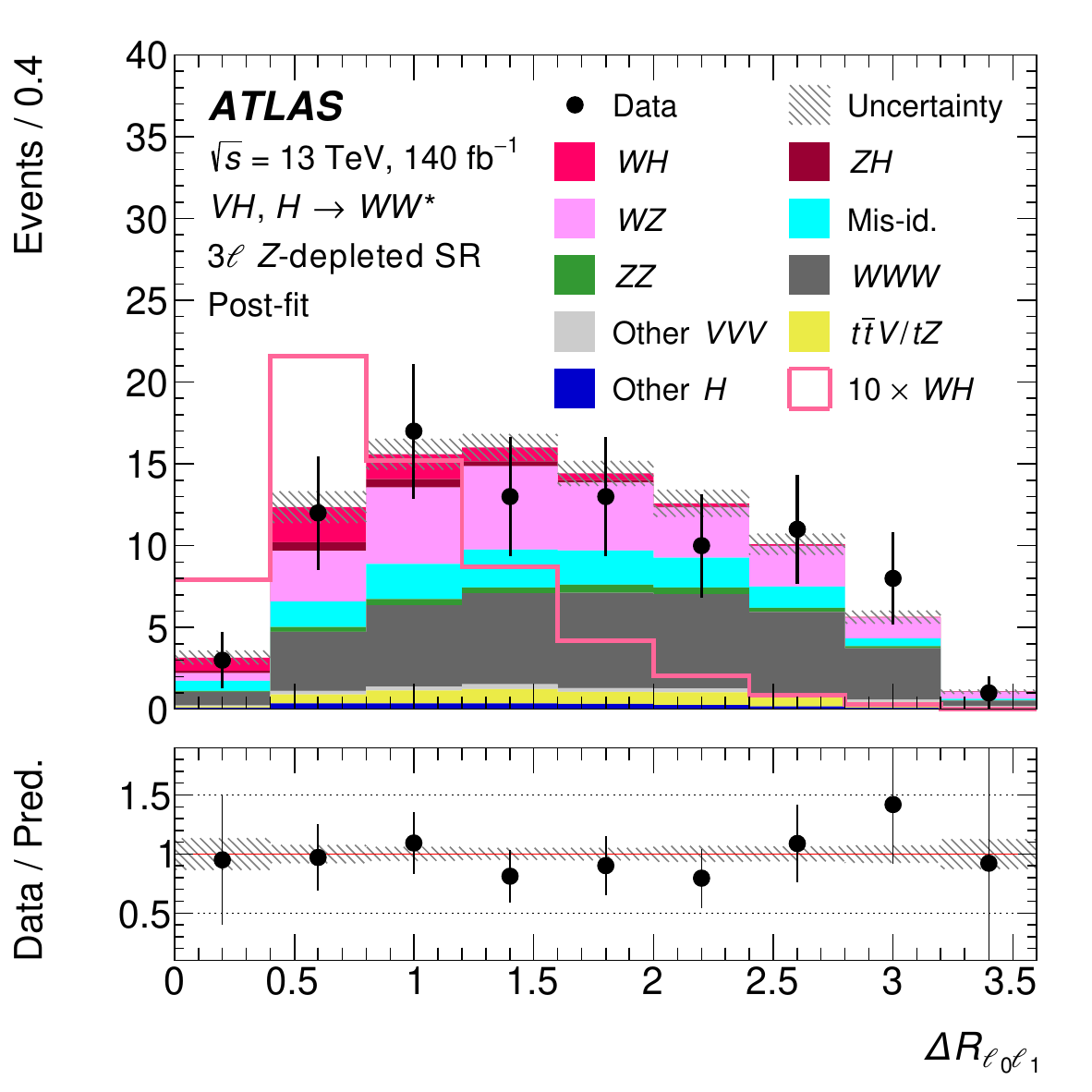}
} \\
\caption{Post-fit distributions of (a) \ANNZdom and (c) the azimuthal separation between lepton~0 and the \met, $\Delta\phi_{\ell_0,\textrm{miss}}$, in the \Z-dominated SR as well as (b) $\ANNZdep^\Delta$ and (d) the angular distance between leptons~0 and 1, $\Delta{R}_{\ell_0\ell_1}$, in the \Z-depleted SR. The post-fit \WH signal yields are additionally overlaid and upscaled in all distributions. The lower panel shows the ratio of the data to the sum of the fitted signal and background. The last bin includes overflow, where applicable. The post-fit results are obtained from the combined 2-POI fit described in Section~\ref{sec:fit}.}
\label{fig:postfit_MVA_3l}
\end{figure}

Figure~\ref{fig:postfit_MVA_3l} shows the post-fit distributions of the ANN discriminants and signal-sensitive kinematic variables in both the \Z-dominated and \Z-depleted SRs.

% End of text imported from the .//sections/chan3l.tex input file

% The next lines are included from the .//sections/chan4l.tex input file
%
\subsection{\texorpdfstring{$4\ell$}{4l} channel}
\label{sec:4l}

The \ZH channel includes events with four isolated leptons with $\pt > \qty{10}{\GeV}$ and total electric charge of zero. Events containing an SFOS lepton pair with $\mll < \qty{12}{\GeV}$ or a DFOS lepton pair with $\mll < \qty{10}{\GeV}$ are rejected to suppress the contamination from heavy-flavour quarkonia. Selected events are classified into channels according to the number of SFOS lepton pairs: 1-SFOS and 2-SFOS. Events with no SFOS lepton pairs are excluded.

The most prominent background process in the $4\ell$ channel is \ZZg production, with smaller contributions from top quark processes (e.g., \ttZ), processes with mis-identified leptons (e.g., \ttW), and \WWZ production.

The reconstruction of the \ZH process proceeds through the identification of the leptons from the \Z boson, called $\ell_2$ and $\ell_3$, as the SFOS lepton pair with invariant mass closest to the mass of the \Z boson. The remaining two leptons, labelled $\ell_0$ and $\ell_1$, are candidates for originating from the Higgs boson decay. This labelling scheme is demonstrated pictorially in the leading-order Feynman diagram, Figure~\ref{fig:channels_4l}. To suppress the \ttZ process, events containing $b$-tagged jets are rejected. To reduce the \ZZg background process in 2-SFOS events, the invariant mass of $\ell_0$ and $\ell_1$, \mHc, is required to be below \qty{50}{\GeV}; to ensure orthogonality with the ATLAS $H \to ZZ\virt \to 4\ell$ measurement~\cite{HIGG-2018-28}, the 4-lepton invariant mass, \mfl, is required to be below \qty{115}{\GeV} or above \qty{130}{\GeV}. Finally, the angular separation between $\ell_0$ and $\ell_1$, $\Delta{R}_{\ell_0\ell_1}$, is required to be larger than 0.2 to remove overlapping leptons.

\begin{figure}[htbp]
\centering
\subfloat[]{
\includegraphics[width=0.48\textwidth]{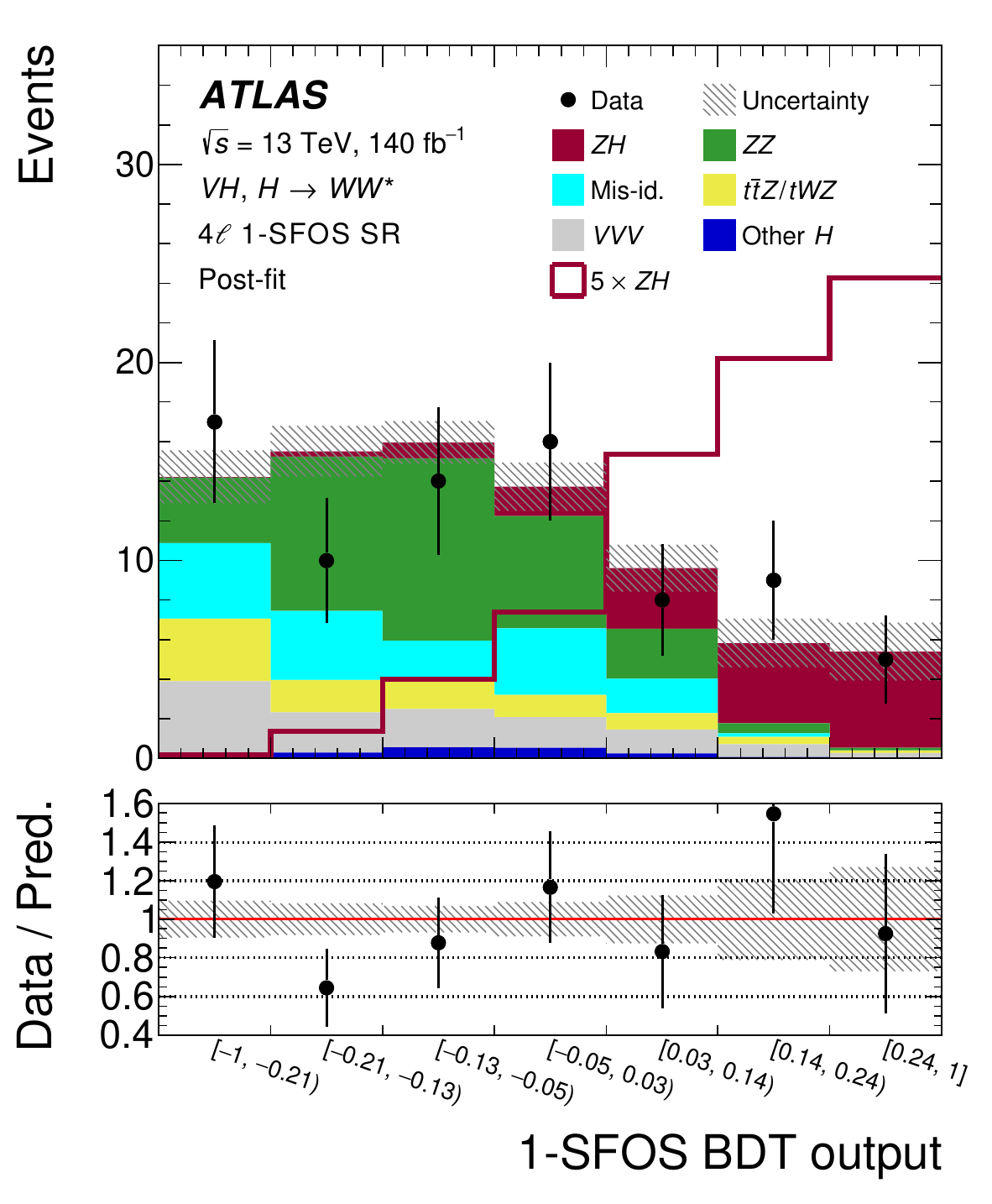}
\label{fig:postfit_MVA_4l_1SFOS}
}
\subfloat[]{
\includegraphics[width=0.48\textwidth]{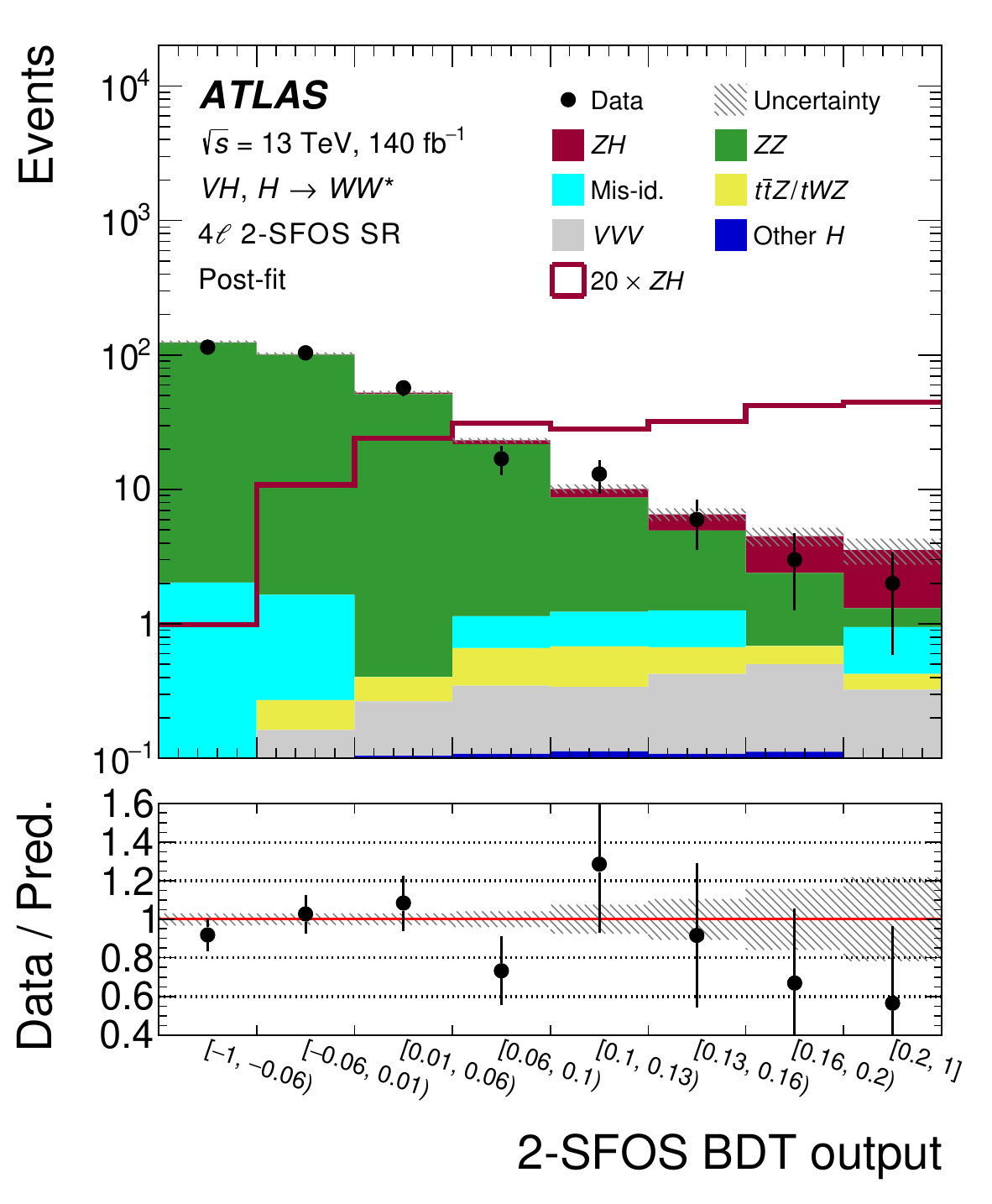}
\label{fig:postfit_MVA_4l_2SFOS}
} \\
\subfloat[]{
\includegraphics[width=0.48\textwidth]{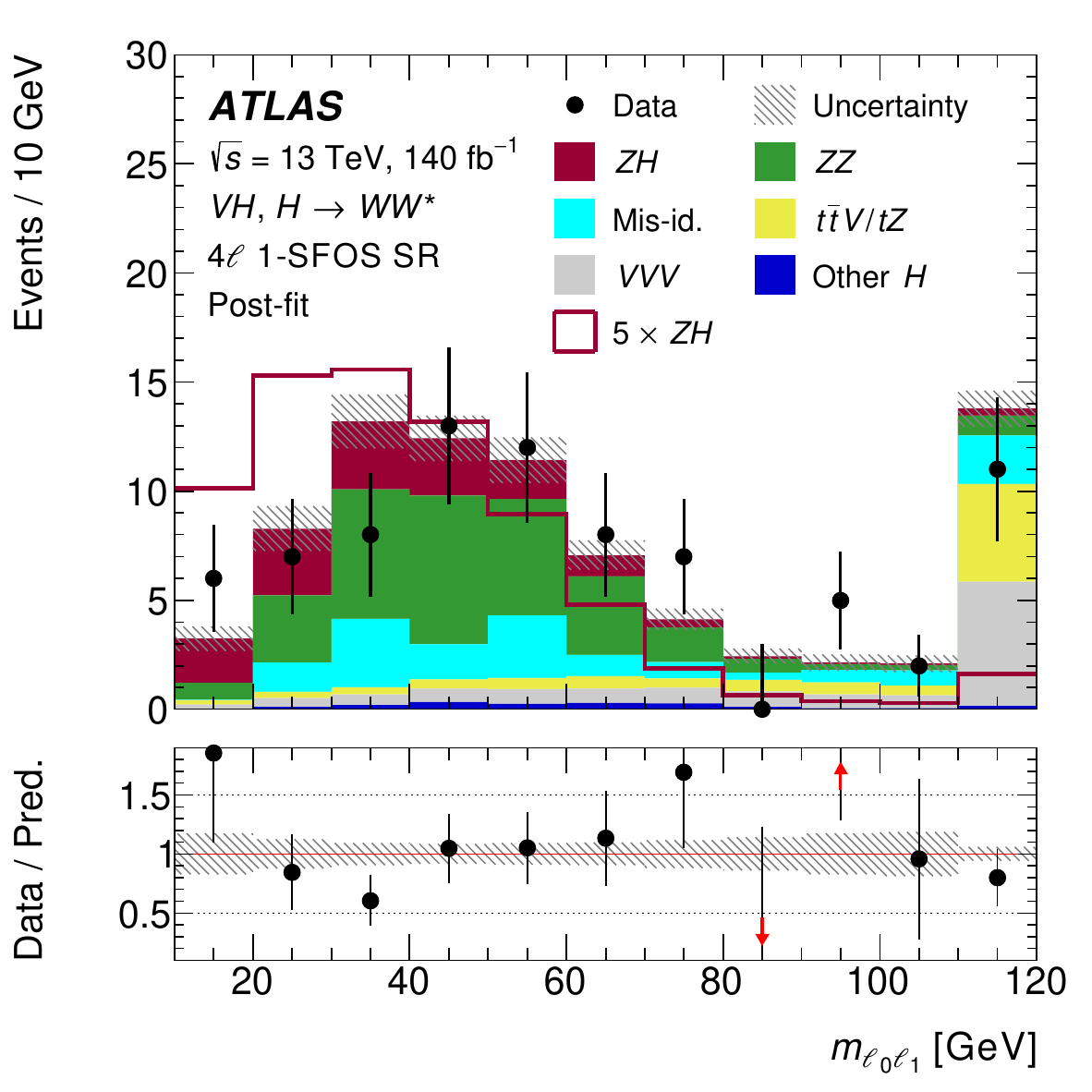}
}
\subfloat[]{
\includegraphics[width=0.48\textwidth]{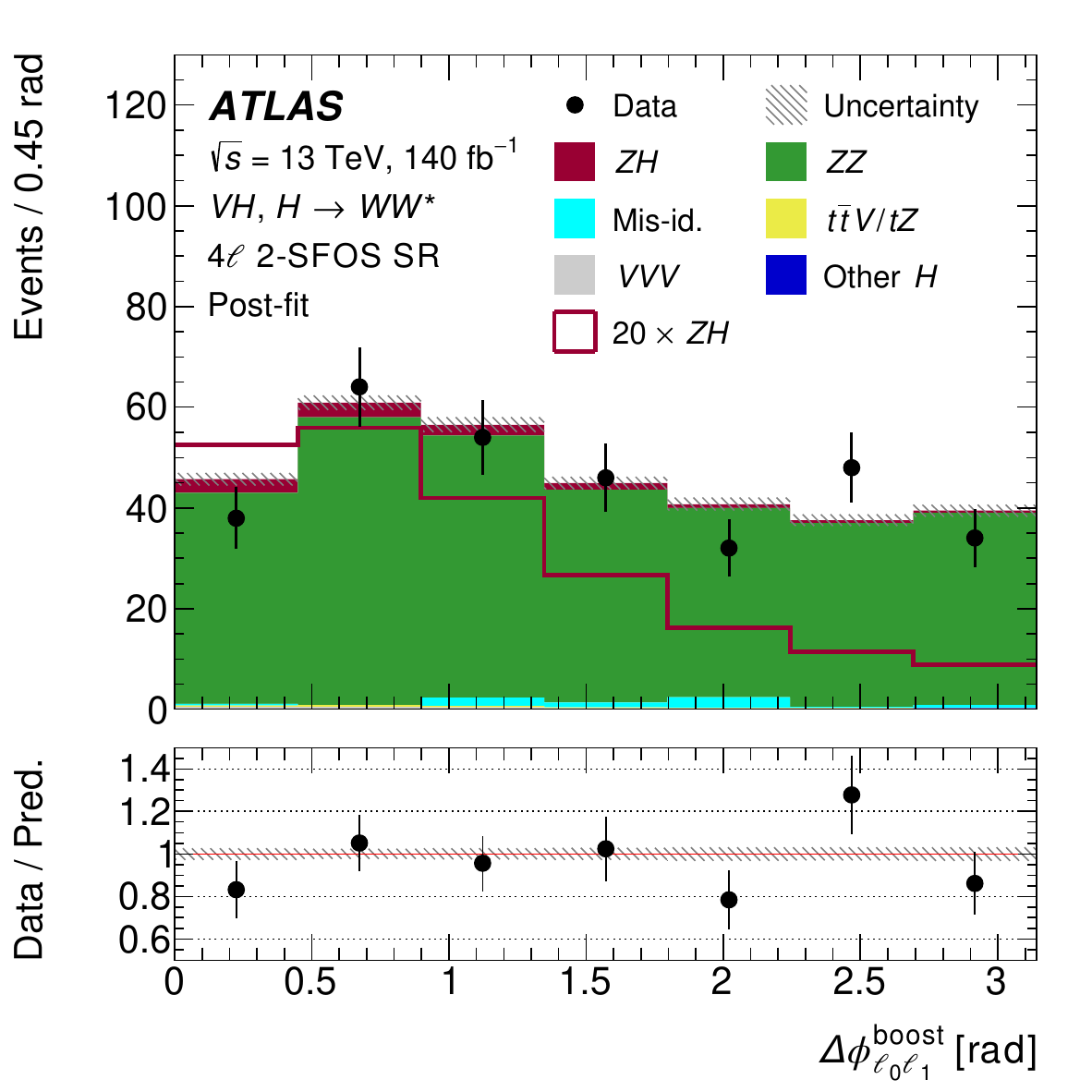}
} \\
\caption{Post-fit distributions of (a) the BDT output and (c) the invariant mass of the Higgs boson candidate, \mHc, in the 1-SFOS SR and (b) the BDT output and (d) the azimuthal separation between the leptons from the Higgs boson candidate in the rest frame of the Higgs boson, \DPhiB, in the 2-SFOS SR. The post-fit \ZH signal yields are additionally overlaid and upscaled in all distributions. The lower panel shows the ratio of the data to the sum of the fitted signal and background, and the arrow in (c) indicates the position of a point outside the vertical axis range. The hatched band shows the total uncertainty. The last bin includes overflow, where applicable. The post-fit results are obtained from the combined 2-POI fit described in Section~\ref{sec:fit}.}
\label{fig:postfit_MVA_4l}
\end{figure}

The selections described above constitute the 1-SFOS and 2-SFOS SRs. For each SR, a discriminant based on a BDT is used to achieve a further separation between the \ZH and \ZZg processes. Independent BDTs are trained for each SR but sharing a common set of input variables, which are summarised in Table~\ref{tab:inputsFourLepton} of Appendix~\ref{app:inputs}. In the 1-SFOS SR, the most important variables for discriminating between \ZH and \ZZg include \mHc, the number of jets, and the azimuthal separation between the leptons from the Higgs boson candidate in the rest frame of the Higgs boson, \DPhiB; in the 2-SFOS SR, the most important variables include \mHc, \met, and \DPhiB. Figure~\ref{fig:postfit_MVA_4l} shows the post-fit distributions of the BDT discriminants and signal-sensitive kinematic variables in the two $4\ell$ SRs.

% End of text imported from the .//sections/chan4l.tex input file

% The next lines are included from the .//sections/STXS.tex input file
%
\subsection{STXS categorisation}
\label{sec:stxs}

The Stage~1.2 STXS categorisation scheme for \VH production is split by the production mode (\qqWH, \qqZH, and \ggZH), the \pt of the associated vector boson, and the number of jets, as shown in Figure~\ref{fig:STXSStage1_2} of Appendix~\ref{app:stxs}. Based on the expected sensitivity of the measurement, kinematic fiducial regions are merged to ensure each cross-section is measured with adequate precision, yielding a reduced categorisation schemes, \ptV and STXS. In both the schemes, there is no splitting of events by jet multiplicity. The fiducial regions for each are shown in Table~\ref{tab:fiducial}.

For the \ptV scheme, the \qqWH, \qqZH, and \ggZH categories are merged into a single \VH category which is inclusive in the decay of the vector boson and which is defined in the fiducial region $|y_H| < 2.5$ where $y_H$ is the rapidity of the Higgs boson. The $250 \leq \ptV < \qty{400}{\GeV}$ and $\ptV \geq \qty{400}{\GeV}$ categories are merged into a single $\ptV \geq \qty{250}{\GeV}$ category. This yields four measured fiducial cross-sections.

In the STXS scheme, the $\qq\to\W(\to\ell\nu)\H$ category, written as \lvH for brevity, is measured independently of the $\qq\to\Z(\to\ell\ell/\nu\nu)\H$ and $gg\to\Z(\to\ell\ell/\nu\nu)\H$ categories, which are merged and written as \llH for brevity. The $150 \leq \ptV < \qty{250}{\GeV}$, $250 \leq \ptV < \qty{400}{\GeV}$, and $\ptV \geq \qty{400}{\GeV}$ categories are merged into a single $\ptV \geq \qty{150}{\GeV}$ category. The EW \qqH category with $60 \leq \mjj < \qty{120}{\GeV}$ is measured using the opposite-sign $2\ell$ channel. This yields seven measured fiducial cross-sections.

\begin{table}[htbp]
\centering
\caption{Summary the differential cross-sections measured in the \ptV and STXS schemes.}
\label{tab:fiducial}
\renewcommand\arraystretch{1.25}
\begin{tabular}{l|l}
\hline\hline
\ptV scheme & STXS scheme \\
\hline
\VH, $0 \leq \ptV < \qty{75}{\GeV}$    & \lvH and \llH, $0 \leq \ptV < \qty{75}{\GeV}$ \\
\VH, $75 \leq \ptV < \qty{150}{\GeV}$  & \lvH and \llH, $75 \leq \ptV < \qty{150}{\GeV}$ \\
\VH, $150 \leq \ptV < \qty{250}{\GeV}$ & \lvH and \llH, $\ptV \geq \qty{150}{\GeV}$ \\
\VH, $\ptV \geq \qty{250}{\GeV}$       & EW \qqH, $60 \leq \mjj < \qty{120}{\GeV}$ \\
\hline\hline
\end{tabular}
\end{table}

For each of the channels considered by the analysis, subregions of the corresponding SRs are defined which target the individual or pairs of the fiducial cross-sections above. The subregions are split using selections on reconstructed proxies of \ptV; the boundaries of these selections were scanned and chosen to ensure adequate expected sensitivity for all POIs measured by the corresponding channel. Table~\ref{tab:STXS_SRs} summarises the \ptV proxies and selections for each channel and the relevant \ptV range for each subregion.

For the opposite-sign $2\ell$ channel, the chosen proxy is the dijet transverse momentum, \ptjj. For the same-sign $2\ell$ channel, the chosen proxy is the scalar sum of the lepton, jet, and missing transverse momenta, \sumpt. The same splitting is shared between the \SSMM, \SSEE, and \SSDF SRs.

For the $3\ell$ channel, a regression ANN is used to reconstruct the transverse momentum of the \W boson. The regression ANN receives a subset of the input variables used by the multi-classifier of the \Z-depleted SR --- these input variables are summarised in Table~\ref{tab:inputsRegression} of Appendix~\ref{app:inputs}. Its output node, \ptR, is chosen as a proxy for \ptV and is utilised in both the \Z-dominated and \Z-depleted SRs.

For the $4\ell$ channel, the chosen proxy is the transverse momentum of the \Z boson candidate, $\ptZ \coloneqq \pt^{\ell_2\ell_3}$. The same splitting is applied to both the 1-SFOS and 2-SFOS SRs.

The performance of the proxies for several channels is shown in Figure~\ref{fig:proxies} of Appendix~\ref{app:stxs}. Each channel uses the trained multivariate discriminant from the inclusive SR in each of its subregions; retraining of the discriminant was tested for the $3\ell$ channel without producing appreciable improvements in sensitivity. The binning of the discriminant in each subregion is necessarily different from that of the inclusive SR, owing to smaller sample sizes. These distributions are provided as input to the combined fit. In the case of the STXS scheme, the opposite-sign $2\ell$ SR is still split according to \ptV; however, only a single fiducial cross-section of EW \qqH is measured.

\begin{table}[htbp]
\centering
\caption{Selections used to define the SRs entering the differential cross-section measurement. For each channel, the \ptV proxy and the selections applied to it are shown. In the context of the Stage~1.2 STXS framework, the \ptV range relevant to each SR is also shown. In the case of the same-sign $2\ell$ channel, the selections are the same for each of the \SSMM, \SSEE, and \SSDF SRs; in the case of the $4\ell$ channel, the selections are the same for each of the 1-SFOS and 2-SFOS SRs.}
\label{tab:STXS_SRs}
\renewcommand\arraystretch{1.25}
\resizebox{\textwidth}{!}{
\begin{tabular}{l|c|c|c}
\hline\hline
Channel & \ptV proxy & Reconstructed SR & Relevant \ptV range \\
\hline
\multirow{3}{*}{Opposite-sign $2\ell$} & \multirowcell{3}{Dijet transverse \\ momentum, \ptjj} & $0 \leq \ptjj < \qty{160}{\GeV}$ & $0 \leq \ptV < \qty{150}{\GeV}$ \\
& & $160 \leq \ptjj < \qty{260}{\GeV}$ & $150 \leq \ptV < \qty{250}{\GeV}$ \\
& & $\ptjj \geq \qty{260}{\GeV}$ & $\ptV \geq \qty{250}{\GeV}$ \\
\hline
\multirow{4}{*}{Same-sign $2\ell$} & \multirowcell{4}{Scalar sum of lepton, jet, \\ and missing transverse \\ momenta, \sumpt} & $0 \leq \sumpt < \qty{200}{\GeV}$ & $0 \leq \ptV < \qty{75}{\GeV}$ \\
& & $200 \leq \sumpt < \qty{320}{\GeV}$ & $75 \leq \ptV < \qty{150}{\GeV}$ \\
& & $320 \leq \sumpt < \qty{460}{\GeV}$ & $150 \leq \ptV < \qty{250}{\GeV}$ \\
& & $\sumpt \geq \qty{460}{\GeV}$ & $\ptV \geq \qty{250}{\GeV}$ \\
\hline
\multirow{3}{*}{$3\ell$ \Z-dominated} & \multirowcell{3}{Regression ANN \\ for \W transverse \\ momentum, \ptR} & $0 \leq \ptR < \qty{90}{\GeV}$ & $0 \leq \ptV < \qty{75}{\GeV}$ \\
& & $90 \leq \ptR < \qty{180}{\GeV}$ & $75 \leq \ptV < \qty{150}{\GeV}$ \\
& & $\ptR \geq \qty{180}{\GeV}$ & $\ptV \geq \qty{150}{\GeV}$ \\
\hline
\multirow{4}{*}{$3\ell$ \Z-depleted} & \multirowcell{4}{Regression ANN \\ for \W transverse \\ momentum, \ptR} & $0 \leq \ptR < \qty{90}{\GeV}$ & $0 \leq \ptV < \qty{75}{\GeV}$ \\
& & $90 \leq \ptR < \qty{180}{\GeV}$ & $75 \leq \ptV < \qty{150}{\GeV}$ \\
& & $180 \leq \ptR < \qty{270}{\GeV}$ & $150 \leq \ptV < \qty{250}{\GeV}$ \\
& & $\ptR \geq \qty{270}{\GeV}$ & $\ptV \geq \qty{250}{\GeV}$ \\
\hline
\multirow{4}{*}{$4\ell$} & \multirowcell{4}{\Z boson transverse \\ momentum, \ptZ} & $0 \leq \ptZ < \qty{75}{\GeV}$ & $0 \leq  \ptV < \qty{75}{\GeV}$ \\
& & $75 \leq  \ptZ < \qty{150}{\GeV}$ & $75 \leq  \ptV < \qty{150}{\GeV}$ \\
& & $150 \leq  \ptZ < \qty{250}{\GeV}$ & $150 \leq  \ptV < \qty{250}{\GeV}$ \\
& & $\ptZ \geq \qty{250}{\GeV}$ & $\ptV \geq \qty{250}{\GeV}$ \\
\hline\hline
\end{tabular}
}
\end{table}

\begin{figure}[htbp]
\centering
\subfloat[]{
\includegraphics[width=0.65\textwidth]{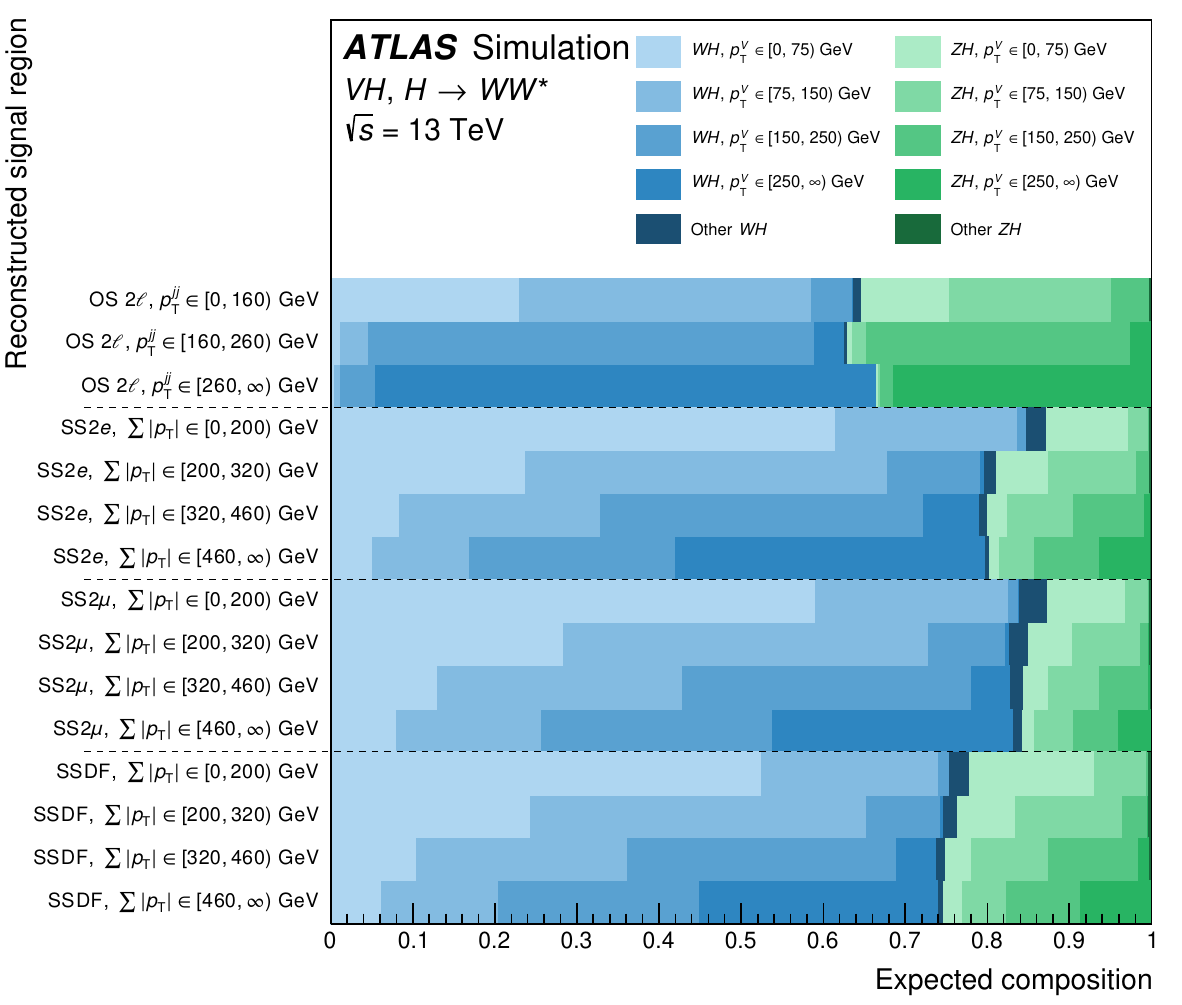}
} \\
\subfloat[]{
\includegraphics[width=0.65\textwidth]{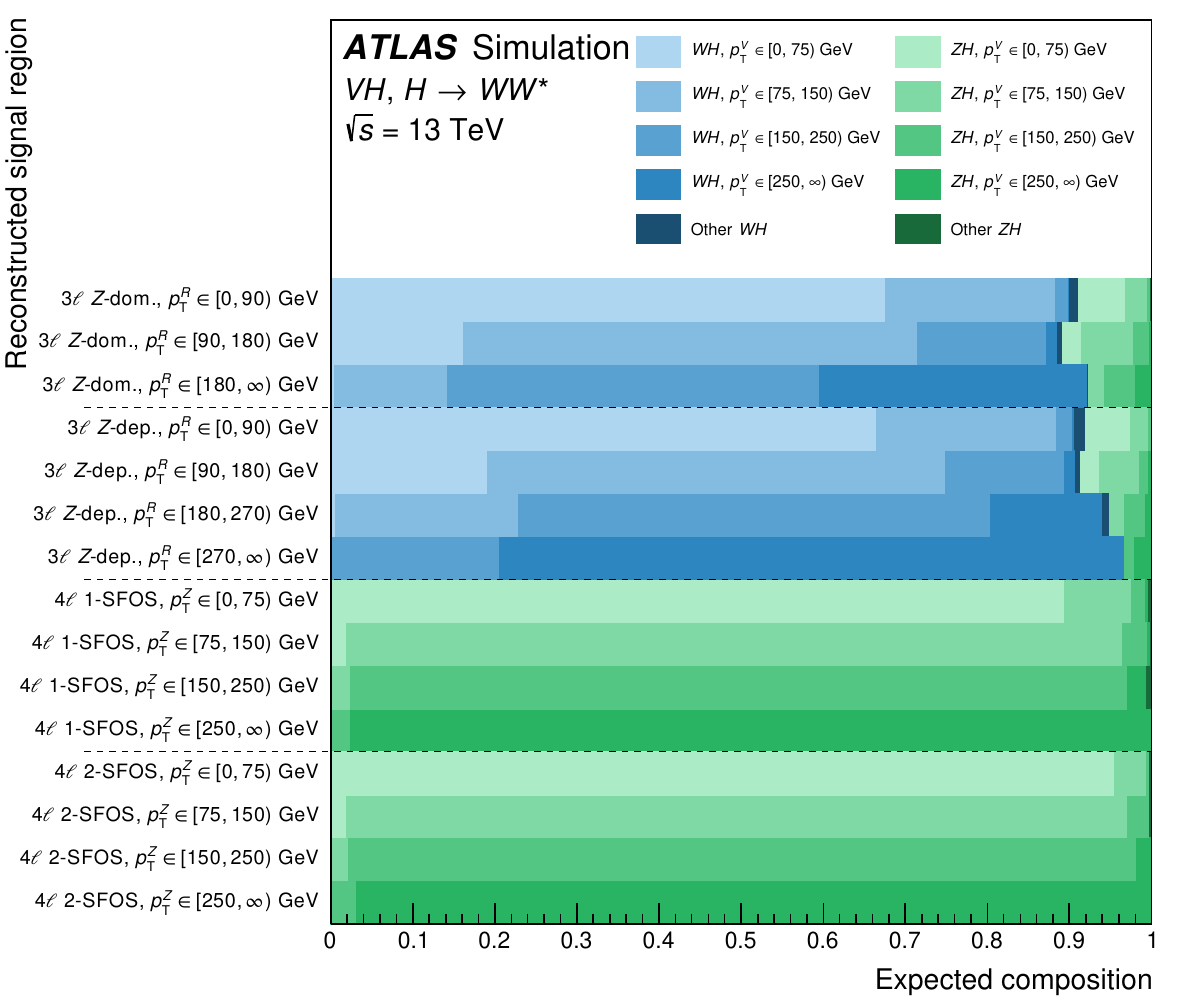}
} \\
\caption{Relative SM signal composition in terms of the reduced categories for each reconstructed SR of the (a) $2\ell$ and (b) $3\ell$/$4\ell$ channels. The categorisation follows the \ptV scheme; however, the \WH and \ZH templates are resolved to show their relative fraction in each SR. ``Other \WH/\ZH'' includes forward Higgs bosons (i.e., Higgs bosons produced with rapidity $|y| > 2.5$) and --- for the $3\ell$ channels only --- Higgs boson processes not included in the categorisation.}
\label{fig:STXSCat}
\end{figure}

\begin{figure}[htbp]
\centering
\subfloat[]{
\includegraphics[width=0.45\textwidth]{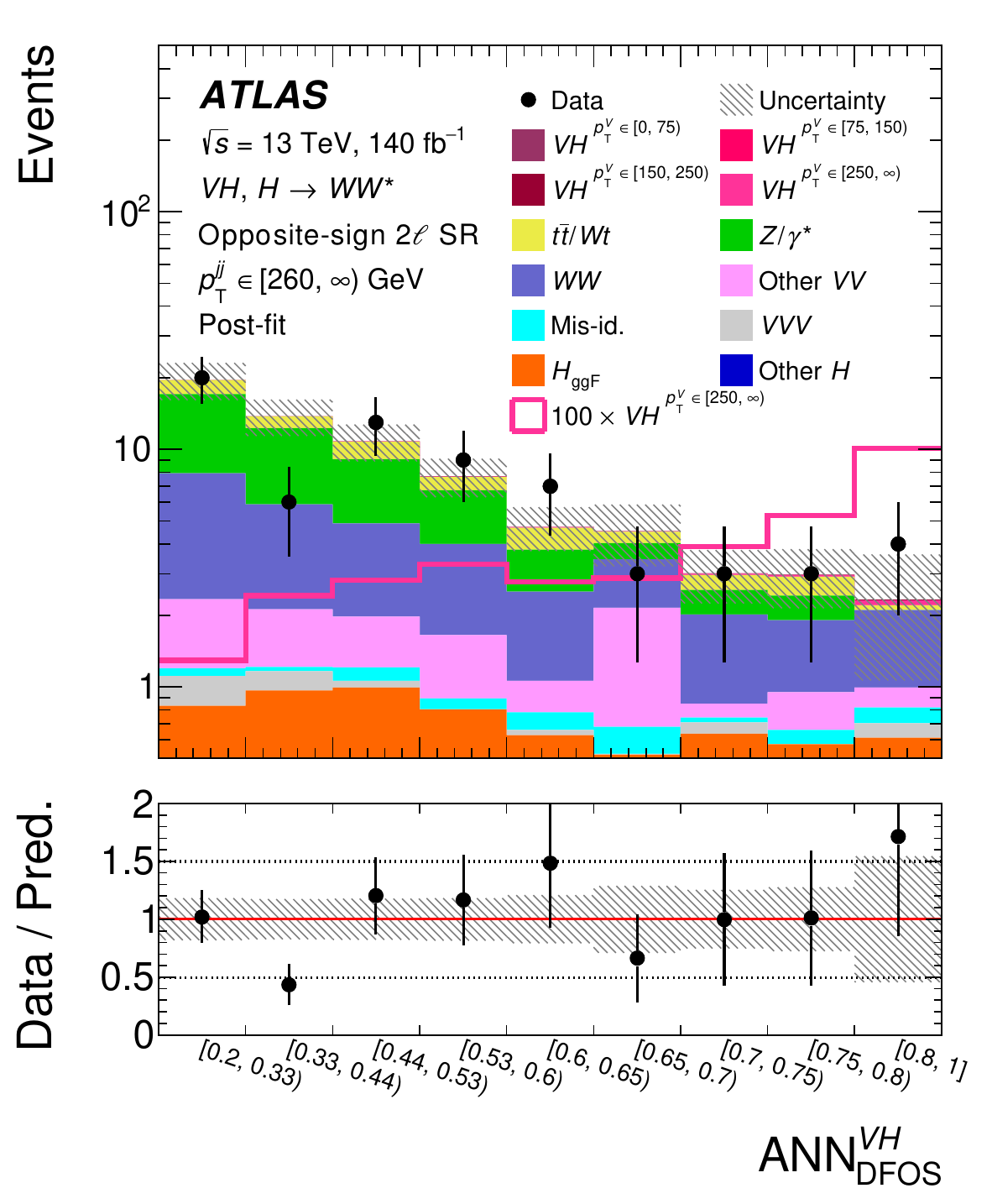}
}
\subfloat[]{
\includegraphics[width=0.45\textwidth]{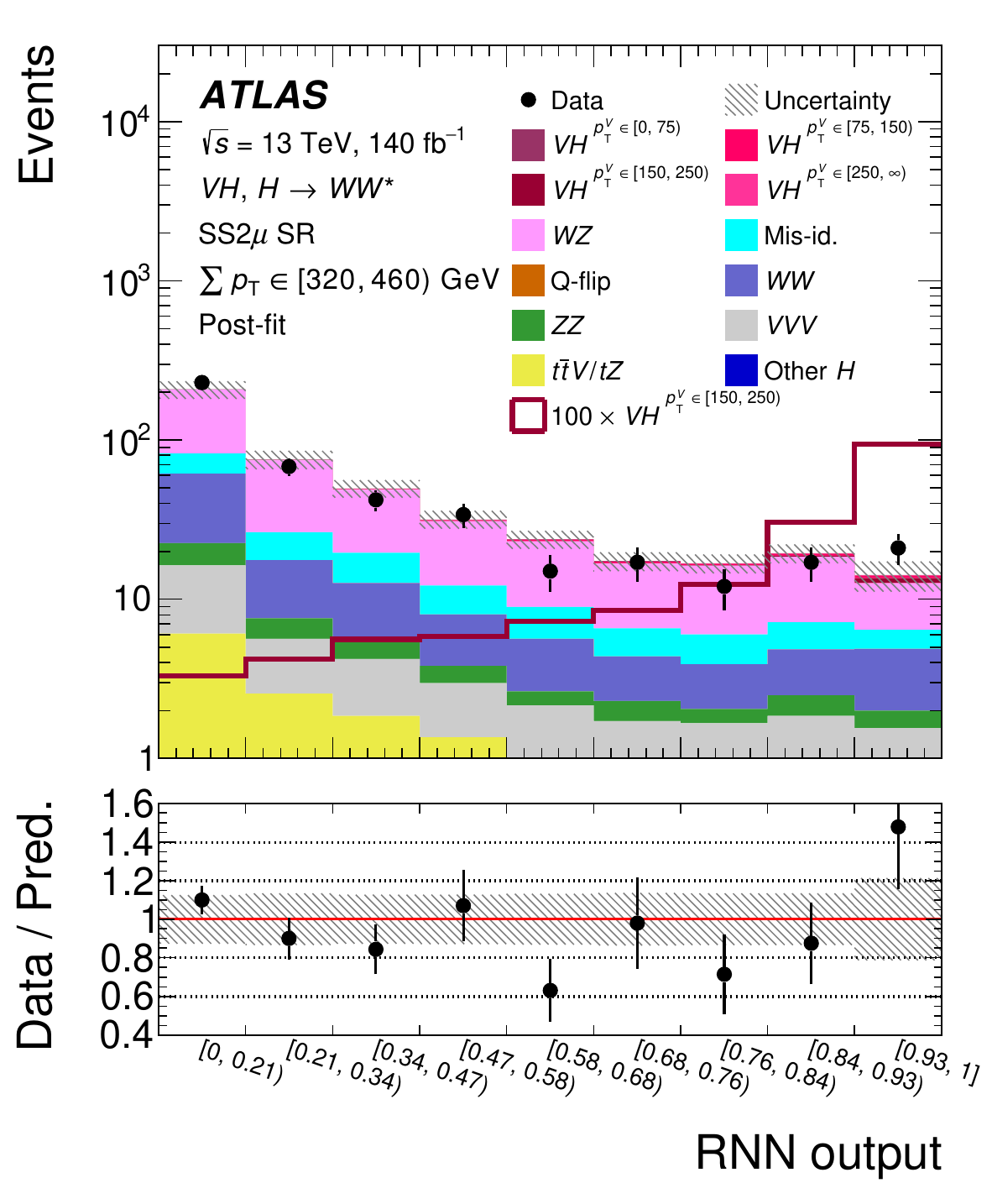}
} \\
\subfloat[]{
\includegraphics[width=0.45\textwidth]{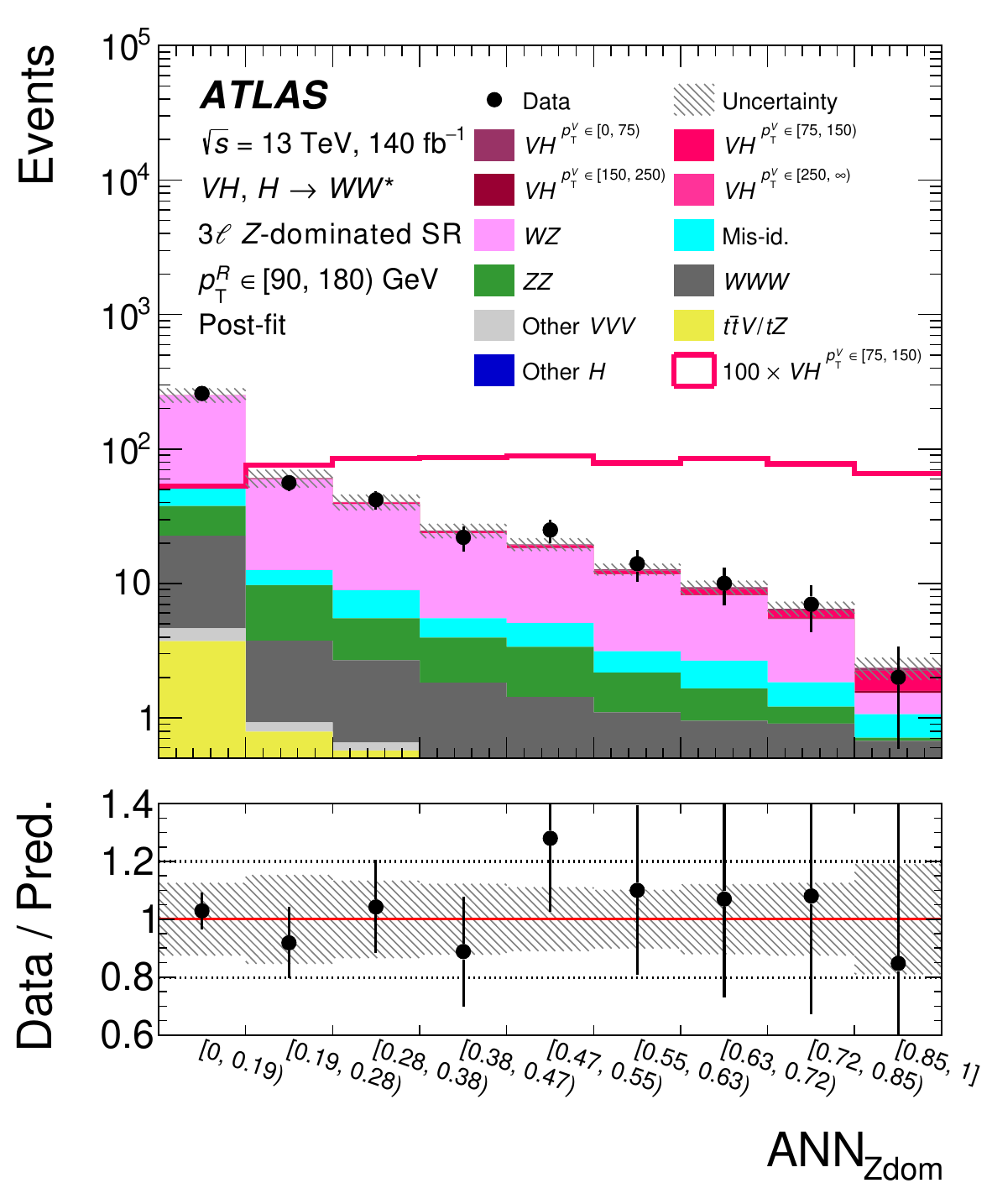}
}
\subfloat[]{
\includegraphics[width=0.45\textwidth]{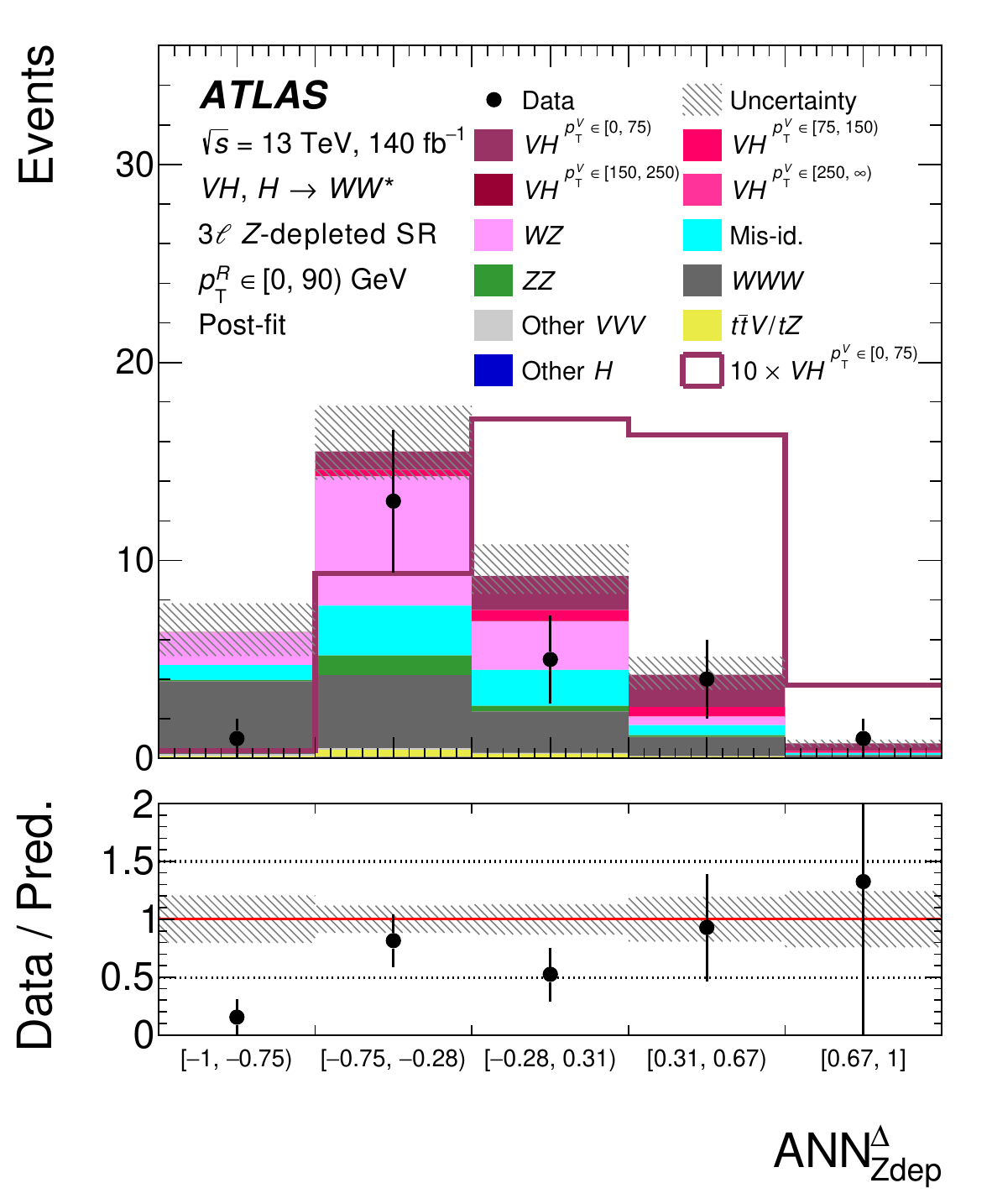}
} \\
\caption{Post-fit ANN/RNN discriminants in a subset of the SRs considered by the analysis: (a) opposite-sign $2\ell$ SR in the $\ptjj \geq \qty{260}{\GeV}$ subregion, (b) \SSMM SR in the $230 \leq \sumpt < \qty{460}{\GeV}$ subregion, (c) $3\ell$ \Z-dominated SR in the $90 \leq \ptR < \qty{180}{\GeV}$ subregion, and (d) $3\ell$ \Z-depleted SR in the $0 \leq \ptR < \qty{90}{\GeV}$ subregion. The lower panel shows the ratio of the data to the sum of the fitted signal and background. In each distribution, the post-fit \VH signal template targeted by that SR is overlaid and upscaled. The hatched band shows the total uncertainty. The post-fit results are obtained from the combined fit for the \ptV scheme, described in Section~\ref{sec:fit}.}
\label{fig:STXS_WH_postFit}
\end{figure}

\begin{figure}[htbp]
\centering
\subfloat[]{
\includegraphics[width=0.45\textwidth]{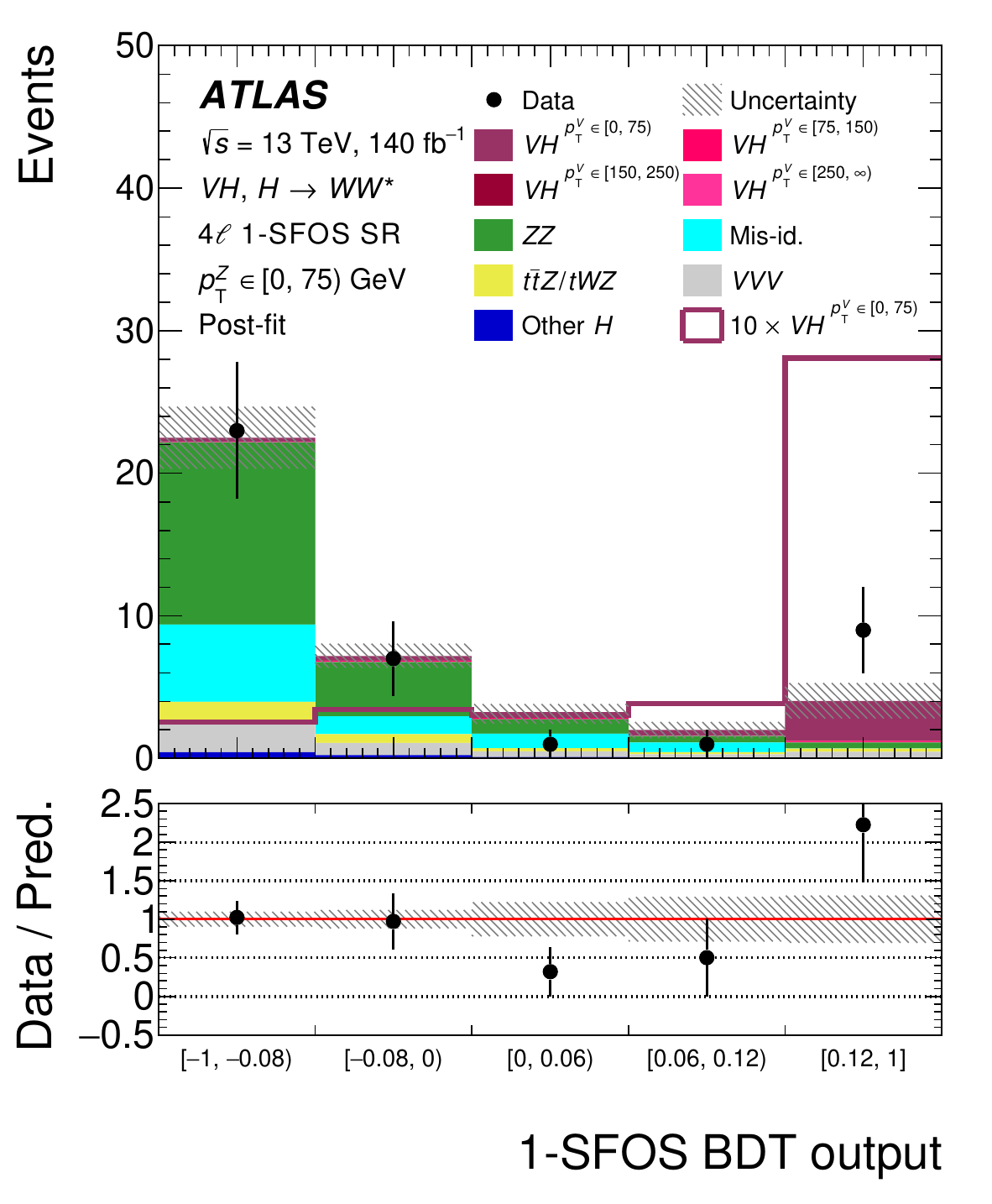}
}
\subfloat[]{
\includegraphics[width=0.45\textwidth]{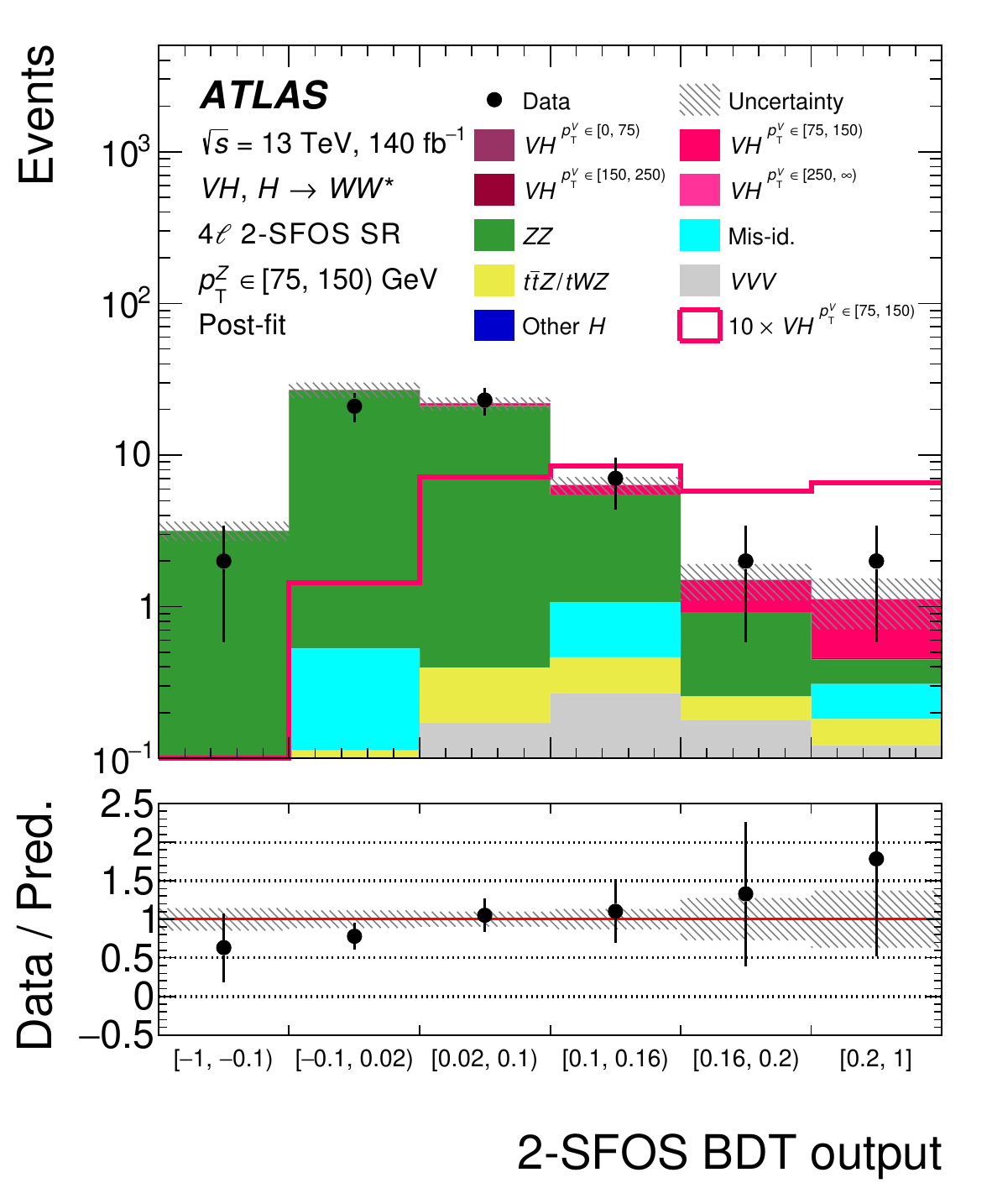}
}
\caption{Post-fit BDT discriminants in a subset of the SRs considered by the analysis: (a) $4\ell$ 1-SFOS SR in the $0 \leq \ptZ < \qty{75}{\GeV}$ subregion and (b) $4\ell$ 2-SFOS SR in the $75 \leq \ptZ < \qty{150}{\GeV}$ subregion. In each distribution, the post-fit \VH signal template targeted by that SR is overlaid and upscaled. The lower panel shows the ratio of the data to the sum of the fitted signal and background. The hatched band shows the total uncertainty. The post-fit results are obtained from the combined fit for the \ptV scheme, described in Section~\ref{sec:fit}.}
\label{fig:STXS_ZH_postFit}
\end{figure}

Figure~\ref{fig:STXSCat} shows the relative contributions of the reduced categories in all reconstructed SRs. In each case, the reduced categories provide the largest contributions in the corresponding SRs which aim to select them. Post-fit distributions for a subset of the SRs are shown in Figures~\ref{fig:STXS_WH_postFit} and \ref{fig:STXS_ZH_postFit} for the $2\ell$/$3\ell$ and $4\ell$ channels, respectively.

% End of text imported from the .//sections/STXS.tex input file

% End of text imported from the .//sections/analyses.tex input file

% The next lines are included from the .//sections/background.tex input file
%
\section{Background modelling}
\label{sec:background}

The background contamination in the SRs results from various physics processes, each modelled by pure data-driven prediction, pure MC prediction, or MC prediction normalised to data either using dedicated CRs or using only the SRs. The first method --- pure data-driven prediction --- is used to estimate the backgrounds with mis-identified leptons and electrons with mis-identified charge. In this estimate, rates and differential distributions (``shapes'') are extracted from data exploiting the method described in Section~\ref{sec:background_fake}. In the second method, the rates and shapes are extracted from simulation and normalised to the predicted cross-sections. In the third method, the rates are fit to data using the CRs --- which are orthogonal to the SRs --- in addition to the SRs themselves, while the shapes are extracted from simulation. Table~\ref{tab:bkg_summary} summarises the method adopted for each background process in each signal region. The same methods are adopted for the STXS measurement with no change of the CR definitions.

\begin{table}[htbp]
\centering
\caption{Summary of background modelling for each channel. ``Normalised in the fit'' are the background processes which are normalised in the fit but which do not have dedicated control regions; instead, their normalisation is determined by the signal discriminant. Data-driven background processes have at least one reconstructed mis-identified lepton.}
\label{tab:bkg_summary}
\renewcommand\arraystretch{1.25}
\resizebox{\textwidth}{!}{
\begin{tabular}{l|c|c|c|c}
\hline\hline
Channel & Pure MC & Normalised in the fit & Control region & Data-driven \\
\hline
Opposite-sign $2\ell$ & \WZg, \ZZg, \VVV, \H (non-\VH)                & --   & \ttbar/\Wt, \Zjets, \WW & \Wy, \Wjets \\
Same-sign $2\ell$     & \WW, \ZZg, \VVV, \ttV/\tZ                     & \WZg & --                      & \Vy, \Vjets \\
$3\ell$               & \ZZg, \VVV (non-\WWW), \ttV/\tZ, \H (non-\VH) & \WWW & \WZg                    & \Zy, \Zjets, \ttbar/\Wt, \WW \\
$4\ell$               & \VVV, \ttV/\tWZ, \H (non-\VH)                 & --   & \ZZg                    & \WZg, \ttW/\tZ, \Zjets, \ttbar/\Wt \\
\hline\hline
\end{tabular}
}
\end{table}

In the opposite-sign $2\ell$ channel, different CRs are defined to normalise the top, \Zjets, and \WW background processes. The \VH output node of the ANN is required to be less than 0.2 to define a region orthogonal to the SR, while dedicated selections on each of the ANN background nodes --- $\ANNDFOS^\textrm{top}$, $\ANNDFOS^{\Zjets}$, and $\ANNDFOS^{\WW}$ --- are used to define three CRs. The top CR is used to normalise both the \ttbar and \Wt background processes. The purities for the top, \Zjets, and \WW CRs are 72\%, 77\%, and 58\%, respectively. In this section, a CR purity is defined as the number of events targeted by that CR relative to the total number of events, each calculated within that CR.

In the same-sign $2\ell$ channel, the \WZg background in the SR is normalised using a free fit parameter. The \ttbar and \Zjets background processes contribute when the charge of one lepton is wrongly assigned. An electron with a hard bremsstrahlung or a mis-measured track curvature has a large charge mis-identification probability, which is measured in data using a control sample of electrons from \Z boson decays. In this sample, the \WZg background is obtained from MC and the mis-identified lepton background is obtained from the data-driven procedure described in Section~\ref{sec:background_fake}. The charge mis-identification probability is parametrised as a function of electron \pt with six bins and $|\eta|$ with six bins. The bins were chosen in accordance with the size of the event sample and the geometry of the detector. The charge mis-identification probability varies from $\mathcal{O}(10^{-5})$ for the electrons with $15 < \pt < \qty{60}{\GeV}$ and $|\eta| < 0.6$ to $\mathcal{O}(10^{-1})$ for electrons with $\pt > \qty{300}{\GeV}$ and $|\eta| > 2.3$. The measured charge mis-identification probabilities are applied to a sample of opposite-sign leptons that satisfies the requirements of same-sign $2\ell$ SRs, resulting in an estimate of the charge mis-identification background in each SR. The fraction of the charge mis-identification background is 26\% in the \SSEE SR and 1.6\% in the \SSDF SR. The muon charge mis-identification probability is negligible in the \pt range relevant to this analysis.

\begin{figure}[htbp]
\centering
\subfloat[]{
\includegraphics[width=0.48\textwidth]{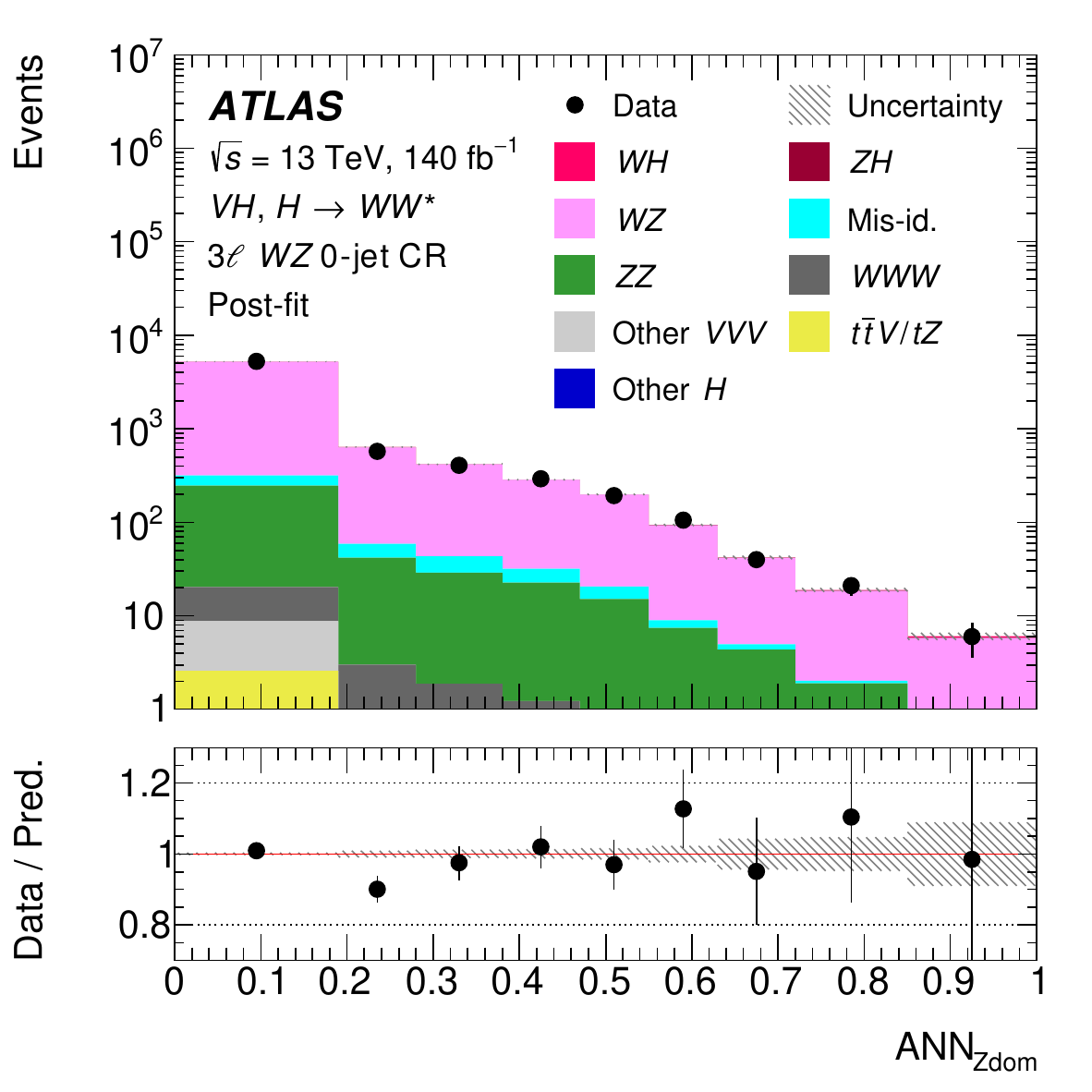}
}
\subfloat[]{
\includegraphics[width=0.48\textwidth]{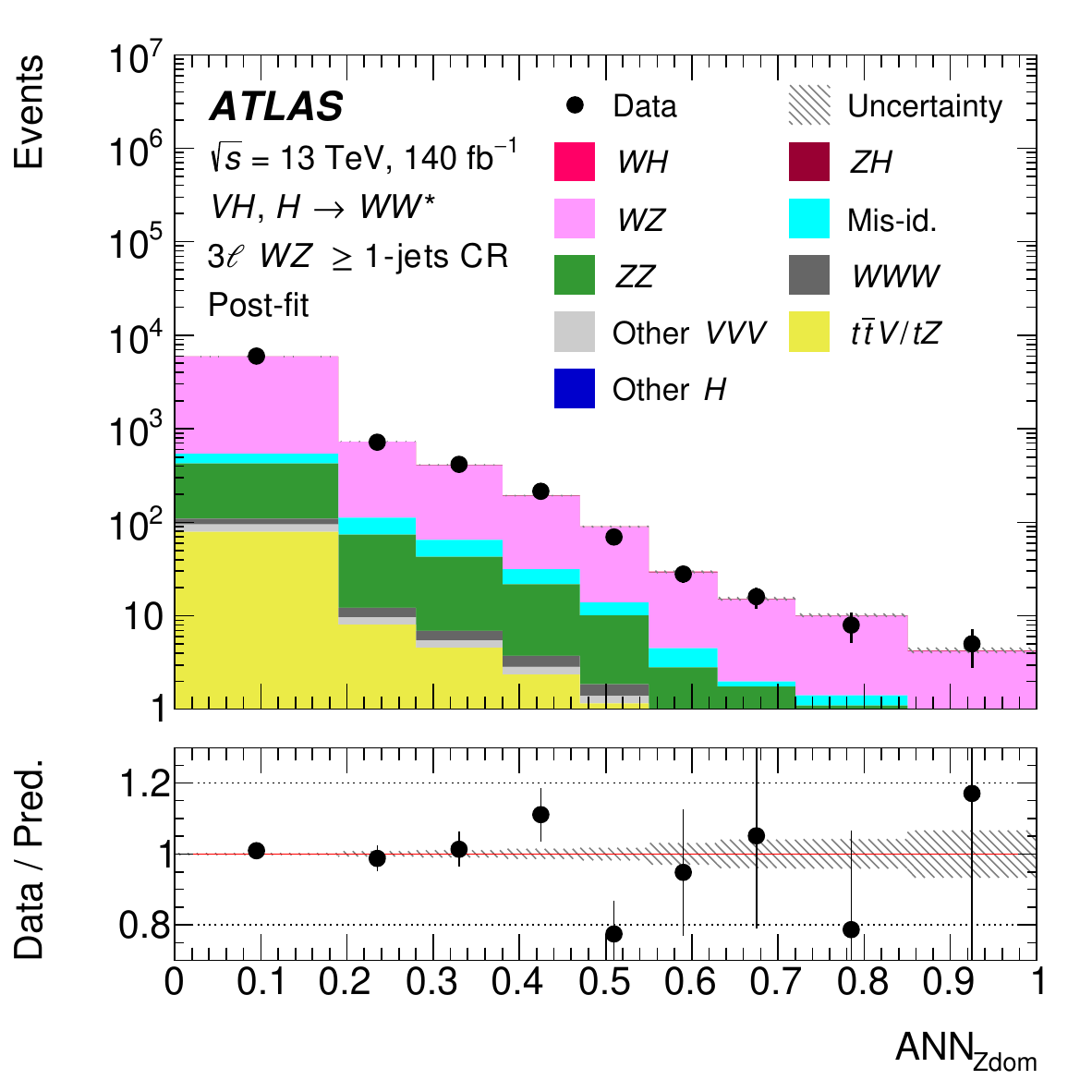}
} \\
\subfloat[]{
\includegraphics[width=0.48\textwidth]{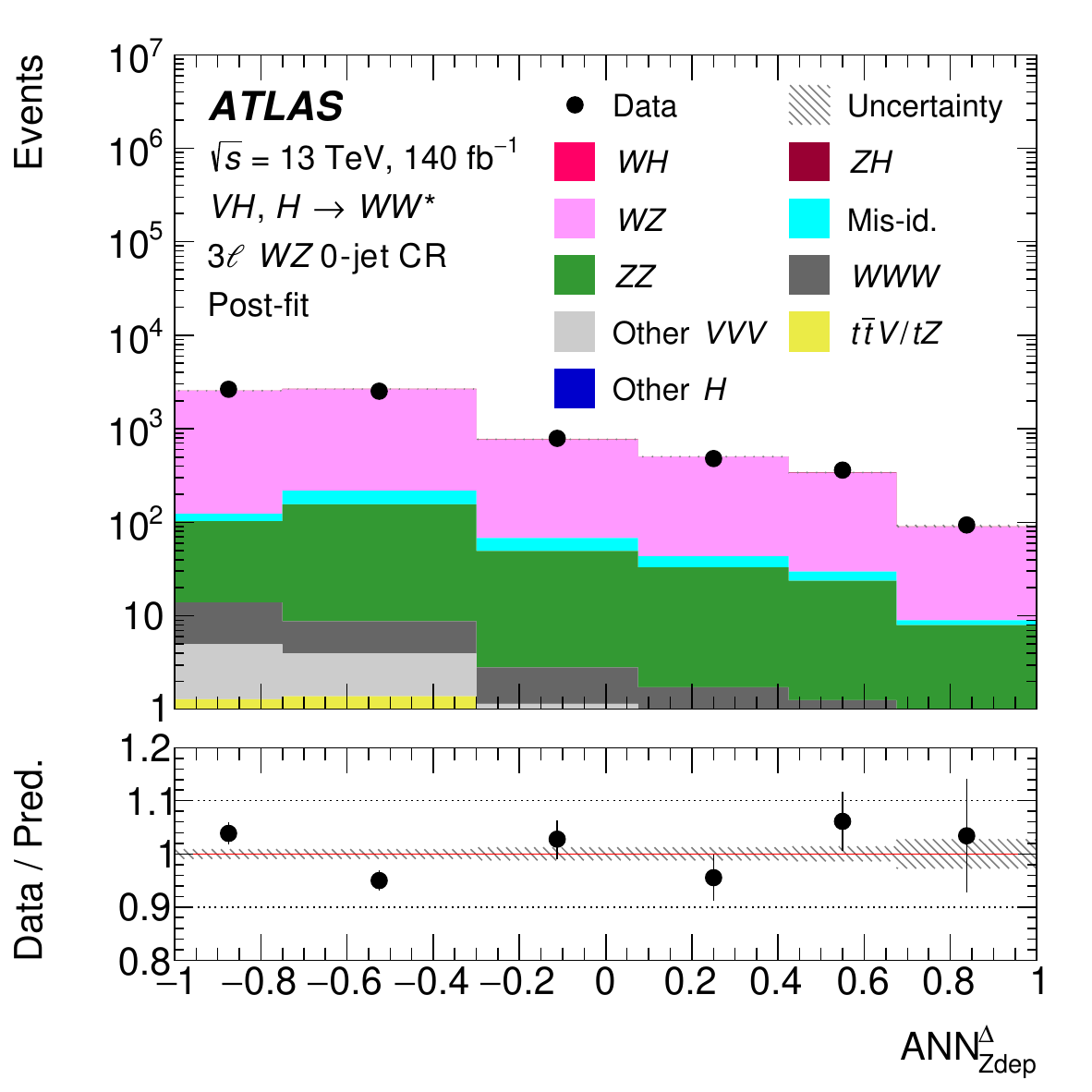}
}
\subfloat[]{
\includegraphics[width=0.48\textwidth]{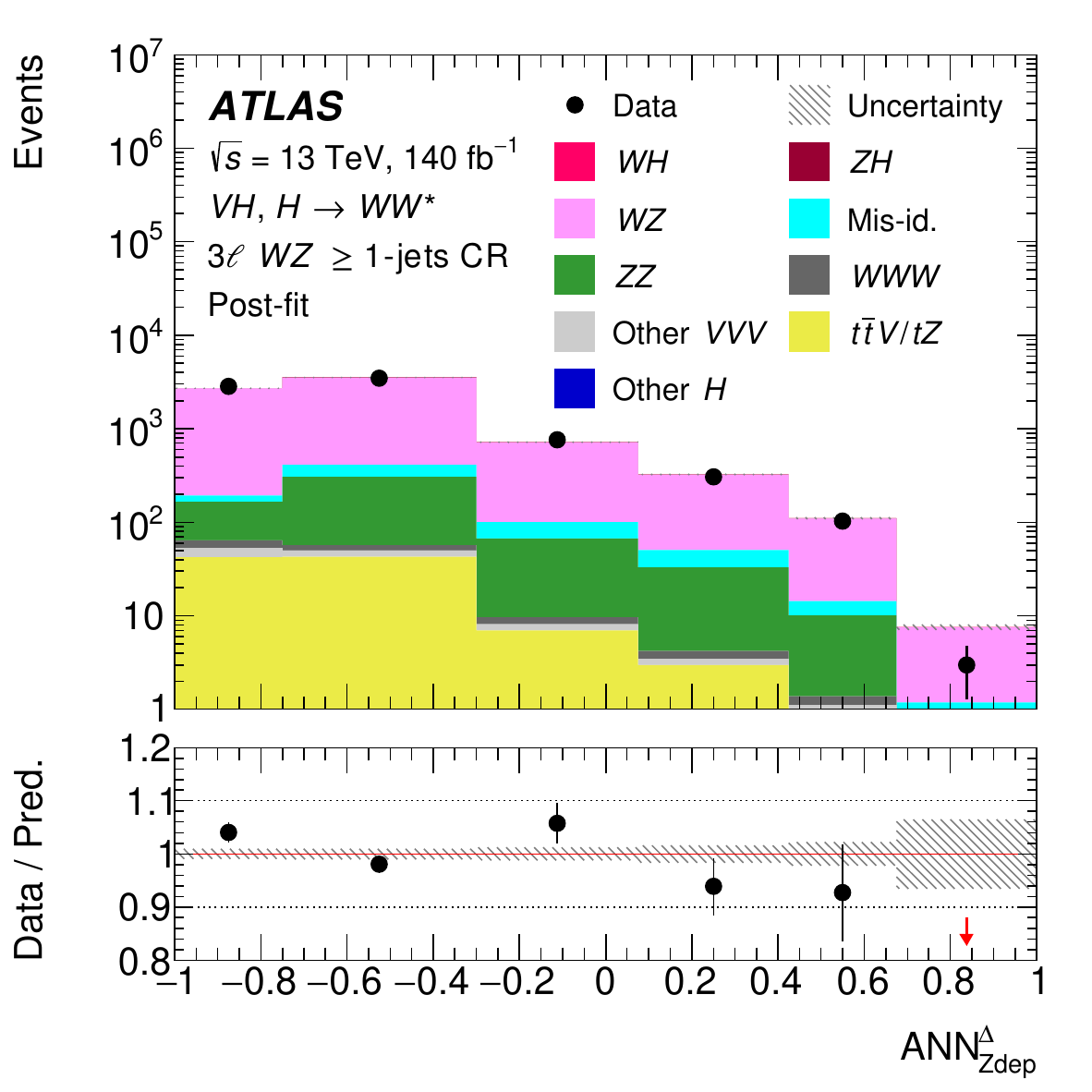}
} \\
\caption{Post-fit distributions of ANN$_\textrm{Zdom}$ in the (a) \WZg CR with no jets and (b) \WZg CR with at least one jet and of ANN$_\textrm{Zdep}^{\Delta}$ in the (c) \WZg CR with no jets and (d) \WZg CR with at least one jet. The binning of each distribution matches that used in the corresponding SR where that distribution is relevant. The lower panel shows the ratio of the data to the sum of the fitted signal and background, and the arrow in (d) indicates the position of a point outside the vertical axis range. The hatched band shows the total uncertainty. The post-fit results are obtained from the combined 2-POI fit described in Section~\ref{sec:fit}.}
\label{fig:dnnThreeLepton_CR}
\end{figure}

The quality of the \WZg modelling of the RNN for the same-sign $2\ell$ channel was evaluated in a \WZg-enriched validation region, defined using the selections common to each of the \SSMM, \SSEE, and \SSDF SRs but inclusive in the flavours of the leading isolated leptons. This validation region also requires an additional reconstructed lepton with $\pt > \qty{15}{\GeV}$ and meeting looser quality criteria than those applied to the two leading isolated leptons. These looser leptons are reconstructed imperfectly and are similar to those missed in \WZg events in which only two leptons are reconstructed. As a consequence, the RNN distribution in this region is sensitive to the quality of simulating events with real leptons from \WZg. This validation region is also depleted in signal, with relative signal fractions of $<1\%$ inclusively and $<2\%$ in the signal-sensitive bin. A shape mis-modelling of the RNN shape for \WZg events is observed, amounting to 30\% less data than predicted in the bins of higher value of the RNN output. The source of this mis-modelling was found to be particularly enhanced at small values of the $\Delta{R}$ distribution between a lepton and its nearest jet as well as at small values of the distribution of the minimum lepton-jet invariant mass considering all leptons and jets. As the signal events tend to be at low values of this variable, the high-score regions of the RNN also exhibit this mis-modelling. A dedicated correction to the RNN shape in the \WZg background is derived by fitting an exponential function to the ratio of data --- with the non-\WZg background subtracted --- to the \WZg background. The correction is then applied to the \WZg background in each of the \SSMM, \SSEE, and \SSDF signal regions, with a conservative 100\% systematic uncertainty assigned to the correction and decorrelated among the signal regions. This uncertainty reduces the expected sensitivity of the same-sign $2\ell$ channel by $\sim$20\% with respect to what is obtained neglecting it. Good modelling is observed for events with three reconstructed leptons that are required to meet stringent quality criteria --- as shown in Figure~\ref{fig:dnnThreeLepton_CR}. Consequently, no additional correction or uncertainty is needed for the $3\ell$ channels.

In the $3\ell$ channel, \WZg CRs are defined by requiring at least one SFOS lepton pair and by inverting the \Z-veto selection with respect to the \Z-dominated SR. The \WZg background normalisation depends on the jet multiplicity of the events, so the CR is split into two regions containing events with no reconstructed jets and events with at least one reconstructed jet, respectively. The purity of the regions without jets is 93\% and the purity of the region with one or more jets is 88\%.

Motivated by an excess in the measurement of the $\WWW + \WH$ rate reported in Ref.~\cite{STDM-2019-09}, the \WWW background is normalised in the fit procedure through a floating parameter and without a dedicated CR. The \WWW process is the largest background in the $3\ell$ \Z-depleted channel followed by \WZg production, and thus the value of the \WWW normalisation factor is entirely determined by this channel. The measured normalisation factor, $2.2^{+0.7}_{-0.6}$, is consistent with the measured signal strength, $1.61 \pm 0.25$, from the ATLAS \WWW analysis~\cite{STDM-2019-09}.

In the $4\ell$ channel, a dedicated CR is used to estimate the normalisation factor of the \ZZg process. It is defined in the 2-SFOS channel after the $b$-jet veto by adding a set of requirements on the invariant mass of each of the lepton pairs. To select on-shell \Z boson leptonic decays, $m_{\ell_2\ell_3}$ is required to be within $\pm$\qty{10}{\GeV} of the \Z boson mass. The invariant mass $m_{\ell_0\ell_1}$ must be above \qty{50}{\GeV}. The purity of this region is about 97\%.

\begin{figure}[htbp]
\centering
\subfloat[]{
\includegraphics[width=0.32\textwidth]{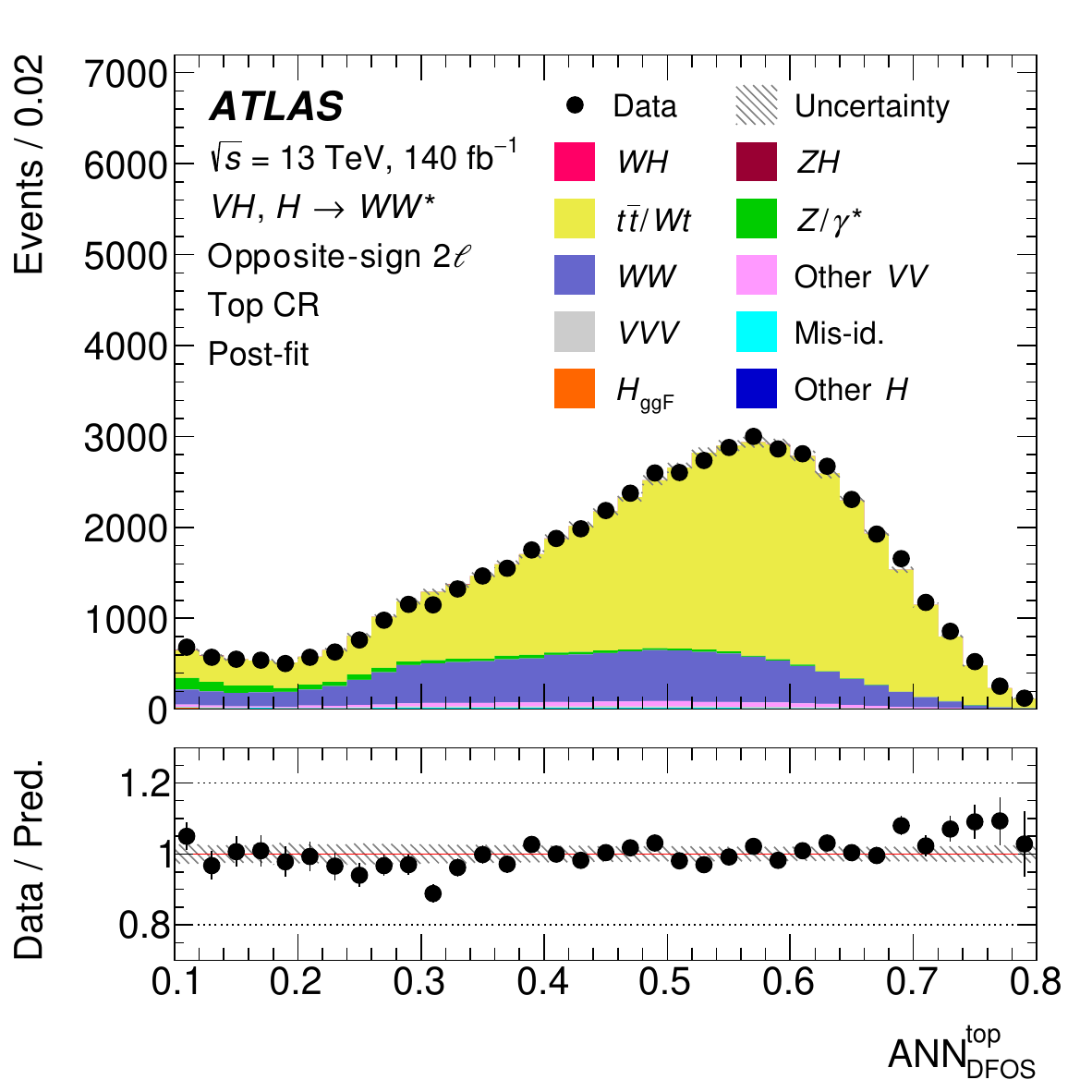}
}
\subfloat[]{
\includegraphics[width=0.32\textwidth]{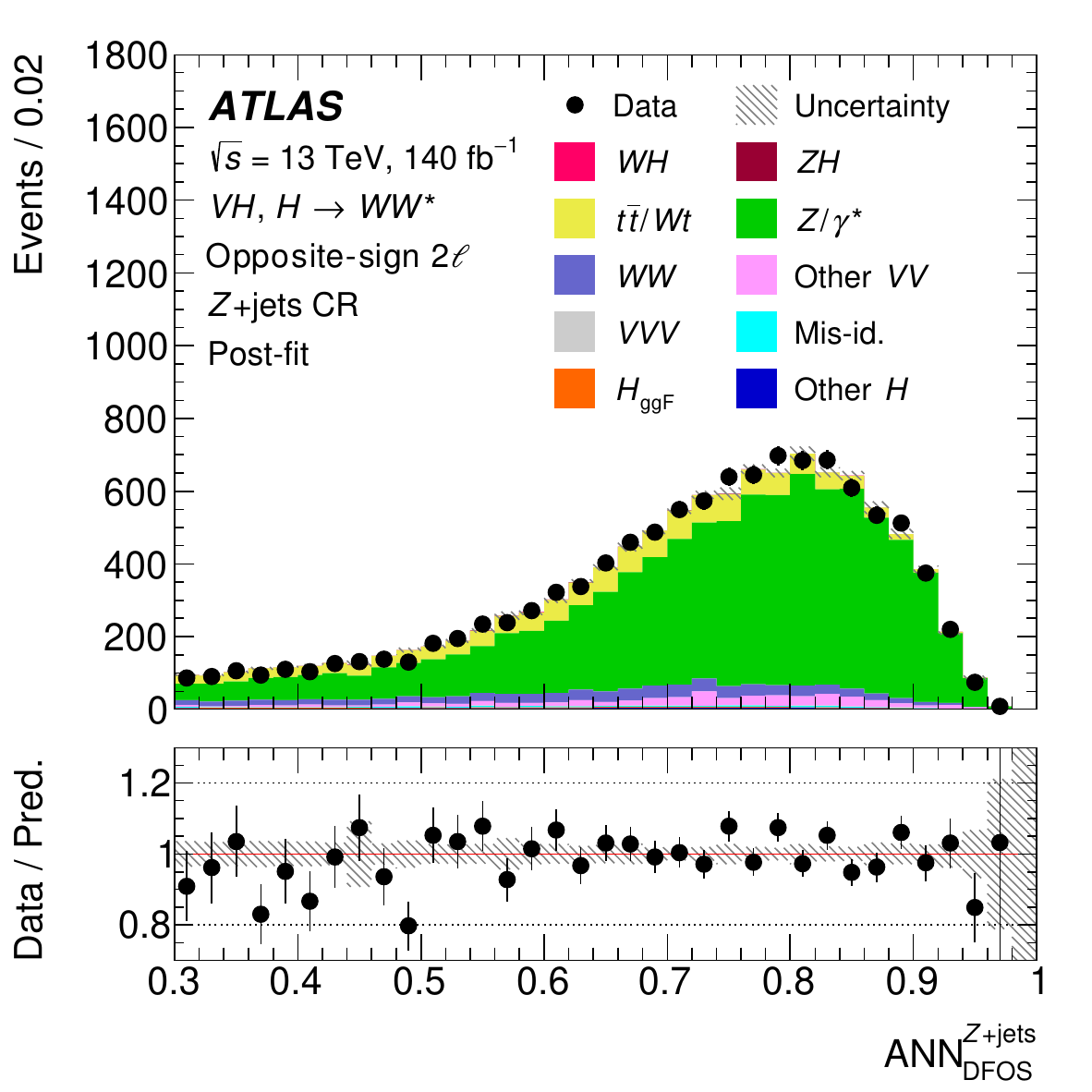}
}
\subfloat[]{
\includegraphics[width=0.32\textwidth]{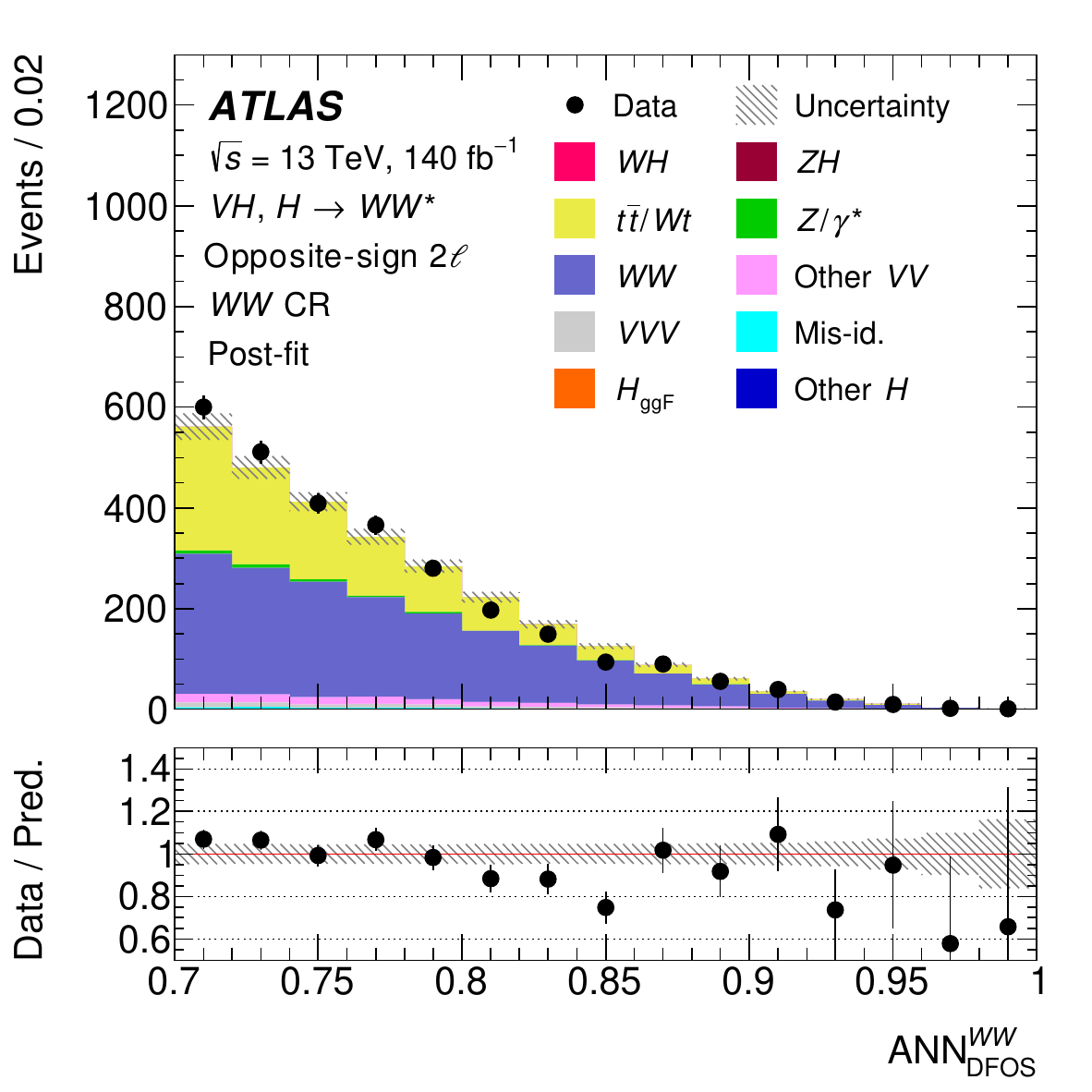}
}
\caption{Post-fit distributions of (a) $\ANNDFOS^\textrm{top}$ in the top CR, (b) $\ANNDFOS^{\Zjets}$ in the \Zjets CR, and (c) $\ANNDFOS^{WW}$ in the \WW CR of the opposite-sign $2\ell$ channel. The lower panel shows the ratio of the data to the sum of the fitted signal and background. The hatched band shows the total uncertainty. The last bin includes overflow, where applicable. The post-fit results are obtained from the combined 2-POI fit described in Section~\ref{sec:fit}.}
\label{fig:dnnTwoLeptonOS_CR}
\end{figure}

\begin{figure}[htbp]
\centering
\subfloat[]{
\includegraphics[width=0.48\textwidth]{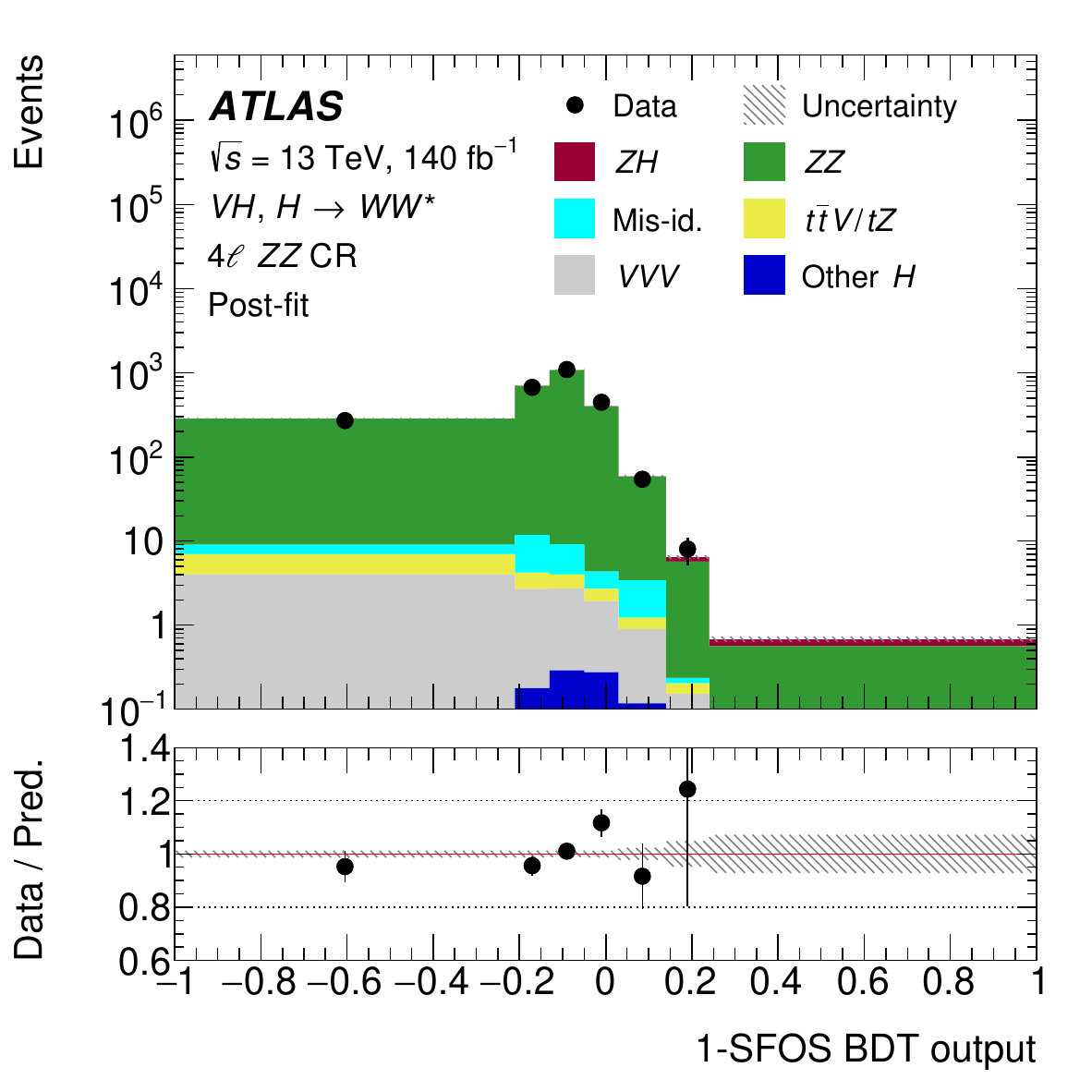}
}
\subfloat[]{
\includegraphics[width=0.48\textwidth]{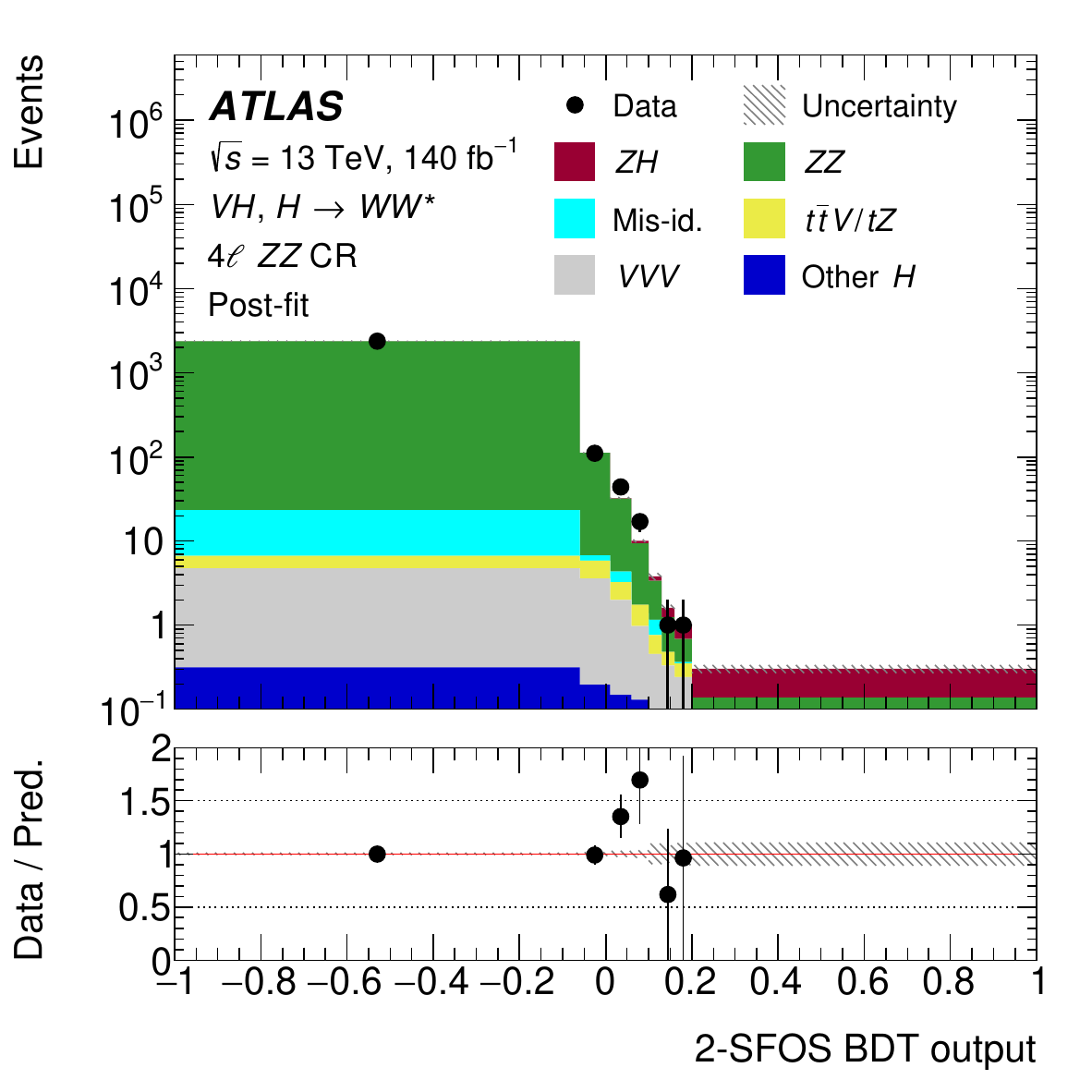}
}
\caption{Post-fit distributions of the (a) 1-SFOS BDT output and (b) 2-SFOS BDT output in the \ZZg CR. The binning of each distribution matches that used in the corresponding SR where that distribution is relevant. The lower panel shows the ratio of the data to the sum of the fitted signal and background. The hatched band shows the total uncertainty. The post-fit results are obtained from the combined 2-POI fit described in Section~\ref{sec:fit}.}
\label{fig:bdtFourLepton_CR}
\end{figure}

Figures~\ref{fig:dnnTwoLeptonOS_CR}, \ref{fig:dnnThreeLepton_CR}, and \ref{fig:bdtFourLepton_CR} show distributions of the multivariate discriminants, described in Section~\ref{sec:analyses}, in the relevant CRs. Good modelling is observed in the CRs for all of the distributions shown.

The background processes which are not normalised in the fit procedure or not measured via data-driven methods are estimated via pure MC prediction and normalised to their theoretical cross-sections.

% The next lines are included from the .//sections/background_fake.tex input file
%
\subsection{Backgrounds with non-prompt leptons}
\label{sec:background_fake}

Non-prompt electrons can originate from the decays of heavy-flavour hadrons, mis-identification of hadronic jets, and photon conversions in the detector material, with an admixture depending on the lepton quality requirements and event categories. Non-prompt muons only originate from decays of heavy-flavour hadrons. The physical processes which contribute non-prompt leptons in each channel are summarised in Table~\ref{tab:bkg_summary}.

All backgrounds with non-prompt leptons are estimated by weighting events from dedicated control samples with extrapolation factors. Except for requiring one of the lepton candidates to be an anti-identified (``anti-ID'') lepton, these control samples satisfy all of the nominal selections given in Table~\ref{tab:lepton_selections}. An anti-ID lepton fails to meet the full identification criteria used to select identified (``ID'') leptons but satisfies a looser set of criteria. The extrapolation factors are defined as the ratios between the numbers of events with non-prompt lepton candidates being ID and anti-ID. Using the control samples, a maximum-likelihood fit measures the extrapolation factors such that the total number of weighted events matches the number of observed data events with only ID leptons after subtracting the expected contribution from processes with only prompt leptons.

The non-prompt electron background in $2\ell$ and $3\ell$ channels is estimated from simulation with data-driven corrections. The corresponding extrapolation factors are measured separately for three sources --- heavy-flavour jets, light-flavour jets, and $\gamma$ conversions --- using three sets of samples enriched in \ttbar, \Zjets, and \Zy events, respectively. The simulated events with a non-prompt anti-ID electron are counted according to the source of the non-prompt electron and weighted by  extrapolation factors. This yields three sets of extrapolation factors which are simultaneously measured using a maximum-likelihood fit. In contrast, the non-prompt muon background in $2\ell$ and $3\ell$ channels is estimated by using a fully data-driven technique. From a sample enriched in \ttbar, the number of events with a non-prompt muon is calculated from the observed data after subtracting the expected contribution from processes with only prompt muons. The corresponding extrapolation factors are then obtained from a maximum-likelihood fit. The non-prompt electron and muon extrapolation factors are both measured in bins of lepton \pt.

The measurement of the non-prompt electron and muon extrapolation factors for the $4\ell$ channel follows a similar procedure to the measurement of the non-prompt muon extrapolation factors for the $2\ell$ and $3\ell$ channels. The differences between the $2\ell$ and $3\ell$ channels are due to the looser lepton selection and the lower sample size available in the control samples. The extrapolation factors are measured in a sample enriched in \Zjets events and are inclusive in lepton \pt. Processes with two non-prompt leptons are accounted for in the extrapolation by applying a correction term evaluated in a sample where two of the lepton candidates are anti-ID.

% End of text imported from the .//sections/background_fake.tex input file

% End of text imported from the .//sections/background.tex input file

% The next lines are included from the .//sections/systematics.tex input file
%
\section{Systematic uncertainties}
\label{sec:systematics}

The experimental and theoretical sources of systematic uncertainties are evaluated using MC samples and their effects are included in the statistical analysis. The impact of each uncertainty is estimated per analysis region and per bin. A list of the systematic uncertainty sources with corresponding impact on the measurement is shown in Section~\ref{sec:results}.

The theoretical and experimental uncertainties are varied in a correlated way for all MC processes across all signal and control region bins, which considers the normalisation extrapolation from control to signal regions.

\subsection{Experimental uncertainties}

Experimental uncertainties associated with leptons originate from the reconstruction, identification, and isolation efficiencies~\cite{EGAM-2021-01, MUON-2018-03} and from the scale and resolution of the energy or momentum~\cite{EGAM-2021-02, MUON-2022-01}. For jets, uncertainties arise from the jet energy scale and resolution~\cite{JETM-2018-05}, the performance of the pile-up jet tagger~\cite{PERF-2014-03, ATL-PHYS-PUB-2019-026}, and the $b$-jet identification~\cite{FTAG-2018-01}. Furthermore, uncertainties due to the trigger selection~\cite{TRIG-2018-05, TRIG-2018-01} and the soft term in the reconstruction of the \met~\cite{JETM-2020-03} are estimated. The uncertainty in the modelling of pile-up for simulated samples is estimated by varying the reweighting to the profile in data within its uncertainties. The uncertainty in the integrated luminosity is 0.83\%~\cite{DAPR-2021-01}, which is measured using the LUCID-2 detector~\cite{LUCID2}. The luminosity uncertainty is applied to the signal and background processes which are normalised to theoretical predictions.

For the data-driven estimate of the backgrounds attributed to mis-identified leptons and electrons with mis-identified charge, uncertainties are considered for the availability of data and MC statistics, variations on the theoretical predictions of the prompt lepton backgrounds, variations in the analysis selections, and the self-consistency of the method. The latter source of uncertainty is estimated purely using MC by comparing the mis-identified lepton estimate obtained from MC to that obtained by applying the data-driven method to the MC sample, using extrapolation factors evaluated from MC.

The largest sources of experimental uncertainties are the following: jet energy scale uncertainties~(9.2\%)\footnote{The numbers quoted for the respective channels here and at the end of Section~\ref{sec:systematics_theory} correspond to the average relative contribution of that source of uncertainty to the total uncertainty on the \VH/\WH/\ZH cross-section times the \HWW branching ratio as measured individually by that channel.} for the opposite-sign $2\ell$ channel; uncertainties relating to the RNN shape correction for \WZg~(200\%), the estimate of the mis-identified lepton background~(150\%), and the estimate of the electron background with mis-identified charge~(56\%) for the same-sign $2\ell$ channel; and uncertainties relating to the estimate of the mis-identified lepton background~(11\% and 2.9\%) and the isolation of leptons~(3.0\% and 3.9\%) for the $3\ell$ and $4\ell$ channels.

% The next lines are included from the .//sections/systematics_theory.tex input file
%
\subsection{Theoretical uncertainties}
\label{sec:systematics_theory}

The impact of the theoretical sources of uncertainty is evaluated by reweighting MC events or by using alternative MC samples, which are detailed in Table~\ref{tab:mcsamples}. Uncertainties computed as differences between two samples are symmetrised. Uncertainties in background processes with negligible contributions in the signal regions are not considered. All uncertainties are computed using detector-level events unless specified otherwise.

For all processes, the uncertainty in missing higher-order corrections is computed as the maximum variation of the envelope resulting from simultaneous variations of the renormalisation and factorisation scales by factors of 0.5 and 2. The uncertainty in the central value of the PDF set is evaluated by comparing the nominal weight to alternative weights, the latter encapsulating the experimental and model-related uncertainties entering the PDF fit. The uncertainty in the central value of the strong coupling constant, $\alpha_s = 0.1180 \pm 0.0015$, is evaluated by varying it up and down by its uncertainty and recalculating the event weight for each case. The midpoint between these varied weights is symmetrised and assigned as the corresponding uncertainty in the nominal weight.

The uncertainty in the PS modelling for the Higgs boson processes is assigned as the difference between the nominal sample showered with \PYTHIA[8] and an alternative sample showered with \HERWIG[7]. For the \VH signal process, this uncertainty is computed using particle-level events with selections similar to those on detector-level and the same multivariate discriminants.

The merging of the NLO and LO matrix elements for the \VV processes, which are modelled by \SHERPA[2.2.2], is performed using the \MEPSatNLO prescription~\cite{Hoeche:2012yf}. A jet merging uncertainty (``CKKW'') is assigned by varying the nominal threshold (\qty{20}{\GeV}) separating the ME and the PS up (\qty{30}{\GeV}) and down (\qty{15}{\GeV}). Similarly, a PS resummation uncertainty (``QSF'') is assigned by varying the resummation scale up and down by factors of 2 and 0.5, respectively. The uncertainty due to the choice of the PS momentum recoil scheme (``CSSKIN'') is computed by comparing the nominal sample, which uses the recoil scheme described in Ref.~\cite{Schumann:2007mg}, to an alternative sample, which uses the recoil scheme described in Ref.~\cite{Hoeche:2009xc}. In a manner analogous to the PS modelling uncertainty in \VH, the CKKW, QSF, and CSSKIN uncertainties in \VV are all computed using particle-level events.

The uncertainty in the ME and PS modelling for the \VVV and \Zjets processes are assigned as the difference between the nominal \SHERPA samples and alternative samples generated with \MGNLO and showered with \PYTHIA[8].

The uncertainty in the hadronisation and fragmentation modelling for the \ttbar sample is assigned as the difference between the nominal sample showered with \PYTHIA[8] and an alternative sample showered with \HERWIG[7]. The uncertainty in the choice of generator and matching algorithm is assigned as the difference between the nominal sample generated with \POWHEG and an alternative sample generated with \MGNLO. Uncertainties in the modelling of initial- and final-state radiation are assigned by varying scale, resummation, and showering parameters.

In the total cross-section measurement, signal modelling uncertainties are computed separately for the \WH and \ZH processes and decorrelated in the fit. The MC weights are varied using the inclusive samples and the uncertainties are computed in each signal and control region. In the differential cross-section measurement, signal modelling uncertainties are computed for each fiducial bin of the corresponding scheme, using the binning defined in the STXS Stage~1.2 convention and without applying any merging. For the missing higher-order QCD corrections to $V(\to\ell\nu/\ell\ell)H$, the uncertainty model described in Ref.~\cite{ATL-PHYS-PUB-2018-035} is used, resulting in a set of uncertainties in the total cross-section correlated across fiducial bins and a set of migration uncertainties across the fiducial bin boundaries. Seven independent QCD uncertainty components are considered. One component affects the total production cross-section, while the other six are pure migration uncertainties. Four of these components are designed to model uncertainties across \ptV boundaries (75, 150, 250, and \qty{400}{\GeV}) while the other two affect jet bin migrations across the $n_\textrm{jets} = 1$ and 2 boundaries. A similar set of uncertainties are considered for $V(\to\qq)H$: one affecting the total production cross-section and nine migration uncertainties relating to the fiducial selection criteria on the transverse momentum of the Higgs boson, the transverse momentum of the Higgs boson and leading dijet system, the invariant mass of the leading dijet system, and the number of jets. The uncertainties affect both the analysis acceptance, defined as the fraction of events in a fiducial region passing the analysis selection, and the predicted cross-section per fiducial region.

The largest sources of theoretical uncertainties are the following: uncertainties in the hadronisation/fragmentation modelling of \ttbar production~(19\%) and on the ME/PS modelling of \Zjets production~(17\%) for the opposite-sign $2\ell$ channel; uncertainties in the choice of PS recoil scheme for \WZg production~(99\%) for the same-sign $2\ell$ channel; uncertainties in the choice of PS recoil scheme for \WZg production~(9.7\%) and on the ME/PS modelling of \WWW production~(9.4\%) for the $3\ell$ channel; and uncertainties in the choice of PS algorithm for \ZH production~(2.1\%) for the $4\ell$ channel.

% End of text imported from the .//sections/systematics_theory.tex input file

% End of text imported from the .//sections/systematics.tex input file

% The next lines are included from the .//sections/results.tex input file
%
\section{Fit procedure and results}
\label{sec:fit_and_results}

\subsection{Fit procedure}
\label{sec:fit}

A binned likelihood function is constructed as a product of Poisson probability terms over the bins of the different SRs defined in Section~\ref{sec:analyses}. The binned likelihood function is parameterised in terms of a signal strength, $\mu$, defined as the ratio of the observed signal yield to that predicted by the SM. The signal strength constitutes the POI of the measurement. Additionally, a Poisson probability term is added for each CR and used to fit the normalisation of its corresponding background process in the combined measurement via a floating normalisation factor. Systematic uncertainties enter as nuisance parameters in the likelihood function --- primarily in the form of Gaussian distributed constraints --- and their correlations are taken into account. The final results are obtained using the profile likelihood method~\cite{Cowan:2010js}.

For the measurement of \WH and \ZH production, two fit scenarios are considered: a combined 1-POI fit, where the \WH and \ZH yields are simultaneously scaled by a single POI, \muVH, and a combined 2-POI fit, where the \WH and \ZH yields are independently scaled by two POIs, \muWH and \muZH, respectively. Table~\ref{tab:fit_summary} summarises the regions entering the combined fits. In the case of the combined 1-POI fit, the SM expectation is assumed for the cross-section ratio of the \WH and \ZH production processes; for both combined fits, the cross-sections for non-\VH Higgs boson processes are fixed to their SM expectations.

A similar prescription is followed for the differential measurements, where the SRs in Table~\ref{tab:fit_summary} are split according to Table~\ref{tab:STXS_SRs} and the four (\ptV scheme) or seven (STXS scheme) fiducial cross-sections of interest are independently scaled by an equal number of POIs.

For both the total and differential cross-section measurements, the pure normalisation components of the signal theory uncertainties are factorised from the pure acceptance components --- only the latter components enter the cross-section fits. Thus, the effect of theoretical uncertainties on the \VH signal in a cross-section measurement should always be as small or smaller that in the corresponding signal strength measurement. For the cross-section measurement, the POIs are written as cross-sections times the \HWW branching ratio, $\XS\times\BR_{\HWW}$, or those same quantities relative their SM expectations, $(\XS\times\BR_{\HWW})/(\XS\times\BR_{\HWW})_\textrm{SM}$, to differentiate them from the POIs of the signal strength measurements, $\mu$.

\begin{table}[htbp]
\centering
\caption{Summary of SRs --- including the chosen SR discriminants --- and CRs in each channel entering the combined total cross-section fit. The SR discriminants enter the combined fit as histograms, and the corresponding figures showing the post-fit histograms are referred in parentheses after each SR discriminant; the CRs enter the combined fit as counters.}
\label{tab:fit_summary}
\renewcommand\arraystretch{1.25}
\begin{tabular}{l|l|l|l}
\hline\hline
Channel & SR & SR discriminant & Relevant CR(s) \\
\hline
Opposite-sign $2\ell$ & -- & $\ANNDFOS^{\VH}$ (\ref{fig:postfit_MVA_2los_fit}) & Top, \Zjets, \WW \\
\hline
\multirow{3}{*}{Same-sign $2\ell$} & \SSMM & RNN output (\ref{fig:postfit_MVA_2lss_MM}) & \multirow{3}{*}{--} \\
& \SSEE & RNN output (\ref{fig:postfit_MVA_2lss_EE}) & \\
& \SSDF & RNN output (\ref{fig:postfit_MVA_2lss_DF}) & \\
\hline
\multirow{2}{*}{$3\ell$} & \Z-dominated & \ANNZdom (\ref{fig:postfit_MVA_3l_Zdom}) & \multirow{2}{*}{\WZ 0-jet, \WZ $\geq$1-jets} \\
& \Z-depleted & $\ANNZdep^\Delta$ (\ref{fig:postfit_MVA_3l_Zdep}) & \\
\hline
\multirow{2}{*}{$4\ell$} & 1-SFOS & 1-SFOS BDT output (\ref{fig:postfit_MVA_4l_1SFOS}) & \multirow{2}{*}{\ZZg} \\
& 2-SFOS & 2-SFOS BDT output (\ref{fig:postfit_MVA_4l_2SFOS}) & \\
\hline\hline
\end{tabular}
\end{table}

\subsection{Results}
\label{sec:results}

\subsubsection{Total cross-section results}
\label{sec:results_total_xs}

The post-fit MC and data yields in each SR are shown in Table~\ref{tab:postfit_yields}. The observed data yields agree, in both the rate and shape, within uncertainties with the expected yields from MC in all SRs.

% The next lines are included from the .//sections/yields_table.tex input file
\begin{table}[htbp]
\centering
\caption{Post-fit signal, background, and observed data yields in each SR as measured by the 2-POI fit. The \Zjets and top processes for a given channel correspond to those with only prompt leptons, as described in Section~\ref{sec:background}. The uncertainties correspond to the total of all statistical and systematic sources. The quadrature sum of the individual sources may differ from the total uncertainty due to correlations.}
\label{tab:postfit_yields}
\resizebox{\textwidth}{!}{
\begin{tabular}{l|S[table-format=3.3,table-column-width=1.4cm]@{$\,\pm\,$}S[table-format=3.3,table-column-width=1.2cm] S[table-format=3.3,table-column-width=1.4cm]@{$\,\pm\,$}S[table-format=3.3,table-column-width=1.2cm] S[table-format=3.3,table-column-width=1.4cm]@{$\,\pm\,$}S[table-format=3.3,table-column-width=1.2cm] S[table-format=3.3,table-column-width=1.4cm]@{$\,\pm\,$}S[table-format=3.3,table-column-width=1.2cm]}
\hline\hline
& \multicolumn{2}{c}{OS $2\ell$} & \multicolumn{6}{c}{SS $2\ell$}  \\
Process & \multicolumn{2}{c}{\phantom{--}} & \multicolumn{2}{c}{\SSMM} & \multicolumn{2}{c}{\SSEE} & \multicolumn{2}{c}{\SSDF} \\
\hline
\WH & 11 & 7 & 33 & 21 & 12 & 8 & 48 & 31 \\
\ZH & 20 & 6 & 7.2 & 2.2 & 4.1 & 1.2 & 14 & 4 \\
\hline
Other Higgs & 61 & 14 & 19.0 & 2.2 & 9.3 & 1.1 & 33 & 4 \\
\WW & 260 & 50 & 260 & 40 & 109 & 15 & 380 & 50 \\
\WZg & 41.5 & 3.1 & 1420 & 150 & 850 & 90 & 2630 & 270 \\
\ZZg & 7.8 & 0.7 & 144 & 17 & 101 & 8 & 309 & 28 \\
\Zy & 10 & 7 & \multicolumn{2}{c}{----} & \multicolumn{2}{c}{----} & \multicolumn{2}{c}{----} \\
\VVV & 19 & 6 & 190 & 50 & 86 & 24 & 290 & 80 \\
\Zjets & 302 & 27 & \multicolumn{2}{c}{----} & \multicolumn{2}{c}{----} & \multicolumn{2}{c}{----} \\
Top & 1020 & 70 & 33.0 & 2.7 & 16.2 & 1.3 & 54 & 5 \\
Mis-identified leptons & 25 & 6 & 320 & 130 & 1090 & 100 & 2080 & 260 \\
Charge-flip electrons & \multicolumn{2}{c}{----} & \multicolumn{2}{c}{----} & 930 & 70 & 109 & 8 \\
\hline
Total & 1780 & 40 & 2420 & 50 & 3200 & 50 & 5950 & 70 \\
Observed & \multicolumn{2}{c}{1788} & \multicolumn{2}{c}{2438} & \multicolumn{2}{c}{3233} & \multicolumn{2}{c}{5906} \\
\hline\hline
& \multicolumn{4}{c}{$3\ell$} & \multicolumn{4}{c}{$4\ell$} \\
Process & \multicolumn{2}{c}{\Z-dominated} & \multicolumn{2}{c}{\Z-depleted} & \multicolumn{2}{c}{1-SFOS} & \multicolumn{2}{c}{2-SFOS} \\
\hline
\WH & 12 & 7 & 6 & 4 & \multicolumn{2}{c}{----} & \multicolumn{2}{c}{----} \\
\ZH & 4.4 & 1.3 & 1.7 & 0.5 & 15 & 4 & 10.7 & 3.2 \\
Other Higgs & 1.80 & 0.17 & 2.83 & 0.31 & 1.88 & 0.22 & 0.68 & 0.08 \\
\WZg & 877 & 20 & 24.6 & 1.6 & \multicolumn{2}{c}{----} & \multicolumn{2}{c}{----} \\
\ZZg & 130 & 17 & 2.63 & 0.28 & 29.2 & 1.5 & 306 & 9 \\
\WWW & 64 & 16 & 35 & 9 & \multicolumn{2}{c}{----} & \multicolumn{2}{c}{----} \\
\WWZ & 3.81 & 0.09 & 1.64 & 0.05 & 10.9 & 1.0 & 1.58 & 0.14 \\
\WZZ & 0.310 & 0.008 & \multicolumn{2}{c}{$<0.1$} & 0.447 & 0.034 & \multicolumn{2}{c}{$<0.1$} \\
Top & 14.5 & 1.3 & 4.8 & 0.4 & 8.8 & 0.9 & 1.45 & 0.14 \\
Mis-identified leptons & 98 & 14 & 12.5 & 2.9 & 14.5 & 3.2 & 5.5 & 1.2 \\
\hline
Total & 1205 & 23 & 92 & 8 & 80 & 5 & 326 & 9 \\
Observed & \multicolumn{2}{c}{1237} & \multicolumn{2}{c}{88} & \multicolumn{2}{c}{79} & \multicolumn{2}{c}{316} \\
\hline\hline
\end{tabular}
}
\end{table}

% End of text imported from the .//sections/yields_table.tex input file

\begin{figure}[htbp]
\centering
\includegraphics[width=0.65\textwidth]{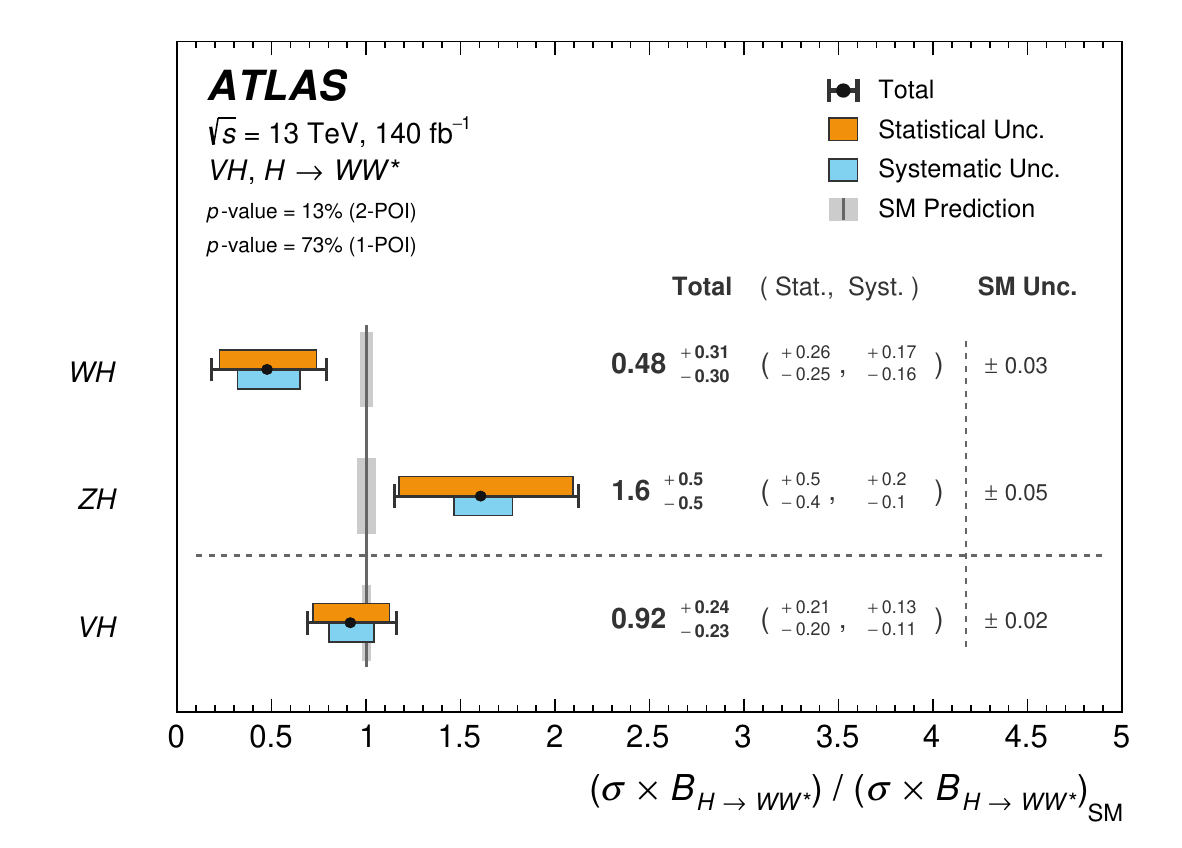}
\caption{Best-fit values of the total \WH, \ZH, and \VH cross-sections times the \HWW branching ratio. The \WH/\ZH and \VH results are obtained from the combined 2- and 1-POI cross-section fits, respectively. Each measurement is normalised to its SM prediction. The black error bars, orange boxes, and blue boxes show the total, statistical, and systematic uncertainties in the measurements, respectively. The grey bands represent the theory uncertainty of the corresponding Higgs boson production mode, calculated from the sources of uncertainty described in Section~\ref{sec:systematics_theory}.}
\label{fig:POIPlot}
\end{figure}

\begin{figure}[htbp]
\centering
\subfloat[]{
\includegraphics[width=0.48\textwidth]{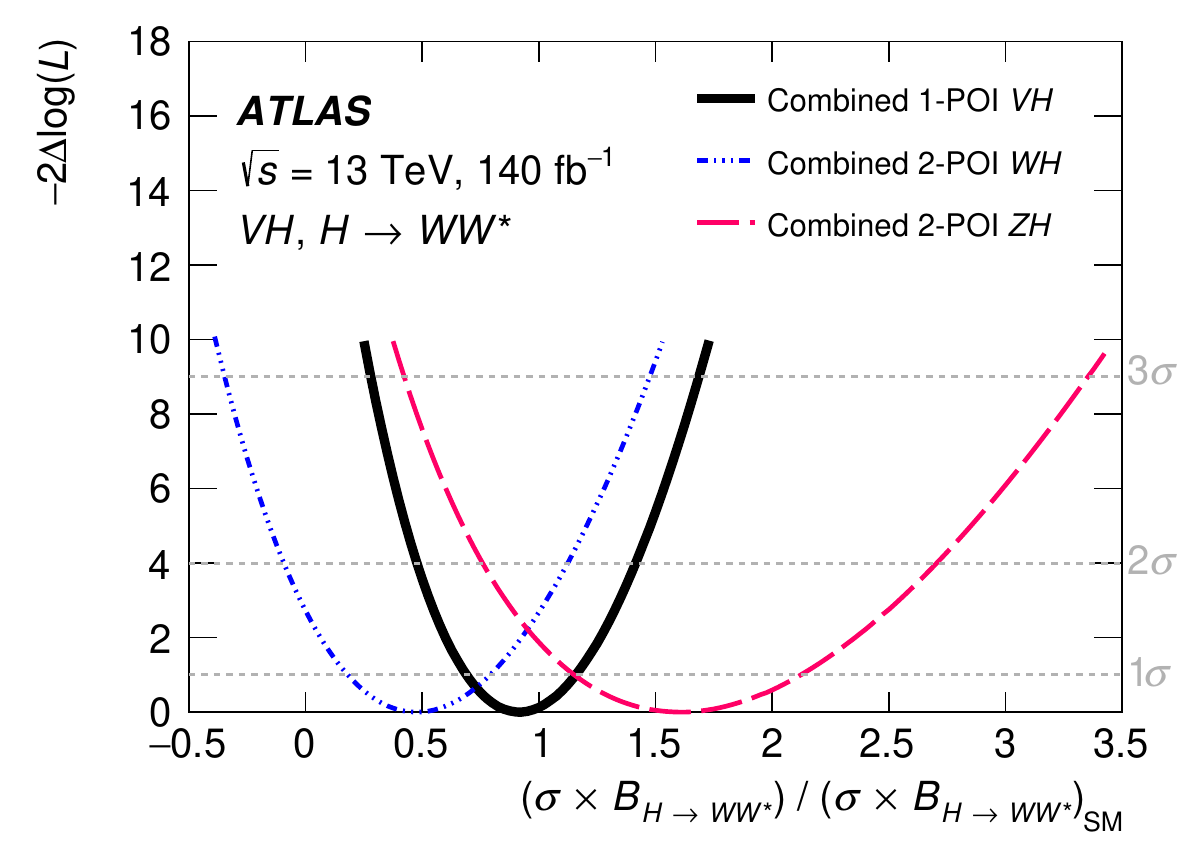}
}
\subfloat[]{
\includegraphics[width=0.48\textwidth]{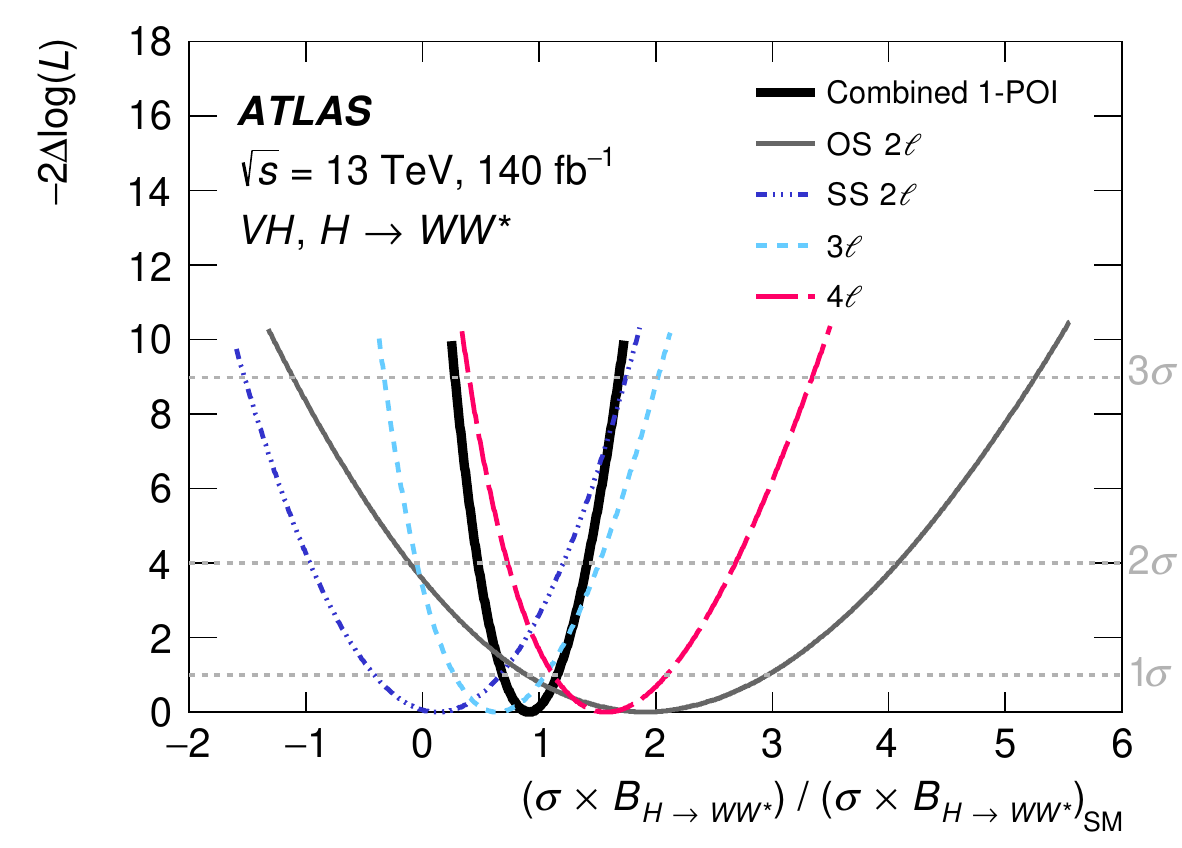}
}
\caption{Observed profile likelihood as a function of $\XS \times \BR_{\HWW}$ normalised by the SM expectation for (a) the \VH and \WH/\ZH measurements from the combined 1- and 2-POI fits, respectively, and (b) the single-channel measurements.}
\label{fig:LHScan}
\end{figure}

\begin{figure}[htbp]
\centering
\includegraphics[width=0.75\textwidth]{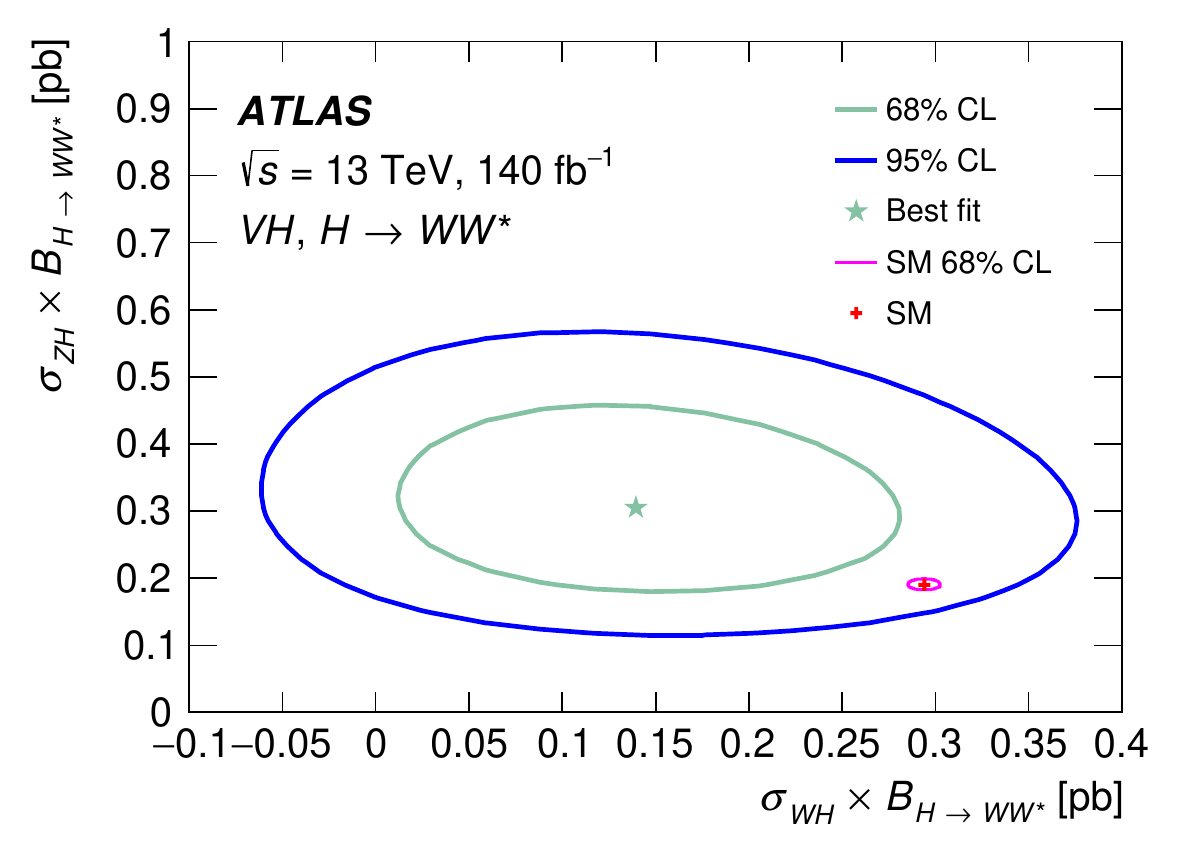}
\caption{Two-dimensional likelihood contours of the observed values of $\XS_{\ZH} \times \BR_{\HWW}$ vs. $\XS_{\WH} \times \BR_{\HWW}$ at the 68\% and 95\% confidence levels (CLs) compared with the predictions from the SM. The 68\% confidence level on the SM predictions for the \ZH and \WH cross-sections times branching fraction is indicated by the magenta ellipse.}
\label{fig:2DCL}
\end{figure}

Table~\ref{tab:fit_results} shows the expected and observed values of the signal strengths for the single-channel, combined 1-POI, and combined 2-POI fits, and Table~\ref{tab:bkg_NFs} shows the observed values of the normalisation factors for the combined 2-POI fit. Figure~\ref{fig:POIPlot} shows the observed values of the total \WH, \ZH, and \VH cross-sections times the \HWW branching ratio, normalised to the SM predictions, and Figure~\ref{fig:LHScan} shows the observed profile likelihood for the combined and single-channel results. Figure~\ref{fig:2DCL} shows the two-dimensional likelihood contours of the observed values of $\XS_{\ZH} \times \BR_{\HWW}$ vs. $\XS_{\WH} \times \BR_{\HWW}$ compared with the SM predictions.

For the combined 1-POI fit, the \VH signal strength is measured to be:

\begin{equation*}
\muVH = 0.92^{+0.21}_{-0.20}\stat^{+0.13}_{-0.12}\syst \,,
\end{equation*}

\noindent corresponding to a $4.5\sigma$ significance over the background-only hypothesis and consistent with the SM expectation with a $p$-value of 73\%. The total \VH cross-section times the \HWW branching ratio is measured to be:

\begin{equation*}
\XS_{\VH} \times \BR_{\HWW} = 0.44^{+0.10}_{-0.09}\stat^{+0.06}_{-0.05}\syst\,\si{pb} \,,
\end{equation*}

\noindent in agreement with the SM expectation of $0.48 \pm 0.01\,\si{pb}$~\cite{deFlorian:2016spz}.

For the combined 2-POI fit, the \WH and \ZH signal strengths are measured to be:

\begin{equation}
\begin{split}
\muWH & = 0.48^{+0.26}_{-0.25}\stat^{+0.18}_{-0.16}\syst \,, \\
\muZH & = 1.6^{+0.5}_{-0.4}\stat\pm0.2\syst \,. \\
\end{split}
\end{equation}

\noindent The observed values of \muWH and \muZH are compatible at a level of $2.0\sigma$, and they are consistent with the SM expectations with a $p$-value of 13\%. The observed \ZH result exceeds the SM prediction due to an excess in the $4\ell$ 1-SFOS SR. The observed \muWH is smaller than the SM prediction due to deficits in the $3\ell$ \Z-depleted SR and the same-sign $2\ell$ channel. The total \WH and \ZH cross-sections times the \HWW branching ratio are measured to be:

\begin{equation}
\begin{split}
\XS_{\WH} \times \BR_{\HWW} & = 0.14^{+0.08}_{-0.07}\stat\pm0.05\syst\,\si{pb} \,, \\
\XS_{\ZH} \times \BR_{\HWW} & = 0.31^{+0.09}_{-0.08}\stat\pm0.03\syst\,\si{pb} \,, \\
\end{split}
\end{equation}

\noindent while the SM expectations are $0.294 \pm 0.009\,\si{pb}$ (\WH) and $0.190^{+0.009}_{-0.008}\,\si{pb}$ (\ZH), respectively~\cite{deFlorian:2016spz}.

\FloatBarrier

\begin{table}[htbp]
\centering
\caption{The observed values of the signal strengths and the corresponding statistical significances, $Z_0$, for the single-channel fits and the combined 1- and 2-POI fits. For each fit, both the expected and observed results are shown. The uncertainties correspond to the total of all statistical and systematic sources. The statistical significances are quoted in units of standard deviations above the background-only hypothesis.}
\label{tab:fit_results}
\renewcommand\arraystretch{1.25}
\begin{tabular}{l | l r r}
\hline\hline
Channel & POI / $Z_0$ & Expected & Observed \\
\hline
\multirow{2}{*}{Opposite-sign $2\ell$} & \muVH & $1.0 \pm 1.0$ & $1.9^{+1.1}_{-1.0}$ \\
& $Z_0$ & 1.0 & 1.9 \\
\hline
\multirow{2}{*}{Same-sign $2\ell$} & \muWH & $1.0 \pm 0.5$ & $0.1 \pm 0.5$ \\
& $Z_0$ & 1.8 & 0.2 \\
\hline
\multirow{2}{*}{$3\ell$} & \muWH & $1.0 \pm 0.4$ & $0.6 \pm 0.4$ \\
& $Z_0$ & 2.8 & 1.8 \\
\hline
\multirow{2}{*}{$4\ell$} & \muZH & $1.0^{+0.5}_{-0.4}$ & $1.6 \pm 0.5$ \\
& $Z_0$ & 3.1 & 4.4 \\
\hline
\multirow{2}{*}{Combined 1-POI} & \muVH & $1.00^{+0.26}_{-0.24}$ & $0.92^{+0.25}_{-0.23}$ \\
& $Z_0$ & 4.8 & 4.5 \\
\hline
\multirow{4}{*}{Combined 2-POI} & \muWH & $1.00^{+0.35}_{-0.33}$ & $0.48^{+0.32}_{-0.30}$ \\
& \muZH & $1.0^{+0.5}_{-0.4}$ & $1.6 \pm 0.5$ \\
& $Z_0^{WH}$ & 3.3 & 1.6 \\
& $Z_0^{ZH}$ & 3.1 & 4.5 \\
\hline\hline
\end{tabular}
\end{table}

\begin{table}[htbp]
\centering
\caption{The observed values of the background normalisation factors for the combined 2-POI fit. The uncertainties correspond to the total of all statistical and systematic sources.}
\label{tab:bkg_NFs}
\renewcommand\arraystretch{1.25}
\begin{tabular}{c|l|c}
\hline\hline
Channel                                & Background        & Normalisation factor \\
\hline
\multirow{3}{*}{Opposite-sign $2\ell$} & Top               & $1.0^{+0.3}_{-0.2}$ \\
& \Zjets            & $0.86^{+0.15}_{-0.14}$ \\
& \WW               & $0.9^{+0.3}_{-0.2}$ \\
\hline
Same-sign $2\ell$                      & \WZg              & $0.90^{+0.17}_{-0.16}$ \\
\hline
\multirow{3}{*}{$3\ell$}               & \WZg 0-jet        & $1.03 \pm 0.06$ \\
& \WZg $\geq$1-jets & $0.88^{+0.16}_{-0.15}$ \\
& \WWW              & $2.2^{+0.7}_{-0.6}$ \\
\hline
$4\ell$                                & \ZZg              & $0.98 \pm 0.07$ \\
\hline\hline
\end{tabular}
\end{table}

\FloatBarrier

Table~\ref{tab:fit_uncertainties} shows the relative impact of the different sources of uncertainty in the observed values of the total cross-sections. For all fit scenarios, the observed results are dominated by statistical uncertainties in data. For the \WH measurement, the RNN shape uncertainty for \WZg (23\%) and the \WZg 0-jet (11\%) and \VVV (12\%) background uncertainties are the dominant systematic uncertainties; for the \ZH measurement, muon experimental uncertainties (4.1\%), primarily on the efficiency of the isolation selection, are the dominant systematic uncertainties.

\begin{table}[htbp]
\centering
\caption{Breakdown of the average contributions to the total uncertainties (in percentage) in the observed values of the cross-sections for the combined 1-POI ($\XS_{\VH} \times \BR_{\HWW}$) and 2-POI ($\XS_{\WH} \times \BR_{\HWW}$ and $\XS_{\ZH} \times \BR_{\HWW}$) fits. Indentation is used to denote subcategories. The quadrature sum of the individual sources may differ from the total uncertainty due to correlations.}
\label{tab:fit_uncertainties}
\resizebox{\textwidth}{!}{
\begin{tabular}{l | S[table-format=3.2] S[table-format=3.2] S[table-format=3.2]}
\hline\hline
\Gape[3mm][3mm]{Source} & $\frac{\Delta(\XS_{\VH} \times \BR_{\HWW})}{\XS_{\VH} \times \BR_{\HWW}}$ [\%] & $\frac{\Delta(\XS_{\WH} \times \BR_{\HWW})}{\XS_{\WH} \times \BR_{\HWW}}$ [\%] & $\frac{\Delta(\XS_{\ZH} \times \BR_{\HWW})}{\XS_{\ZH} \times \BR_{\HWW}}$ [\%] \\
\hline
Statistical uncertainties in data & 22 & 54 & 29 \\
\,\,\,\, Statistical uncertainties in SR data & 20 & 46 & 28 \\
\,\,\,\, Statistical uncertainties in CR data & 10 & 29 & 6.3 \\
\hline
Systematic uncertainties & 13 & 35 & 10 \\
\,\,\,\, Statistical uncertainties in simulation & 6.2 & 13 & 5.9 \\
\,\,\,\, Experimental systematic uncertainties & 5.4 & 12 & 5.7 \\
\,\,\,\,\,\,\,\, Electrons & 1.1 & 1.6 & 1.6 \\
\,\,\,\,\,\,\,\, Muons & 2.7 & 3.0 & 4.1 \\
\,\,\,\,\,\,\,\, Jet energy scale & 1.0 & 3.1 & 0.5 \\
\,\,\,\,\,\,\,\, Jet energy resolution & 0.5 & 2.4 & 0.6 \\
\,\,\,\,\,\,\,\, Flavour tagging & 1.0 & 1.5 & 0.8 \\
\,\,\,\,\,\,\,\, Missing transverse momentum & 0.6 & 0.2 & 0.9 \\
\,\,\,\,\,\,\,\, Pile-up & 1.0 & 1.3 & 0.8 \\
\,\,\,\,\,\,\,\, Luminosity & 1.1 & 1.2 & 1.1 \\
\,\,\,\,\,\,\,\, Mis-identified leptons & 3.7 & 10 & 2.8 \\
\,\,\,\,\,\,\,\, Charge-flip electrons & 1.7 & 5.0 & 0.0 \\
\,\,\,\, Theoretical uncertainties & 6.9 & 20 & 4.2 \\
\,\,\,\,\,\,\,\, \WH & 2.1 & 2.3 & 0.1 \\
\,\,\,\,\,\,\,\, \ZH & 0.5 & 0.3 & 2.4 \\
\,\,\,\,\,\,\,\, Other $H$ & 1.2 & 2.4 & 0.9 \\
\,\,\,\,\,\,\,\, \WW & 1.2 & 3.7 & 0.2 \\
\,\,\,\,\,\,\,\, \WZg 0-jet & 3.2 & 11 & 0.2 \\
\,\,\,\,\,\,\,\, \WZg $\geq$1-jets & 3.2 & 9.7 & 0.4 \\
\,\,\,\,\,\,\,\, \ZZg & 1.2 & 2.1 & 0.8 \\
\,\,\,\,\,\,\,\, \VVV & 2.7 & 12 & 1.0 \\
\,\,\,\,\,\,\,\, Top & 2.8 & 5.0 & 2.4 \\
\,\,\,\,\,\,\,\, \Zjets & 1.6 & 2.9 & 1.5 \\
\,\,\,\, RNN shape uncertainty for \WZg & 7.5 & 23 & 0.7 \\
\hline
Total & 26 & 64 & 30 \\
\hline\hline
\end{tabular}
}
\end{table}

\FloatBarrier

\subsubsection{Differential cross-section results}
\label{sec:results_stxs}

For the differential cross-section measurement, two fits are performed: one for the \ptV scheme and one for the STXS scheme. The analysis regions are the same for both the fits. The signal samples are split into signal templates following the STXS Stage~1.2 categorisation. The cross-section for each template is scaled by its own POI, $(\XS\times\BR_{\HWW})/(\XS\times\BR_{\HWW})_\textrm{SM}$. The POIs are correlated (i.e., coherently tuned via a single POI) to provide the results in merged bins of the STXS categorisation; for example, the POIs for different jet multiplicity bins are correlated as are the POIs for the $250 \leq \ptV < \qty{400}{\GeV}$ and $\ptV \geq \qty{400}{\GeV}$ bins.

For the \ptV scheme, the POIs for the \WH and \ZH production modes are correlated to provide combined \VH production mode POIs; for the $\V\to\qq$ signal template fit, the signal is split into four \ptV bins and the POIs are correlated with those of the $\V\to\ell\ell/\nu\nu$ channels.

\begin{figure}[htbp]
\centering
\subfloat[]{
\includegraphics[width=0.7\textwidth]{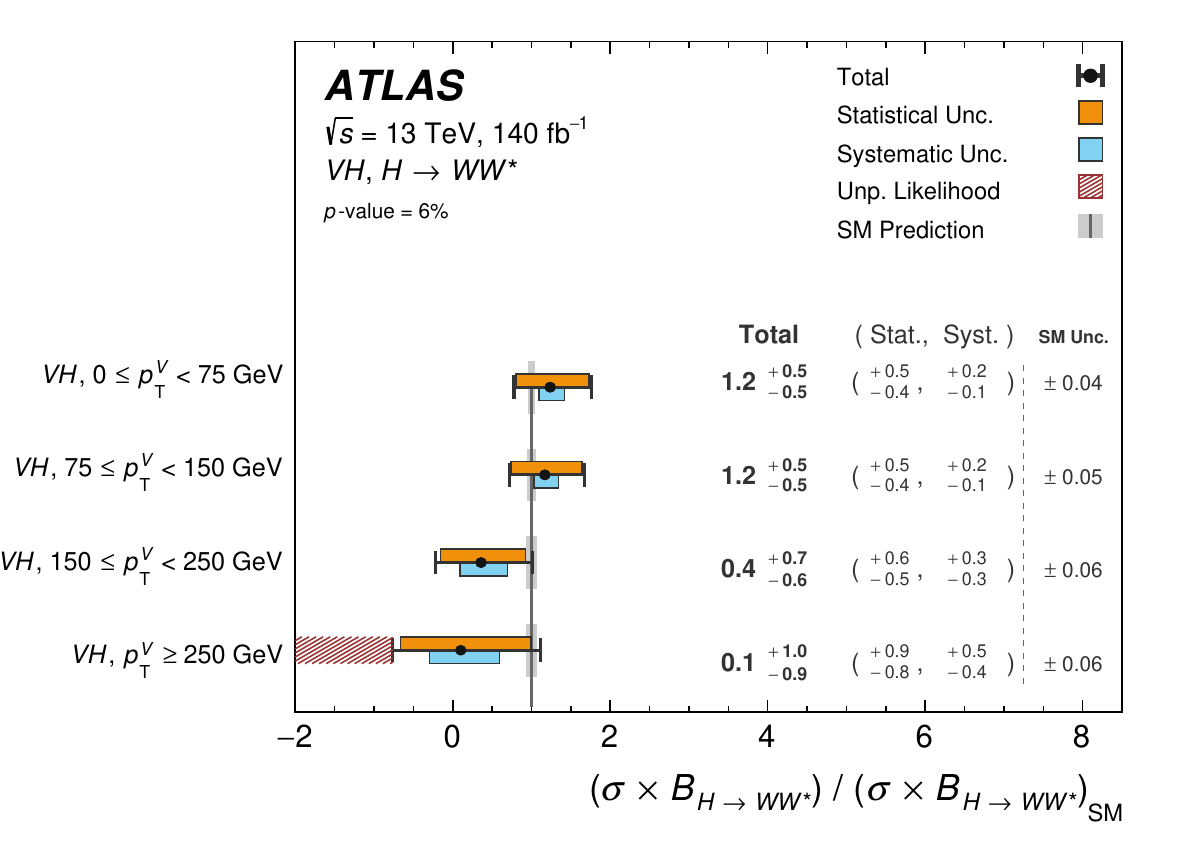}
} \\
\subfloat[]{
\includegraphics[width=0.7\textwidth]{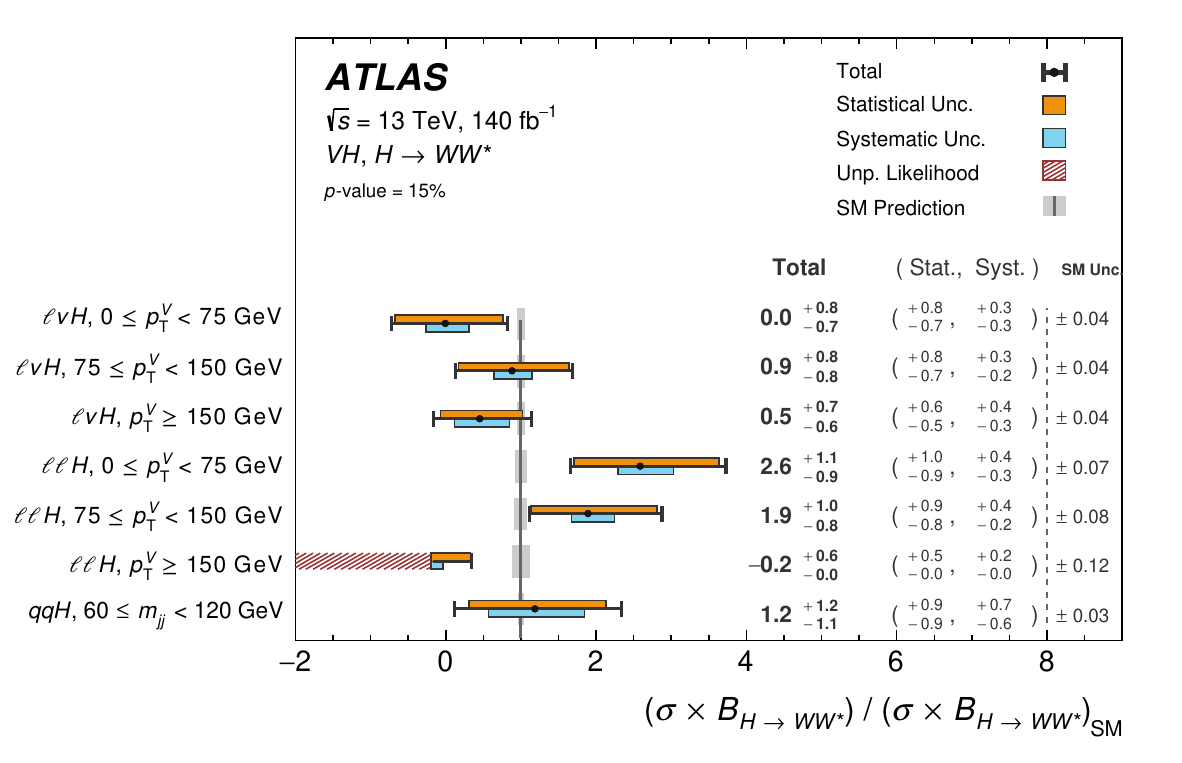}
}
\caption{Best-fit values of the cross-sections times the \HWW branching ratio measured in the (a) \ptV and (b) STXS schemes. Each cross-section is normalised to its SM prediction. The black error bars, orange boxes, and blue boxes show the total, statistical, and systematic uncertainties in the measurements, respectively. The grey bands represent the theory uncertainty of the corresponding Higgs boson production mode, calculated from the sources of uncertainty described in Section~\ref{sec:systematics_theory}. The hatched band indicated with ``Unp. Likelihood'' shows the region of the parameter space where the likelihood becomes unphysical.}
\label{fig:STXS_mu_plots_summary}
\end{figure}

The fitted POIs and normalisation factors for the \ptV and STXS schemes are summarised in Tables~\ref{tab:pTVscheme} and \ref{tab:STXSscheme}, respectively. The POIs are plotted in Figure~\ref{fig:STXS_mu_plots_summary}.

\begin{table}[htbp]
\centering
\caption{Observed fit results for the \ptV scheme. ``Stat. unc.'' and ``Syst. unc.'' indicate the statistical and systematic components, respectively, of the total uncertainty in the corresponding POI. For the normalisation factors, only the total uncertainties are quoted.  A lower limit on the POI for the $\VH, \ptV \geq \qty{250}{\GeV}$ category was set at $-0.76$ to prevent the likelihood from reaching an unphysical region with zero observed and zero expected events in any bin of the analysis. The label ``LL'' indicates that the interval reaches its minimum allowed value. A 95\% confidence level upper limit on the POI for the $\VH, \ptV \geq \qty{250}{\GeV}$ category was computed and found to be 2.1. In the limit computation, the other POIs were left free to float.}
\label{tab:pTVscheme}
\renewcommand\arraystretch{1.25}
\resizebox{\textwidth}{!}{
\begin{tabular}{l|r||l|r|r|r}
\hline\hline
\multicolumn{6}{c}{Fit results for the \ptV scheme} \\
\hline
\multicolumn{2}{c||}{Normalisation factors} & \multicolumn{4}{c}{$(\XS_\textrm{category}\times\BR_{\HWW})/(\XS_\textrm{category}\times\BR_{\HWW})_\textrm{SM}$} \\
\hline
Parameter & Fit result & Category & Fit result & Stat. unc. & Syst. unc. \\
\hline
Top & $0.9^{+0.3}_{-0.2}$ & $\VH, 0 \leq \ptV < \qty{75}{\GeV}$ & $1.2 \pm 0.5$ & ${}^{+0.5}_{-0.4}$ & ${}^{+0.2}_{-0.1}$ \\
\Zjets & $0.86^{+0.12}_{-0.11}$ & $\VH, 75 \leq \ptV < \qty{150}{\GeV}$ & $1.2 \pm 0.5$ & ${}^{+0.5}_{-0.4}$ & ${}^{+0.2}_{-0.1}$ \\
\WW & $0.9^{+0.3}_{-0.2}$ & $\VH, 150 \leq \ptV < \qty{250}{\GeV}$ & $0.4^{+0.7}_{-0.6}$ & ${}^{+0.6}_{-0.5}$ & $\pm 0.3$ \\
\WZg (SS $2\ell$) & $1.12^{+0.17}_{-0.15}$ & $\VH, \ptV \geq \qty{250}{\GeV}$ & $0.1^{+1.0}_{-0.9\,\textrm{(LL)}}$ & ${}^{+0.9}_{-0.8}$ & ${}^{+0.5}_{-0.4}$ \\
\WZg 0-jet & $1.00 \pm 0.06$ & & & & \\
\WZg $\geq$1-jets & $0.89^{+0.16}_{-0.15}$ & & & & \\
\WWW & $2.0^{+0.6}_{-0.5}$ & & & & \\
\ZZg & $1.00 \pm 0.03$ & & & & \\
\hline\hline
\end{tabular}
}
\end{table}

\begin{table}[htbp]
\centering
\caption{Observed fit results for the STXS scheme. ``Stat. unc.'' and ``Syst. unc.'' indicate the statistical and systematic components, respectively, of the total uncertainty in the corresponding POI. For the normalisation factors, only the total uncertainties are quoted. A lower limit on the POI for the $\llH, \ptV \geq \qty{150}{\GeV}$ category was set at $-0.2$ to prevent the likelihood from reaching an unphysical region with zero observed and zero expected events in any bin of the analysis. The label ``LL'' indicates that the interval reaches its minimum allowed value. A 95\% confidence level upper limit on the POI for the $\llH, \ptV \geq \qty{150}{\GeV}$ category was computed and found to be 1.6. In the limit computation, the other POIs were left free to float.}
\label{tab:STXSscheme}
\renewcommand\arraystretch{1.25}
\begin{tabular}{l|r||l|r|r|r}
\hline\hline
\multicolumn{6}{c}{Fit results for the STXS scheme} \\
\hline
\multicolumn{2}{c||}{Normalisation factors} & \multicolumn{4}{c}{$(\XS_\textrm{STXS category}\times\BR_{\HWW})/(\XS_\textrm{STXS category}\times\BR_{\HWW})_\textrm{SM}$} \\
\hline
Parameter & Fit result & STXS category & Fit result & Stat. unc. & Syst. unc.  \\
\hline
Top & $0.9^{+0.3}_{-0.2}$ & $\lvH, 0 \leq \ptV < \qty{75}{\GeV}$ & $0.0^{+0.8}_{-0.7}$ & ${}^{+0.8}_{-0.7}$ & $\pm 0.3$ \\
\Zjets & $0.85^{+0.12}_{-0.11}$ & $\lvH, 75 \leq \ptV < \qty{150}{\GeV}$ & $0.9 \pm 0.8$ & ${}^{+0.8}_{-0.7}$ & ${}^{+0.3}_{-0.2}$ \\
\WW & $0.9^{+0.3}_{-0.2}$ & $\lvH, \ptV \geq \qty{150}{\GeV}$ & $0.5^{+0.7}_{-0.6}$ & ${}^{+0.6}_{-0.5}$ & ${}^{+0.4}_{-0.3}$ \\
\WZg (SS $2\ell$) & $1.14^{+0.17}_{-0.15}$ & $\llH, 0 \leq \ptV < \qty{75}{\GeV}$ & $2.6^{+1.1}_{-0.9}$ & ${}^{+1.0}_{-0.9}$ & ${}^{+0.4}_{-0.3}$ \\
\WZg 0-jet & $1.00 \pm 0.06$ & $\llH, 75 \leq \ptV < \qty{150}{\GeV}$ & $1.9^{+1.0}_{-0.8}$ & ${}^{+0.9}_{-0.8}$ & ${}^{+0.4}_{-0.2}$ \\
\WZg $\geq$1-jets & $0.88^{+0.16}_{-0.15}$ & $\llH, \ptV \geq \qty{150}{\GeV}$ & $-0.2^{+0.6}_{-0.0\,\textrm{(LL)}}$ & ${}^{+0.5}_{-0.0\,\textrm{(LL)}}$ & ${}^{+0.2}_{-0.0\,\textrm{(LL)}}$ \\
\WWW & $2.2^{+0.7}_{-0.6}$ & $qqH, 60 \leq \mjj < \qty{120}{\GeV}$ & $1.2^{+1.2}_{-1.1}$ & $\pm 0.9$ & ${}^{+0.7}_{-0.6} $ \\
\ZZg & $0.99 \pm 0.03$ & & & & \\
\hline\hline
\end{tabular}
\end{table}

\FloatBarrier

Due to deficits in data in several bins of the analysis --- in particular the \SSMM region --- some of observed POIs are close to zero. Moreover, the expected total yield becomes negative for large negative values of the POIs; therefore, a constraint was imposed on the POI for the $VH, \ptV \geq \qty{250}{\GeV}$ category in the \ptV scheme and on the POI for the $\llH, \ptV \geq \qty{150}{\GeV}$ category in the STXS scheme. In both the cases, the lower limit of the 68\% confidence interval cannot be reached before the likelihood becomes ill-defined due to negative expected yields in some bins. The value at which this happens is specified in Tables~\ref{tab:pTVscheme} and \ref{tab:STXSscheme}. Given that this indicates a very low expected signal yield, a test of the validity of the asymptotic approximation was performed on such parameters. A small degree of over-coverage (70\% with respect to 68\%) of the confidence interval was observed for the POI for the $\llH, \ptV \geq \qty{150}{\GeV}$ category. The confidence interval is defined as in Ref.~\cite{Cowan:2010js}.

There are also some notable differences in the normalisation factors obtained from the total and differential cross-section measurements. For the opposite-sign $2\ell$ channel, the top normalisation factor in the differential cross-section measurement is smaller than that obtained in the total cross-section measurement. The nuisance parameter which tunes the \ttbar parton-shower modelling uncertainty is sensitive to the \pTV modelling; therefore, it can be pulled differently in the two measurements. In the case of the differential cross-section measurement, the best-fit value of this nuisance parameter increases the \ttbar yield in the top CR. To compensate for such an increase, the top normalisation factor decreases. For the same-sign $2\ell$ channel, the \WZ normalisation factor in the differential cross-section measurement is larger than that obtained in the total cross-section measurement. The \WZ normalisation factor is driven by the background-like bins of the signal regions in the same-sign $2\ell$ channel. As those signal regions are split differently in the two measurement scenarios, the resulting \WZ normalisation factors are also different.

\begin{figure}[htbp]
\centering
\includegraphics[width=0.7\textwidth]{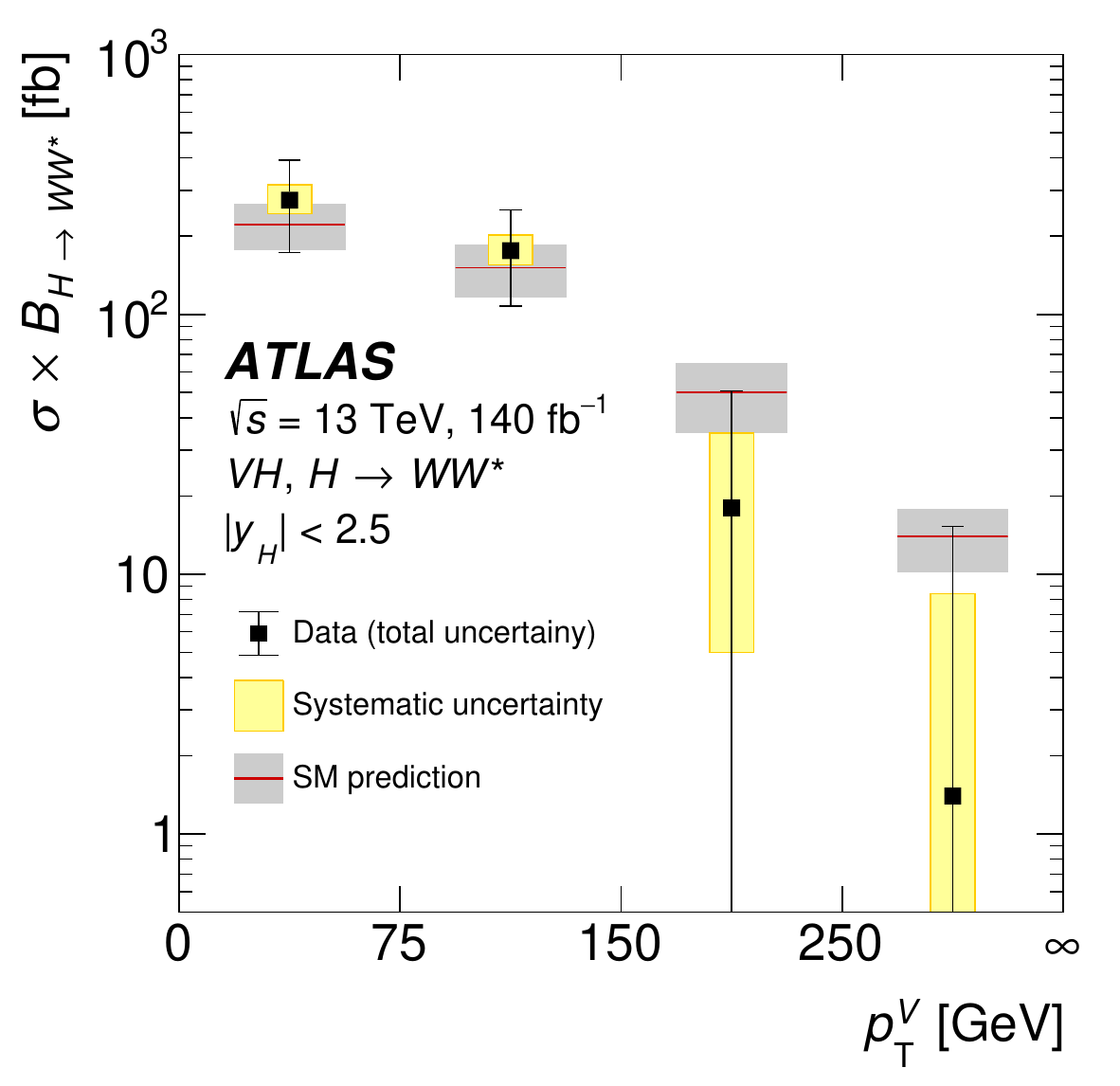}
\caption{Differential \VH production cross-section as a function of the \pt of the associated vector boson. The theoretical expectation is also shown, whose uncertainty is calculated from the sources described in Section~\ref{sec:systematics_theory}.}
\label{fig:diff_PTV}
\end{figure}

The POIs were converted to fiducial cross-sections by multiplying the POIs by the cross-sections used to normalise the signal templates in each bin of the \ptV and STXS schemes. The resulting values are shown in Table~\ref{tab:STXS_XS}. The differential cross-section as a function of \ptV is shown in Figure~\ref{fig:diff_PTV}.

\begin{table}[htbp]
\centering
\caption{Observed cross-sections times the \HWW branching ratio for the \ptV and STXS schemes. ``Stat. unc.'' and ``Syst. unc.'' indicate the statistical and systematic components, respectively, of the total uncertainty on the corresponding POI. ``SM'' indicates the SM expectation, whose uncertainty is calculated from the sources described in Section~\ref{sec:systematics_theory}. Lower limits on $(\XS\times\BR_{\HWW}) / (\XS\times\BR_{\HWW})_\textrm{SM}$ for the $\VH, \ptV \geq \qty{250}{\GeV}$ category in the \pTV scheme and the $\llH, \ptV \geq \qty{150}{\GeV}$ category in the STXS scheme were set at $-0.76$ and $-0.2$, respectively, to prevent the likelihoods from reaching unphysical regions. The label ``LL'' indicates that the interval reaches its minimum allowed value.}
\label{tab:STXS_XS}
\renewcommand\arraystretch{1.25}
\resizebox{\textwidth}{!}{
\begin{tabular}{l|r|r|r|r||l|r|r|r|r}
\hline\hline
\multicolumn{10}{c}{$\XS\times\BR_{\HWW}$ [fb]} \\
\hline
\multicolumn{5}{c||}{\ptV scheme ($|y_H| < 2.5$)} & \multicolumn{5}{c}{STXS scheme} \\
\hline
\ptV interval [$\GeV\xspace$] & Value & Stat. unc. & Syst. unc. & SM & STXS category [\ptV and \mjj in $\GeV\xspace$] & Value & Stat. unc. & Syst. unc. & SM \\
\hline
$\VH\,(0 \leq \ptV < 75)$ & 270 & ${}^{+110}_{-100}$ & ${}^{+40}_{-30}$ & $220 \pm 40$ & $\lvH\,(0 \leq \ptV < 75)$ & 0 & ${}^{+40}_{-30}$ & $\pm 10$ & $46.3 \pm 1.8$ \\
$\VH\,(75 \leq \ptV < 150)$ & 180 & ${}^{+70}_{-60}$ & ${}^{+30}_{-20}$& $151 \pm 34$ & $\lvH\,(75 \leq \ptV < 150)$ & 26 & ${}^{+22}_{-21}$ & ${}^{+8}_{-7}$ & $28.9 \pm 1.2$ \\
$\VH\,(150 \leq \ptV < 250)$ & 18 & ${}^{+28}_{-26}$ & ${}^{+17}_{-13}$ & $50 \pm 15$ & $\lvH\,(\ptV \geq 150)$ & 5 & $\pm 6$ & $\pm 4$ & $11.5 \pm 0.5$ \\
$\VH\,(\ptV \geq 250)$ & 1 & ${}^{+12}_{-11}$ & ${}^{+7}_{-6}$ & $14 \pm 4$ & $\llH\,(0 \leq \ptV < 75)$ & 62 & ${}^{+25}_{-21}$ & ${}^{+11}_{-7}$ & $24.1 \pm 1.7$ \\
& & & & & $\llH\,(75 \leq \ptV < 150)$ & 35 & ${}^{+17}_{-14}$ & ${}^{+7}_{-4}$ & $18.7 \pm 1.5$ \\
& & & & & $\llH\,(\ptV \geq 150)$ & $-2$ & ${}^{+5}_{-0\,\textrm{(LL)}}$ & ${}^{+1}_{-0\,\textrm{(LL)}}$ & $8.7 \pm 1.0$ \\
& & & & & $qqH\,(60 \leq \mjj < 120)$ & 110 & ${}^{+90}_{-80}$ & $\pm 60$ & $91.8 \pm 3.3$ \\
\hline\hline
\end{tabular}
}
\end{table}

Finally, Tables~\ref{tab:breakdown_ptV} and \ref{tab:breakdown_STXS} of Appendix~\ref{app:uncertainties} show the fractional contributions from statistical and systematic sources to the total uncertainties reported in \ptV and STXS schemes, respectively. As for the total cross-section measurement, the differential results are dominated by statistical uncertainties in data.

% End of text imported from the .//sections/results.tex input file

%
%
\clearpage

% The next lines are included from the .//sections/conclusions.tex input file
%
\section{Conclusions}
\label{sec:conclusions}

A measurement of the total Higgs boson production cross-sections via associated \WH and \ZH production using \HWWlnln and \HWWlnjj decays is presented. Results for combined \WH and \ZH production are also presented. The analysis uses proton--proton events delivered by the Large Hadron Collider at a centre-of-mass energy of \qty{13}{\TeV} and recorded by the ATLAS detector between 2015 and 2018. The data correspond to an integrated luminosity of \qty{140}{\ifb}. The \VH signal strength is measured to be $0.92^{+0.21}_{-0.20}\stat^{+0.13}_{-0.12}\syst$, corresponding to a $4.5\sigma$ significance over the background-only hypothesis. The products of the \HWW branching fraction times the \WH and \ZH cross-sections are measured to be $0.14^{+0.08}_{-0.07}\stat\pm0.05\syst\,\si{pb}$ and $0.31^{+0.09}_{-0.08}\stat\pm0.03\syst\,\si{pb}$, respectively, in agreement with the Standard Model predictions. Differential cross-sections have also been measured, $\sigma_{\VH}\times\BR_{\HWW}$ as a function of the \pt of the associated vector boson and Simplified Template Cross-Sections for \VH and EW \qqH production. The results obtained are in agreement with their Standard Model expectations.

% End of text imported from the .//sections/conclusions.tex input file

%
\section*{Acknowledgements}
%

% The next lines are included from the .//acknowledgements/Acknowledgements.tex input file
%
%

%
%

We thank CERN for the very successful operation of the LHC and its injectors, as well as the support staff at
CERN and at our institutions worldwide without whom ATLAS could not be operated efficiently.

The crucial computing support from all WLCG partners is acknowledged gratefully, in particular from CERN, the ATLAS Tier-1 facilities at TRIUMF/SFU (Canada), NDGF (Denmark, Norway, Sweden), CC-IN2P3 (France), KIT/GridKA (Germany), INFN-CNAF (Italy), NL-T1 (Netherlands), PIC (Spain), RAL (UK) and BNL (USA), the Tier-2 facilities worldwide and large non-WLCG resource providers. Major contributors of computing resources are listed in Ref.~\cite{ATL-SOFT-PUB-2025-001}.

We gratefully acknowledge the support of ANPCyT, Argentina; YerPhI, Armenia; ARC, Australia; BMWFW and FWF, Austria; ANAS, Azerbaijan; CNPq and FAPESP, Brazil; NSERC, NRC and CFI, Canada; CERN; ANID, Chile; CAS, MOST and NSFC, China; Minciencias, Colombia; MEYS CR, Czech Republic; DNRF and DNSRC, Denmark; IN2P3-CNRS and CEA-DRF/IRFU, France; SRNSFG, Georgia; BMBF, HGF and MPG, Germany; GSRI, Greece; RGC and Hong Kong SAR, China; ISF and Benoziyo Center, Israel; INFN, Italy; MEXT and JSPS, Japan; CNRST, Morocco; NWO, Netherlands; RCN, Norway; MNiSW, Poland; FCT, Portugal; MNE/IFA, Romania; MSTDI, Serbia; MSSR, Slovakia; ARIS and MVZI, Slovenia; DSI/NRF, South Africa; MICIU/AEI, Spain; SRC and Wallenberg Foundation, Sweden; SERI, SNSF and Cantons of Bern and Geneva, Switzerland; NSTC, Taipei; TENMAK, T\"urkiye; STFC/UKRI, United Kingdom; DOE and NSF, United States of America.

Individual groups and members have received support from BCKDF, CANARIE, CRC and DRAC, Canada; CERN-CZ, FORTE and PRIMUS, Czech Republic; COST, ERC, ERDF, Horizon 2020, ICSC-NextGenerationEU and Marie Sk{\l}odowska-Curie Actions, European Union; Investissements d'Avenir Labex, Investissements d'Avenir Idex and ANR, France; DFG and AvH Foundation, Germany; Herakleitos, Thales and Aristeia programmes co-financed by EU-ESF and the Greek NSRF, Greece; BSF-NSF and MINERVA, Israel; NCN and NAWA, Poland; La Caixa Banking Foundation, CERCA Programme Generalitat de Catalunya and PROMETEO and GenT Programmes Generalitat Valenciana, Spain; G\"{o}ran Gustafssons Stiftelse, Sweden; The Royal Society and Leverhulme Trust, United Kingdom.

In addition, individual members wish to acknowledge support from Armenia: Yerevan Physics Institute (FAPERJ); CERN: European Organization for Nuclear Research (CERN PJAS); Chile: Agencia Nacional de Investigaci\'on y Desarrollo (FONDECYT 1230812, FONDECYT 1230987, FONDECYT 1240864); China: Chinese Ministry of Science and Technology (MOST-2023YFA1605700, MOST-2023YFA1609300), National Natural Science Foundation of China (NSFC - 12175119, NSFC 12275265, NSFC-12075060); Czech Republic: Czech Science Foundation (GACR - 24-11373S), Ministry of Education Youth and Sports (ERC-CZ-LL2327, FORTE CZ.02.01.01/00/22\_008/0004632), PRIMUS Research Programme (PRIMUS/21/SCI/017); EU: H2020 European Research Council (ERC - 101002463); European Union: European Research Council (ERC - 948254, ERC 101089007), Horizon 2020 Framework Programme (MUCCA - CHIST-ERA-19-XAI-00), European Union, Future Artificial Intelligence Research (FAIR-NextGenerationEU PE00000013), Italian Center for High Performance Computing, Big Data and Quantum Computing (ICSC, NextGenerationEU); France: Agence Nationale de la Recherche (ANR-20-CE31-0013, ANR-21-CE31-0013, ANR-21-CE31-0022, ANR-22-EDIR-0002), Investissements d'Avenir Labex (ANR-11-LABX-0012); Germany: Baden-Württemberg Stiftung (BW Stiftung-Postdoc Eliteprogramme), Deutsche Forschungsgemeinschaft (DFG - 469666862, DFG - CR 312/5-2); Italy: Istituto Nazionale di Fisica Nucleare (ICSC, NextGenerationEU), Ministero dell'Università e della Ricerca (NextGenEU I53D23000820006 M4C2.1.1); Japan: Japan Society for the Promotion of Science (JSPS KAKENHI JP22H01227, JSPS KAKENHI JP22H04944, JSPS KAKENHI JP22KK0227, JSPS KAKENHI JP23KK0245); Netherlands: Netherlands Organisation for Scientific Research (NWO Veni 2020 - VI.Veni.202.179); Norway: Research Council of Norway (RCN-314472); Poland: Ministry of Science and Higher Education (IDUB AGH, POB8, D4 no 9722), Polish National Agency for Academic Exchange (PPN/PPO/2020/1/00002/U/00001), Polish National Science Centre (NCN 2021/42/E/ST2/00350, NCN OPUS nr 2022/47/B/ST2/03059, NCN UMO-2019/34/E/ST2/00393, NCN \& H2020 MSCA 945339, UMO-2020/37/B/ST2/01043, UMO-2021/40/C/ST2/00187, UMO-2022/47/O/ST2/00148, UMO-2023/49/B/ST2/04085, UMO-2023/51/B/ST2/00920); Portugal: Foundation for Science and Technology (FCT); Slovenia: Slovenian Research Agency (ARIS grant J1-3010); Spain: Generalitat Valenciana (Artemisa, FEDER, IDIFEDER/2018/048), Ministry of Science and Innovation (MCIN \& NextGenEU PCI2022-135018-2, MICIN \& FEDER PID2021-125273NB, RYC2019-028510-I, RYC2020-030254-I, RYC2021-031273-I, RYC2022-038164-I); Sweden: Carl Trygger Foundation (Carl Trygger Foundation CTS 22:2312), Swedish Research Council (Swedish Research Council 2023-04654, VR 2018-00482, VR 2021-03651, VR 2022-03845, VR 2022-04683, VR 2023-03403), Knut and Alice Wallenberg Foundation (KAW 2018.0458, KAW 2019.0447, KAW 2022.0358); Switzerland: Swiss National Science Foundation (SNSF - PCEFP2\_194658); United Kingdom: Leverhulme Trust (Leverhulme Trust RPG-2020-004), Royal Society (NIF-R1-231091); United States of America: U.S. Department of Energy (ECA DE-AC02-76SF00515), Neubauer Family Foundation.

%
%

% End of text imported from the .//acknowledgements/Acknowledgements.tex input file

%
%
%
%
%
%
%

%
\clearpage
\appendix
\part*{Appendix}
\addcontentsline{toc}{part}{Appendix}
%

%
%

% The next lines are included from the .//sections/app_inputs.tex input file
%
\section{Input variables for multivariate discriminants}
\label{app:inputs}

This appendix describes the input variables for each of the multivariate discriminants utilised by this measurement.

\paragraph{Opposite-sign $2\ell$ channel:} The input variables for the ANN utilised by the opposite-sign $2\ell$ channel --- see Section~\ref{sec:os2l} --- are shown in Table~\ref{tab:inputsTwoLeptonOS}.

\begin{table}[htbp]
\centering
\caption{Summary of input variables for the ANN in the opposite-sign $2\ell$ channel.}
\label{tab:inputsTwoLeptonOS}
\renewcommand\arraystretch{1.25}
\resizebox{\textwidth}{!}{
\begin{tabular}{l | l}
\hline\hline
Variable(s) & Description \\
\hline
$\pt^{\ell_0}$, $\pt^{\ell_1}$ & Leading and subleading lepton transverse momentum \\
\mll & Dilepton invariant mass \\
\dphill & Dilepton azimuthal separation \\
$|\dyll|$ & Dilepton rapidity separation \\
\mt & Transverse mass (Eq.~\ref{eqn:mT}) \\
$\pt^{j_0}$, $\pt^{j_1}$ & Leading and subleading jet transverse momentum \\
\mjj & Dijet invariant mass \\
\dphijj & Dijet azimuthal separation \\
$|\dyjj|$ & Dijet rapidity separation \\
$m_{\ell_0 j_0}$, $m_{\ell_0 j_1}$, $m_{\ell_1 j_0}$, $m_{\ell_1 j_1}$ & All lepton-jet invariant mass combinations \\
\mtt & Invariant mass of the $\tau$-lepton pair using the collinear approximation~\cite{Plehn:1999xi}, \\
& assuming the electrons and muons result from $\tau$-lepton decays \\
\met & Missing transverse momentum \\
\metsig & Object-based \met significance \\
$H_\mathrm{T}$ & Transverse momentum sum of all hard objects \\
\hline\hline
\end{tabular}
}
\end{table}

\paragraph{$3\ell$ channel:} The input variables for the ANN utilised by the $3\ell$ \Z-dominated channel --- see Section~\ref{sec:3l} --- are shown in Table~\ref{tab:inputsZdom}. The transverse mass (Eq.~\ref{eqn:mT}) of the \W boson, $\mt^W$. It is constructed from the \met and the lepton not belonging to the SFOS pair with an invariant mass closest to the mass of the \Z boson, either $\ell_1$ or $\ell_2$.

\begin{table}[htbp]
\centering
\caption{Summary of input variables for the ANN in the $3\ell$ \Z-dominated channel.}
\label{tab:inputsZdom}
\renewcommand\arraystretch{1.25}
\begin{tabular}{l | l}
\hline\hline
Variable(s) & Description \\
\hline
$\pt^{\ell_0}$ & Transverse momentum of $\ell_0$ \\
$\left|\sum_{i=0}^2\vec{p}_\textrm{T}^{\,\ell_i}\right|$ & Magnitude of the vectorial sum of the lepton transverse momenta \\
$|\Delta\eta_{\ell_0\ell_1}|$, $|\Delta\eta_{\ell_1\ell_2}|$ & Pseudorapidity separation between $\ell_0$ and $\ell_1$ and between $\ell_1$ and $\ell_2$ \\
$\Delta\phi_{\ell_0\ell_2}$ & Azimuthal separation between $\ell_0$ and $\ell_2$ \\
$\Delta{R}_{\ell_0\ell_1}$, $\Delta{R}_{\ell_0\ell_2}$ & Angular separation between $\ell_0$ and $\ell_1$ and between $\ell_0$ and $\ell_2$ \\
$m_{\ell_0\ell_1}$, $m_{\ell_0\ell_2}$, $m_{\ell_1\ell_2}$ & Dilepton invariant mass for each combination of leptons \\
\met & Missing transverse momentum \\
$\Delta\phi_{\ell_0,\textrm{miss}}$, $\Delta\phi_{\ell_1,\textrm{miss}}$, $\Delta\phi_{\ell_2,\textrm{miss}}$ & Azimuthal separation between each lepton and the \met \\
$\mt^W$ & Transverse mass of the \W boson (Eq.~\ref{eqn:mT}) \\
\hline\hline
\end{tabular}
\end{table}

The input variables for the ANN utilised by the $3\ell$ \Z-depleted channel are shown in Table~\ref{tab:inputsZdep}. An input deserving additional explanation is the compatibility of the event with the \WZg hypothesis, $F_\alpha$. Given the reconstructed charged lepton momenta and the \pmet, the event kinematics can be calculated under the \WZg with $Z\to\tau\tau$ hypothesis and using the collinear approximation for the $\tau$-lepton  decays with one remaining unknown --- for example, the ratio of one $\tau$-lepton's energy to the energy of the lepton from the same $\tau$-lepton's decay. This unknown is varied, and the number of physical kinematic solutions is taken as a measure of the compatibility with the \WZg hypothesis.

\begin{table}[htbp]
\centering
\caption{Summary of input variables for the ANN in the $3\ell$ \Z-depleted channel.}
\label{tab:inputsZdep}
\renewcommand\arraystretch{1.25}
\begin{tabular}{l | l}
\hline\hline
Variable(s) & Description \\
\hline
$\pt^{\ell_0}$, $\pt^{\ell_1}$, $\pt^{\ell_2}$ & Transverse momentum for each lepton \\
$|\Delta\eta_{\ell_0\ell_1}|$, $|\Delta\eta_{\ell_0\ell_2}|$, $|\Delta\eta_{\ell_1\ell_2}|$ & Dilepton pseudorapidity separation for each combination of leptons \\
$\Delta{R}_{\ell_0\ell_1}$, $\Delta{R}_{\ell_0\ell_2}$, $\Delta{R}_{\ell_1\ell_2}$ & Dilepton angular separation for each combination of leptons \\
$m_{\ell_0\ell_1}$, $m_{\ell_0\ell_2}$, $m_{\ell_1\ell_2}$ & Dilepton invariant mass for each combination of leptons \\
$\mt^{\ell_0\ell_1}$, $\mt^{\ell_0\ell_2}$, $\mt^{\ell_1\ell_2}$ & Dilepton transverse mass for each combination of leptons \\
$\left|\sum_{i=0}^2\vec{p}_\textrm{T}^{\,\ell_i}\right|$ & Magnitude of the vectorial sum of the lepton transverse momenta \\
$m_{\ell\ell\ell}$ & Trilepton invariant mass \\
$n_\textrm{jets}$ & Number of jets \\
$n_{b\textrm{-jets}}$ & Number of $b$-tagged jets \\
$\pt^{j_0}$ & Transverse momentum of the leading jet \\
\met & Missing transverse momentum \\
$\Delta\phi_{\ell_0,\textrm{miss}}$, $\Delta\phi_{\ell_1,\textrm{miss}}$, $\Delta\phi_{\ell_2,\textrm{miss}}$ & Azimuthal separation between each lepton and the \met \\
$\metsig\,/ \met$ & Ratio of the \met significance to the \met \\
$F_\alpha$ & Compatibility of the event with the \WZg hypothesis \\
\hline\hline
\end{tabular}
\end{table}

\paragraph{$4\ell$ channel:} The input variables for the BDTs utilised by the $4\ell$ channel --- see Section~\ref{sec:4l} --- are shown in Table~\ref{tab:inputsFourLepton}. The azimuthal separation between the leptons from the Higgs boson candidate in the frame where the Higgs boson \pt is zero is denoted by \DPhiB. The Higgs boson transverse momentum is approximated with $\vec{p}_\textrm{T}^{\,H} \approx - \vec{p}_\textrm{T}^{\,Z}$ or $\vec{p}_\textrm{T}^{\,H} \approx - \vec{p}_\textrm{T}^{\,Z} - \sum \vec{p}_\textrm{T}^\textrm{\,jet}$, if at least one jet is in the event.

\begin{table}[htbp]
\centering
\caption{Summary of input variables for the BDTs in the $4\ell$ channel.}
\label{tab:inputsFourLepton}
\renewcommand\arraystretch{1.25}
\resizebox{\textwidth}{!}{
\begin{tabular}{l | l}
\hline\hline
Variable(s) & Description \\
\hline
$\pt^{\ell_0}$, $\pt^{\ell_1}$, $\pt^{\ell_2}$, $\pt^{\ell_3}$ & Transverse momentum for each lepton \\
\mHc & Invariant mass of the Higgs boson candidate \\
\mZc & Invariant mass of the \Z candidate \\
\ptfl & Transverse momentum of the 4-lepton system \\
\mfl & Invariant mass of the 4-lepton system \\
\mtt & Invariant mass of the $\tau$-lepton pair using the collinear approximation~\cite{Plehn:1999xi}, \\
& assuming the electrons and muons result from $\tau$-lepton decays \\
$n_\textrm{jets}$ & Number of jets \\
\met & Missing transverse momentum \\
\DPhillMET & Azimuthal separation between the dilepton system composed of $\ell_0$ and $\ell_1$ and the \met \\
\DPhiB & Azimuthal separation between $\ell_0$ and $\ell_1$ in the Higgs boson candidate's frame \\
\hline\hline
\end{tabular}
}
\end{table}

\paragraph{STXS categorisation:} Table~\ref{tab:inputsRegression} shows the input variables for the ANN regressing the \W boson's \pt. This ANN is utilised by the $3\ell$ channels for their differential measurement --- see Section~\ref{sec:stxs}.

\begin{table}[htbp]
\centering
\caption{Summary of input variables for the regression ANN in the $3\ell$ channel.}
\label{tab:inputsRegression}
\renewcommand\arraystretch{1.25}
\begin{tabular}{l | l}
\hline\hline
Variable(s) & Description \\
\hline
$\pt^{\ell_0}$, $\pt^{\ell_1}$, $\pt^{\ell_2}$ & Transverse momentum for each lepton \\
$\Delta{R}_{\ell_0\ell_1}$, $\Delta{R}_{\ell_0\ell_2}$, $\Delta{R}_{\ell_1\ell_2}$ & Dilepton angular separation for each combination of leptons \\
$m_{\ell_0\ell_1}$, $m_{\ell_0\ell_2}$, $m_{\ell_1\ell_2}$ & Dilepton invariant mass for each combination of leptons \\
$\mt^{\ell_0\ell_1}$, $\mt^{\ell_0\ell_2}$, $\mt^{\ell_1\ell_2}$ & Dilepton transverse mass for each combination of leptons \\
$\sum_{i=0}^2\pt^{\ell_i}$ & Scalar sum of the lepton transverse momenta \\
\met & Missing transverse momentum \\
$\Delta\phi_{\ell_0,\textrm{miss}}$, $\Delta\phi_{\ell_1,\textrm{miss}}$, $\Delta\phi_{\ell_2,\textrm{miss}}$ & Azimuthal separation between each lepton and the \met \\
\hline\hline
\end{tabular}
\end{table}

% End of text imported from the .//sections/app_inputs.tex input file

% The next lines are included from the .//sections/app_stxs.tex input file
%
\section{Additional plots for the STXS categorisation}
\label{app:stxs}

This appendix includes additional plots supporting the STXS categorisation scheme described in Section~\ref{sec:stxs}.

\begin{figure}[htbp]
\centering
\includegraphics[width=0.75\textwidth]{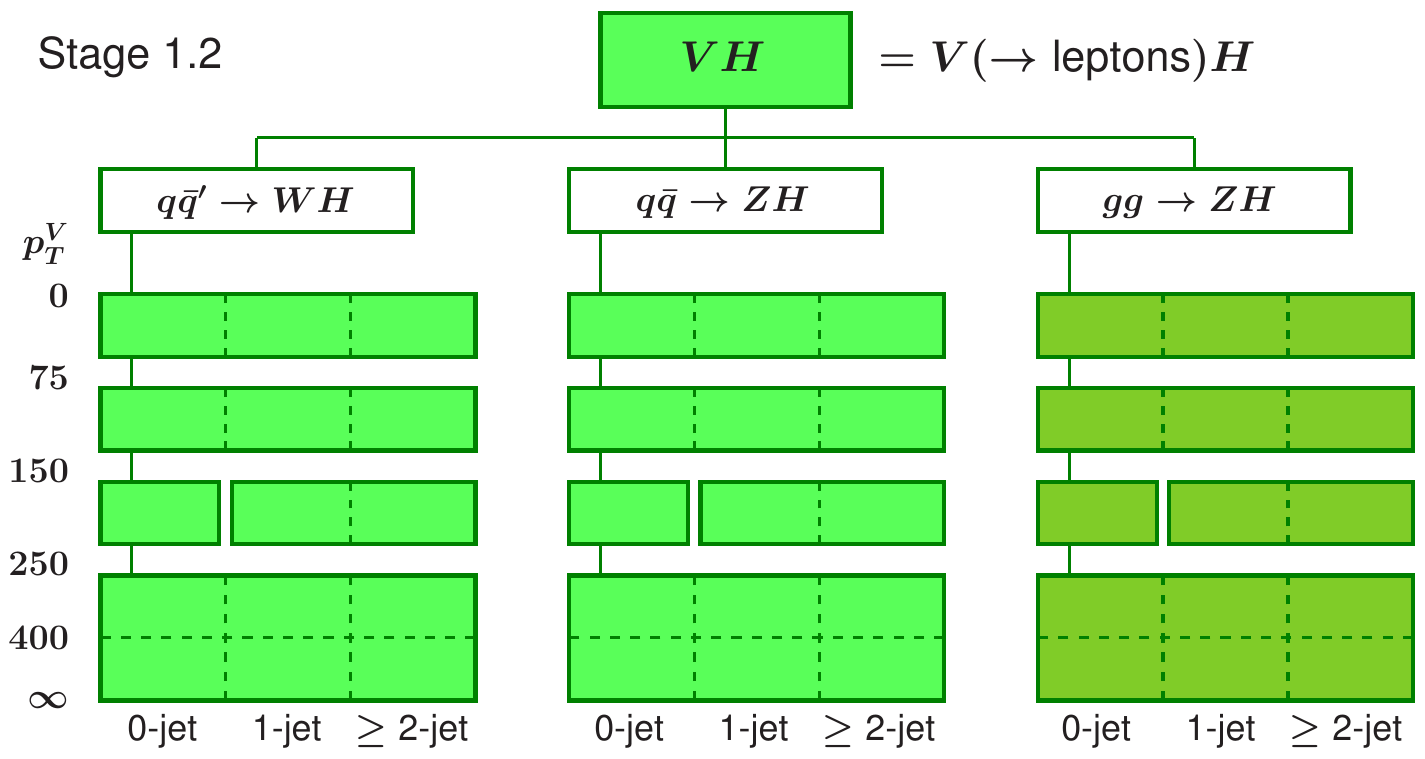}
\caption{The Stage~1.2 STXS categorisation scheme for \VH production~\cite{Berger:2019wnu}.}
\label{fig:STXSStage1_2}
\end{figure}

\begin{figure}[htbp]
\centering
\subfloat[]{
\includegraphics[width=0.32\textwidth]{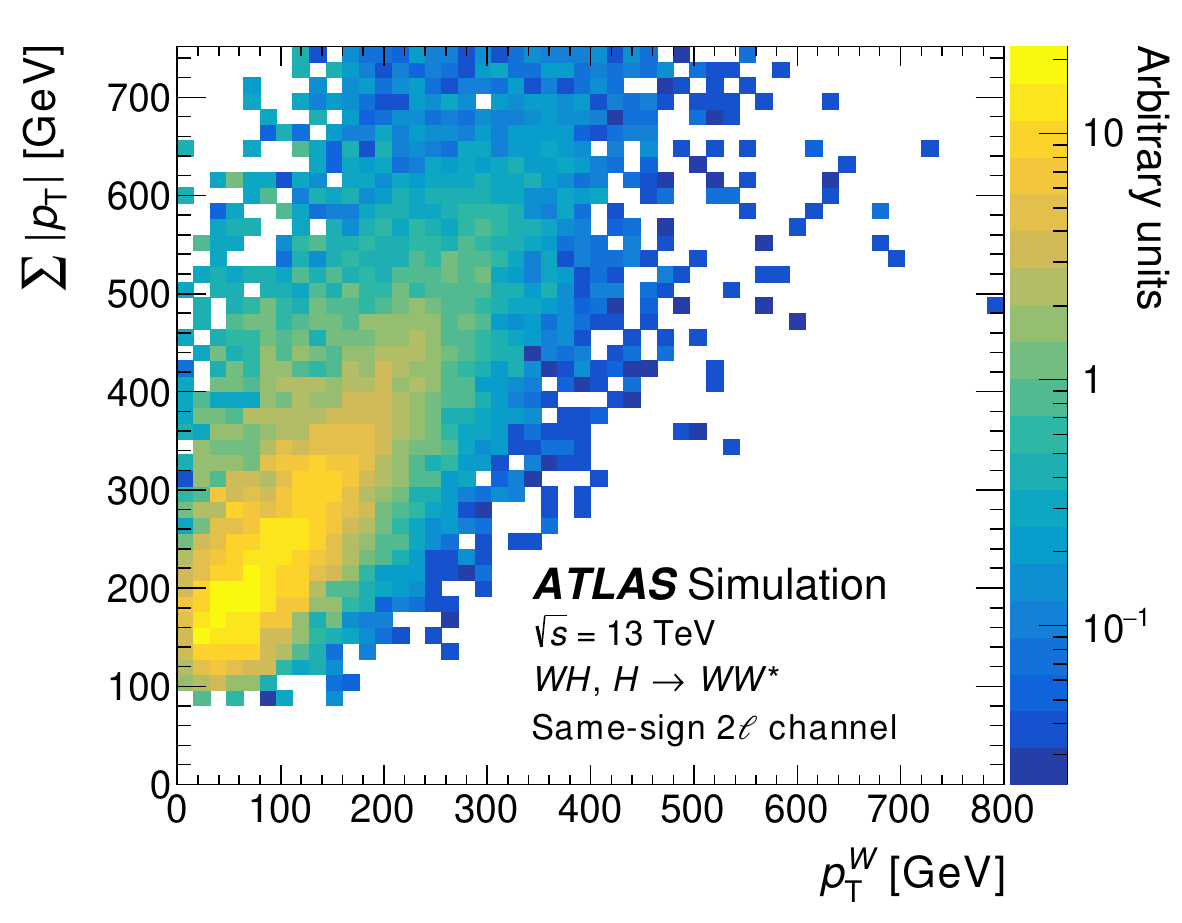}}
\subfloat[]{
\includegraphics[width=0.32\textwidth]{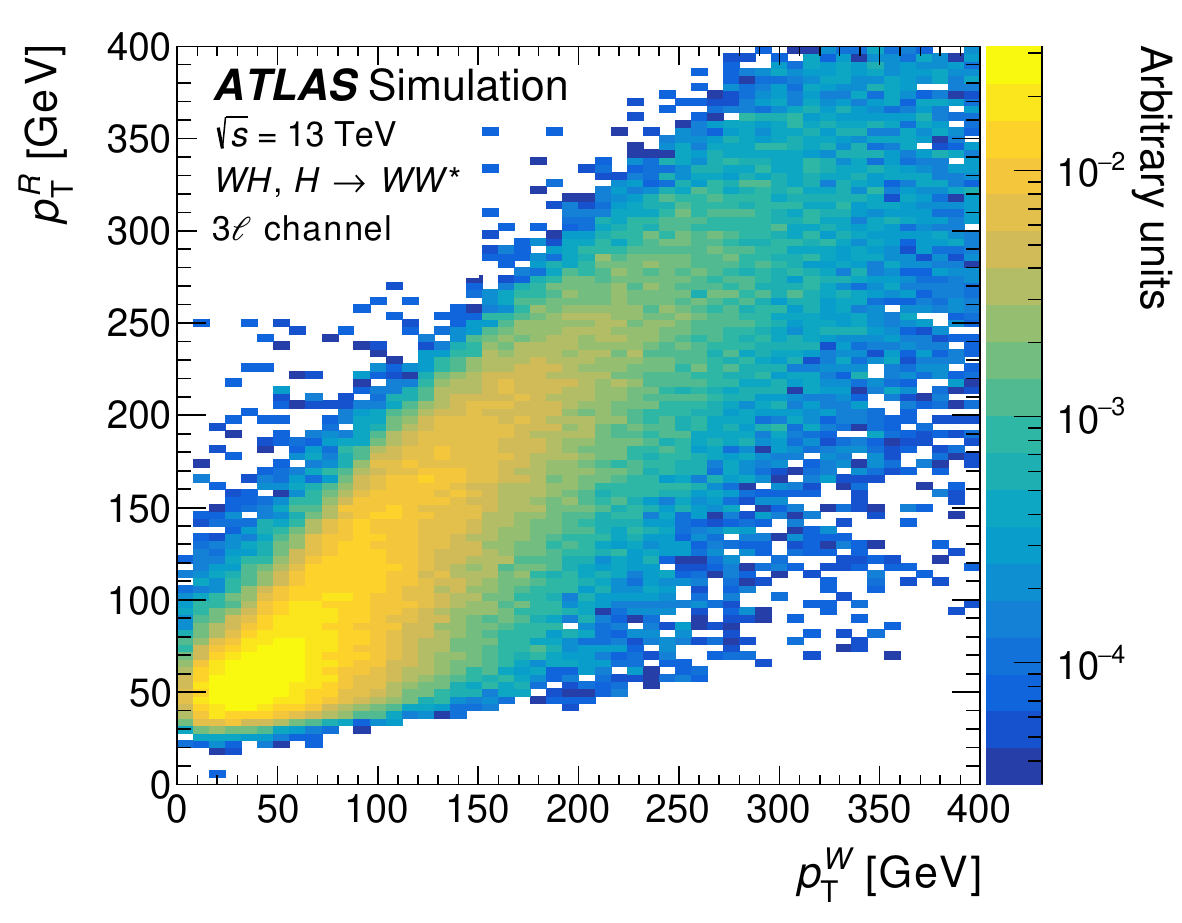}}
\subfloat[]{
\includegraphics[width=0.32\textwidth]{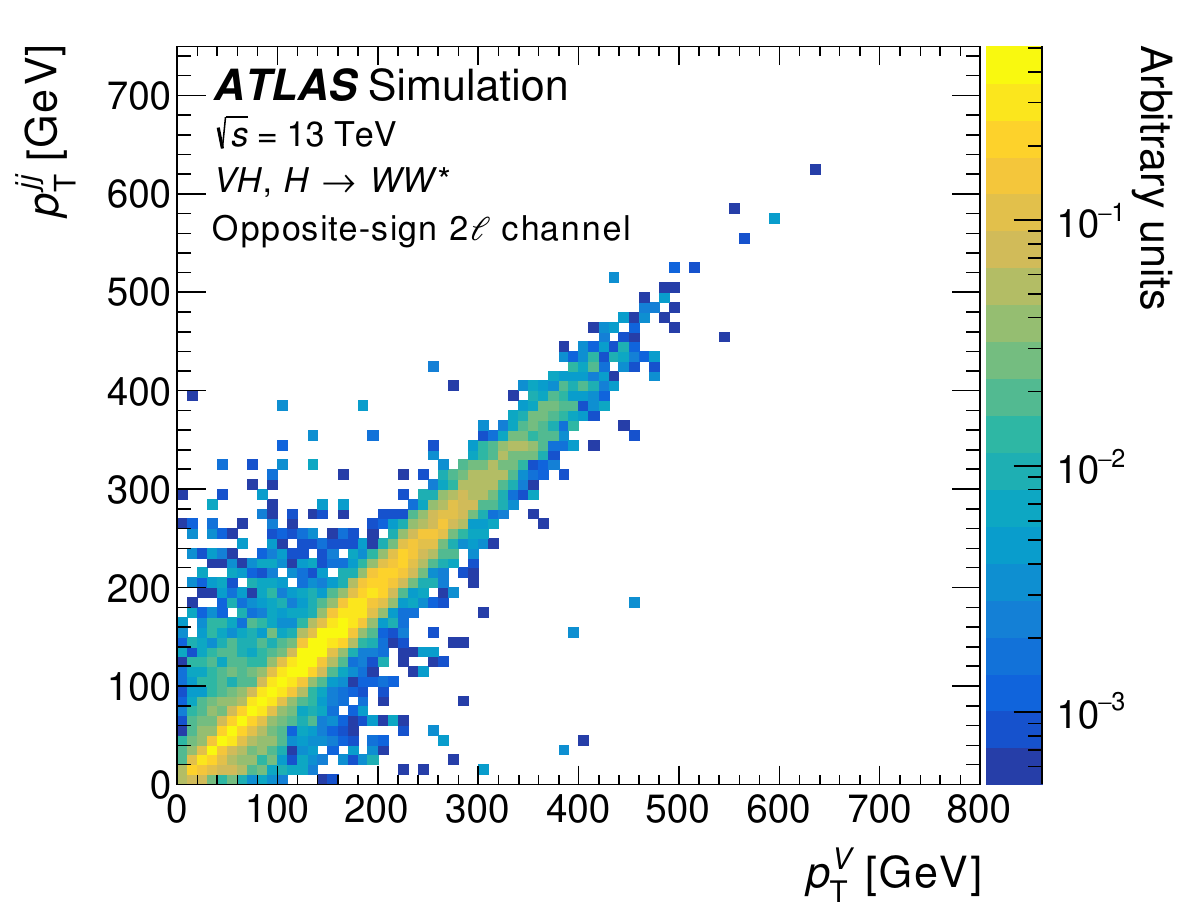}}
\\
\subfloat[]{
\includegraphics[width=0.32\textwidth]{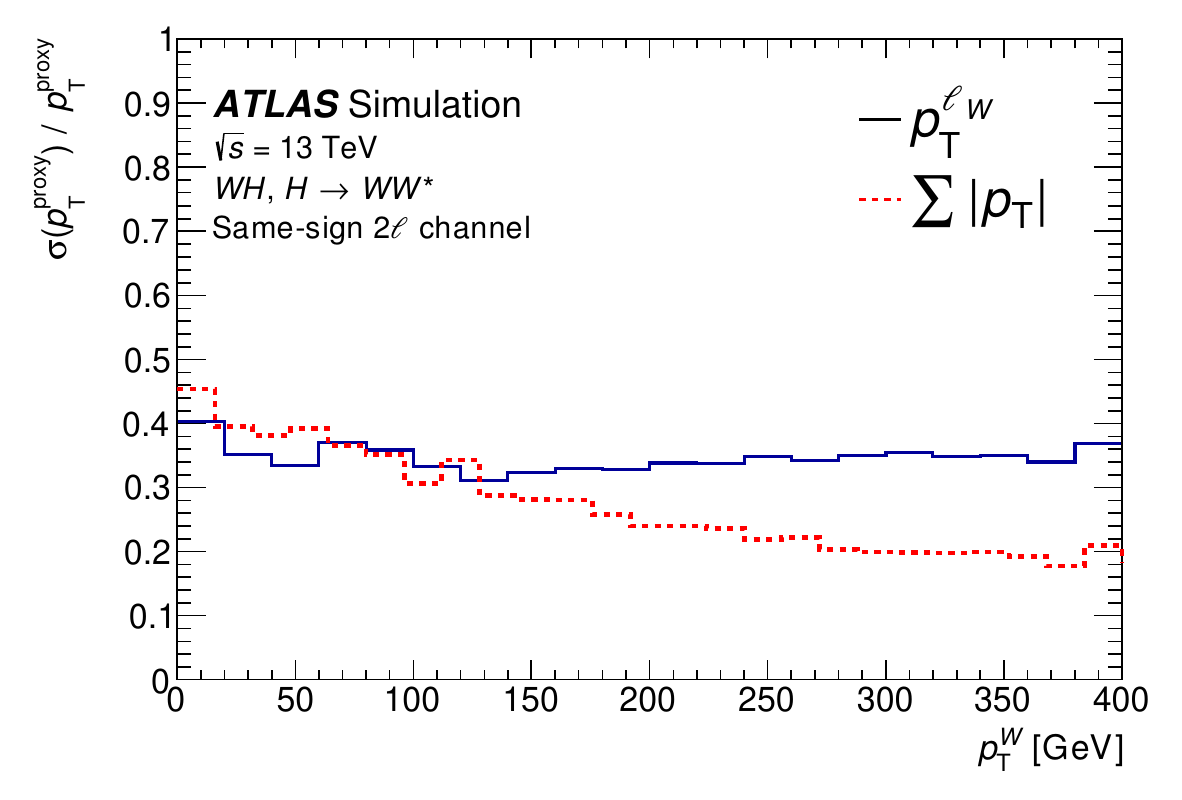}}
\subfloat[]{
\includegraphics[width=0.32\textwidth]{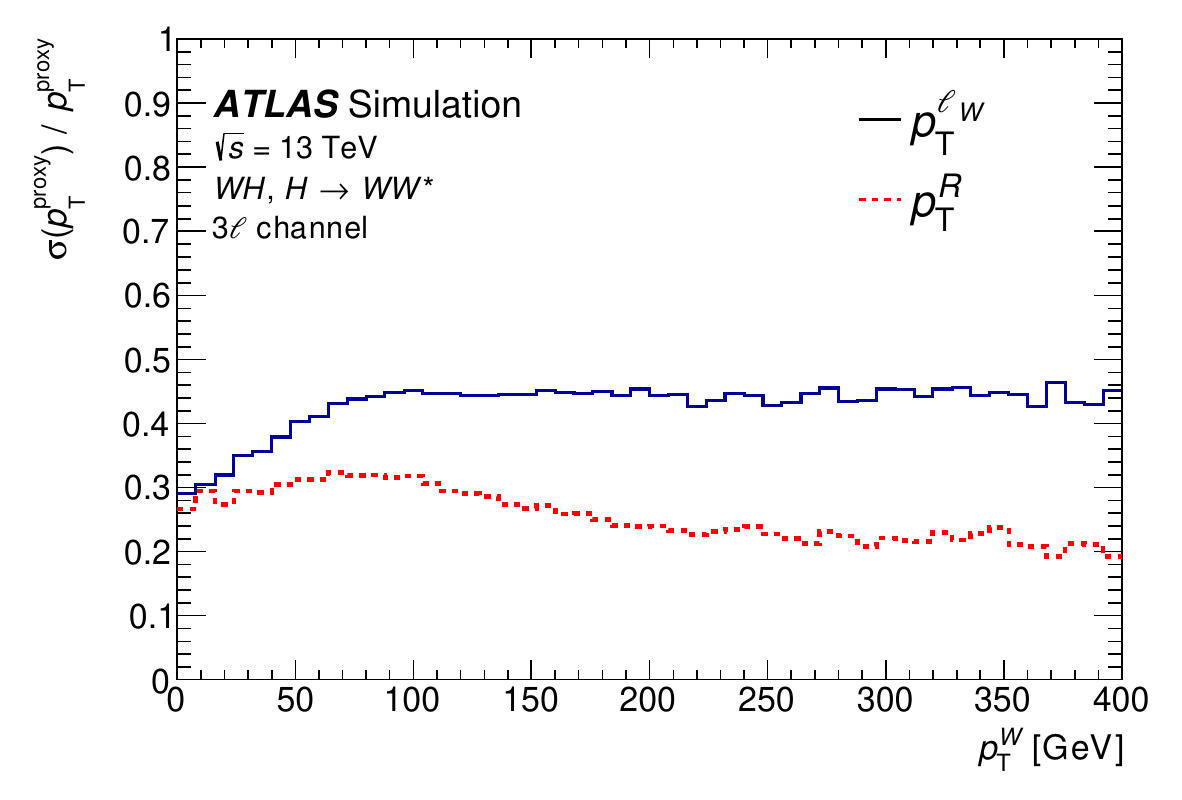}}
\subfloat[]{
\includegraphics[width=0.32\textwidth]{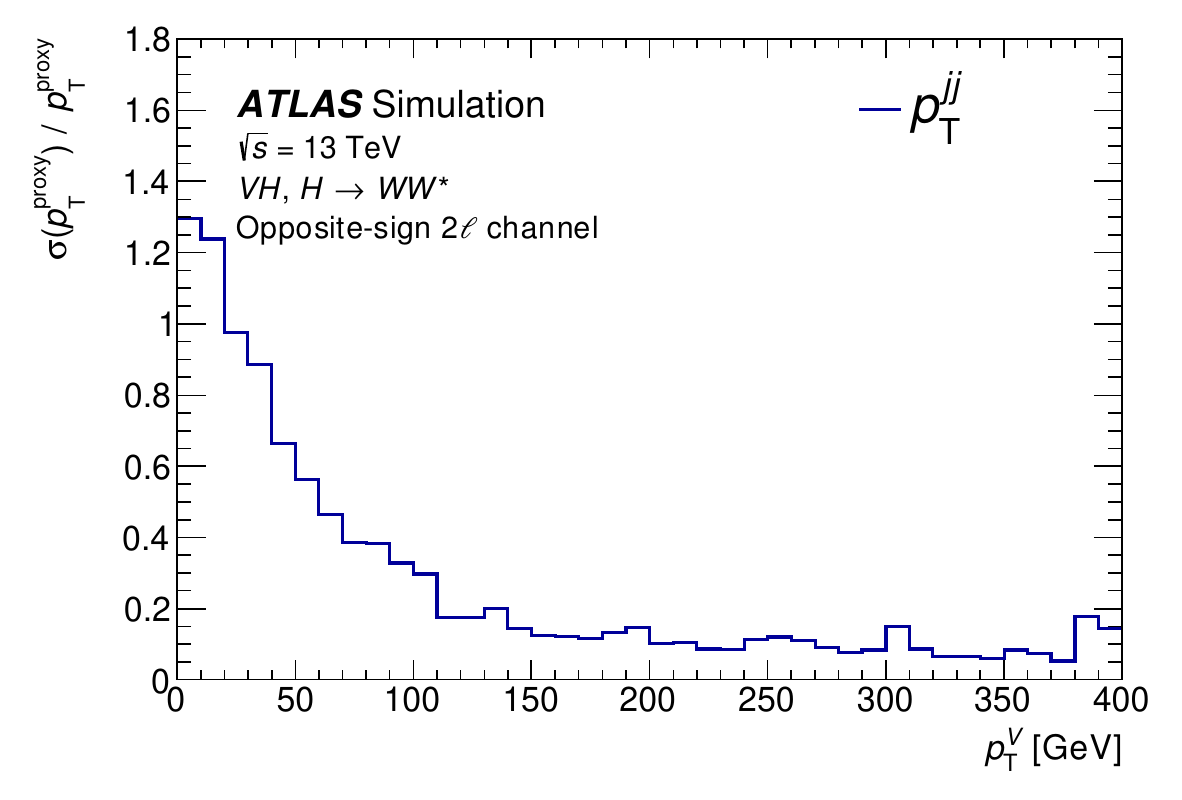}}
\caption{Performance of the \ptV proxies for several channels: (a), (b), (c) the correlation between the \pt of the associated boson and the proxy and (d), (e), (f) the fractional resolution of the proxy as a function of \ptV. Shown are (a), (d) the \sumpt of all objects in the \SSEE channel, (b), (e) the regression neural network output, \ptR, in the $3\ell$ \Z-dominated channel, and (c), (f) the dijet \pt, \ptjj, in the opposite-sign $2\ell$ channel. The \ptR resolution is compared with that obtained using as a proxy the \pt of the lepton identified as that from the \W boson decay, while the \sumpt resolution is compared with that obtained using as a proxy the highest \pt lepton.}
\label{fig:proxies}
\end{figure}

Figure~\ref{fig:STXSStage1_2} shows the Stage~1.2 STXS categorisation scheme for \VH production. Figure~\ref{fig:proxies} shows the correlation between \ptV and the proxy used for its reconstruction for several channels.

% End of text imported from the .//sections/app_stxs.tex input file

% The next lines are included from the .//sections/app_uncertainties.tex input file
%
\section{Breakdown of uncertainties contributing to observed results}
\label{app:uncertainties}

This appendix includes tables which show the breakdown of uncertainties contributing to the observed results presented in Section~\ref{sec:results}.

Tables~\ref{tab:breakdown_ptV} and \ref{tab:breakdown_STXS} show the fractional contributions from statistical and systematic sources to the total uncertainties in the observed values of the differential cross-sections, presented in Section~\ref{sec:results_stxs}.

\begin{table}[htbp]
\centering
\caption{Breakdown of the average contributions to the total uncertainties in the differential cross-sections measured in the \ptV scheme. The uncertainties of a single source, $\Delta_i$, are expressed in percentage of the total uncertainty, $\Delta$. Indentation is used to denote subcategories. Also shown is the total relative (non-fractional) uncertainty in the cross-section, $\Delta\,(\XS\times\BR_{\HWW}) / (\XS\times\BR_{\HWW})$, in percent. The quadrature sum of the individual sources may differ from the total uncertainty due to correlations.}
\label{tab:breakdown_ptV}
\resizebox{\textwidth}{!}{
\begin{tabular}{l | S[table-format=3.2] S[table-format=3.2] S[table-format=3.2] S[table-format=3.2]}
\hline\hline
& \multicolumn{4}{c}{\Gape[1mm][1mm]{$\frac{\Delta_i\,(\XS_{\VH}\,(X \leq \ptV < Y)\times\BR_{\HWW})}{\Delta\,(\XS_{\VH}(X \leq \ptV < Y)\times\BR_{\HWW})}$ [\%], $[X, Y)$: $X \leq \ptV < Y\,\si{\GeV}$}} \\
\cline{2-5}
\Gape[1mm][1mm]{Source} & {$\VH\,[0, 75)$} & {$\VH\,[75, 150)$} & {$\VH\,[150, 250)$} & {$\VH\,[250, \infty)$} \\
\hline
Statistical uncertainties in data & 95 & 95 & 87 & 81 \\
\,\,\,\, Statistical uncertainties in SR data & 92 & 93 & 87 & 81 \\
\,\,\,\, Statistical uncertainties in CR data & 22 & 18 & 5.9 & 5.7 \\
\hline
Systematic uncertainties & 33 & 32 & 49 & 44 \\
\,\,\,\, Statistical uncertainties in simulation & 20 & 22 & 24 & 33 \\
\,\,\,\, Experimental systematic uncertainties & 17 & 13 & 13 & 10 \\
\,\,\,\,\,\,\,\, Electrons & 3.8 & 1.6 & 1.5 & 1.2 \\
\,\,\,\,\,\,\,\, Muons & 6.6 & 3.8 & 6.6 & 4.9 \\
\,\,\,\,\,\,\,\, Jet energy scale & 4.0 & 2.9 & 6.5 & 5.6 \\
\,\,\,\,\,\,\,\, Jet energy resolution & 7.1 & 5.7 & 7.4 & 5.1 \\
\,\,\,\,\,\,\,\, Flavour tagging & 1.3 & 1.4 & 1.7 & 1.6 \\
\,\,\,\,\,\,\,\, Missing transverse momentum & 5.1 & 1.6 & 1.2 & 0.9 \\
\,\,\,\,\,\,\,\, Pile-up & 4.4 & 1.2 & 0.8 & 0.6 \\
\,\,\,\,\,\,\,\, Luminosity & 3.1 & 2.5 & 1.3 & 1.1 \\
\,\,\,\,\,\,\,\, Mis-identified leptons & 9.2 & 9.0 & 4.4 & 3.7 \\
\,\,\,\,\,\,\,\, Charge-flip electrons & 2.1 & 2.2 & 4.0 & 3.1 \\
\,\,\,\, Theoretical uncertainties & 16 & 16 & 21 & 11 \\
\,\,\,\,\,\,\,\, \WH & 0.6 & 0.7 & 1.1 & 0.6 \\
\,\,\,\,\,\,\,\, \ZH & 0.6 & 0.7 & 1.2 & 0.6 \\
\,\,\,\,\,\,\,\, STXS bin migration & 2.1 & 4.2 & 3.2 & 1.7 \\
\,\,\,\,\,\,\,\, Other $H$ & 0.6 & 1.3 & 7.7 & 7.3 \\
\,\,\,\,\,\,\,\, \WW & 1.2 & 4.4 & 12.1 & 2.2 \\
\,\,\,\,\,\,\,\, \WZg 0-jet & 2.5 & 5.3 & 5.0 & 4.0 \\
\,\,\,\,\,\,\,\, \WZg $\geq$1-jets & 2.2 & 3.1 & 3.6 & 2.5 \\
\,\,\,\,\,\,\,\, \ZZg & 14 & 11 & 6.3 & 2.0 \\
\,\,\,\,\,\,\,\, \VVV & 4.5 & 2.8 & 4.5 & 2.7 \\
\,\,\,\,\,\,\,\, Top & 2.3 & 8.4 & 4.8 & 5.0 \\
\,\,\,\,\,\,\,\, \Zjets & 0.6 & 3.0 & 1.8 & 2.9 \\
\,\,\,\, RNN shape uncertainty for \WZg & 1.1 & 8.0 & 36 & 7.7 \\
\hline
Total relative uncertainty (non-fractional) & 40 & 41 & 170 & 1000 \\
\hline\hline
\end{tabular}
}
\end{table}

\begin{table}[htbp]
\centering
\caption{Breakdown of the average contributions to the total uncertainties in the differential cross-sections measured in the STXS scheme. The uncertainties of a single source, $\Delta_i$, are expressed in percentage of the total uncertainty, $\Delta$. Indentation is used to denote subcategories. Also shown is the total relative (non-fractional) uncertainty in the cross-section, $\Delta\,(\XS\times\BR_{\HWW}) / (\XS\times\BR_{\HWW})$, in percent. The quadrature sum of the individual sources may differ from the total uncertainty due to correlations.}
\label{tab:breakdown_STXS}
\resizebox{\textwidth}{!}{
\begin{tabular}{l | S[table-format=4.2] S[table-format=3.2] S[table-format=3.2] S[table-format=3.2] S[table-format=3.2] S[table-format=3.2,table-align-comparator=false,table-comparator=true] S[table-format=3.2]}
\hline\hline
& \multicolumn{7}{c}{\Gape[1mm][1mm]{$\frac{\Delta_i\,(\XS_\textrm{STXS category}\times\BR_{\HWW})}{\Delta\,(\XS_\textrm{STXS category}\times\BR_{\HWW})}$ [\%], $[X, Y)$: $X \leq (\ptV\textrm{ or }\mjj) < Y\,\si{\GeV}$}} \\
\cline{2-8}
\Gape[1mm][1mm]{Source} & {$\lvH\,[0, 75)$} & {$\lvH\,[75, 150)$} & {$\lvH\,[150, \infty)$} & {$\llH\,[0, 75)$} & {$\llH\,[75, 150)$} & {$\llH\,[150, \infty)$} & {$\qqH\,[60, 120)$} \\
\hline
Statistical uncertainties in data & 93 & 95 & 83 & 94 & 95 & 96 & 82 \\
\,\,\,\, Statistical uncertainties in SR data & 90 & 94 & 83 & 92 & 94 & 95 & 82 \\
\,\,\,\, Statistical uncertainties in CR data & 23 & 15 & 8.1 & 14 & 12 & 7.6 & 5.7 \\
\hline
Systematic uncertainties & 36 & 32 & 56 & 35 & 32 & 30 & 57 \\
\,\,\,\, Statistical uncertainties in simulation & 22 & 24 & 18 & 31 & 28 & 20 & 28 \\
\,\,\,\, Experimental systematic uncertainties & 23 & 18 & 11 & 8.7 & 8.3 & 20 & 23 \\
\,\,\,\,\,\,\,\, Electrons & 2.6 & 1.2 & 1.6 & 3.6 & 2.0 & 0.9 & 2.1 \\
\,\,\,\,\,\,\,\, Muons & 2.6 & 1.7 & 2.0 & 3.1 & 2.5 & 19 & 2.4 \\
\,\,\,\,\,\,\,\, Jet energy scale & 4.8 & 3.7 & 2.4 & 1.7 & 2.2 & 1.3 & 20 \\
\,\,\,\,\,\,\,\, Jet energy resolution & 11 & 9.4 & 2.2 & 1.9 & 1.0 & 0.8 & 10 \\
\,\,\,\,\,\,\,\, Flavour tagging & 0.9 & 1.5 & 2.3 & 0.8 & 1.3 & 1.2 & 3.5 \\
\,\,\,\,\,\,\,\, Missing transverse momentum & 3.4 & 1.9 & 0.9 & 4.0 & 1.7 & 1.2 & 1.1 \\
\,\,\,\,\,\,\,\, Pile-up & 3.2 & 1.1 & 0.4 & 1.4 & 1.1 & 0.3 & 0.8 \\
\,\,\,\,\,\,\,\, Luminosity & 1.2 & 0.9 & 1.3 & 2.3 & 2.3 & 0.9 & 2.0 \\
\,\,\,\,\,\,\,\, Mis-identified leptons & 18 & 13 & 7.3 & 4.2 & 6.2 & 4.2 & 7.1 \\
\,\,\,\,\,\,\,\, Charge-flip electrons & 3.8 & 4.2 & 6.3 & 0.9 & 0.8 & 0.1 & 1.3 \\
\,\,\,\, Theoretical uncertainties & 12 & 12 & 19 & 15 & 13 & 11 & 44 \\
\,\,\,\,\,\,\,\, \WH & 0.5 & 0.8 & 0.7 & 0.9 & 0.8 & <0.1 & 0.7 \\
\,\,\,\,\,\,\,\, \ZH & 0.3 & 0.3 & 0.5 & 1.0 & 0.7 & 0.3 & 0.7 \\
\,\,\,\,\,\,\,\, STXS bin migration & 1.6 & 2.5 & 2.0 & 2.5 & 3.6 & 4.7 & 1.0 \\
\,\,\,\,\,\,\,\, Other $H$ & 0.3 & 0.3 & 0.5 & 0.9 & 0.8 & <0.1 & 14.2 \\
\,\,\,\,\,\,\,\, \WW & 0.6 & 3.7 & 0.7 & 0.9 & 0.8 & 0.1 & 19 \\
\,\,\,\,\,\,\,\, \WZg 0-jet & 5.7 & 6.8 & 11 & 1.1 & 0.9 & 0.4 & 0.7 \\
\,\,\,\,\,\,\,\, \WZg $\geq$1-jets & 3.8 & 6.2 & 9.5 & 1.0 & 0.8 & 0.5 & 1.9 \\
\,\,\,\,\,\,\,\, \ZZg & 3.9 & 1.8 & 2.7 & 14 & 12 & 9.6 & 0.6 \\
\,\,\,\,\,\,\,\, \VVV & 8.4 & 2.2 & 10 & 1.8 & 0.8 & 1.1 & 0.6 \\
\,\,\,\,\,\,\,\, Top & 2.1 & 1.3 & 0.7 & 1.0 & 0.8 & 0.5 & 32 \\
\,\,\,\,\,\,\,\, \Zjets & 0.7 & 0.9 & 0.5 & 0.9 & 0.8 & <0.1 & 4.5 \\
\,\,\,\, RNN shape uncertainty for \WZg & 2.2 & 6.9 & 49 & 0.9 & 0.9 & 0.9 & 0.6 \\
\hline
Total relative uncertainty (non-fractional) & 19000 & 88 & 140 & 40 & 47 & 140 & 93 \\
\hline\hline
\end{tabular}
}
\end{table}

% End of text imported from the .//sections/app_uncertainties.tex input file

%
%
%
\clearpage
\printbibliography

@Article{SOFT-2022-02_updated,
    author         = "{ATLAS Collaboration}",
    title          = "{Software and computing for Run~3 of the ATLAS experiment at the LHC}",
    journal        = "Eur. Phys. J. C",
    volume         = "85",
    year           = "2024",
    pages          = "234",
    doi            = "10.1140/epjc/s10052-024-13701-w",
    reportNumber   = "CERN-EP-2024-100",
    eprint         = "2404.06335",
    archivePrefix  = "arXiv",
    primaryClass   = "hep-ex",
}

@article{Bredenstein:2006nk_fixed, 
      author         = "Bredenstein, A. and Denner, Ansgar and Dittmaier, S. and Weber, M. M.",
      title          = "{Precision calculations for the Higgs decays \(H \to ZZ/WW \to 4\) leptons}",
      booktitle      = "{Proceedings, 8th DESY Workshop on Elementary Particle
                        Theory: Loops and Legs in Quantum Field Theory}",
      venue          = "Eisenach, Germany",
      eventdate      = "2006-04-23/2006-04-28",
      journal        = "Nucl. Phys. B Proc. Suppl.",
      volume         = "160",
      year           = "2006",
      pages          = "131-135",
      doi            = "10.1016/j.nuclphysbps.2006.09.104",
      eprint         = "hep-ph/0607060",
      archivePrefix  = "arXiv",
      primaryClass   = "hep-ph",
      reportNumber   = "PSI-PR-06-09",
      SLACcitation   = "%%CITATION = HEP-PH/0607060;%%"
}

@article{Lee:1973iz,
    author = "Lee, T. D.",
    editor = "Feinberg, G.",
    title = "{A Theory of Spontaneous $T$ Violation}",
    doi = "10.1103/PhysRevD.8.1226",
    journal = "Phys. Rev. D",
    volume = "8",
    pages = "1226--1239",
    year = "1973"
}

@article{Branco:2011iw,
    author = "Branco, G. C. and Ferreira, P. M. and Lavoura, L. and Rebelo, M. N. and Sher, Marc and Silva, Joao P.",
    title = "{Theory and phenomenology of two-Higgs-doublet models}",
    eprint = "1106.0034",
    archivePrefix = "arXiv",
    primaryClass = "hep-ph",
    doi = "10.1016/j.physrep.2012.02.002",
    journal = "Phys. Rept.",
    volume = "516",
    pages = "1--102",
    year = "2012"
}

@article{deBlas:2012qp,
    author = "de Blas, J. and Lizana, J. M. and P{\'e}rez-Victoria, M.",
    title = "{Combining searches of Z$^\prime$ and W$^\prime$ bosons}",
    eprint = "1211.2229",
    archivePrefix = "arXiv",
    primaryClass = "hep-ph",
    doi = "10.1007/JHEP01(2013)166",
    journal = "JHEP",
    volume = "01",
    pages = "166",
    year = "2013"
}

@article{Pappadopulo:2014qza,
    author = "Pappadopulo, Duccio and Thamm, Andrea and Torre, Riccardo and Wulzer, Andrea",
    title = "{Heavy Vector Triplets: Bridging Theory and Data}",
    eprint = "1402.4431",
    archivePrefix = "arXiv",
    primaryClass = "hep-ph",
    doi = "10.1007/JHEP09(2014)060",
    journal = "JHEP",
    volume = "09",
    pages = "060",
    year = "2014"
}

@inproceedings{Andersen:2016qtm,
    author = "Andersen, J. R. and others",
    title = "{Les Houches 2015: Physics at TeV Colliders Standard Model Working Group Report}",
    booktitle = "{9th Les Houches Workshop on Physics at TeV Colliders}",
    eprint = "1605.04692",
    archivePrefix = "arXiv",
    primaryClass = "hep-ph",
    reportNumber = "FERMILAB-CONF-16-175-PPD-T",
    month = "5",
    year = "2016"
}

@article{Berger:2019wnu,
    author = "Berger, Nicolas and others",
    title = "{Simplified Template Cross Sections - Stage 1.1}",
    eprint = "1906.02754",
    archivePrefix = "arXiv",
    primaryClass = "hep-ph",
    reportNumber = "LHCHXSWG-2019-003, DESY-19-070",
    month = "6",
    year = "2019"
}

@article{Abadi:2016kxi,
    author = "Abadi, Mart\'\i{}n and others",
    title = "{TensorFlow: A system for large-scale machine learning}",
    eprint = "1605.08695",
    archivePrefix = "arXiv",
    primaryClass = "cs.DC",
    month = "5",
    year = "2016"
}

@article{Cascioli:2013gfa,
    author = {Cascioli, F. and H\"oche, S. and Krauss, F. and Maierh\"ofer, P. and Pozzorini, S. and Siegert, F.},
    title = "{Precise Higgs-background predictions: merging NLO QCD and squared quark-loop corrections to four-lepton + 0,1 jet production}",
    eprint = "1309.0500",
    archivePrefix = "arXiv",
    primaryClass = "hep-ph",
    reportNumber = "IPPP-13-66, DCPT-13-132, MCNET-13-12, SLAC-PUB-15714, ZU-TH-15-13, LPN13-056, FR-PHENO-2013-007",
    doi = "10.1007/JHEP01(2014)046",
    journal = "JHEP",
    volume = "01",
    pages = "046",
    year = "2014"
}

@article{Caola:2015rqy,
    author = {Caola, Fabrizio and Melnikov, Kirill and R\"ontsch, Raoul and Tancredi, Lorenzo},
    title = "{QCD corrections to $W^+W^-$ production through gluon fusion}",
    eprint = "1511.08617",
    archivePrefix = "arXiv",
    primaryClass = "hep-ph",
    reportNumber = "CERN-PH-TH-2015-276, TTP15-041",
    doi = "10.1016/j.physletb.2016.01.046",
    journal = "Phys. Lett. B",
    volume = "754",
    pages = "275--280",
    year = "2016"
}

@article{Grazzini:2018owa,
    author = "Grazzini, Massimiliano and Kallweit, Stefan and Wiesemann, Marius and Yook, Jeong Yeon",
    title = "{$ZZ$ production at the LHC: NLO QCD corrections to the loop-induced gluon fusion channel}",
    eprint = "1811.09593",
    archivePrefix = "arXiv",
    primaryClass = "hep-ph",
    reportNumber = "CERN-TH-2018-240, ZU-TH 42/18",
    doi = "10.1007/JHEP03(2019)070",
    journal = "JHEP",
    volume = "03",
    pages = "070",
    year = "2019"
}

@article{Binoth:2008kt,
    author = "Binoth, T. and Ossola, G. and Papadopoulos, C. G. and Pittau, R.",
    title = "{NLO QCD corrections to tri-boson production}",
    eprint = "0804.0350",
    archivePrefix = "arXiv",
    primaryClass = "hep-ph",
    doi = "10.1088/1126-6708/2008/06/082",
    journal = "JHEP",
    volume = "06",
    pages = "082",
    year = "2008"
}

@article{Broggio:2019ewu,
    author = "Broggio, Alessandro and Ferroglia, Andrea and Frederix, Rikkert and Pagani, Davide and Pecjak, Benjamin D. and Tsinikos, Ioannis",
    title = "{Top-quark pair hadroproduction in association with a heavy boson at NLO+NNLL including EW corrections}",
    eprint = "1907.04343",
    archivePrefix = "arXiv",
    primaryClass = "hep-ph",
    reportNumber = "TUM-HEP-1208/19, LU-TP 19-30, IPPP/19/57",
    doi = "10.1007/JHEP08(2019)039",
    journal = "JHEP",
    volume = "08",
    pages = "039",
    year = "2019"
}

@article{Campbell:2013yla,
    author = {Campbell, John and Ellis, R. Keith and R\"ontsch, Raoul},
    title = "{Single top production in association with a Z boson at the LHC}",
    eprint = "1302.3856",
    archivePrefix = "arXiv",
    primaryClass = "hep-ph",
    reportNumber = "FERMILAB-PUB-13-043-T",
    doi = "10.1103/PhysRevD.87.114006",
    journal = "Phys. Rev. D",
    volume = "87",
    pages = "114006",
    year = "2013"
}

@article{Maltoni:2019aot,
    author = "Maltoni, Fabio and Mantani, Luca and Mimasu, Ken",
    title = "{Top-quark electroweak interactions at high energy}",
    eprint = "1904.05637",
    archivePrefix = "arXiv",
    primaryClass = "hep-ph",
    reportNumber = "CP3-19-16",
    doi = "10.1007/JHEP10(2019)004",
    journal = "JHEP",
    volume = "10",
    pages = "004",
    year = "2019"
}

@article{Hochreiter:1997yld,
    author = {Hochreiter, Sepp and Schmidhuber, J\"urgen},
    title = "{Long Short-Term Memory}",
    doi = "10.1162/neco.1997.9.8.1735",
    journal = "Neural Comput.",
    volume = "9",
    number = "8",
    pages = "1735--1780",
    year = "1997"
}

@article{Plehn:1999xi,
    author = "Plehn, T. and Rainwater, D. and Zeppenfeld, D.",
    title = "{A Method for identifying $H\to\tau^+\tau^-\to e^\pm \mu^\mp p_T$ at the CERN LHC}",
    eprint = "hep-ph/9911385",
    archivePrefix = "arXiv",
    reportNumber = "MADPH-99-1142, FERMILAB-PUB-99-290-T",
    doi = "10.1103/PhysRevD.61.093005",
    journal = "Phys. Rev. D",
    volume = "61",
    pages = "093005",
    year = "2000"
}

@article{Hoeche:2009xc,
    author = "H{\"o}che, Stefan and Schumann, Steffen and Siegert, Frank",
    title = "{Hard photon production and matrix-element parton-shower merging}",
    eprint = "0912.3501",
    archivePrefix = "arXiv",
    primaryClass = "hep-ph",
    reportNumber = "ZU-TH-19-09, IPPP-09-94, DCPT-09-188, HD-THEP-09-28, MCNET-09-18",
    doi = "10.1103/PhysRevD.81.034026",
    journal = "Phys. Rev. D",
    volume = "81",
    pages = "034026",
    year = "2010"
}

@Article{Evans:2008zzb,
      author         = "Evans, Lyndon and Bryant, Philip",
      title          = "{LHC Machine}",
      journal        = "JINST",
      volume         = "3",
      pages          = "S08001",
      doi            = "10.1088/1748-0221/3/08/S08001",
      year           = "2008",
      SLACcitation   = "%%CITATION = JINST,3,S08001;%%",
}

@Article{Cacciari:2008gp,
     author    = "Cacciari, Matteo and Salam, Gavin P. and Soyez, Gregory",
     title     = "{The anti-\(k_{t}\) jet clustering algorithm}",
     journal   = "JHEP",
     volume    = "04",
     year      = "2008",
     pages     = "063",
     eprint    = "0802.1189",
     archivePrefix = "arXiv",
     primaryClass  =  "hep-ph",
     doi       = "10.1088/1126-6708/2008/04/063",
     SLACcitation  = "%%CITATION = 0802.1189;%%"
}

@Article{Fastjet,
      author         = "Cacciari, Matteo and Salam, Gavin P. and Soyez, Gregory",
      title          = "{FastJet user manual}",
      journal        = "Eur. Phys. J. C",
      volume         = "72",
      year           = "2012",
      pages          = "1896",
      doi            = "10.1140/epjc/s10052-012-1896-2",
      eprint         = "1111.6097",
      archivePrefix  = "arXiv",
      primaryClass   = "hep-ph",
      reportNumber   = "CERN-PH-TH-2011-297",
      SLACcitation   = "%%CITATION = ARXIV:1111.6097;%%"
}

@Article{Butterworth:2015oua,
      author         = "Butterworth, Jon and others",
      title          = "{PDF4LHC recommendations for LHC Run II}",
      journal        = "J. Phys. G",
      volume         = "43",
      year           = "2016",
      pages          = "023001",
      doi            = "10.1088/0954-3899/43/2/023001",
      eprint         = "1510.03865",
      archivePrefix  = "arXiv",
      primaryClass   = "hep-ph",
      reportNumber   = "OUTP-15-17P, SMU-HEP-15-12, TIF-UNIMI-2015-14,
                        LCTS-2015-27, CERN-PH-TH-2015-249",
      SLACcitation   = "%%CITATION = ARXIV:1510.03865;%%"
}

@article{Ball:2012cx,
      author         = "{NNPDF Collaboration} and Ball, Richard D. and others",
      title          = "{Parton distributions with LHC data}",
      journal        = "Nucl. Phys. B",
      volume         = "867",
      year           = "2013",
      pages          = "244",
      doi            = "10.1016/j.nuclphysb.2012.10.003",
      eprint         = "1207.1303",
      archivePrefix  = "arXiv",
      primaryClass   = "hep-ph",
      reportNumber   = "EDINBURGH-2012-08, IFUM-FT-997, FR-PHENO-2012-014,
                        RWTH-TTK-12-25, CERN-PH-TH-2012-037, SFB-CPP-12-47\,
                        --CERN-PH-TH-2012-037",
      SLACcitation   = "%%CITATION = ARXIV:1207.1303;%%"
}

@article{Ball:2014uwa,
      author         = "{NNPDF Collaboration} and Ball, Richard D. and others",
      title          = "{Parton distributions for the LHC run II}",
      journal        = "JHEP",
      volume         = "04",
      year           = "2015",
      pages          = "040",
      doi            = "10.1007/JHEP04(2015)040",
      eprint         = "1410.8849",
      archivePrefix  = "arXiv",
      primaryClass   = "hep-ph",
      reportNumber   = "EDINBURGH-2014-15, IFUM-1034-FT, CERN-PH-TH-2013-253,
                        OUTP-14-11P, CAVENDISH-HEP-14-11",
      SLACcitation   = "%%CITATION = ARXIV:1410.8849;%%"
}

@article{Czakon:2012pz,
      author         = "Czakon, Michal and Mitov, Alexander",
      title          = "{NNLO corrections to top pair production at hadron
                        colliders: the quark-gluon reaction}",
      journal        = "JHEP",
      volume         = "01",
      pages          = "080",
      doi            = "10.1007/JHEP01(2013)080",
      year           = "2013",
      eprint         = "1210.6832",
      archivePrefix  = "arXiv",
      primaryClass   = "hep-ph",
      SLACcitation   = "%%CITATION = ARXIV:1210.6832;%%",
}

@Article{Sjostrand:2007gs,
	author    = "Sj{\"o}strand, T. and Mrenna, S. and Skands, P.",
	title     = "{A brief introduction to PYTHIA 8.1}",
	journal   = "Comput. Phys. Commun.",
	volume    = "178",
	year      = "2008",
	pages     = "852-867",
	eprint    = "0710.3820",
	archivePrefix = "arXiv",
	primaryClass  =  "hep-ph",
	doi       = "10.1016/j.cpc.2008.01.036",
	SLACcitation  = "%%CITATION = 0710.3820;%%"
}

@article{Sjostrand:2014zea,
      author         = "Sj{\"o}strand, Torbj{\"o}rn and Ask, Stefan and Christiansen,
                        Jesper R. and Corke, Richard and Desai, Nishita and Ilten,
                        Philip and Mrenna, Stephen and Prestel, Stefan and
                        Rasmussen, Christine O. and Skands, Peter Z.",
      title          = "{An introduction to PYTHIA 8.2}",
      journal        = "Comput. Phys. Commun.",
      volume         = "191",
      year           = "2015",
      pages          = "159",
      doi            = "10.1016/j.cpc.2015.01.024",
      eprint         = "1410.3012",
      archivePrefix  = "arXiv",
      primaryClass   = "hep-ph",
      reportNumber   = "LU-TP-14-36, MCNET-14-22, CERN-PH-TH-2014-190,
                        FERMILAB-PUB-14-316-CD, DESY-14-178, SLAC-PUB-16122,
                        --FERMILAB-PUB-14-316-CD",
      SLACcitation   = "%%CITATION = ARXIV:1410.3012;%%"
}

@Article{Lange:2001uf,
      author         = "Lange, D. J.",
      title          = "{The EvtGen particle decay simulation package}",
      booktitle      = "{Proceedings, 7th International Conference on B physics
                        at hadron machines (BEAUTY 2000)}",
      journal        = "Nucl. Instrum. Meth. A",
      volume         = "462",
      year           = "2001",
      pages          = "152",
      doi            = "10.1016/S0168-9002(01)00089-4",
      SLACcitation   = "%%CITATION = NUIMA,A462,152;%%"
}

@Article{Frixione:2008yi,
     author    = "Frixione, Stefano and Laenen, Eric and Motylinski, Patrick and White, Chris and Webber, Bryan R.",
     title     = "{Single-top hadroproduction in association with a \(W\) boson}",
     journal   = "JHEP",
     volume    = "07",
     year      = "2008",
     pages     = "029",
     eprint    = "0805.3067",
     archivePrefix = "arXiv",
     primaryClass  =  "hep-ph",
     doi       = "10.1088/1126-6708/2008/07/029"
}

@Article{Alwall:2014hca,
      author         = "Alwall, J. and Frederix, R. and Frixione, S. and Hirschi,
                        V. and Maltoni, F. and Mattelaer, O. and Shao, H. -S. and
                        Stelzer, T. and Torrielli, P. and Zaro, M.",
      title          = "{The automated computation of tree-level and
                        next-to-leading order differential cross sections, and
                        their matching to parton shower simulations}",
      journal        = "JHEP",
      volume         = "07",
      year           = "2014",
      pages          = "079",
      doi            = "10.1007/JHEP07(2014)079",
      eprint         = "1405.0301",
      archivePrefix  = "arXiv",
      primaryClass   = "hep-ph",
      reportNumber   = "CERN-PH-TH-2014-064, CP3-14-18, LPN14-066, MCNET-14-09,
                        ZU-TH-14-14",
      SLACcitation   = "%%CITATION = ARXIV:1405.0301;%%"
}

@Article{Bahr:2008pv,
      author         = "B{\"a}hr, M. and others",
      title          = "{Herwig++ physics and manual}",
      journal        = "Eur. Phys. J. C",
      volume         = "58",
      year           = "2008",
      pages          = "639",
      doi            = "10.1140/epjc/s10052-008-0798-9",
      eprint         = "0803.0883",
      archivePrefix  = "arXiv",
      primaryClass   = "hep-ph",
      reportNumber   = "CERN-PH-TH-2008-038, CAVENDISH-HEP-08-03, KA-TP-05-2008,
                        DCPT-08-22, IPPP-08-11, CP3-08-05",
      SLACcitation   = "%%CITATION = ARXIV:0803.0883;%%"
}

@Article{Bellm:2015jjp,
      author         = "Bellm, Johannes and others",
      title          = "{Herwig 7.0/Herwig++ 3.0 release note}",
      journal        = "Eur. Phys. J. C",
      volume         = "76",
      year           = "2016",
      number         = "4",
      pages          = "196",
      doi            = "10.1140/epjc/s10052-016-4018-8",
      eprint         = "1512.01178",
      archivePrefix  = "arXiv",
      primaryClass   = "hep-ph",
      reportNumber   = "CERN-PH-TH-2015-289, MAN-HEP-2015-15, IFJPAN-IV-2015-13,
                        HERWIG-2015-01, KA-TP-18-2015, DCPT-15-142, MCNET-15-28,
                        IPPP-15-71, --HERWIG-2015-01",
      SLACcitation   = "%%CITATION = ARXIV:1512.01178;%%"
}

@Article{Nason:2004rx,
      author         = "Nason, Paolo",
      title          = "{A new method for combining NLO QCD with shower Monte Carlo algorithms}",
      journal        = "JHEP",
      volume         = "11",
      pages          = "040",
      doi            = "10.1088/1126-6708/2004/11/040",
      year           = "2004",
      eprint         = "hep-ph/0409146",
      archivePrefix  = "arXiv",
}

@Article{Frixione:2007vw,
      author         = "Frixione, Stefano and Nason, Paolo and Oleari, Carlo",
      title          = "{Matching NLO QCD computations with parton shower
                        simulations: the POWHEG method}",
      journal        = "JHEP",
      volume         = "11",
      pages          = "070",
      doi            = "10.1088/1126-6708/2007/11/070",
      year           = "2007",
      eprint         = "0709.2092",
      archivePrefix  = "arXiv",
      primaryClass   = "hep-ph",
}

@Article{Alioli:2010xd,
      author         = "Alioli, Simone and Nason, Paolo and Oleari, Carlo and Re,
                        Emanuele",
      title          = "{A general framework for implementing NLO calculations in
                        shower Monte Carlo programs: the POWHEG BOX}",
      journal        = "JHEP",
      volume         = "06",
      pages          = "043",
      doi            = "10.1007/JHEP06(2010)043",
      year           = "2010",
      eprint         = "1002.2581",
      archivePrefix  = "arXiv",
      primaryClass   = "hep-ph",
}

@article{Hamilton:2012rf,
      author         = "Hamilton, Keith and Nason, Paolo and Oleari, Carlo and
                        Zanderighi, Giulia",
      title          = "{Merging H/W/Z + 0 and 1 jet at NLO with no merging
                        scale: a path to parton shower + NNLO matching}",
      journal        = "JHEP",
      volume         = "05",
      year           = "2013",
      pages          = "082",
      doi            = "10.1007/JHEP05(2013)082",
      eprint         = "1212.4504",
      archivePrefix  = "arXiv",
      primaryClass   = "hep-ph",
      reportNumber   = "CERN-PH-TH-2012-356",
      SLACcitation   = "%%CITATION = ARXIV:1212.4504;%%"
}

@article{Hamilton:2012np,
      author         = "Hamilton, Keith and Nason, Paolo and Zanderighi, Giulia",
      title          = "{MINLO: multi-scale improved NLO}",
      journal        = "JHEP",
      volume         = "10",
      year           = "2012",
      pages          = "155",
      doi            = "10.1007/JHEP10(2012)155",
      eprint         = "1206.3572",
      archivePrefix  = "arXiv",
      primaryClass   = "hep-ph",
      reportNumber   = "CERN-PH-TH-2012-166, OUTP-12-11P, MCNET-12-07",
      SLACcitation   = "%%CITATION = ARXIV:1206.3572;%%"
}

@article{Hartanto:2015uka,
      author         = "Hartanto, Heribertus B. and J{\"a}ger, Barbara and Reina,
                        Laura and Wackeroth, Doreen",
      title          = "{Higgs boson production in association with top quarks in
                        the POWHEG BOX}",
      journal        = "Phys. Rev. D",
      volume         = "91",
      year           = "2015",
      number         = "9",
      pages          = "094003",
      doi            = "10.1103/PhysRevD.91.094003",
      eprint         = "1501.04498",
      archivePrefix  = "arXiv",
      primaryClass   = "hep-ph",
      SLACcitation   = "%%CITATION = ARXIV:1501.04498;%%"
}

@article{Re:2010bp,
      author         = "Re, Emanuele",
      title          = "{Single-top \(Wt\)-channel production matched with parton
                        showers using the POWHEG method}",
      journal        = "Eur. Phys. J. C",
      volume         = "71",
      year           = "2011",
      pages          = "1547",
      doi            = "10.1140/epjc/s10052-011-1547-z",
      eprint         = "1009.2450",
      archivePrefix  = "arXiv",
      primaryClass   = "hep-ph",
      reportNumber   = "IPPP-10-74, DCPT-10-148",
      SLACcitation   = "%%CITATION = ARXIV:1009.2450;%%"
}

@article{Kidonakis:2010ux,
      author         = "Kidonakis, Nikolaos",
      title          = "{Two-loop soft anomalous dimensions for single top quark
                        associated production with a \(W^{-}\) or \(H^{-}\)}",
      journal        = "Phys. Rev. D",
      volume         = "82",
      year           = "2010",
      pages          = "054018",
      doi            = "10.1103/PhysRevD.82.054018",
      eprint         = "1005.4451",
      archivePrefix  = "arXiv",
      primaryClass   = "hep-ph",
      SLACcitation   = "%%CITATION = ARXIV:1005.4451;%%"
}

@inproceedings{Kidonakis:2013zqa,
      author         = "Kidonakis, Nikolaos",
      title          = "{Top Quark Production}",
      booktitle      = "{Proceedings, Helmholtz International Summer School on Physics of Heavy Quarks and Hadrons (HQ2013)}",
      eventdate      = "2013-07-15/2013-07-28",
      venue          = "JINR, Dubna, Russia",
      pages          = "139-168",
      doi            = "10.3204/DESY-PROC-2013-03/Kidonakis",
      eprint         = "1311.0283",
      archivePrefix  = "arXiv",
      primaryClass   = "hep-ph",
      reportNumber   = "KSU-HEP-110113",
      SLACcitation   = "%%CITATION = ARXIV:1311.0283;%%"
}

@article{Beneke:2011mq,
      author         = "Beneke, M. and Falgari, P. and Klein, S. and Schwinn, C.",
      title          = "{Hadronic top-quark pair production with NNLL threshold
                        resummation}",
      journal        = "Nucl. Phys. B",
      volume         = "855",
      year           = "2012",
      pages          = "695-741",
      doi            = "10.1016/j.nuclphysb.2011.10.021",
      eprint         = "1109.1536",
      archivePrefix  = "arXiv",
      primaryClass   = "hep-ph",
      reportNumber   = "TTK-11-38, ITP-UU-11-26, SPIN-11-19, FR-PHENO-2011-015,
                        SFB-CPP-11-49",
      SLACcitation   = "%%CITATION = ARXIV:1109.1536;%%"
}

@article{Anastasiou:2003ds,
      author         = "Anastasiou, Charalampos and Dixon, Lance and Melnikov,
                        Kirill and Petriello, Frank",
      title          = "{High-precision QCD at hadron colliders: Electroweak
                        gauge boson rapidity distributions at next-to-next-to leading order}",
      journal        = "Phys. Rev. D",
      volume         = "69",
      year           = "2004",
      pages          = "094008",
      doi            = "10.1103/PhysRevD.69.094008",
      eprint         = "hep-ph/0312266",
      archivePrefix  = "arXiv",
      reportNumber   = "SLAC-PUB-10288, UH-511-1042-03",
      SLACcitation   = "%%CITATION = HEP-PH/0312266;%%"
}

@article{Cacciari:2011hy,
      author         = "Cacciari, Matteo and Czakon, Michal and Mangano,
                        Michelangelo and Mitov, Alexander and Nason, Paolo",
      title          = "{Top-pair production at hadron colliders with
                        next-to-next-to-leading logarithmic soft-gluon
                        resummation}",
      journal        = "Phys. Lett. B",
      volume         = "710",
      pages          = "612-622",
      doi            = "10.1016/j.physletb.2012.03.013",
      year           = "2012",
      eprint         = "1111.5869",
      primaryClass   = "hep-ph",
      archivePrefix  = "arXiv",
      reportNumber   = "CERN-PH-TH-2011-277, TTK-11-54",
      SLACcitation   = "%%CITATION = ARXIV:1111.5869;%%",
}

@article{Czakon:2012zr,
      author         = "Czakon, Michal and Mitov, Alexander",
      title          = "{NNLO corrections to top-pair production at hadron
                        colliders: the all-fermionic scattering channels}",
      journal        = "JHEP",
      volume         = "12",
      pages          = "054",
      doi            = "10.1007/JHEP12(2012)054",
      year           = "2012",
      eprint         = "1207.0236",
      archivePrefix  = "arXiv",
      primaryClass   = "hep-ph",
      SLACcitation   = "%%CITATION = ARXIV:1207.0236;%%",
}

@article{Czakon:2013goa,
      author         = "Czakon, Michal and Fiedler, Paul and Mitov, Alexander",
      title          = "{Total Top-Quark Pair-Production Cross Section at
                        Hadron Colliders Through \(\mathcal{O}(\alpha_S^4)\)}",
      journal        = "Phys. Rev. Lett.",
      volume         = "110",
      pages          = "252004",
      doi            = "10.1103/PhysRevLett.110.252004",
      year           = "2013",
      eprint         = "1303.6254",
      archivePrefix  = "arXiv",
      primaryClass   = "hep-ph",
      reportNumber   = "CERN-PH-TH-2013-056, TTK-13-08",
      SLACcitation   = "%%CITATION = ARXIV:1303.6254;%%",
}

@article{Baernreuther:2012ws,
      author         = "B{\"a}rnreuther, Peter and Czakon, Michal and Mitov,
                        Alexander",
      title          = "{Percent-Level-Precision Physics at the
                  Tevatron: Next-to-Next-to-Leading Order QCD
                  Corrections to \(q \bar{q} \to t \bar{t} + X\)}",
      journal        = "Phys. Rev. Lett.",
      volume         = "109",
      pages          = "132001",
      doi            = "10.1103/PhysRevLett.109.132001",
      year           = "2012",
      eprint         = "1204.5201",
      archivePrefix  = "arXiv",
      primaryClass   = "hep-ph",
      SLACcitation   = "%%CITATION = ARXIV:1204.5201;%%",
}

@article{Frixione:2007nw,
      author         = "Frixione, Stefano and Ridolfi, Giovanni and Nason, Paolo",
      title          = "{A positive-weight next-to-leading-order Monte Carlo for
                        heavy flavour hadroproduction}",
      journal        = "JHEP",
      volume         = "09",
      pages          = "126",
      doi            = "10.1088/1126-6708/2007/09/126",
      year           = "2007",
      eprint         = "0707.3088",
      archivePrefix  = "arXiv",
      primaryClass   = "hep-ph",
      SLACcitation   = "%%CITATION = ARXIV:0707.3088;%%",
}

@article{deFlorian:2016spz,
      author         = "de Florian, D. and others",
      title          = "{Handbook of LHC Higgs Cross Sections: 4. Deciphering the Nature of the Higgs Sector}",
      collaboration  = "LHC Higgs Cross Section Working Group",
      doi            = "10.23731/CYRM-2017-002",
      year           = "2017",
      eprint         = "1610.07922",
      archivePrefix  = "arXiv",
      primaryClass   = "hep-ph",
      reportNumber   = "FERMILAB-FN-1025-T, CERN-2017-002-M",
      SLACcitation   = "%%CITATION = ARXIV:1610.07922;%%"
}

@Article{Hoeche:2009rj,
      author         = "H{\"o}che, Stefan and Krauss, Frank and Schumann, Steffen and Siegert, Frank",
      title          = "{QCD matrix elements and truncated showers}",
      journal        = "JHEP",
      volume         = "05",
      pages          = "053",
      doi            = "10.1088/1126-6708/2009/05/053",
      year           = "2009",
      eprint         = "0903.1219",
      archivePrefix  = "arXiv",
      primaryClass   = "hep-ph",
}

@Article{Gleisberg:2008fv,
      author         = "Gleisberg, Tanju and H{\"o}che, Stefan",
      title          = "{Comix, a new matrix element generator}",
      journal        = "JHEP",
      volume         = "12",
      pages          = "039",
      doi            = "10.1088/1126-6708/2008/12/039",
      year           = "2008",
      eprint         = "0808.3674",
      archivePrefix  = "arXiv",
      primaryClass   = "hep-ph",
}

@Article{Schumann:2007mg,
      author         = "Schumann, Steffen and Krauss, Frank",
      title          = "{A parton shower algorithm based on Catani--Seymour dipole factorisation}",
      journal        = "JHEP",
      volume         = "03",
      pages          = "038",
      doi            = "10.1088/1126-6708/2008/03/038",
      year           = "2008",
      eprint         = "0709.1027",
      archivePrefix  = "arXiv",
      primaryClass   = "hep-ph",
}

@article{Hoeche:2012yf,
      author         = "H{\"o}che, Stefan and Krauss, Frank and Sch{\"o}nherr, Marek and
                        Siegert, Frank",
      title          = "{QCD matrix elements + parton showers. The NLO case}",
      journal        = "JHEP",
      volume         = "04",
      year           = "2013",
      pages          = "027",
      doi            = "10.1007/JHEP04(2013)027",
      eprint         = "1207.5030",
      archivePrefix  = "arXiv",
      primaryClass   = "hep-ph",
      reportNumber   = "SLAC-PUB-15191, IPPP-12-52, DCPT-12-104, LPN12-081,
                        FR-PHENO-2012-017, MCNET-12-09, --FR-PHENO-2012-017",
      SLACcitation   = "%%CITATION = ARXIV:1207.5030;%%"
}

@article{Bothmann:2019yzt,
      author = "Bothmann, Enrico and others",
      title = "{Event generation with Sherpa 2.2}",
      journal = "SciPost Phys.",
      volume = "7",
      year = "2019",
      number = "3",
      pages = "034",
      doi = "10.21468/SciPostPhys.7.3.034",
      reportNumber = "FERMILAB-PUB-19-218-T, SLAC-PUB-17433, IPPP/19/42, MCNET-19-11",
      eprint = "1905.09127",
      archivePrefix = "arXiv",
      primaryClass = "hep-ph",
}

@article{Hoeche:2011fd,
      author         = "H{\"o}che, Stefan and Krauss, Frank and Sch{\"o}nherr, Marek and
                        Siegert, Frank",
      title          = "{A critical appraisal of NLO+PS matching methods}",
      journal        = "JHEP",
      volume         = "09",
      year           = "2012",
      pages          = "049",
      doi            = "10.1007/JHEP09(2012)049",
      eprint         = "1111.1220",
      archivePrefix  = "arXiv",
      primaryClass   = "hep-ph",
      reportNumber   = "SLAC-PUB-14661, IPPP-11-67, DCPT-11-134, LPN11-58,
                        FR-PHENO-2011-019, MCNET-11-24",
      SLACcitation   = "%%CITATION = ARXIV:1111.1220;%%"
}

@article{Catani:2001cc,
      author         = "Catani, S. and Krauss, F. and Webber, B. R. and Kuhn, R.",
      title          = "{QCD Matrix Elements + Parton Showers}",
      journal        = "JHEP",
      volume         = "11",
      year           = "2001",
      pages          = "063",
      doi            = "10.1088/1126-6708/2001/11/063",
      eprint         = "hep-ph/0109231",
      archivePrefix  = "arXiv",
      reportNumber   = "CERN-TH-2000-367, CAVENDISH-HEP-00-03",
      SLACcitation   = "%%CITATION = HEP-PH/0109231;%%"
}

@article{Cascioli:2011va,
      author         = "Cascioli, Fabio and Maierh{\"o}fer, Philipp and Pozzorini, Stefano",
      title          = "{Scattering Amplitudes with Open Loops}",
      journal        = "Phys. Rev. Lett.",
      volume         = "108",
      year           = "2012",
      pages          = "111601",
      doi            = "10.1103/PhysRevLett.108.111601",
      eprint         = "1111.5206",
      archivePrefix  = "arXiv",
      primaryClass   = "hep-ph",
      reportNumber   = "ZU-TH-23-11, LPN11-66",
      SLACcitation   = "%%CITATION = ARXIV:1111.5206;%%"
}

@article{Buccioni:2017yxi,
    author = "Buccioni, Federico and Pozzorini, Stefano and Zoller, Max",
    title = "{On-the-fly reduction of open loops}",
    eprint = "1710.11452",
    archivePrefix = "arXiv",
    primaryClass = "hep-ph",
    reportNumber = "ZU-TH-29-17",
    doi = "10.1140/epjc/s10052-018-5562-1",
    journal = "Eur. Phys. J. C",
    volume = "78",
    number = "1",
    pages = "70",
    year = "2018"
}

@article{Buccioni:2019sur,
    author = "Buccioni, Federico and Lang, Jean-Nicolas and Lindert, Jonas M. and Maierh{\"o}fer, Philipp and Pozzorini, Stefano and Zhang, Hantian and Zoller, Max F.",
    title = "{OpenLoops 2}",
    eprint = "1907.13071",
    archivePrefix = "arXiv",
    primaryClass = "hep-ph",
    reportNumber = "IPPP/19/62, FR-PHENO-2019-12, PSI-PR-19-15, ZU-TH 37/19",
    doi = "10.1140/epjc/s10052-019-7306-2",
    journal = "Eur. Phys. J. C",
    volume = "79",
    number = "10",
    pages = "866",
    year = "2019"
}

@article{Denner:2016kdg,
      author         = "Denner, Ansgar and Dittmaier, Stefan and Hofer, Lars",
      title          = "{\textsc{Collier}: A fortran-based complex one-loop library in
                        extended regularizations}",
      journal        = "Comput. Phys. Commun.",
      volume         = "212",
      year           = "2017",
      pages          = "220-238",
      doi            = "10.1016/j.cpc.2016.10.013",
      eprint         = "1604.06792",
      archivePrefix  = "arXiv",
      primaryClass   = "hep-ph",
      reportNumber   = "FR-PHENO-2016-003, ICCUB-16-016",
      SLACcitation   = "%%CITATION = ARXIV:1604.06792;%%"
}

@article{Czakon:2011xx,
  author         = "Czakon, Michal and Mitov, Alexander",
  title          = "{Top++: A program for the calculation of the top-pair cross-section at hadron colliders}",
  journal        = "Comput. Phys. Commun.",
  volume         = "185",
  year           = "2014",
  pages          = "2930",
  doi            = "10.1016/j.cpc.2014.06.021",
  eprint         = "1112.5675",
  archivePrefix  = "arXiv",
  primaryClass   = "hep-ph",
  reportNumber   = "CERN-PH-TH-2011-303, TTK-11-58",
}

@article{Hamilton:2013fea,
      author         = "Hamilton, Keith and Nason, Paolo and Re, Emanuele and
                        Zanderighi, Giulia",
      title          = "{NNLOPS simulation of Higgs boson production}",
      journal        = "JHEP",
      volume         = "10",
      year           = "2013",
      pages          = "222",
      doi            = "10.1007/JHEP10(2013)222",
      eprint         = "1309.0017",
      archivePrefix  = "arXiv",
      primaryClass   = "hep-ph",
      reportNumber   = "MCNET-13-11, CERN-PH-TH-2013-205, OUTP-13-18P",
      SLACcitation   = "%%CITATION = ARXIV:1309.0017;%%"
}

@article{Campbell:2012am,
      author         = "Campbell, John M. and Ellis, R. Keith and Frederix,
                        Rikkert and Nason, Paolo and Oleari, Carlo and Williams,
                        Ciaran",
      title          = "{NLO Higgs boson production plus one and two jets using
                        the POWHEG BOX, MadGraph4 and MCFM}",
      journal        = "JHEP",
      volume         = "07",
      year           = "2012",
      pages          = "092",
      doi            = "10.1007/JHEP07(2012)092",
      eprint         = "1202.5475",
      archivePrefix  = "arXiv",
      primaryClass   = "hep-ph",
      reportNumber   = "FERMILAB-PUB-12-040-T, CERN-PH-TH-2012-048",
      SLACcitation   = "%%CITATION = ARXIV:1202.5475;%%"
}

@article{Catani:2007vq,
      author         = "Catani, Stefano and Grazzini, Massimiliano",
      title          = "{Next-to-Next-to-Leading-Order Subtraction Formalism in Hadron Collisions and
                        its Application to Higgs-Boson Production at the Large Hadron Collider}",
      journal        = "Phys. Rev. Lett.",
      volume         = "98",
      year           = "2007",
      pages          = "222002",
      doi            = "10.1103/PhysRevLett.98.222002",
      eprint         = "hep-ph/0703012",
      archivePrefix  = "arXiv",
      SLACcitation   = "%%CITATION = HEP-PH/0703012;%%"
}

@article{Nason:2009ai,
      author         = "Nason, Paolo and Oleari, Carlo",
      title          = "{NLO Higgs boson production via vector-boson fusion
                        matched with shower in POWHEG}",
      journal        = "JHEP",
      volume         = "02",
      year           = "2010",
      pages          = "037",
      doi            = "10.1007/JHEP02(2010)037",
      eprint         = "0911.5299",
      archivePrefix  = "arXiv",
      primaryClass   = "hep-ph",
      SLACcitation   = "%%CITATION = ARXIV:0911.5299;%%"
}

@article{Anastasiou:2016cez,
      author         = "Anastasiou, Charalampos and Duhr, Claude and Dulat, Falko
                        and Furlan, Elisabetta and Gehrmann, Thomas and Herzog,
                        Franz and Lazopoulos, Achilleas and Mistlberger, Bernhard",
      title          = "{High precision determination of the gluon fusion Higgs
                        boson cross-section at the LHC}",
      journal        = "JHEP",
      volume         = "05",
      year           = "2016",
      pages          = "058",
      doi            = "10.1007/JHEP05(2016)058",
      eprint         = "1602.00695",
      archivePrefix  = "arXiv",
      primaryClass   = "hep-ph",
      reportNumber   = "CP3-16-01, ZU-TH-27-15, NIKHEF-2016-004,
                        CERN-TH-2016-006",
      SLACcitation   = "%%CITATION = ARXIV:1602.00695;%%"
}

@article{Anastasiou:2015ema,
      author         = "Anastasiou, Charalampos and Duhr, Claude and Dulat, Falko
                        and Herzog, Franz and Mistlberger, Bernhard",
      title          = "{Higgs Boson Gluon-Fusion Production in QCD at Three
                        Loops}",
      journal        = "Phys. Rev. Lett.",
      volume         = "114",
      year           = "2015",
      pages          = "212001",
      doi            = "10.1103/PhysRevLett.114.212001",
      eprint         = "1503.06056",
      archivePrefix  = "arXiv",
      primaryClass   = "hep-ph",
      reportNumber   = "CERN-PH-TH-2015-055, CP3-15-07",
      SLACcitation   = "%%CITATION = ARXIV:1503.06056;%%"
}

@article{Dulat:2018rbf,
      author         = "Dulat, Falko and Lazopoulos, Achilleas and Mistlberger,
                        Bernhard",
      title          = "{iHixs 2 -- Inclusive Higgs cross sections}",
      journal        = "Comput. Phys. Commun.",
      volume         = "233",
      year           = "2018",
      pages          = "243-260",
      doi            = "10.1016/j.cpc.2018.06.025",
      eprint         = "1802.00827",
      archivePrefix  = "arXiv",
      primaryClass   = "hep-ph",
      reportNumber   = "CERN-TH-2018-019, SLAC-PUB-17222",
      SLACcitation   = "%%CITATION = ARXIV:1802.00827;%%"
}

@article{Harlander:2009mq,
      author         = "Harlander, Robert V. and Ozeren, Kemal J.",
      title          = "{Finite top mass effects for hadronic Higgs production at
                        next-to-next-to-leading order}",
      journal        = "JHEP",
      volume         = "11",
      year           = "2009",
      pages          = "088",
      doi            = "10.1088/1126-6708/2009/11/088",
      eprint         = "0909.3420",
      archivePrefix  = "arXiv",
      primaryClass   = "hep-ph",
      reportNumber   = "WUB-09-10",
      SLACcitation   = "%%CITATION = ARXIV:0909.3420;%%"
}

@article{Harlander:2009bw,
      author         = "Harlander, Robert V. and Ozeren, Kemal J.",
      title          = "{Top mass effects in Higgs production at
                        next-to-next-to-leading order QCD: Virtual corrections}",
      journal        = "Phys. Lett. B",
      volume         = "679",
      year           = "2009",
      pages          = "467-472",
      doi            = "10.1016/j.physletb.2009.08.012",
      eprint         = "0907.2997",
      archivePrefix  = "arXiv",
      primaryClass   = "hep-ph",
      reportNumber   = "WUB-09-08",
      SLACcitation   = "%%CITATION = ARXIV:0907.2997;%%"
}

@article{Harlander:2009my,
      author         = "Harlander, Robert V. and Mantler, Hendrik and Marzani,
                        Simone and Ozeren, Kemal J.",
      title          = "{Higgs production in gluon fusion at
                        next-to-next-to-leading order QCD for finite top mass}",
      journal        = "Eur. Phys. J. C",
      volume         = "66",
      year           = "2010",
      pages          = "359-372",
      doi            = "10.1140/epjc/s10052-010-1258-x",
      eprint         = "0912.2104",
      archivePrefix  = "arXiv",
      primaryClass   = "hep-ph",
      reportNumber   = "WUB-09-18, MAN-HEP-2009-44",
      SLACcitation   = "%%CITATION = ARXIV:0912.2104;%%"
}

@article{Pak:2009dg,
      author         = "Pak, Alexey and Rogal, Mikhail and Steinhauser, Matthias",
      title          = "{Finite top quark mass effects in NNLO Higgs boson
                        production at LHC}",
      journal        = "JHEP",
      volume         = "02",
      year           = "2010",
      pages          = "025",
      doi            = "10.1007/JHEP02(2010)025",
      eprint         = "0911.4662",
      archivePrefix  = "arXiv",
      primaryClass   = "hep-ph",
      reportNumber   = "SFB-CPP-09-116, TTP09-43",
      SLACcitation   = "%%CITATION = ARXIV:0911.4662;%%"
}

@article{Actis:2008ug,
      author         = "Actis, Stefano and Passarino, Giampiero and Sturm,
                        Christian and Uccirati, Sandro",
      title          = "{NLO electroweak corrections to Higgs boson production at
                        hadron colliders}",
      journal        = "Phys. Lett. B",
      volume         = "670",
      year           = "2008",
      pages          = "12-17",
      doi            = "10.1016/j.physletb.2008.10.018",
      eprint         = "0809.1301",
      archivePrefix  = "arXiv",
      primaryClass   = "hep-ph",
      reportNumber   = "PITHA-08-20, SFB-CPP-08-62, TTP08-38",
      SLACcitation   = "%%CITATION = ARXIV:0809.1301;%%"
}

@article{Actis:2008ts,
      author         = "Actis, Stefano and Passarino, Giampiero and Sturm,
                        Christian and Uccirati, Sandro",
      title          = "{NNLO computational techniques: The cases \(H \to \gamma\gamma\) and \(H \to gg\)}",
      journal        = "Nucl. Phys. B",
      volume         = "811",
      year           = "2009",
      pages          = "182-273",
      doi            = "10.1016/j.nuclphysb.2008.11.024",
      eprint         = "0809.3667",
      archivePrefix  = "arXiv",
      primaryClass   = "hep-ph",
      reportNumber   = "PITHA-08-24, SFB-CPP-08-73, TTP08-42",
      SLACcitation   = "%%CITATION = ARXIV:0809.3667;%%"
}

@article{Bonetti:2018ukf,
      author         = "Bonetti, Marco and Melnikov, Kirill and Tancredi,
                        Lorenzo",
      title          = "{Higher order corrections to mixed QCD-EW contributions
                        to Higgs boson production in gluon fusion}",
      journal        = "Phys. Rev. D",
      volume         = "97",
      year           = "2018",
      number         = "5",
      pages          = "056017",
      doi            = "10.1103/PhysRevD.97.056017",
      eprint         = "1801.10403",
      archivePrefix  = "arXiv",
      primaryClass   = "hep-ph",
      reportNumber   = "CERN-TH-2018-011, TTP18-004",
      SLACcitation   = "%%CITATION = ARXIV:1801.10403;%%",
      related = "Bonetti:2018ukf-err",
      relatedstring  = "Erratum:",
}

@article{Ciccolini:2007jr,
     author         = "Ciccolini, M. and Denner, Ansgar and Dittmaier, S.",
     title          = "{Strong and Electroweak Corrections to the Production of a Higgs Boson + 2 Jets via Weak Interactions at the Large Hadron Collider}",
     journal        = "Phys. Rev. Lett.",
     volume         = "99",
     pages          = "161803",
     doi            = "10.1103/PhysRevLett.99.161803",
     year           = "2007",
     eprint         = "0707.0381",
     archivePrefix  = "arXiv",
     primaryClass   = "hep-ph",
}

@Article{Ciccolini:2007ec,
    author    = "Ciccolini, Mariano and Denner, Ansgar and Dittmaier, Stefan",
    title     = "{Electroweak and QCD corrections to Higgs production via
                 vector-boson fusion at the CERN LHC}",
    journal   = "Phys. Rev. D",
    volume    = "77",
    year      = "2008",
    pages     = "013002",
    eprint    = "0710.4749",
    archivePrefix = "arXiv",
    primaryClass  =  "hep-ph",
    doi       = "10.1103/PhysRevD.77.013002",
    SLACcitation  = "%%CITATION = 0710.4749;%%"
}

@Article{Bolzoni:2010xr,
    author    = "Bolzoni, Paolo and Maltoni, Fabio and Moch, Sven-Olaf and
                 Zaro, Marco",
    title     = "{Higgs Boson Production via Vector-Boson Fusion at Next-to-Next-to-Leading Order in QCD}",
    journal   = "Phys. Rev. Lett.",
    volume    = "105",
    year      = "2010",
    pages     = "011801",
    eprint    = "1003.4451",
    archivePrefix = "arXiv",
    primaryClass  =  "hep-ph",
    doi       = "10.1103/PhysRevLett.105.011801",
    SLACcitation  = "%%CITATION = 1003.4451;%%"
}

@article{Djouadi:1997yw,
    author    = "Djouadi, A. and Kalinowski, J. and Spira, M.",
    title     = "{HDECAY: a program for Higgs boson decays in the Standard
                 Model and its  supersymmetric extension}",
    journal   = "Comput. Phys. Commun.",
    volume    = "108",
    year      = "1998",
    pages     = "56",
    eprint    = "hep-ph/9704448",
    archivePrefix = "arXiv",
    doi       = "10.1016/S0010-4655(97)00123-9",
    reportNumber   = "DESY-97-079, IFT-96-29, PM-97-04",
    SLACcitation  = "%%CITATION = HEP-PH/9704448;%%"
}

@article{Spira:1997dg,
      author         = "Spira, Michael",
      title          = "{QCD Effects in Higgs Physics}",
      journal        = "Fortsch. Phys.",
      volume         = "46",
      year           = "1998",
      pages          = "203-284",
      doi            = "10.1002/(SICI)1521-3978(199804)46:3<203::AID-PROP203>3.0.CO;2-4",
      eprint         = "hep-ph/9705337",
      archivePrefix  = "arXiv",
      reportNumber   = "CERN-TH-97-068, CERN-TH-97-68",
      SLACcitation   = "%%CITATION = HEP-PH/9705337;%%"
}

@article{Luisoni:2013cuh,
    author = "Luisoni, Gionata and Nason, Paolo and Oleari, Carlo and Tramontano, Francesco",
    title = "{\(HW^{\pm}/HZ\) + 0 and 1 jet at NLO with the POWHEG BOX interfaced to GoSam and their merging within MiNLO}",
    eprint = "1306.2542",
    archivePrefix = "arXiv",
    primaryClass = "hep-ph",
    doi = "10.1007/JHEP10(2013)083",
    journal = "JHEP",
    volume = "10",
    pages = "083",
    year = "2013"
}

@article{Djouadi:2006bz,
      author         = "Djouadi, A. and M{\"u}hlleitner, M. M. and Spira, M.",
      title          = "{Decays of Supersymmetric Particles: the Program SUSY-HIT
                        (SUspect-SdecaY-Hdecay-InTerface)}",
      booktitle      = "{Proceedings, Physics at LHC, 3rd Conference}",
      venue          = "Cracow, Poland",
      eventdate      = "2006-07-03/2006-07-08",
      journal        = "Acta Phys. Polon. B",
      volume         = "38",
      year           = "2007",
      pages          = "635-644",
      eprint         = "hep-ph/0609292",
      archivePrefix  = "arXiv",
      XprimaryClass   = "hep-ph",
      reportNumber   = "CERN-PH-TH-2006-200",
      SLACcitation   = "%%CITATION = HEP-PH/0609292;%%"
}

@article{Bredenstein:2006ha,
      author         = "Bredenstein, A. and Denner, Ansgar and Dittmaier, S. and Weber, M. M.",
      title          = "{Radiative corrections to the semileptonic and hadronic
                        Higgs-boson decays \(H \to WW / ZZ \to 4\) fermions}",
      journal        = "JHEP",
      volume         = "02",
      year           = "2007",
      pages          = "080",
      doi            = "10.1088/1126-6708/2007/02/080",
      eprint         = "hep-ph/0611234",
      archivePrefix  = "arXiv",
      xprimaryClass   = "hep-ph",
      reportNumber   = "KEK-CP-187, KEK-PREPRINT-2006-47, MPP-2006-138,
                        PSI-PR-06-11, WUB-06-08",
      SLACcitation   = "%%CITATION = HEP-PH/0611234;%%"
}

@article{Bredenstein:2006rh,
      author         = "Bredenstein, A. and Denner, Ansgar and Dittmaier, S. and Weber, M. M.",
      title          = "{Precise predictions for the Higgs-boson decay $H \to WW/ZZ \to 4$ leptons}",
      journal        = "Phys. Rev. D",
      volume         = "74",
      year           = "2006",
      pages          = "013004",
      doi            = "10.1103/PhysRevD.74.013004",
      eprint         = "hep-ph/0604011",
      archivePrefix  = "arXiv",
      primaryClass   = "hep-ph",
      reportNumber   = "MPP-2005-24, PSI-PR-06-05",
      SLACcitation   = "%%CITATION = HEP-PH/0604011;%%"
}

@article{Ciccolini:2003jy,
      author         = "Ciccolini, M. L. and Dittmaier, S. and Kr{\"a}mer, M.",
      title          = "{Electroweak radiative corrections to associated \(WH\) and
                        \(ZH\) production at hadron colliders}",
      journal        = "Phys. Rev. D",
      volume         = "68",
      year           = "2003",
      pages          = "073003",
      doi            = "10.1103/PhysRevD.68.073003",
      eprint         = "hep-ph/0306234",
      archivePrefix  = "arXiv",
      reportNumber   = "EDINBURGH-2003-05, MPI-PHT-2003-24",
      SLACcitation   = "%%CITATION = HEP-PH/0306234;%%"
}

@article{Brein:2003wg,
      author         = "Brein, Oliver and Djouadi, Abdelhak and Harlander, Robert",
      title          = "{NNLO QCD corrections to the Higgs-strahlung processes at
                        hadron colliders}",
      journal        = "Phys. Lett. B",
      volume         = "579",
      year           = "2004",
      pages          = "149-156",
      doi            = "10.1016/j.physletb.2003.10.112",
      eprint         = "hep-ph/0307206",
      archivePrefix  = "arXiv",
      XprimaryClass   = "hep-ph",
      reportNumber   = "MPP-2003-35, CERN-TH-2003-161, PM-03-16",
      SLACcitation   = "%%CITATION = HEP-PH/0307206;%%"
}

@article{Brein:2011vx,
      author         = "Brein, Oliver and Harlander, Robert V. and Wiesemann, Marius and Zirke, Tom",
      title          = "{Top-quark mediated effects in hadronic Higgs-Strahlung}",
      journal        = "Eur. Phys. J. C",
      volume         = "72",
      year           = "2012",
      pages          = "1868",
      doi            = "10.1140/epjc/s10052-012-1868-6",
      eprint         = "1111.0761",
      archivePrefix  = "arXiv",
      primaryClass   = "hep-ph",
      reportNumber   = "CERN-PH-TH-2011-268, FR-PHENO-2011-016, WUB-11-15",
      SLACcitation   = "%%CITATION = ARXIV:1111.0761;%%"
}

@article{Altenkamp:2012sx,
      author         = "Altenkamp, Lukas and Dittmaier, Stefan and Harlander,
                        Robert V. and Rzehak, Heidi and Zirke, Tom J. E.",
      title          = "{Gluon-induced Higgs-strahlung at next-to-leading order
                        QCD}",
      journal        = "JHEP",
      volume         = "02",
      year           = "2013",
      pages          = "078",
      doi            = "10.1007/JHEP02(2013)078",
      eprint         = "1211.5015",
      archivePrefix  = "arXiv",
      primaryClass   = "hep-ph",
      reportNumber   = "CERN-PH-TH-2012-312, FR-PHENO-2012-023, WUB-12-21",
      SLACcitation   = "%%CITATION = ARXIV:1211.5015;%%"
}

@article{Denner:2014cla,
      author         = "Denner, Ansgar and Dittmaier, Stefan and Kallweit, Stefan
                        and M{\"u}ck, Alexander",
      title          = "{HAWK 2.0: A Monte Carlo program for Higgs production
                        in vector-boson fusion and Higgs strahlung at hadron
                        colliders}",
      journal        = "Comput. Phys. Commun.",
      volume         = "195",
      year           = "2015",
      pages          = "161-171",
      doi            = "10.1016/j.cpc.2015.04.021",
      eprint         = "1412.5390",
      archivePrefix  = "arXiv",
      primaryClass   = "hep-ph",
      reportNumber   = "FR-PHENO-2014-013, MITP-14-101, TTK-14-36",
      SLACcitation   = "%%CITATION = ARXIV:1412.5390;%%"
}

@article{Brein:2012ne,
      author         = "Brein, Oliver and Harlander, Robert V. and Zirke, Tom J.
                        E.",
      title          = "{vh@nnlo -- Higgs Strahlung at hadron colliders}",
      journal        = "Comput. Phys. Commun.",
      volume         = "184",
      year           = "2013",
      pages          = "998-1003",
      doi            = "10.1016/j.cpc.2012.11.002",
      eprint         = "1210.5347",
      archivePrefix  = "arXiv",
      primaryClass   = "hep-ph",
      SLACcitation   = "%%CITATION = ARXIV:1210.5347;%%"
}

@article{Harlander:2014wda,
      author         = "Harlander, Robert V. and Kulesza, Anna and Theeuwes,
                        Vincent and Zirke, Tom",
      title          = "{Soft gluon resummation for gluon-induced Higgs
                        Strahlung}",
      journal        = "JHEP",
      volume         = "11",
      year           = "2014",
      pages          = "082",
      doi            = "10.1007/JHEP11(2014)082",
      eprint         = "1410.0217",
      archivePrefix  = "arXiv",
      primaryClass   = "hep-ph",
      reportNumber   = "MS-TP-14-17, WUB-14-10, LPN14-112",
      SLACcitation   = "%%CITATION = ARXIV:1410.0217;%%"
}

@article{Agostinelli:2002hh,
      author         = "Agostinelli, S. and others",
      title          = "{\textsc{Geant4} -- a simulation toolkit}",
      journal        = "Nucl. Instrum. Meth. A",
      volume         = "506",
      year           = "2003",
      pages          = "250",
      doi            = "10.1016/S0168-9002(03)01368-8",
      reportNumber   = "SLAC-PUB-9350, FERMILAB-PUB-03-339",
      SLACcitation   = "%%CITATION = NUIMA,A506,250;%%"
}

@Article{Cowan:2010js,
   Author = {Cowan, Glen and Cranmer, Kyle and Gross, Eilam and Vitells, Ofer},
   Title = {{Asymptotic formulae for likelihood-based tests of new physics}},
   Journal = {Eur. Phys. J. C},
   Volume = {71},
   Year = {2011},
   Pages = {1554},
   doi = "10.1140/epjc/s10052-011-1554-0",
   Eprint = "1007.1727",
   Archiveprefix = {arXiv},
   Primaryclass = {physics.data-an},
   related = "Cowan:2010js-err",
   relatedstring  = "Erratum:",
}

@Booklet{TMVA,
      author         = "Hoecker, Andreas and others",
      title          = "{TMVA - Toolkit for Multivariate Data Analysis}",
      year           = "2009",
      eprint         = "physics/0703039",
      archivePrefix  = "arXiv",
      primaryClass   = "physics.data-an",
}

@online{chollet2015keras,
      author         = {Chollet, Fran\c{c}ois and others},
      title          = {Keras},
      year           = {2015},
      url            = {https://keras.io},
}

@Article{PERF-2007-01,
    author         = "{ATLAS Collaboration}",
    title          = "{The ATLAS Experiment at the CERN Large Hadron Collider}",
    journal        = "JINST",
    volume         = "3",
    year           = "2008",
    pages          = "S08003",
    doi            = "10.1088/1748-0221/3/08/S08003",
    primaryClass   = "hep-ex",
}

@Article{SOFT-2010-01,
    author         = "{ATLAS Collaboration}",
    title          = "{The ATLAS Simulation Infrastructure}",
    journal        = "Eur. Phys. J. C",
    volume         = "70",
    year           = "2010",
    pages          = "823",
    doi            = "10.1140/epjc/s10052-010-1429-9",
    eprint         = "1005.4568",
    archivePrefix  = "arXiv",
    primaryClass   = "physics.ins-det",
}

@Article{STDM-2012-23,
    author         = "{ATLAS Collaboration}",
    title          = "{Measurement of the \(Z/\gamma^*\) boson transverse momentum distribution in \(pp\) collisions at \(\sqrt{s} = 7\,\text{TeV}\) with the ATLAS detector}",
    journal        = "JHEP",
    volume         = "09",
    year           = "2014",
    pages          = "145",
    doi            = "10.1007/JHEP09(2014)145",
    reportNumber   = "CERN-PH-EP-2014-075",
    eprint         = "1406.3660",
    archivePrefix  = "arXiv",
    primaryClass   = "hep-ex",
}

@Article{HIGG-2013-15,
    author         = "{ATLAS Collaboration}",
    title          = "{Study of \((W/Z)H\) production and Higgs boson couplings using \(H \rightarrow WW^{\ast}\) decays with the ATLAS detector}",
    journal        = "JHEP",
    volume         = "08",
    year           = "2015",
    pages          = "137",
    doi            = "10.1007/JHEP08(2015)137",
    reportNumber   = "CERN-PH-EP-2015-104",
    eprint         = "1506.06641",
    archivePrefix  = "arXiv",
    primaryClass   = "hep-ex",
}

@Article{HIGG-2014-14,
    author         = "{ATLAS and CMS Collaborations}",
    title          = "{Combined Measurement of the Higgs Boson Mass in \(pp\) Collisions at \(\sqrt{s} = 7\) and \(8\,\text{TeV}\) with the ATLAS and CMS Experiments}",
    journal        = "Phys. Rev. Lett.",
    volume         = "114",
    year           = "2015",
    pages          = "191803",
    doi            = "10.1103/PhysRevLett.114.191803",
    reportNumber   = "CERN-PH-EP-2015-075",
    eprint         = "1503.07589",
    archivePrefix  = "arXiv",
    primaryClass   = "hep-ex",
}

@Article{PERF-2014-03,
    author         = "{ATLAS Collaboration}",
    title          = "{Performance of pile-up mitigation techniques for jets in \(pp\) collisions at \(\sqrt{s} = 8\,\text{TeV}\) using the ATLAS detector}",
    journal        = "Eur. Phys. J. C",
    volume         = "76",
    year           = "2016",
    pages          = "581",
    doi            = "10.1140/epjc/s10052-016-4395-z",
    reportNumber   = "CERN-PH-EP-2015-206",
    eprint         = "1510.03823",
    archivePrefix  = "arXiv",
    primaryClass   = "hep-ex",
}

@Article{PERF-2015-09,
    author         = "{ATLAS Collaboration}",
    title          = "{Jet reconstruction and performance using particle flow with the ATLAS Detector}",
    journal        = "Eur. Phys. J. C",
    volume         = "77",
    year           = "2017",
    pages          = "466",
    doi            = "10.1140/epjc/s10052-017-5031-2",
    reportNumber   = "CERN-EP-2017-024",
    eprint         = "1703.10485",
    archivePrefix  = "arXiv",
    primaryClass   = "hep-ex",
}

@Article{PERF-2016-06,
    author         = "{ATLAS Collaboration}",
    title          = "{Identification and rejection of pile-up jets at high pseudorapidity with the ATLAS detector}",
    journal        = "Eur. Phys. J. C",
    volume         = "77",
    year           = "2017",
    pages          = "580",
    doi            = "10.1140/epjc/s10052-017-5081-5",
    reportNumber   = "CERN-EP-2017-055",
    eprint         = "1705.02211",
    archivePrefix  = "arXiv",
    primaryClass   = "hep-ex",
    related        = "PERF-2016-06-err",
    relatedstring  = "Erratum:",
}

@Article{TRIG-2016-01,
    author         = "{ATLAS Collaboration}",
    title          = "{Performance of the ATLAS trigger system in 2015}",
    journal        = "Eur. Phys. J. C",
    volume         = "77",
    year           = "2017",
    pages          = "317",
    doi            = "10.1140/epjc/s10052-017-4852-3",
    reportNumber   = "CERN-EP-2016-241",
    eprint         = "1611.09661",
    archivePrefix  = "arXiv",
    primaryClass   = "hep-ex",
}

@Article{HIGG-2017-14,
    author         = "{ATLAS Collaboration}",
    title          = "{Measurement of the production cross section for a Higgs boson in association with a vector boson in the \(H \to WW^{*} \to \ell\nu\ell\nu\) channel in \(pp\) collisions at \(\sqrt{s} = 13\,\text{TeV}\) with the ATLAS detector}",
    journal        = "Phys. Lett. B",
    volume         = "798",
    year           = "2019",
    pages          = "134949",
    doi            = "10.1016/j.physletb.2019.134949",
    reportNumber   = "CERN-EP-2019-038",
    eprint         = "1903.10052",
    archivePrefix  = "arXiv",
    primaryClass   = "hep-ex",
}

@Article{DAPR-2018-01,
    author         = "{ATLAS Collaboration}",
    title          = "{ATLAS data quality operations and performance for 2015--2018 data-taking}",
    journal        = "JINST",
    volume         = "15",
    year           = "2020",
    pages          = "P04003",
    doi            = "10.1088/1748-0221/15/04/P04003",
    reportNumber   = "CERN-EP-2019-207",
    eprint         = "1911.04632",
    archivePrefix  = "arXiv",
    primaryClass   = "physics.ins-det",
}

@Article{EGAM-2018-01,
    author         = "{ATLAS Collaboration}",
    title          = "{Electron and photon performance measurements with the ATLAS detector using the 2015--2017 LHC proton--proton collision data}",
    journal        = "JINST",
    volume         = "14",
    year           = "2019",
    pages          = "P12006",
    doi            = "10.1088/1748-0221/14/12/P12006",
    reportNumber   = "CERN-EP-2019-145",
    eprint         = "1908.00005",
    archivePrefix  = "arXiv",
    primaryClass   = "hep-ex",
}

@Article{FTAG-2018-01,
    author         = "{ATLAS Collaboration}",
    title          = "{ATLAS \(b\)-jet identification performance and efficiency measurement with \(t\bar{t}\) events in \(pp\) collisions at \(\sqrt{s} = 13\,\text{TeV}\)}",
    journal        = "Eur. Phys. J. C",
    volume         = "79",
    year           = "2019",
    pages          = "970",
    doi            = "10.1140/epjc/s10052-019-7450-8",
    reportNumber   = "CERN-EP-2019-132",
    eprint         = "1907.05120",
    archivePrefix  = "arXiv",
    primaryClass   = "hep-ex",
}

@Article{HIGG-2018-28,
    author         = "{ATLAS Collaboration}",
    title          = "{Higgs boson production cross-section measurements and their EFT interpretation in the \(4\ell\) decay channel at \(\sqrt{s} = 13\,\text{TeV}\) with the ATLAS detector}",
    journal        = "Eur. Phys. J. C",
    volume         = "80",
    year           = "2020",
    pages          = "957",
    doi            = "10.1140/epjc/s10052-020-8227-9",
    reportNumber   = "CERN-EP-2020-034",
    eprint         = "2004.03447",
    archivePrefix  = "arXiv",
    primaryClass   = "hep-ex",
    related        = "HIGG-2018-28-err",
    relatedstring  = "Erratum:",
}

@Article{HIGG-2018-51,
    author         = "{ATLAS Collaboration}",
    title          = "{Measurements of \(WH\) and \(ZH\) production in the \(H \to b\bar{b}\) decay channel in \(pp\) collisions at \(13\,\text{TeV}\) with the ATLAS detector}",
    journal        = "Eur. Phys. J. C",
    volume         = "81",
    year           = "2021",
    pages          = "178",
    doi            = "10.1140/epjc/s10052-020-08677-2",
    reportNumber   = "CERN-EP-2020-087",
    eprint         = "2007.02873",
    archivePrefix  = "arXiv",
    primaryClass   = "hep-ex",
}

@Article{JETM-2018-05,
    author         = "{ATLAS Collaboration}",
    title          = "{Jet energy scale and resolution measured in proton--proton collisions at \(\sqrt{s} = 13\,\text{TeV}\) with the ATLAS detector}",
    journal        = "Eur. Phys. J. C",
    volume         = "81",
    year           = "2021",
    pages          = "689",
    doi            = "10.1140/epjc/s10052-021-09402-3",
    reportNumber   = "CERN-EP-2020-083",
    eprint         = "2007.02645",
    archivePrefix  = "arXiv",
    primaryClass   = "hep-ex",
}

@Article{MUON-2018-03,
    author         = "{ATLAS Collaboration}",
    title          = "{Muon reconstruction and identification efficiency in ATLAS using the full Run~2 \(pp\) collision data set at \(\sqrt{s} = 13\,\text{TeV}\)}",
    journal        = "Eur. Phys. J. C",
    volume         = "81",
    year           = "2021",
    pages          = "578",
    doi            = "10.1140/epjc/s10052-021-09233-2",
    reportNumber   = "CERN-EP-2020-199",
    eprint         = "2012.00578",
    archivePrefix  = "arXiv",
    primaryClass   = "hep-ex",
}

@Article{PIX-2018-001,
    author         = "Abbott, B. and others",
    title          = "{Production and integration of the ATLAS Insertable B-Layer}",
    journal        = "JINST",
    volume         = "13",
    year           = "2018",
    pages          = "T05008",
    doi            = "10.1088/1748-0221/13/05/T05008",
    eprint         = "1803.00844",
    archivePrefix  = "arXiv",
    primaryClass   = "physics.ins-det",
}

@Article{STDM-2018-32,
    author         = "{ATLAS Collaboration}",
    title          = "{Measurement and interpretation of same-sign \(W\) boson pair production in association with two jets in \(pp\) collisions at \(\sqrt{s} = 13\,\text{TeV}\) with the ATLAS detector}",
    journal        = "JHEP",
    volume         = "04",
    year           = "2024",
    pages          = "026",
    doi            = "10.1007/JHEP04(2024)026",
    reportNumber   = "CERN-EP-2023-221",
    eprint         = "2312.00420",
    archivePrefix  = "arXiv",
    primaryClass   = "hep-ex",
}

@Article{TRIG-2018-01,
    author         = "{ATLAS Collaboration}",
    title          = "{Performance of the ATLAS muon triggers in Run~2}",
    journal        = "JINST",
    volume         = "15",
    year           = "2020",
    pages          = "P09015",
    doi            = "10.1088/1748-0221/15/09/p09015",
    reportNumber   = "CERN-EP-2020-031",
    eprint         = "2004.13447",
    archivePrefix  = "arXiv",
    primaryClass   = "physics.ins-det",
}

@Article{TRIG-2018-05,
    author         = "{ATLAS Collaboration}",
    title          = "{Performance of electron and photon triggers in ATLAS during LHC Run~2}",
    journal        = "Eur. Phys. J. C",
    volume         = "80",
    year           = "2020",
    pages          = "47",
    doi            = "10.1140/epjc/s10052-019-7500-2",
    reportNumber   = "CERN-EP-2019-169",
    eprint         = "1909.00761",
    archivePrefix  = "arXiv",
    primaryClass   = "hep-ex",
}

@Article{FTAG-2019-07,
    author         = "{ATLAS Collaboration}",
    title          = "{ATLAS flavour-tagging algorithms for the LHC Run~2 \(pp\) collision dataset}",
    journal        = "Eur. Phys. J. C",
    volume         = "83",
    year           = "2023",
    pages          = "681",
    doi            = "10.1140/epjc/s10052-023-11699-1",
    reportNumber   = "CERN-EP-2022-226",
    eprint         = "2211.16345",
    archivePrefix  = "arXiv",
    primaryClass   = "physics.data-an",
}

@Article{HIGG-2019-09,
    author         = "{ATLAS Collaboration}",
    title          = "{Measurements of Higgs boson production cross-sections in the \(H\to\tau^{+}\tau^{-}\) decay channel in \(pp\) collisions at \(\sqrt{s} = 13\,\text{TeV}\) with the ATLAS detector}",
    journal        = "JHEP",
    volume         = "08",
    year           = "2022",
    pages          = "175",
    doi            = "10.1007/JHEP08(2022)175",
    reportNumber   = "CERN-EP-2021-217",
    eprint         = "2201.08269",
    archivePrefix  = "arXiv",
    primaryClass   = "hep-ex",
}

@Article{STDM-2019-09,
    author         = "{ATLAS Collaboration}",
    title          = "{Observation of \(WWW\) Production in \(pp\) Collisions at \(\sqrt{s} = 13\,\text{TeV}\) with the ATLAS Detector}",
    journal        = "Phys. Rev. Lett.",
    volume         = "129",
    year           = "2022",
    pages          = "061803",
    doi            = "10.1103/PhysRevLett.129.061803",
    reportNumber   = "CERN-EP-2021-243",
    eprint         = "2201.13045",
    archivePrefix  = "arXiv",
    primaryClass   = "hep-ex",
}

@Article{HIGG-2020-16,
    author         = "{ATLAS Collaboration}",
    title          = "{Measurement of the properties of Higgs boson production at \(\sqrt{s} = 13\,\text{TeV}\) in the \(H \to \gamma\gamma\) channel using \(139\,\text{fb}^{-1}\) of \(pp\) collision data with the ATLAS experiment}",
    journal        = "JHEP",
    volume         = "07",
    year           = "2023",
    pages          = "088",
    doi            = "10.1007/JHEP07(2023)088",
    reportNumber   = "CERN-EP-2022-094",
    eprint         = "2207.00348",
    archivePrefix  = "arXiv",
    primaryClass   = "hep-ex",
}

@Article{JETM-2020-03,
    author         = "{ATLAS Collaboration}",
    title          = "{The performance of missing transverse momentum reconstruction and its significance with the ATLAS detector using \(140\,\text{fb}^{-1}\) of \(\sqrt{s}=13\,\text{TeV}\) \(pp\) collisions}",
    year           = "2024",
    reportNumber   = "CERN-EP-2024-023",
    eprint         = "2402.05858",
    archivePrefix  = "arXiv",
    primaryClass   = "hep-ex",
}

@Article{DAPR-2021-01,
    author         = "{ATLAS Collaboration}",
    title          = "{Luminosity determination in \(pp\) collisions at \(\sqrt{s} = 13\,\text{TeV}\) using the ATLAS detector at the LHC}",
    journal        = "Eur. Phys. J. C",
    volume         = "83",
    year           = "2023",
    pages          = "982",
    doi            = "10.1140/epjc/s10052-023-11747-w",
    reportNumber   = "CERN-EP-2022-281",
    eprint         = "2212.09379",
    archivePrefix  = "arXiv",
    primaryClass   = "hep-ex",
}

@Article{EGAM-2021-01,
    author         = "{ATLAS Collaboration}",
    title          = "{Electron and photon efficiencies in LHC Run~2 with the ATLAS experiment}",
    journal        = "JHEP",
    volume         = "05",
    year           = "2024",
    pages          = "162",
    doi            = "10.1007/JHEP05(2024)162",
    reportNumber   = "CERN-EP-2023-182",
    eprint         = "2308.13362",
    archivePrefix  = "arXiv",
    primaryClass   = "hep-ex",
}

@Article{EGAM-2021-02,
    author         = "{ATLAS Collaboration}",
    title          = "{Electron and photon energy calibration with the ATLAS detector using LHC Run~2 data}",
    journal        = "JINST",
    volume         = "19",
    year           = "2024",
    pages          = "P02009",
    doi            = "10.1088/1748-0221/19/02/P02009",
    reportNumber   = "CERN-EP-2023-128",
    eprint         = "2309.05471",
    archivePrefix  = "arXiv",
    primaryClass   = "hep-ex",
}

@Article{HIGG-2021-20,
    author         = "{ATLAS Collaboration}",
    title          = "{Measurements of Higgs boson production by gluon--gluon fusion and vector-boson fusion using \(H \to W W^{*} \to e\nu \mu\nu\) decays in \(pp\) collisions at \(\sqrt{s} = 13\,\text{TeV}\) with the ATLAS detector}",
    journal        = "Phys. Rev. D",
    volume         = "108",
    year           = "2023",
    pages          = "032005",
    doi            = "10.1103/PhysRevD.108.032005",
    reportNumber   = "CERN-EP-2022-078",
    eprint         = "2207.00338",
    archivePrefix  = "arXiv",
    primaryClass   = "hep-ex",
}

@Article{MUON-2022-01,
    author         = "{ATLAS Collaboration}",
    title          = "{Studies of the muon momentum calibration and performance of the ATLAS detector with \(pp\) collisions at \(\sqrt{s} = 13\,\text{TeV}\)}",
    journal        = "Eur. Phys. J. C",
    volume         = "83",
    year           = "2023",
    pages          = "686",
    doi            = "10.1140/epjc/s10052-023-11584-x",
    reportNumber   = "CERN-EP-2022-274",
    eprint         = "2212.07338",
    archivePrefix  = "arXiv",
    primaryClass   = "hep-ex",
}

@Article{SOFT-2022-02,
    author         = "{ATLAS Collaboration}",
    title          = "{Software and computing for Run~3 of the ATLAS experiment at the LHC}",
    journal        = "Eur. Phys. J. C",
    volume         = "85",
    year           = "2025",
    pages          = "234",
    doi            = "10.1140/epjc/s10052-024-13701-w",
    reportNumber   = "CERN-EP-2024-100",
    eprint         = "2404.06335",
    archivePrefix  = "arXiv",
    primaryClass   = "hep-ex",
}

@Article{HIGG-2023-09,
    author         = "{ATLAS Collaboration}",
    title          = "{Measurements of the production cross-sections of a Higgs boson in association with a vector boson and decaying into \(WW^{*}\) with the ATLAS detector at \(\sqrt{s} = 13\,\text{TeV}\)}",
    year           = "2025",
    reportNumber   = "CERN-EP-2025-037",
    eprint         = "2503.19420",
    archivePrefix  = "arXiv",
    primaryClass   = "hep-ex",
}

@Booklet{ATL-SOFT-PUB-2025-001,
    author         = "{ATLAS Collaboration}",
    title          = "{ATLAS Computing Acknowledgements}",
    howpublished   = "{ATL-SOFT-PUB-2025-001}",
    url            = "https://cds.cern.ch/record/2922210",
    year           = "2025",
}

@Report{ATLAS-TDR-19,
    author         = "{ATLAS Collaboration}",
    title          = "{ATLAS Insertable B-Layer: Technical Design Report}",
    type           = "ATLAS-TDR-19; CERN-LHCC-2010-013",
    year           = "2010",
    url            = "https://cds.cern.ch/record/1291633",
    related        = "ATLAS-TDR-19-addm",
    relatedstring  = "Addendum:",
}

@Article{LUCID2,
    author         = "Avoni, G. and others",
    title          = "{The new LUCID-2 detector for luminosity measurement and monitoring in ATLAS}",
    journal        = "JINST",
    volume         = "13",
    year           = "2018",
    pages          = "P07017",
    doi            = "10.1088/1748-0221/13/07/P07017",
}

@Article{CMS-HIG-13-023,
    author         = "{CMS Collaboration}",
    title          = "{Measurement of Higgs boson production and properties in the \(WW\) decay channel with leptonic final states}",
    journal        = "JHEP",
    volume         = "01",
    year           = "2014",
    pages          = "096",
    doi            = "10.1007/JHEP01(2014)096",
    reportNumber   = "CERN-PH-EP-2013-221",
    eprint         = "1312.1129",
    archivePrefix  = "arXiv",
    primaryClass   = "hep-ex",
}

@Article{CMS-HIG-18-016,
    author         = "{CMS Collaboration}",
    title          = "{Observation of Higgs Boson Decay to Bottom Quarks}",
    journal        = "Phys. Rev. Lett.",
    volume         = "121",
    year           = "2018",
    pages          = "121801",
    doi            = "10.1103/PhysRevLett.121.121801",
    reportNumber   = "CERN-EP-2018-223",
    eprint         = "1808.08242",
    archivePrefix  = "arXiv",
    primaryClass   = "hep-ex",
}

@Article{CMS-HIG-19-001,
    author         = "{CMS Collaboration}",
    title          = "{Measurements of production cross sections of the Higgs boson in the four-lepton final state in proton--proton collisions at \(\sqrt{s} = 13\,\text{TeV}\)}",
    journal        = "Eur. Phys. J. C",
    volume         = "81",
    year           = "2021",
    pages          = "488",
    doi            = "10.1140/epjc/s10052-021-09200-x",
    reportNumber   = "CERN-EP-2021-016",
    eprint         = "2103.04956",
    archivePrefix  = "arXiv",
    primaryClass   = "hep-ex",
}

@Article{CMS-HIG-19-010,
    author         = "{CMS Collaboration}",
    title          = "{Measurements of Higgs boson production in the decay channel with a pair of \(\tau\) leptons in proton--proton collisions at \(\sqrt{s} = 13\,\text{TeV}\)}",
    journal        = "Eur. Phys. J. C",
    volume         = "83",
    year           = "2023",
    pages          = "562",
    doi            = "10.1140/epjc/s10052-023-11452-8",
    reportNumber   = "CERN-EP-2022-027",
    eprint         = "2204.12957",
    archivePrefix  = "arXiv",
    primaryClass   = "hep-ex",
}

@Article{CMS-HIG-19-015,
    author         = "{CMS Collaboration}",
    title          = "{Measurements of Higgs boson production cross sections and couplings in the diphoton decay channel at \(\sqrt{s} = 13\,\text{TeV}\)}",
    journal        = "JHEP",
    volume         = "07",
    year           = "2021",
    pages          = "027",
    doi            = "10.1007/JHEP07(2021)027",
    reportNumber   = "CERN-EP-2021-038",
    eprint         = "2103.06956",
    archivePrefix  = "arXiv",
    primaryClass   = "hep-ex",
}

@Article{CMS-HIG-20-013,
    author         = "{CMS Collaboration}",
    title          = "{Measurements of the Higgs boson production cross section and couplings in the \(W\) boson pair decay channel in proton--proton collisions at \(\sqrt{s} = 13\,\text{TeV}\)}",
    journal        = "Eur. Phys. J. C",
    volume         = "83",
    year           = "2023",
    pages          = "667",
    doi            = "10.1140/epjc/s10052-023-11632-6",
    reportNumber   = "CERN-EP-2022-120",
    eprint         = "2206.09466",
    archivePrefix  = "arXiv",
    primaryClass   = "hep-ex",
}

@Booklet{ATL-PHYS-PUB-2014-021,
    author         = "{ATLAS Collaboration}",
    title          = "{ATLAS Pythia~8 tunes to \(7~\text{TeV}\) data}",
    howpublished   = "{ATL-PHYS-PUB-2014-021}",
    url            = "https://cds.cern.ch/record/1966419",
    year           = "2014",
}

@Booklet{ATL-PHYS-PUB-2015-026,
    author         = "{ATLAS Collaboration}",
    title          = "{Vertex Reconstruction Performance of the ATLAS Detector at \(\sqrt{s} = 13~\text{TeV}\)}",
    howpublished   = "{ATL-PHYS-PUB-2015-026}",
    url            = "https://cds.cern.ch/record/2037717",
    year           = "2015",
}

@Booklet{ATL-PHYS-PUB-2016-017,
    author         = "{ATLAS Collaboration}",
    title          = "{The Pythia~8 A3 tune description of ATLAS minimum bias and inelastic measurements incorporating the Donnachie--Landshoff diffractive model}",
    howpublished   = "{ATL-PHYS-PUB-2016-017}",
    url            = "https://cds.cern.ch/record/2206965",
    year           = "2016",
}

@Booklet{ATL-PHYS-PUB-2016-020,
    author         = "{ATLAS Collaboration}",
    title          = "{Studies on top-quark Monte Carlo modelling for Top2016}",
    howpublished   = "{ATL-PHYS-PUB-2016-020}",
    url            = "https://cds.cern.ch/record/2216168",
    year           = "2016",
}

@Booklet{ATL-PHYS-PUB-2018-035,
    author         = "{ATLAS Collaboration}",
    title          = "{Evaluation of theoretical uncertainties for simplified template cross section measurements of \(V\)-associated production of the Higgs boson}",
    howpublished   = "{ATL-PHYS-PUB-2018-035}",
    url            = "https://cds.cern.ch/record/2649241",
    year           = "2018",
}

@Booklet{ATL-PHYS-PUB-2019-026,
    author         = "{ATLAS Collaboration}",
    title          = "{Forward jet vertex tagging using the particle flow algorithm}",
    howpublished   = "{ATL-PHYS-PUB-2019-026}",
    url            = "https://cds.cern.ch/record/2683100",
    year           = "2019",
}
\clearpage
% ATLAS Collaboration author list
% Reference date of HIGG-2023-09 is 2024-07-01
% Author list last updated on date 22-MAY-25
% Data extracted on 22-May-2025 for paper reference HIGG-2023-09
% at 8:44am
 
\begin{flushleft}
\hypersetup{urlcolor=black}
{\Large The ATLAS Collaboration}

\bigskip

\AtlasOrcid[0000-0002-6665-4934]{G.~Aad}$^\textrm{\scriptsize 105}$,
\AtlasOrcid[0000-0001-7616-1554]{E.~Aakvaag}$^\textrm{\scriptsize 17}$,
\AtlasOrcid[0000-0002-5888-2734]{B.~Abbott}$^\textrm{\scriptsize 124}$,
\AtlasOrcid[0000-0002-0287-5869]{S.~Abdelhameed}$^\textrm{\scriptsize 120a}$,
\AtlasOrcid[0000-0002-1002-1652]{K.~Abeling}$^\textrm{\scriptsize 57}$,
\AtlasOrcid[0000-0001-5763-2760]{N.J.~Abicht}$^\textrm{\scriptsize 51}$,
\AtlasOrcid[0000-0002-8496-9294]{S.H.~Abidi}$^\textrm{\scriptsize 30}$,
\AtlasOrcid[0009-0003-6578-220X]{M.~Aboelela}$^\textrm{\scriptsize 46}$,
\AtlasOrcid[0000-0002-9987-2292]{A.~Aboulhorma}$^\textrm{\scriptsize 36e}$,
\AtlasOrcid[0000-0001-5329-6640]{H.~Abramowicz}$^\textrm{\scriptsize 156}$
% ...
\bigskip
\\

% ...
$^{17}$Department for Physics and Technology, University of Bergen, Bergen; Norway.\\
% ...
$^{30}$Physics Department, Brookhaven National Laboratory, Upton NY; United States of America.\\
% ...
$^{36}$$^{(a)}$Facult\'e des Sciences Ain Chock, Universit\'e Hassan II de Casablanca;$^{(b)}$Facult\'{e} des Sciences, Universit\'{e} Ibn-Tofail, K\'{e}nitra;$^{(c)}$Facult\'e des Sciences Semlalia, Universit\'e Cadi Ayyad, LPHEA-Marrakech;$^{(d)}$LPMR, Facult\'e des Sciences, Universit\'e Mohamed Premier, Oujda;$^{(e)}$Facult\'e des sciences, Universit\'e Mohammed V, Rabat;$^{(f)}$Institute of Applied Physics, Mohammed VI Polytechnic University, Ben Guerir; Morocco.\\
% ...
$^{46}$Physics Department, Southern Methodist University, Dallas TX; United States of America.\\
% ...
$^{51}$Fakult\"{a}t Physik , Technische Universit{\"a}t Dortmund, Dortmund; Germany.\\
% ...
$^{57}$II. Physikalisches Institut, Georg-August-Universit\"{a}t G\"ottingen, G\"ottingen; Germany.\\
% ...
$^{105}$CPPM, Aix-Marseille Universit\'e, CNRS/IN2P3, Marseille; France.\\
% ...
$^{120}$$^{(a)}$New York University Abu Dhabi, Abu Dhabi;$^{(b)}$United Arab Emirates University, Al Ain; United Arab Emirates.\\
% ...
$^{124}$Homer L. Dodge Department of Physics and Astronomy, University of Oklahoma, Norman OK; United States of America.\\
% ...
$^{156}$Raymond and Beverly Sackler School of Physics and Astronomy, Tel Aviv University, Tel Aviv; Israel.\\
% ...
$^{a}$ Also Affiliated with an institute covered by a cooperation agreement with CERN.\\
$^{b}$ Also at An-Najah National University, Nablus; Palestine.\\
$^{c}$ Also at Borough of Manhattan Community College, City University of New York, New York NY; United States of America.\\
$^{d}$ Also at Center for High Energy Physics, Peking University; China.\\
$^{e}$ Also at Center for Interdisciplinary Research and Innovation (CIRI-AUTH), Thessaloniki; Greece.\\
$^{f}$ Also at CERN, Geneva; Switzerland.\\
% ...

\end{flushleft}

% Created with Glance <Atlas.Glance@cern.ch>

%

%
%

%

%
%
%
%

\end{document}